\documentclass{jfm}

\usepackage{amsmath}
\usepackage{graphicx}
\usepackage{natbib}

\usepackage[nameinlink]{cleveref}
\crefname{paragraph}{\S}{\S\S}
\crefformat{section}{\S#2#1#3}
\crefformat{subsection}{\S#2#1#3}
\crefformat{subsubsection}{\S#2#1#3}
\usepackage{subcaption}
\usepackage{float}
\usepackage[outdir=./]{epstopdf}
\usepackage{placeins}
\usepackage{subfloat}
\usepackage{physics}
\usepackage{empheq}
\usepackage[T1]{fontenc}
\usepackage{amsmath,mathtools}
\usepackage[utf8]{inputenc}
\usepackage{amssymb}
\usepackage{bm}
\usepackage{xspace}
\usepackage{subcaption}
\usepackage{float}
\usepackage{multirow}
\usepackage[table]{xcolor}
\usepackage{caption}
\usepackage{tikz}
\allowdisplaybreaks
\usepackage{amssymb}
\usepackage{pifont}
\newcommand{\cmark}{\ding{51}}%
\newcommand{\xmark}{\ding{55}}%

\begin{document}

\newtheorem{lemma}{Lemma}
\newtheorem{corollary}{Corollary}
\title{Generalised quasilinear approximations of turbulent channel flow: Part 2. Spanwise scale interactions}
\author
{
Carlos G. Hern\'andez\aff{1}\corresp{\email{cg1116@imperial.ac.uk}}, 
Qiang Yang\aff{2} \and Yongyun Hwang\aff{1}
}
\affiliation
{
\aff{1}
Department of Aeronautics, Imperial College London, South Kensington, London SW7 2AZ, UK
\aff{2}
State Key Laboratory of Aerodynamics, China Aerodynamics Research and Development Centre, Mianyang 621000, PR China
}
\maketitle
\begin{abstract}
Continuing from Part 1 (Hernández \emph{et al.}, \emph{arXiv:2108.12395}, 2021), a generalized quasilinear (GQL) approximation is studied in turbulent channel flow using a flow decomposition defined with spanwise Fourier modes: the flow is decomposed into a set of low-wavenumber spanwise Fourier modes and the rest high-wavenumber modes. This decomposition leads to the nonlinear low-wavenumber group that supports the self-sustaining process within the given integral length scales, whereas the linearised high-wavenumber group is not able to do so, unlike the GQL models in Part 1 which place a minimal mathematical description for the self-sustaining process across all integral scales. Despite the important physical difference, it is shown that the GQL models in this study share some similarities with those in Part 1: i.e. the reduced multi-scale behaviour and anisotropic turbulent fluctuations. Furthermore, despite not being able to support the self-sustaining process in the high-wavenumber group, the GQL models in the present study are found to reproduce some key statistical features in the high-wavenumber group solely through the `scattering' mechanism proposed by previous studies. Finally, using the nature of the GQL approximation, a set of numerical experiments suppressing certain triadic nonlinear interactions are further carried out. This unveils some key roles played by the certain types of triadic interactions including energy cascade and inverse energy transfer in the near-wall region. In particular, the inhibition of inverse energy transfer in the spanwise direction leads to suppression of the near-wall positive turbulent transport at large scales. 
\end{abstract}

\section{Introduction}\label{sec:sec1}

The quasilinear (QL) and generalized quasilinear (GQL) approximations have recently been proposed in various forms for several turbulent flows. Common to all various forms, this approach introduces a decomposition of the given flow into two groups: one in which nonlinear terms are kept, and the other in which all self-interactions are ignored or suitably modelled. The resulting equations for the first are unchanged from or close to the original, while those for the second become equivalent to a linearisation around the first group often with an additional model (e.g. stochastic forcing). The early work may be found in \cite{malkus54,malkus56} and \cite{herring63, herring64, herring66} where the equations for the second group are closed using the `marginal stability' ansatz. In the last two decades, several variants of the QL framework have been proposed for many different flows with various types of models for the self-interaction term of the second group (e.g. stochastic forcing, eddy viscosity, etc): for example, stochastic structural stability theory (S3T) \citep{farrell03,farrell07,farrell12}, direct statistical simulation (DSS) \citep{marston08,tobias13prl}, self-consistent approximation for linearly unstable flows \citep{lugo14,lugo16}, minimal quasilinear approximation augmented with an eddy-viscosity model \citep{hwang19,skouloudis}, restricted nonlinear model (RNL) \citep{thomas14,thomas15,farrell17,pausch18,hernandez} and generalised quasilinear approximations (GQL) \citep{constantinou15,marston16,tobias17}. 

Continuing from Part 1 \citep*{paper1}, the starting point of the present study is the RNL model, which we have referred to as the QL model. The important advantage of this model is that it can capture the so-called `self-sustaining process' \citep{hamilton95,waleffe97}, i.e. the two-way interaction between `streamwise elongated' structure of streamwise velocity (streaks) and its `streamwise wavy' instability involving cross-streamwise velocities (waves and rolls), in a minimal manner \citep{thomas14,thomas15,pausch18,hernandez} -- 
in the QL model, the elongated streaks are captured by the nonlinear streamwise mean equations and the streamwise wavy instability is by the linearised equations. The recent examination of this model in uniform shear turbulence \citep{hernandez} revealed that the inhibition of the streamwise energy cascade in the QL model by construction results in highly elevated spectral energy intensity residing only at the streamwise integral length scale, and only a small number of streamwise Fourier modes consequently remain active \cite[]{thomas14,thomas15,farrell16,tobias17}. While the QL model captures the most fundamental physical mechanism in wall-bounded turbulence reasonably well (especially at low Reynolds numbers), it was found that the approximation yields some non-negligible damages to the slow pressure. This consequently inhibits the related pressure strain which transfers the turbulent kinetic energy produced at the streamwise component to the cross-streamwise components, explaining the typical anisotropic structures in the QL model throughout the entire wavenumber space.

In Part 1 \citep{paper1}, the QL model was extended by applying the GQL approximation to the streamwise direction in turbulent channel flow at $Re_\tau \simeq 1700$ to explore a route to improvement of the QL model. 
The GQL approximation employs the flow decomposition in the QL model in a more flexible manner: the nonlinear equations for the streamwise mean flow is replaced with nonlinear equations for a set of low-wavenumber streamwise Fourier modes (i.e. the low-wavenumber group), and the rest of the flow (i.e. the high-wavenumber group) is described by the linearised equations about the set of low-wavenumber streamwise Fourier modes. 
This is a direct extension of the previous QL model \citep{thomas14,thomas15,farrell17,pausch18,hernandez}, which offers the dynamics of the energy-containing streamwise waves in a minimal manner. In Part 1 \citep{paper1}, the GQL approximation was found to improve the QL model effectively. By incorporation of only a few more streamwise Fourier modes into the low-wavenumber group, the GQL model was able to recover the self-similar scaling with the distance from the wall of the turbulence statistics and spectra. This was associated with the mean-fluctuation interaction improved by the enhanced energy transfer in the streamwise wavenumber space. Furthermore, it was found that this energy transfer from the low to the high-wavenumber group in the GQL model depends on the neutrally stable leading Lyapunov spectrum of the linearised equations for the high-wavenumber group. In particular, if the threshold wavenumber distinguishing the two groups is sufficiently high, this energy transfer is inhibited due to the linear nature of the equations for the high-wavenumber group.

The GQL model employed in Part 1 provides useful physical insights into the modelling of wall-bounded turbulence in the context of the self-sustaining process. However, if the Reynolds number is sufficiently high, the self-sustaining processes emerge at all (spanwise) integral length scales that vary continuously with the distance from the wall \citep{flores10,hwang10,hwang11,hwang15,hwangbengana16}. There have been a number of previous investigations on the interactions between the related energy-containing fluid motions in the near-wall and outer regions  \citep{hutchins07,mathis09,duvvuri15,agostini16,zhang16}. In these studies, the velocity field is often decomposed into large and small scales, and their interactions have been modelled and explored in terms of superposition and amplitude modulation in the near-wall region. However, in general, such interactions are expected to exist across almost every wall-normal locations, given the existence of the self-sustaining processes across all integral length scales in the form of the `attached eddies' of \cite{townsend76}. Recent investigations on the statistical features in the spectral energy transfer revealed that the interaction dynamics appears to be highly complex \citep{kawata18,cho18,lee19}, although there were some progresses made for an idealised system in which there are only two integral length scales \citep{doohan21,doohan21b}. The newly-found scale interactions include mediation of the energy cascade from given integral scale by larger energy-containing motions \citep{cho18}, small-scale turbulence production driven by energy cascade from larger scales \citep{doohan21}, the near-wall inverse energy transfer that forms large-scale near-wall inactive motions of \cite{townsend76} \citep{cho18,doohan21,doohan21b}, and the mean-fluctuation interactions across all integral scales \citep{doohan21b}.  

\begin{figure}
\centering
\includegraphics[width=0.8\textwidth]{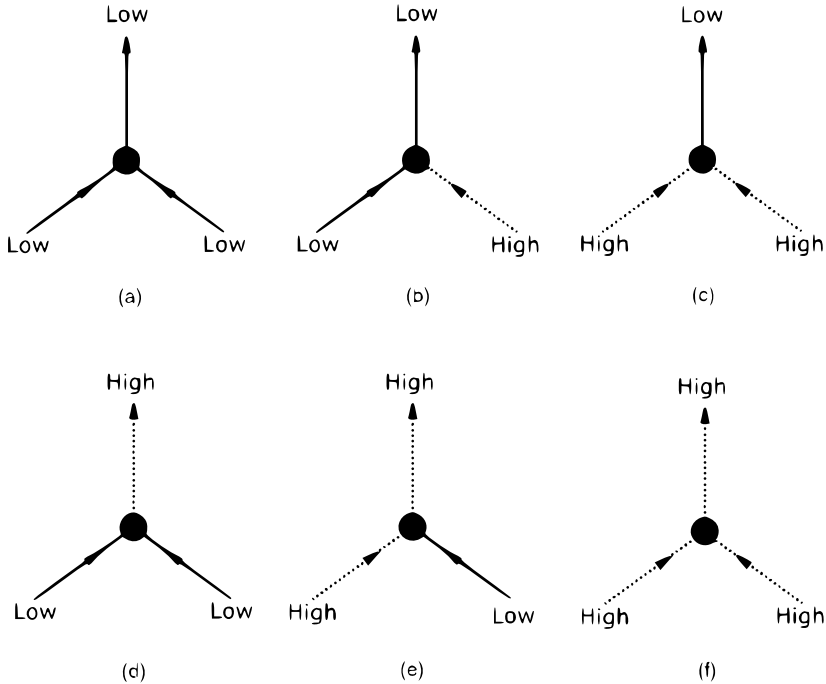}
\label{1}
\caption{Set of triadic interactions. In the GQL approximation, three interactions (a,c,e) are retained out of six in total. The other three triads are discarded \citep{marston16}.}
\label{fig:triad}
\end{figure}

As was briefly discussed in Part 1, the GQL approximation suppresses particular triadic wave interactions between the low and high wavenumber groups. This was sketched in figure \ref{fig:triad}, where the GQL approximations ignore those in (b), (d) and (f). Similarly, this feature can further be exploited to study the scale interactions in wall-bounded turbulence: for example, the other interactions such as (c) can also be suppressed to understand the effect of the inverse energy transfer. From this perspective, a flexible utilisation of the GQL approximation may perhaps provide a unique interventional tool to explore these scale interactions.   

The objective of the present study is to explore the scale interactions in wall-bounded turbulence using the GQL approximation. For this purpose, unlike Part 1, the low wavenumber-mode group will be composed of the plane Fourier modes for $k_z\leq k_{z,c}$ ($k_z$ is the spanwise Fourier wavenumber and $k_{z,c}$ the corresponding threshold wavenumber for the decomposition). The spanwise length scale has been understood to represent the size of coherent structures underpinned by the self-sustaining process, as the spanwise Fourier modes well represent the streaks highly elongated in the streamwise directions and the related quasi-streamwise vortical structures \cite[e.g.][]{hwang15,cho18}. The proposed GQL approximation will then be first examined to present its effect in the spanwise direction. Further to this, a set of numerical experiments, which deliberately suppress the terms presented in figure \ref{fig:triad}, will be performed to understand the scale interaction in wall-bounded turbulence. 

The paper is organized as follows. The GQL approximation together with its spectral energy budget is briefly formulated in \cref{sec:sec2} for completeness. In \cref{sec:sec3}, the statistics and spectra of the GQL model with spanwise threshold wavenumber for channel flow are compared to those of a wall-resolved LES. The energy-budget spectra are also presented here with a detailed analysis to explain the statistical features of the GQL model in the spanwise direction. Comparison of the GQL in the present study is subsequently made with that in Part 1, and various numerical experiments suppressing certain interactions are carried out in \cref{sec:sec4}, highlighting the use of the GQL-approximation oriented idea to analyze scale interactions. Finally, the paper concludes in \cref{sec:sec5} with some remarks.

\section{Problem formulation} \label{sec:sec2}
\subsection{Generalised quasilinear approximation}
We consider a pressure-driven plane channel flow, in which the density and kinematic viscosity are given by $\rho$ and $\nu$, respectively. The time is denoted by $t$, and the space is by $\mathbf{x}=(x,y,z)$ with $x$, $y$ and $z$ being the streamwise, wall-normal and spanwise directions, respectively. The lower and upper walls of the channel are set to be located at $y=0,2h$. For the GQL approximation, 
the velocity $\bf u$ is decomposed into two groups using a discrete Fourier transform in the wall-parallel directions:
\begin{subequations}\label{eq:2.0}
\begin{equation}\label{eq:2.3z}
\bold{u}=\bold{U}_l+\bold{u}_h,
\end{equation}
where 
\begin{equation}\label{eq:2.0b}
    \bold{U}_l=\sum_{n=-M_{z,F}}^{M_{z,F}}\sum_{m=-M_{x,F}}^{M_{x,F}} \hat{\bold{u}}_{m,n} e^{i(m k_{x,0}x+n k_{z,0}z)}
\end{equation}
\end{subequations}
and $\mathbf{u}_h$ is given from (\ref{eq:2.3z}). Here, $\hat{\bold{u}}_{m,n}$ is the discrete Fourier mode of the velocity,  $k_{x,0}$ and $k_{z,0}$ are the fundamental streamwise and spanwise wavenumbers for the given horizontal domain, and $M_{x,F}$ and $M_{z,F}$ define the threshold streamwise and spanwise wavenumbers for the decomposition such that $k_{x,c}=M_{x,F}k_{x,0}$ and $k_{z,c}=M_{z,F}k_{z,0}$. Given the scope of the present study, the main parameter of interest is $k_{z,c}$ (or $M_{z,F}$), while keeping $M_{x,F}$ to be identical to the number of the streamwise Fourier modes used in the full simulation (see \S\ref{sec:sec23} for further details). 

With the decomposition in (\ref{eq:2.0}), two related projection operators are defined as
\begin{equation}\label{eq:2.4}
    \mathcal{P}_l[\bold{u}]\equiv\bold{U}_l,~~\mathcal{P}_h[\bold{u}]\equiv\bold{u}-\bold{U}_l=\bold{u}_h,
\end{equation}
with the following properties: 
\begin{subequations}\label{eq:2.5}
\begin{equation}\label{eq:2.5a}
\mathcal{P}_l[\cdot]+\mathcal{P}_h[\cdot]=\mathcal{I}[\cdot],
\end{equation}
\begin{equation}
\mathcal{P}_l[\mathcal{P}_l[\cdot]]=\mathcal{P}_l[\cdot],~ \mathcal{P}_h[\mathcal{P}_h[\cdot]]=\mathcal{P}_h[\cdot],
\end{equation}
\begin{equation}
\mathcal{P}_l[\mathcal{P}_h[\cdot]]=\mathcal{P}_h[\mathcal{P}_l[\cdot]]=\mathbf{0},
\end{equation}
\end{subequations}
where $\mathcal{I}[\cdot]$ is the identity operator. Using the definition and the properties listed in (\ref{eq:2.4}) and (\ref{eq:2.5}), the Navier-Stokes equations are first projected onto the $\mathcal{P}_l$ (or low wavenumber) and $\mathcal{P}_h$ (or high wavenumber) subspaces. The subsequent linearisation of the equations for $\mathbf{u}_h$ about $\mathbf{U}_l$ leads to the GQL system of interest, i.e.
\begin{subequations}\label{eq:full}
\begin{align}\label{eq:aa}
\pdv{\bold{U}_l}{t} + \mathcal{P}_{l}[(\bold{U}_{l} \cdot \nabla) \bold{U}_{l}]  =-\frac{1}{\rho}\bold{\nabla} P_l + \nu \bold{\nabla}^2  \bold{U}_l +  \mathcal{P}_{l}[ \bold{\nabla} \cdot \bm{\tau}_{SGS} ] -\mathcal{P}_{l}\left[\left(\textbf{u}_{h}\cdot\nabla\right)\textbf{u}_{h}\right],
\end{align}
and 
\begin{equation}\label{eq:bb}
\pdv{\bold{u}_h}{t} + \mathcal{P}_h[(\bold{u}_h \cdot \bold{\nabla}) \bold{U}_l] + \mathcal{P}_h[(\bold{U}_l \cdot \bold{\nabla}) \bold{u}_h] =-\frac{1}{\rho}\bold{\nabla} p_h +  \nu \bold{\nabla}^2  \bold{u}_h +  \mathcal{P}_{h}[ \bold{\nabla} \cdot \bm{\tau}_{SGS} ] , 
\end{equation}
\end{subequations}
where $P_l$ and $p_h$ are given to enforce $\bold{\nabla} \cdot \bold{U}_l=0$ and $\bold{\nabla} \cdot \bold{u}_h=0$, respectively, with $p=P_l+p_h$. As in Part 1, the terms $\mathcal{P}_{l}\left[\left(\bold{U}_{l}\cdot\nabla\right)\bold{u}_{h}\right]$ and $\mathcal{P}_{l}\left[\left(\bold{u}_{h}\cdot\nabla\right)\bold{U}_{l}\right]$ are neglected in (\ref{eq:aa}). The numerical simulations in this study are performed using LES. Therefore, the subgrid-scale stress (SGS) tensor, given by $\bm{\mathbf{\tau}}_{SGS}= - ( \bold{u} \bold{u} - \overline{\bold{u} \bold{u}})$, appears in (\ref{eq:full}), where the overbar $(\overline{\cdot})$ denotes the application of the grid filter given by the numerical discretisation of the equations (see \S\ref{sec:sec23}). The SGS tensor for the present LES here employs a mixing-length type model $\bm{\mathbf{\tau}}_{SGS}= \nu_t (\nabla \bold{u}+ \nabla \bold{u}^T)$, where the eddy viscosity $\nu_t$ is determined with the Vreman model (\citealp{vreman}). Further details are provided in Part 1 (\citealp{paper1}).


\subsection{Spectral energy budget}\label{sec:sec21}
To analyse the turbulence statistics and spectra of the given GQL and original systems, the Reynolds decomposition of the velocity $\mathbf{u}=(u,v,w)$ is employed: i.e.
\begin{equation}\label{eq:2.1}
    \mathbf{u}=\mathbf{U}+\mathbf{u}',
\end{equation}
in which $\mathbf{U}(\equiv\langle \mathbf{u} \rangle_{x,z,t})=(U(y),0,0)$ is the mean velocity with $\langle\cdot\rangle_{x,z,t}$ being an average in $t$-, $x$- and $z$-directions. Similarly to Part 1, this decomposition is only introduced for the statistical analysis of LES and GQL models, and it should not be confused with the one in (\ref{eq:2.0}) introduced for the GQL approximations. 
Using (\ref{eq:2.0}) and (\ref{eq:2.1}), the turbulent velocity fluctuation is further decomposed into the low and high wavenumber components as in (\ref{eq:2.3z}):
\begin{equation}\label{eq:2.3}
\bold{u}'=\bold{u}_l+\bold{u}_h.
\end{equation}

\begin{table}
  \begin{center}
\def~{\hphantom{0}}
  \begin{tabular}{lccccccccc}
\multicolumn{1}{c}{\multirow{2}{*}{Case}} &
\multicolumn{1}{c}{\multirow{2}{*}{$Re$}} &
\multicolumn{1}{c}{\multirow{2}{*}{$Re_{\tau}$}} &
\multicolumn{1}{c}{\multirow{2}{*}{$\Delta_x^+$}}  & \multicolumn{1}{c}{\multirow{2}{*}{$\Delta_z^+$}} & 
\multicolumn{1}{c}{\multirow{2}{*}{$M_{x, F}$}} &
\multicolumn{1}{c}{\multirow{2}{*}{$M_{z, F}$}} &
\multicolumn{1}{c}{\multirow{2}{*}{$\lambda_{z,c}/h$}} &
\multicolumn{1}{c}{\multirow{2}{*}{$\lambda_{z,c}^+$}} &
\multicolumn{1}{c}{\multirow{2}{*}{$N_{x} \times N_y \times N_z$}} \\ \\[3pt]
\multicolumn{1}{c}{LES}  & 66667 & \multicolumn{1}{c}{1974}  &  \multicolumn{1}{c}{64.6} & \multicolumn{1}{c}{32.3} &  \multicolumn{1}{c}{384}&  \multicolumn{1}{c}{96}
 &\multicolumn{1}{c}{-} & \multicolumn{1}{c}{-}   &\multicolumn{1}{c}{$1152 \times 169 \times 288$} \\ [2pt]
\multicolumn{1}{c}{GQLZ1} & 66667  & \multicolumn{1}{c}{1816} &   \multicolumn{1}{c}{59.4} & \multicolumn{1}{c}{29.7} & \multicolumn{1}{c}{384} &  \multicolumn{1}{c}{1}& \multicolumn{1}{c}{3.14}&  \multicolumn{1}{c}{5702} &\multicolumn{1}{c}{$1152 \times 169 \times 288$} \\ [2pt]
\multicolumn{1}{c}{GQLZ5} & 66667  & \multicolumn{1}{c}{1945} &   \multicolumn{1}{c}{63.6} & \multicolumn{1}{c}{31.8} & \multicolumn{1}{c}{384} &  \multicolumn{1}{c}{5}& \multicolumn{1}{c}{0.63}&  \multicolumn{1}{c}{1221} &\multicolumn{1}{c}{$1152 \times 169 \times 288$} \\ [2pt]
\multicolumn{1}{c}{GQLZ25} & 66667  & \multicolumn{1}{c}{1958} &   \multicolumn{1}{c}{64.0} & \multicolumn{1}{c}{32.0} & \multicolumn{1}{c}{384}&  \multicolumn{1}{c}{25}& \multicolumn{1}{c}{0.13}&  \multicolumn{1}{c}{246}  &\multicolumn{1}{c}{$1152 \times 169 \times 288$} \\ [2pt]
\multicolumn{1}{c}{GQLZ64} & 66667  & \multicolumn{1}{c}{1978} &   \multicolumn{1}{c}{64.7} & \multicolumn{1}{c}{32.3} & \multicolumn{1}{c}{384}&  \multicolumn{1}{c}{64}& \multicolumn{1}{c}{0.05}&  \multicolumn{1}{c}{97}  &\multicolumn{1}{c}{$1152 \times 169 \times 288$} \\ [6pt]
\multicolumn{1}{c}{TRIAZ25}& 66667 &  \multicolumn{1}{c}{1981}
&   \multicolumn{1}{c}{64.8} & \multicolumn{1}{c}{32.4} & \multicolumn{1}{c}{384} & \multicolumn{1}{c}{25} &\multicolumn{1}{c}{0.13} &\multicolumn{1}{c}{249} &\multicolumn{1}{c}{$1152 \times 169 \times 288$} \\ [2pt]
\multicolumn{1}{c}{TRIBZ25} & 66667 &  \multicolumn{1}{c}{1860}& \multicolumn{1}{c}{60.8} & \multicolumn{1}{c}{30.4} & \multicolumn{1}{c}{384} & \multicolumn{1}{c}{25} &\multicolumn{1}{c}{0.13} &\multicolumn{1}{c}{234} &\multicolumn{1}{c}{$1152 \times 169 \times 288$} \\ [2pt]
\multicolumn{1}{c}{TRICZ25} & 66667 &  \multicolumn{1}{c}{1721}&   \multicolumn{1}{c}{56.3} & \multicolumn{1}{c}{28.1} & 
\multicolumn{1}{c}{384} & \multicolumn{1}{c}{25} &\multicolumn{1}{c}{0.13} &\multicolumn{1}{c}{216} &\multicolumn{1}{c}{$1152 \times 169 \times 288$} \\ [2pt]
\multicolumn{1}{c}{TRIDZ25} & 66667 &  \multicolumn{1}{c}{2066} & \multicolumn{1}{c}{67.6} & \multicolumn{1}{c}{33.8}
&\multicolumn{1}{c}{384} & \multicolumn{1}{c}{25} &\multicolumn{1}{c}{0.13} &\multicolumn{1}{c}{259} &\multicolumn{1}{c}{$1152 \times 169 \times 288$} \\ [2pt]
\end{tabular}
\caption{Simulation parameters in the present study. The domain size in the $x$-, $y$- and $z$ directions is $L_x/h=8 \pi$, $L_y/h=2$ and $L_z/h=\pi$, respectively. Here, $Re=U_0 h/\nu$ and $Re_\tau=u_\tau h /\nu$, where $U_0$ and $u_\tau$ are the centerline and wall shear velocities, respectively. The grid spacings in the $x$- and $z$-directions are $\Delta_x^+$ and $\Delta_z^+$ (after aliasing). $\lambda_{z,c}$ is the threshold spanwise wavelength. $N_x$, $N_y$, $N_z$ are the number of grid points in the $x$-, $y$- and $z$-directions, respectively.
}
\label{tab:tab}
\end{center}
\end{table}

As in Part 1, the energy balance in the Fourier space for the GQL approximation is given by
\begin{align}\label{eq:2.9}
&0 = \underbrace{\left\langle \Real \left\{-{\widehat{u^{\prime}}}^* (k)\widehat{v^{\prime}}(k) \: \frac{\textup{d} U}{\textup{d} y} \right\} \right\rangle_{r^\perp,t}}_{\widehat{P}\left(y, k \right)}
+\underbrace{\left\langle - \nu \frac{\partial{{\widehat{u_{i}^{\prime}}}^*(k)} }{\partial{x_{j}} } \frac{\partial{{\widehat{u_{i}^{\prime}}}(k)} }{\partial{x_{j}} } \: \right\rangle_{r^\perp,t}}_{\widehat{\varepsilon}\left(y, k \right)} \nonumber\\
&+\underbrace{\left\langle \Real \left\{- {\widehat{u_{i}^{\prime}}}^*(k) \left(\frac{\partial }{\partial{x_{j}} }  \left( \widehat{ \tau_{ij,SGS}^{\prime}}(k) \right) \right)\: \right\} \right\rangle_{r^\perp,t}}_{\widehat{\varepsilon}_{SGS}\left(y, k \right)} +\underbrace{\left\langle \Real \left\{\frac{\textup{d}}{\textup{d} y} \left ( \frac{\widehat{p^{\prime}}(k) {\widehat{v^{\prime}}}^*(k)}{\rho} \right) \right\} \right\rangle_{r^\perp,t}}_{\widehat{T}_{p}\left(y, k \right)} \nonumber \\
&+ \underbrace{\left\langle \Real \left\{-{\widehat{u_{i}^{\prime}}}^*(k) \left(\frac{\partial }{\partial{x_{j}} }  \left( \widehat{ u_{i}^{\prime}u_{j}^{\prime} }(k) -\mathcal{P}_h\left[\widehat{ u_{h,i}^{\prime}u_{h,j}^{\prime} }(k)\right]
-\mathcal{P}_h\left[\widehat{ u_{l,i}^{\prime}u_{l,j}^{\prime} }(k)\right] \right. \right. \right. \right.}_{\widehat{T}_{turb}\left(y, k\right)} \nonumber \\
& \underbrace{\left. \left. \left. \left. -\mathcal{P}_l\left[\widehat{ u_{l,i}^{\prime}u_{h,j}^{\prime} }(k)\right] -\mathcal{P}_l\left[\widehat{ u_{h,i}^{\prime}u_{l,j}^{\prime} }(k)\right] \right) \right)\: \right\} \right\rangle_{r^\perp,t}}_{\widehat{T}_{turb}\left(y, k\right)} +\underbrace{\left\langle \nu \frac{\textup{d}^2{\widehat{e}(k)}}{\textup{d}{y}^2 } \right\rangle_{r^\perp,t}}_{\widehat{T}_{\nu}\left(y, k \right)},
\end{align}
for $i,j=1,2,3$. Here, $(x_1,x_2,x_3)=(x,y,z)$, $(u_1^{\prime},u_2^{\prime},u_3^{\prime})=(u^{\prime},v^{\prime},w^{\prime})$, $(\widehat{\cdot})$ denotes the Fourier-transformed coefficient in the $r$-direction, where $r(=x~\mathrm{or}~z)$ is the streamwise or spanwise coordinate with $r^+(=z~\mathrm{or}~x)$ being the wall-parallel direction orthogonal to $r$, and  $k(=k_x~\mathrm{or}~k_z)$ the spatial wavenumber in the $r$-direction. Also, $\widehat{e}(k)=( |{\widehat{u'}(k)} |^2 + |{\widehat{v'}(k)} |^2 +|{\widehat{w'}(k)} |^2 )/2$, the superscript $(\cdot)^*$ indicates the complex conjugate, and $\Real \left \{\hspace{1mm}\cdot\hspace{1mm}\right \}$ the real part. The terms on the right-hand side are the rate of turbulence production $\widehat{P}(y,k)$, viscous dissipation $\widehat{\varepsilon}(y,k)$, SGS dissipation $\widehat{\varepsilon}_{SGS}(y,k)$, pressure transport $\widehat{T}_p(y,k)$, turbulent transport $\widehat{T}_{turb}(y,k)$ and viscous transport $\widehat{T}_{\nu}(y,k)$, respectively.

\begin{figure*}
\centering
\begin{subfigure}[b]{0.45\textwidth}
\includegraphics[width=\textwidth]{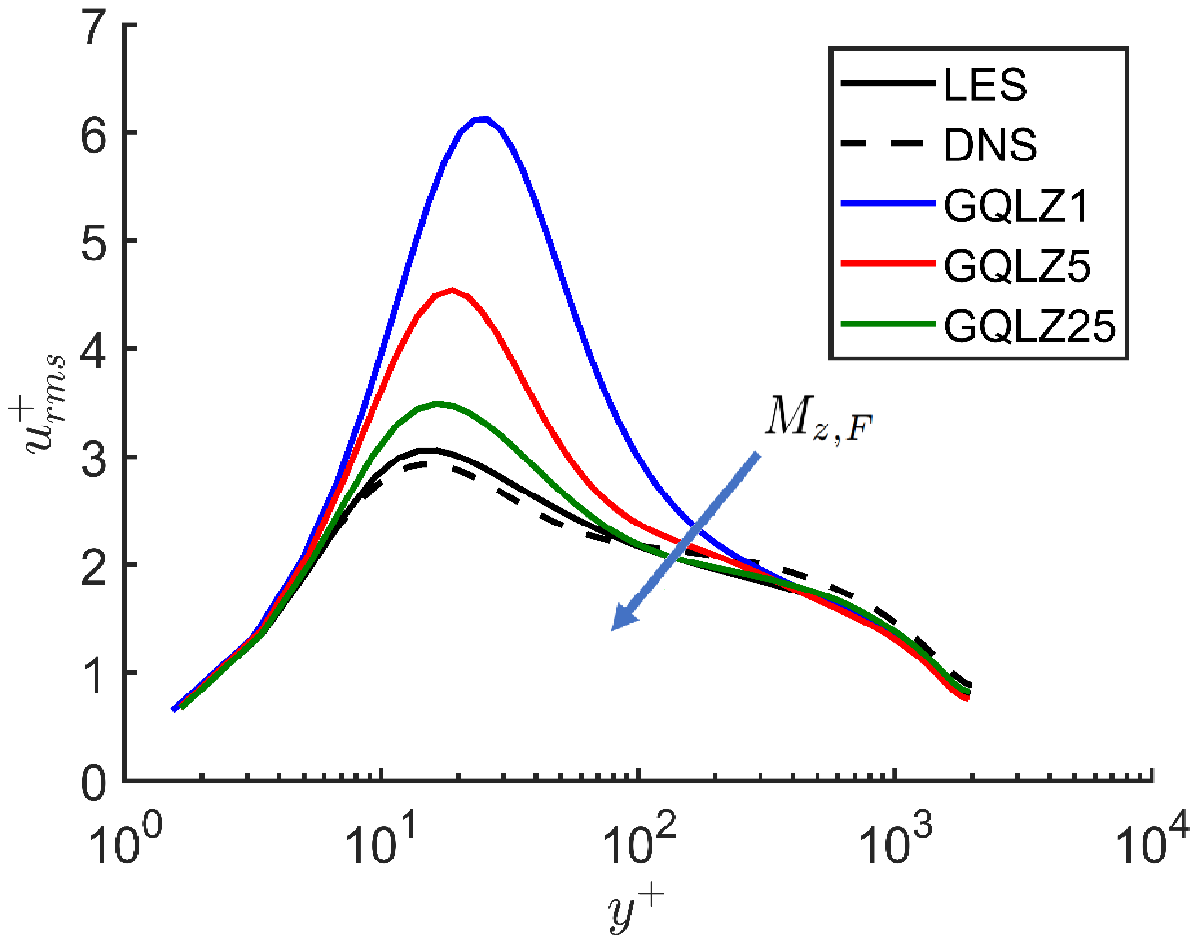}
\caption{$u_{rms}^+(y^+)$}
\label{fig:uu}
\end{subfigure}
\begin{subfigure}[b]{0.45\textwidth}
\includegraphics[width=\textwidth]{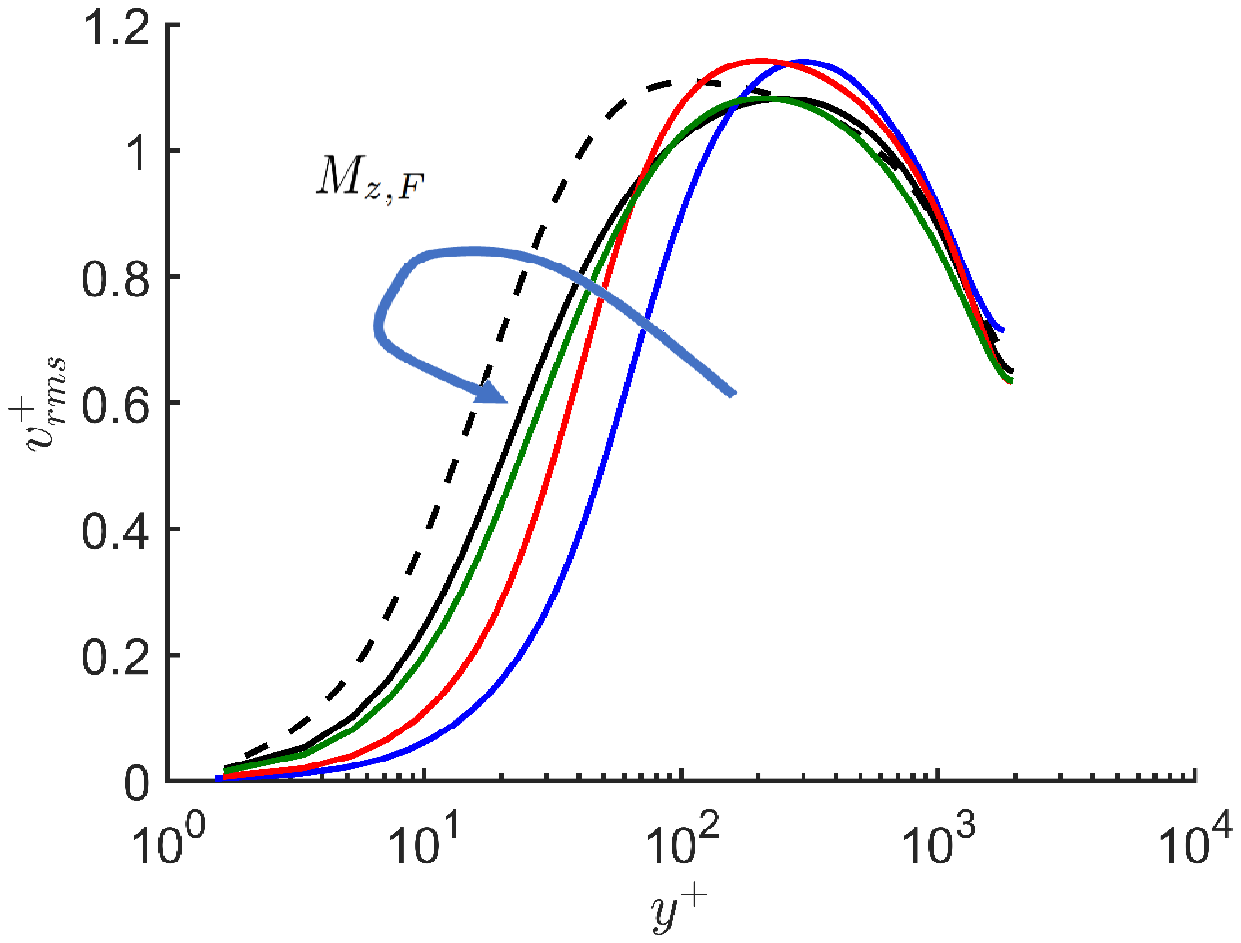}
\caption{$v_{rms}^+(y^+)$}
\label{fig:vv}
\end{subfigure}
\begin{subfigure}[b]{0.45\textwidth}
\includegraphics[width=\textwidth]{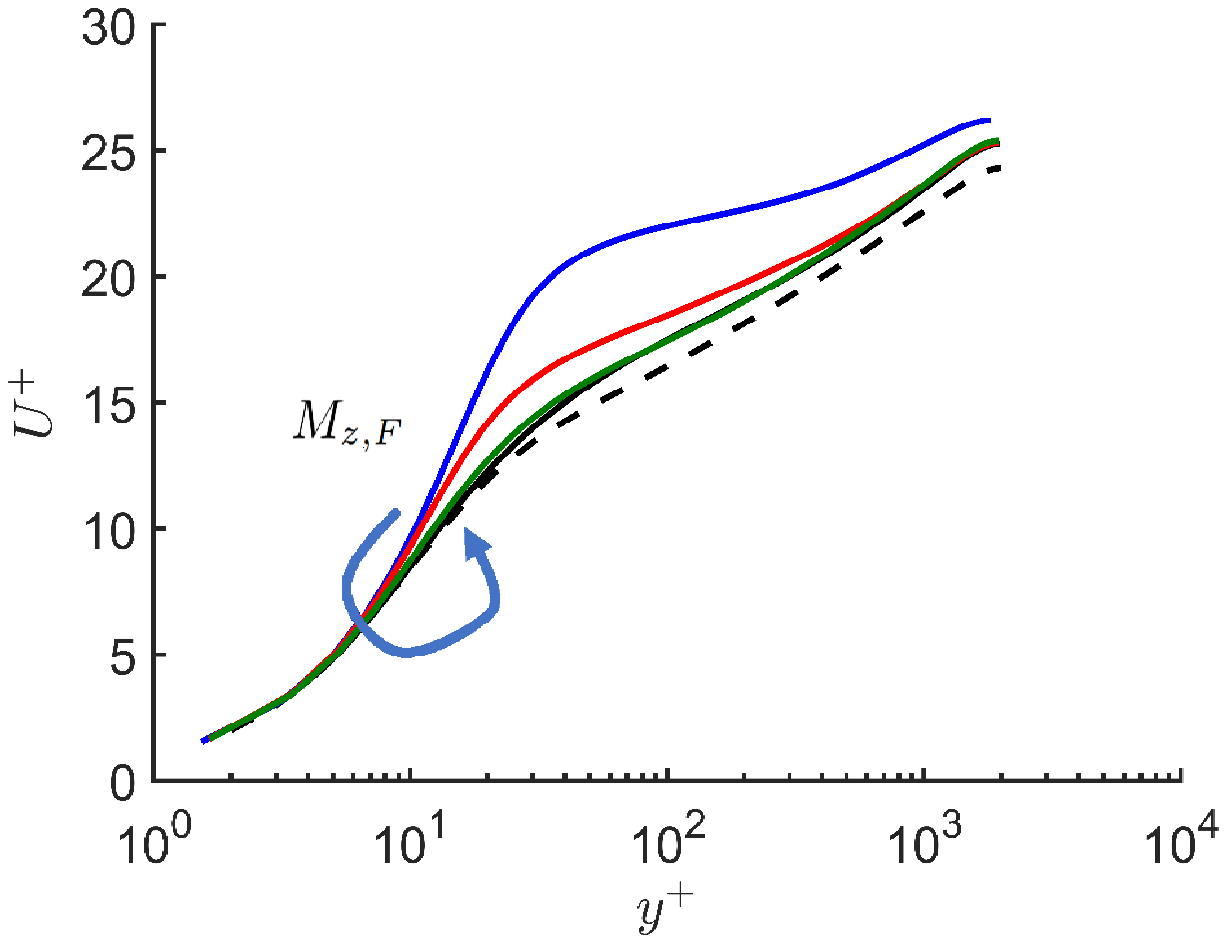}
\caption{$U^+(y^+)$}
\label{fig:ww}
\end{subfigure}
\begin{subfigure}[b]{0.45\textwidth}
\includegraphics[width=\textwidth]{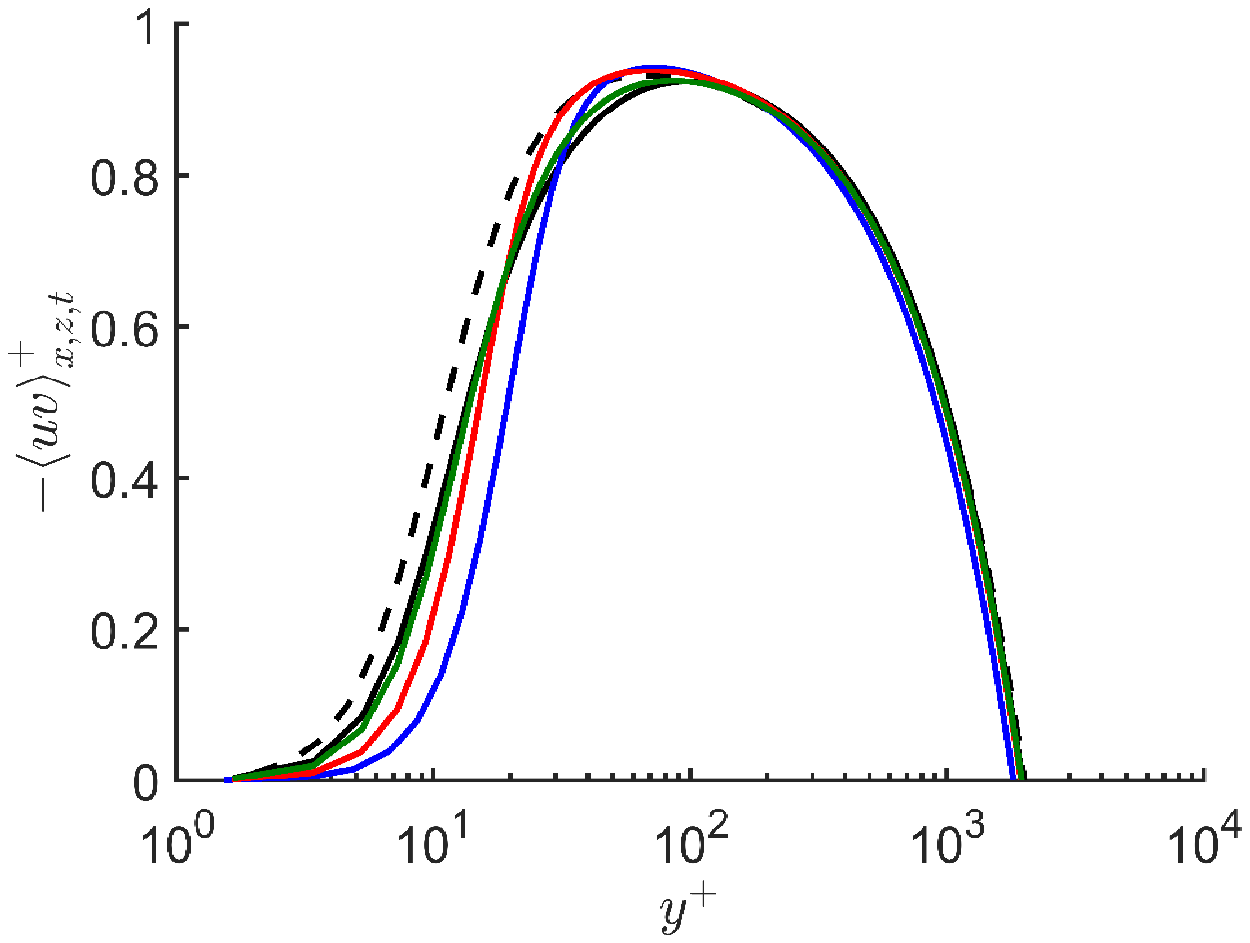}
\caption{$ \langle u^{\prime} v^{\prime}\rangle_{x,z,t} ^+(y^+)$}
\label{fig:uv}
\end{subfigure}
\caption{First- and second-order turbulence statistics for the present LES, DNS at $Re_\tau=2003$ (\citealp{hoyas08}), GQLZ1, GQLZ5 and GQLZ25 cases: (a) $u_{rms}^+(y^+)$; (b) $v_{rms}^+(y^+)$; (c) $U^+(y^+)$; (d) $ \langle u^{\prime} v^{\prime}\rangle_{x,z,t}^+(y^+)$.}
\label{fig:stat}
\end{figure*}

Equation (\ref{eq:2.9}) can further be split into each component. In this case, the energy transports by pressure strain subsequently appear as
\begin{align}
&\widehat{\Pi}_x(y, k)=\left \langle \Real \left \{ \frac{\widehat{p'} (k)}{\rho} \pdv{{\widehat{u'}}^* (k)}{x} \right \} \right \rangle_{r^\perp,t}, \quad \widehat{\Pi}_y(y, k)= \left \langle \Real \left \{\frac{\widehat{p'} (k)}{\rho} \pdv{{\widehat{v'}}^*(k)}{y} \right \} \right \rangle_{r^\perp,t} , \nonumber\\ 
&\widehat{\Pi}_z(y, k)=\left \langle \Real \left \{\frac{\widehat{p'} (k)}{\rho} \pdv{{\widehat{w'}}^* (k)}{z} \right \} \right \rangle_{r^\perp,t},
\end{align}
where $\widehat{\Pi}_x$, $\widehat{\Pi}_y$ and $\widehat{\Pi}_z$ are one-dimensional spectra of the streamwise, wall-normal and spanwise components of pressure strain, respectively. From the incompressibility of the given fluid, the pressure strain terms satisfy the following relation
\begin{equation}
\widehat{\Pi}_x(y,k)+\widehat{\Pi}_y(y,k)+\widehat{\Pi}_z(y,k)=0,
\end{equation}
and a further discussion on this can be found in previous studies \citep[][]{mizuno16,cho18,lee19,hernandez,paper1}.

\subsection{Numerical simulations}\label{sec:sec23}

A large eddy simulation (LES) is first carried out with the same numerical solver used in Part 1 \citep{hernandez}. The main difference of the simulation in the present study from that in Part 1 is the computational domain size. In Part 1, the streamwise computational domain was $L_x/h = \pi$. This size was chosen to match the most energetic streak instability length scale of large-scale outer structures in the streamwise direction \cite[]{degiovanetti17}. However, the scope of the GQL in the present study is to understand the scale interactions. For this purpose, the simulations here are carried out by well resolving all the known coherent structures. Since the largest streamwise and spanwise length scales in turbulent channel flow are $\lambda_x/h \simeq 10$ and $\lambda_z/h \simeq 1$  \cite[]{delalamo03,hwang15}, a sufficiently large computational domain is considered in the streamwise and spanwise directions: i.e. $L_x/h = 8\pi$ and $L_z/h = \pi$. We note that this domain size is 16 times bigger than that in Part 1. Table \ref{tab:tab} summarizes the parameters for the simulations performed in this study. The Reynolds number for all the considered cases is $Re=U_0 h /\nu=66667$. The grid spacings are $\Delta_x^+=54.9-67.3$ and $\Delta_z^+=27.5-33.6$. For each case, the number of spanwise Fourier modes used in the GQL approximation is varied from $M_{z,F}=1$ (GQL1) to the maximum number, the latter of which corresponds to that of the given LES.

\begin{figure}
\begin{minipage}{\textwidth}
\centering

\begin{subfigure}[b]{0.42\textwidth}
  \includegraphics[width=\textwidth]{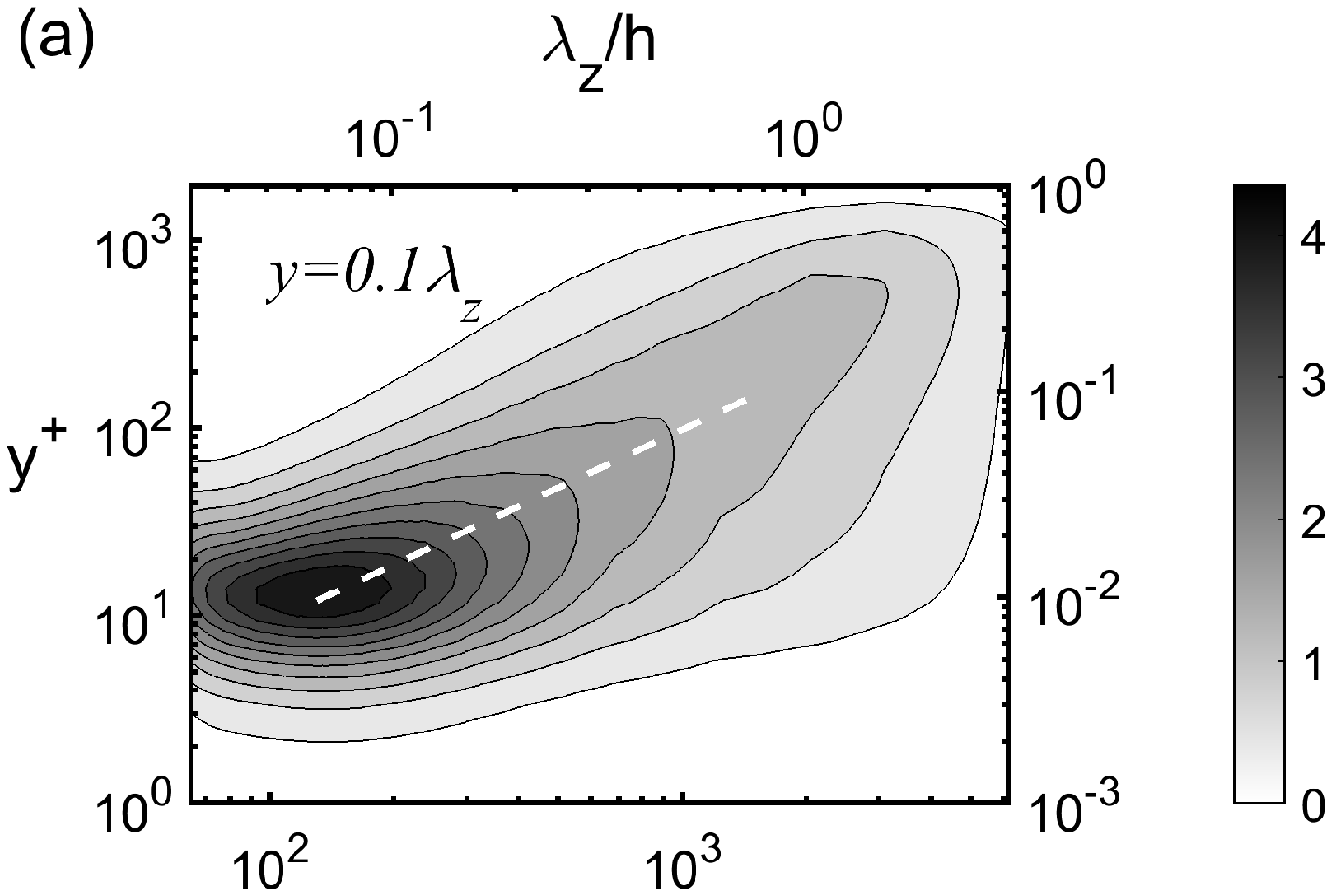}
\label{1}
\end{subfigure}
\vspace{-0.8cm}
\begin{subfigure}[b]{0.42\textwidth}
  \includegraphics[width=\textwidth]{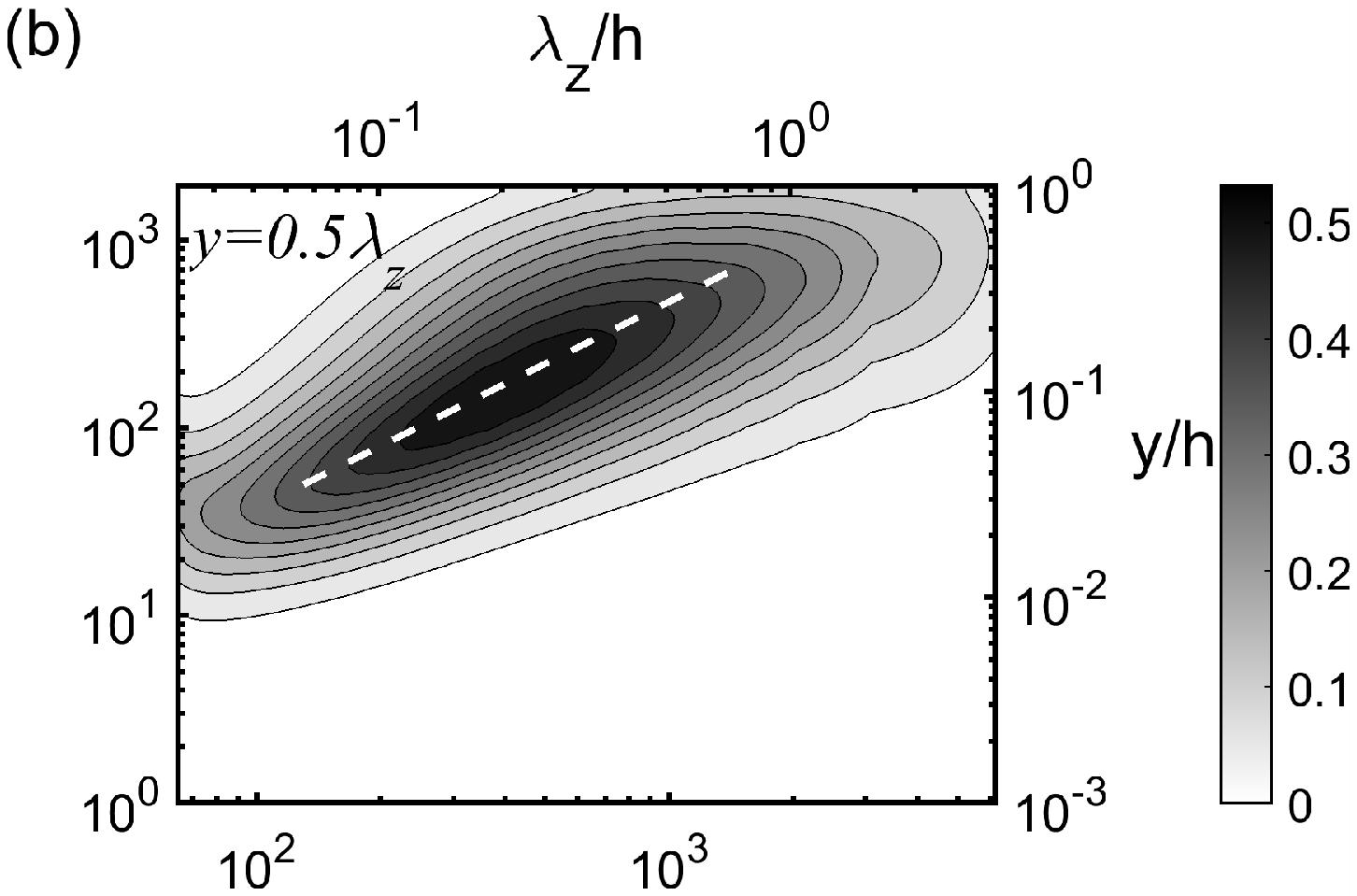}
\label{2}
\end{subfigure}
\vspace{-0.8cm}
\begin{subfigure}[b]{0.42\textwidth}
  \includegraphics[width=\textwidth]{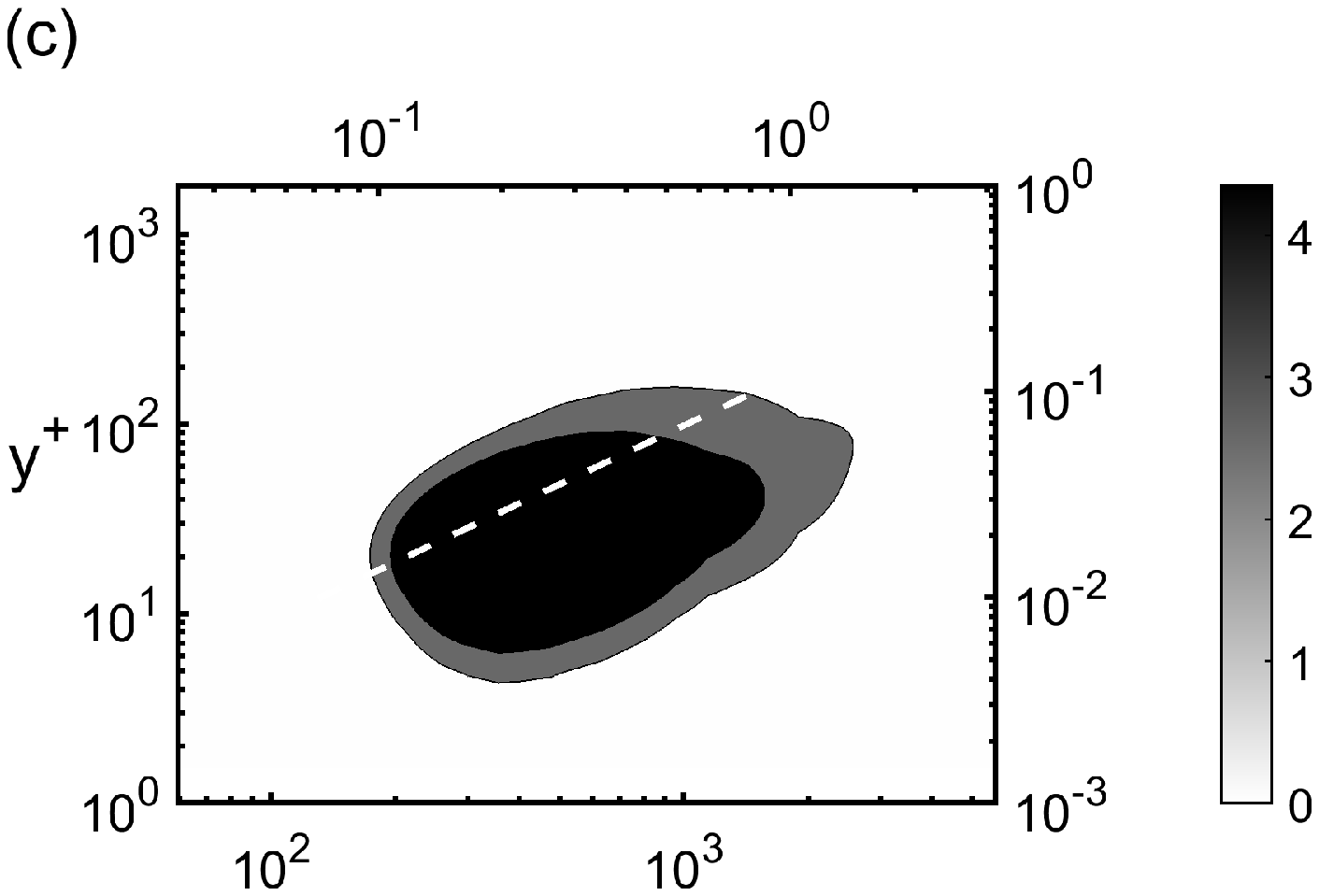}
  \label{3}
\end{subfigure}
\begin{subfigure}[b]{0.42\textwidth}
  \includegraphics[width=\textwidth]{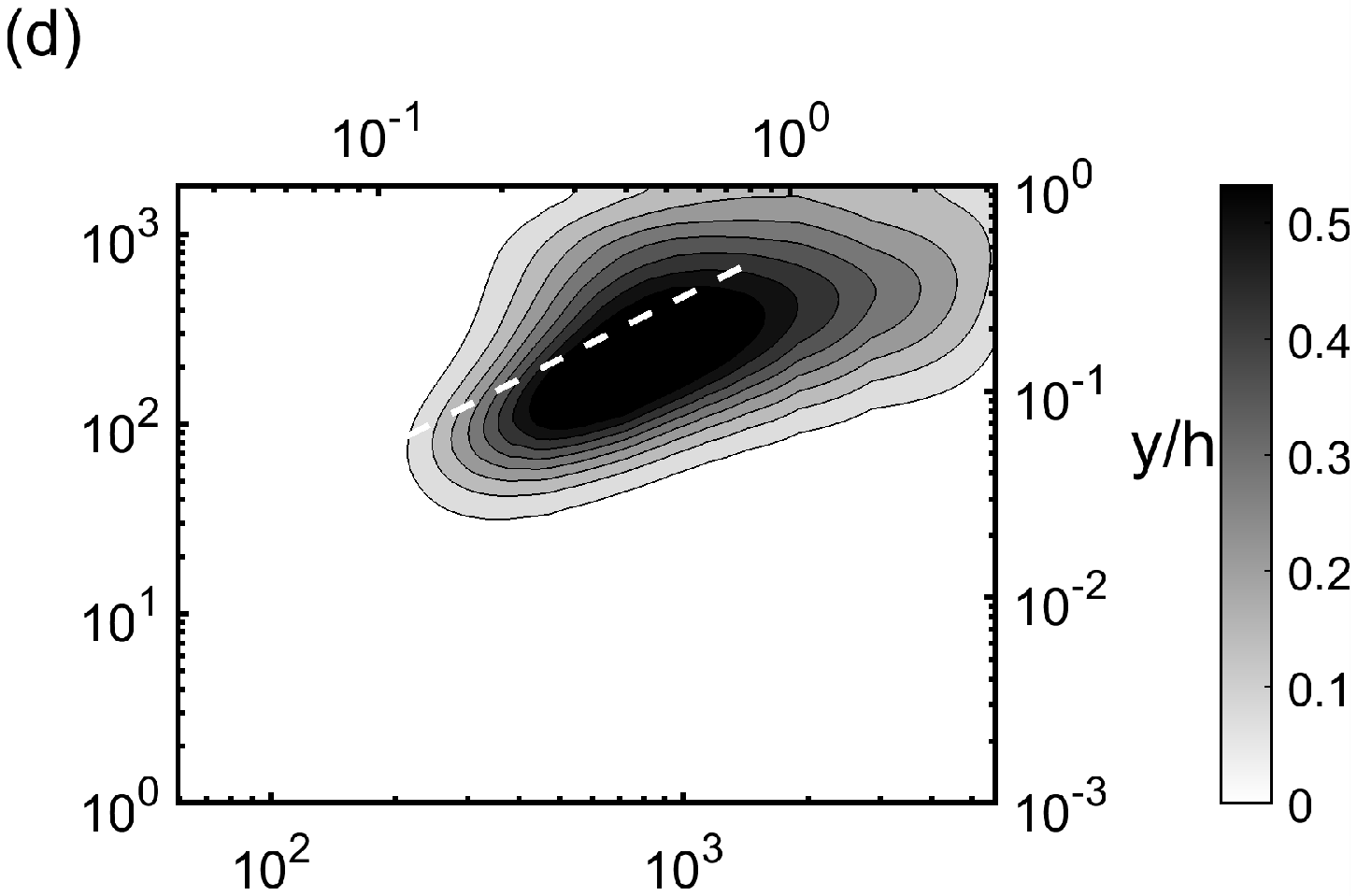}
  \label{4}
\end{subfigure}
\vspace{-0.8cm}
\begin{subfigure}[b]{0.42\textwidth}
  \includegraphics[width=\textwidth]{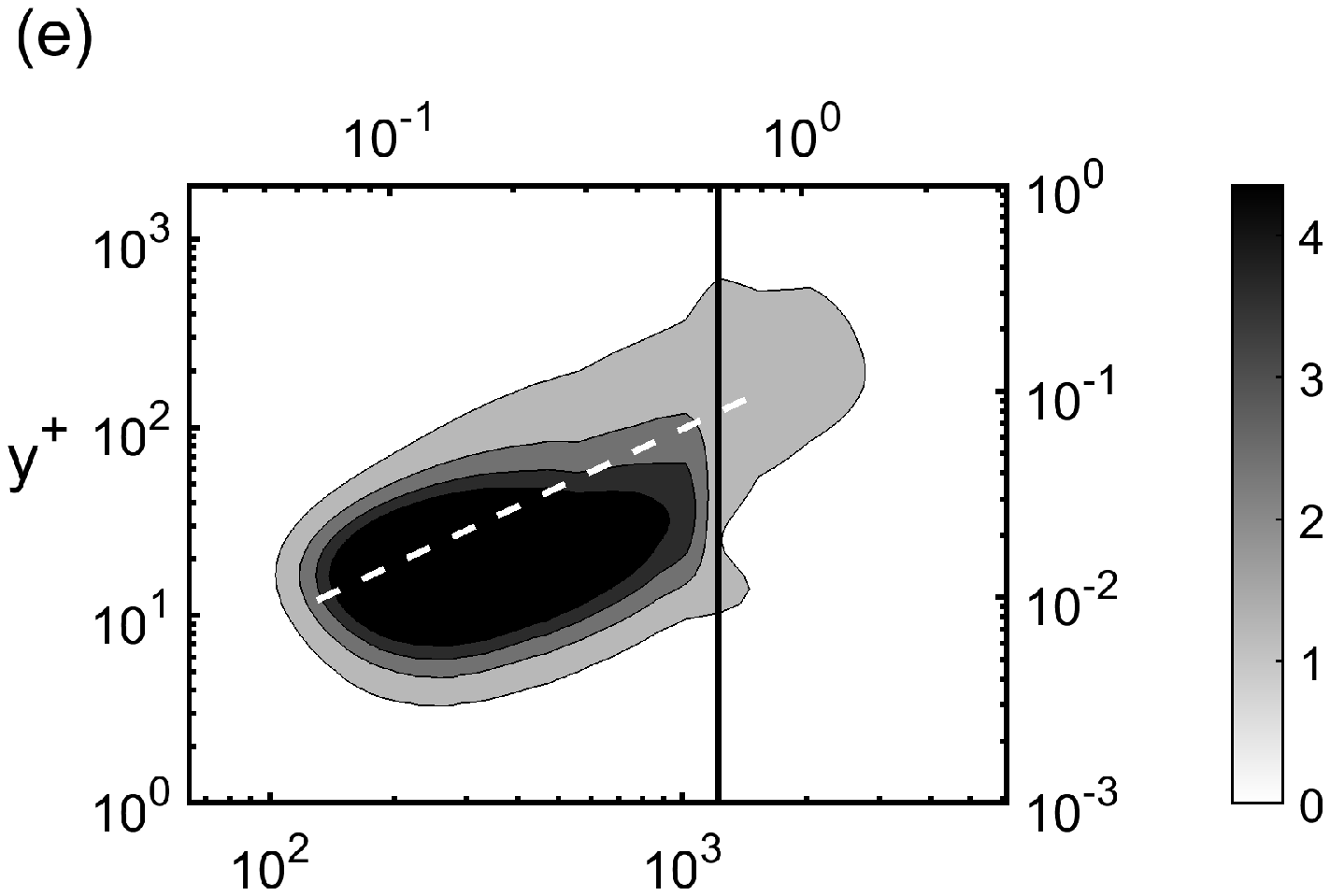}
  \label{5}
\end{subfigure}
\begin{subfigure}[b]{0.42\textwidth}
  \includegraphics[width=\textwidth]{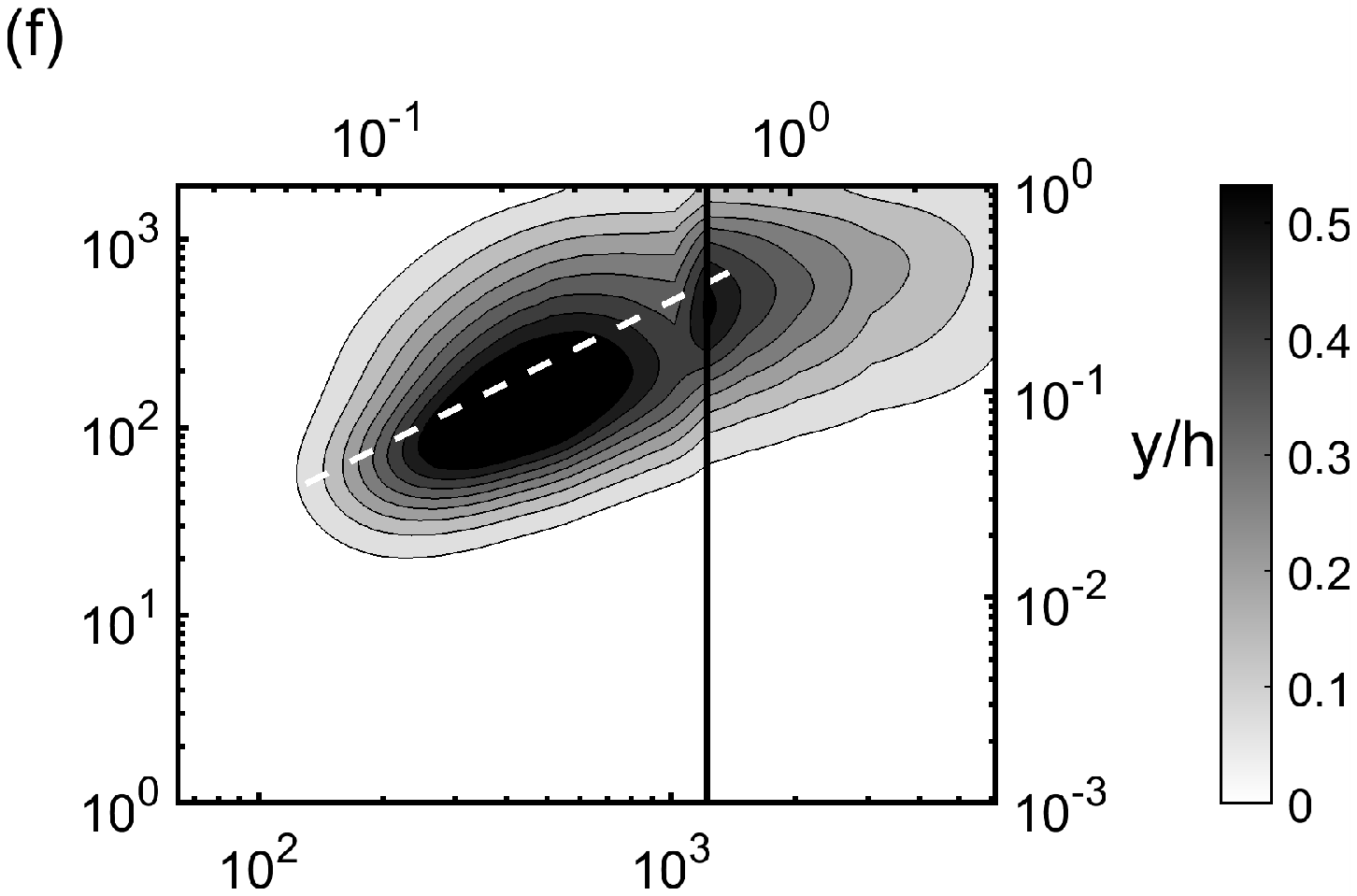}
  \label{6}
\end{subfigure}
\begin{subfigure}[b]{0.42\textwidth}
  \includegraphics[width=\textwidth]{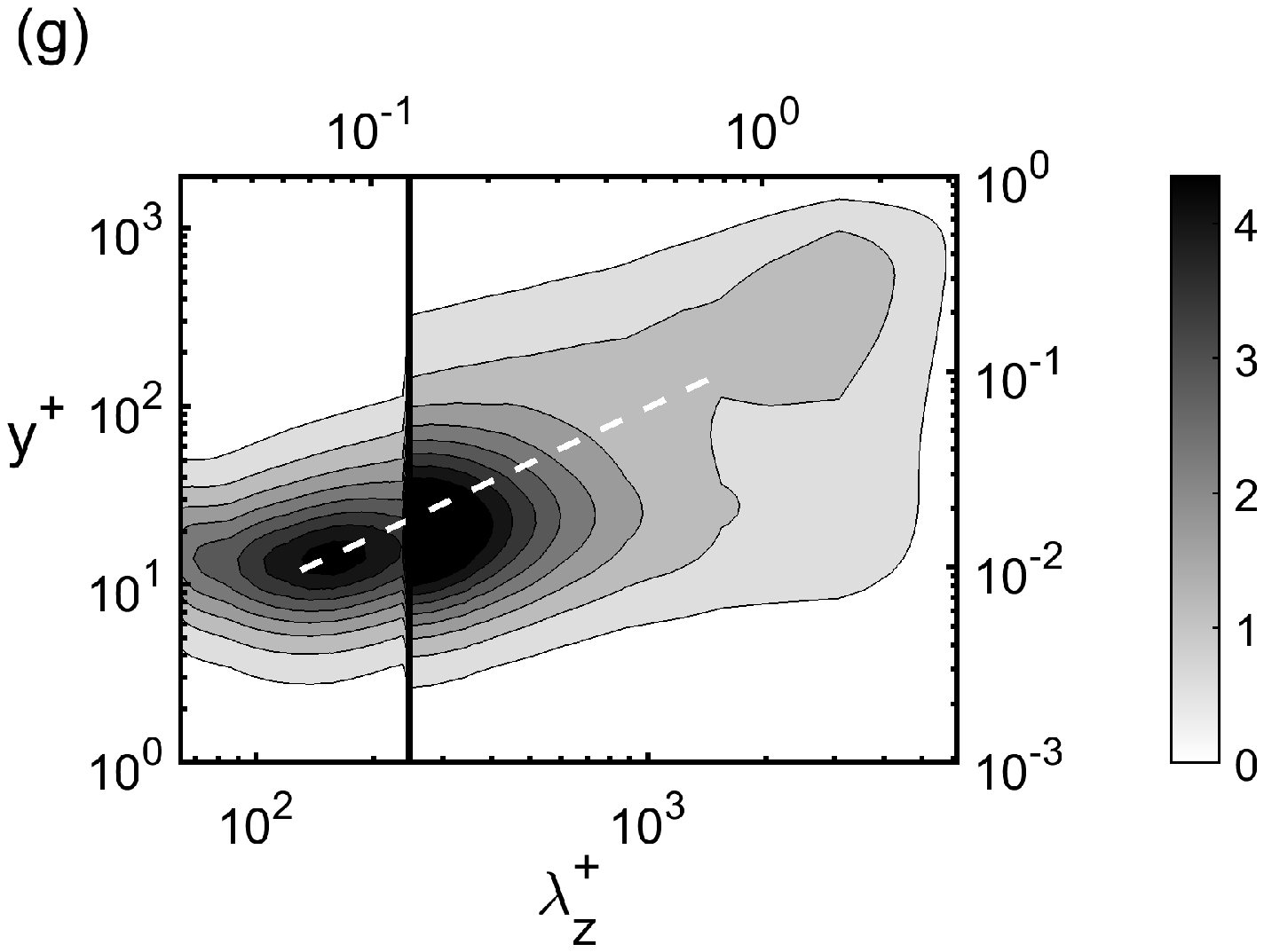}
  \label{5}
\end{subfigure}
\begin{subfigure}[b]{0.42\textwidth}
  \includegraphics[width=\textwidth]{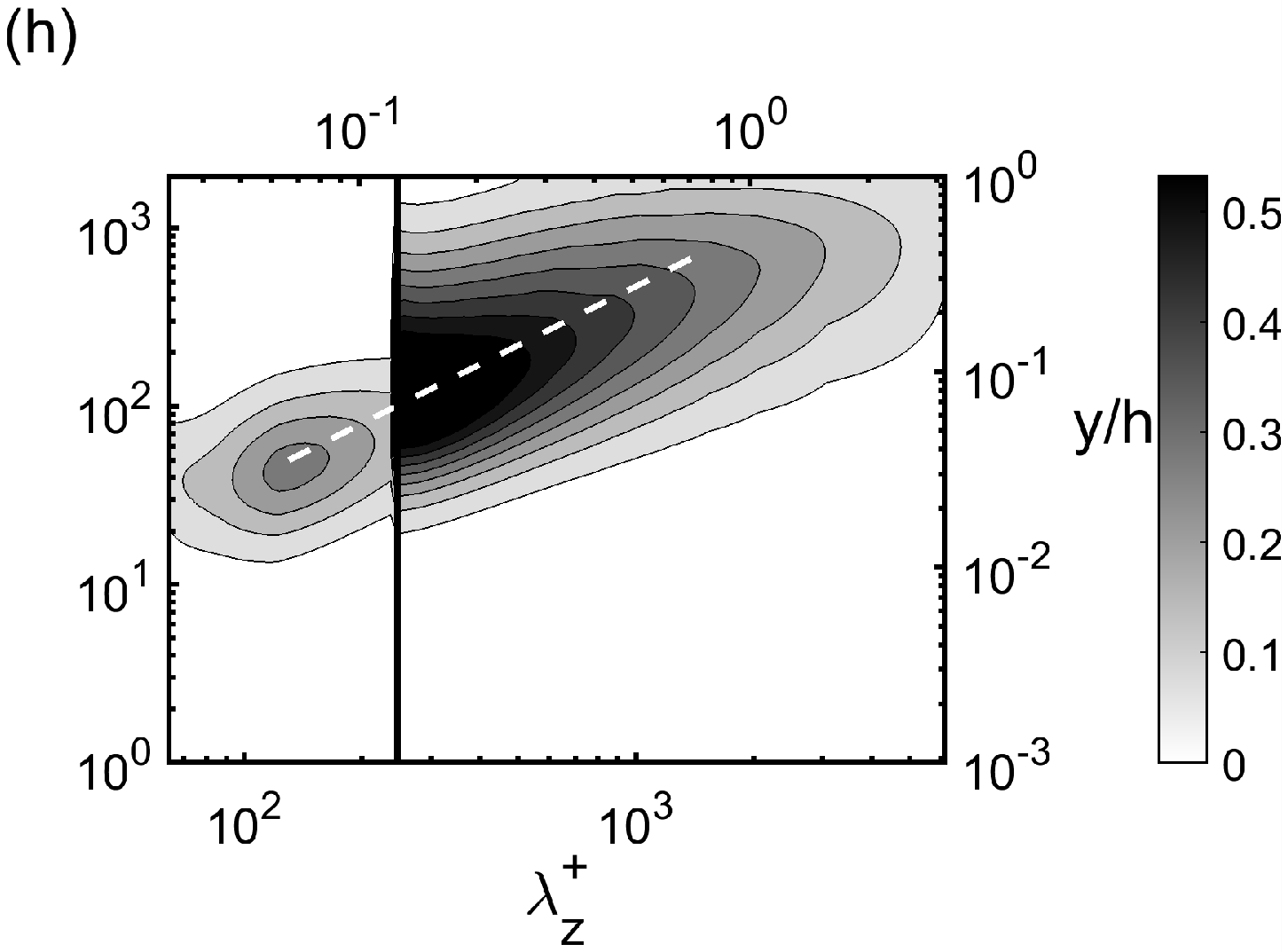}
  \label{6}
\end{subfigure}
\end{minipage}%
\caption{Premultiplied spanwise wavenumber spectra of streamwise $k_z^+ \Phi_{uu}^+(y^+,\lambda_z^+)$ (left column) and wall-normal $k_z^+ \Phi_{vv}^+(y^+,\lambda_z^+)$ (right column) velocity for (a,b) LES, (c,d) GQLZ1, (e,f) GQLZ5 and (g,h) GQLZ25 cases. Here, the vertical line represents the spanwise cut-off wavelength ($\lambda_{z,c}$) dividing the $\mathcal{P}_h$- (left) and $\mathcal{P}_l$-subspace (right) regions.}
\label{fig:zspectra}
\end{figure}

\section{GQL approximations}\label{sec:sec3}

\subsection{Turbulence statistics and spectra}\label{sec:sec31}
The first- and second-order turbulence statistics as a function of the wall-normal direction $y^+$ are plotted in figure \ref{fig:stat} (the superscript $(\cdot)^+$ denotes normalisation by viscous inner scale). There can be observed good agreement between the LES and DNS statistics, plotted with dashed line from \cite{hoyas08} at $Re_\tau=2003$. Similarly to the QL case in Part 1, the statistics of the GQLZ1 case are the most anisotropic in comparison to the reference LES, although the physical nature of the flow decomposition adopted in GQLZ1 is quite different from the QL case (see also \S\ref{sec:sec41} and \S\ref{sec:sec42} for a further discussion). In particular, $u_{rms}^+$ (figure \ref{fig:stat}a) and $U^+$ (figure \ref{fig:stat}c) are larger than those of the reference LES, behaving similarly to the QL model in the previous studies \citep{paper1,thomas14,farrell16} where the quasilinear approximation is made in the streamwise direction. The behaviour of $v_{rms}^+$ (figure \ref{fig:stat}b) and $\langle u^{\prime} v^{\prime}\rangle_{x,z,t}^+$ depends on the $y$ location: they are smaller on and below the log layer and greater above it. The enrichment of the $\mathcal{P}_l$-subspace group by further incorporating more spanwise Fourier modes (i.e. GQLZ5 and GQLZ25 cases) leads to the magnitude of the near-wall streamwise velocity peak being considerably reduced, whilst $v_{rms}^+$, $w_{rms}^+$ and $\langle u^{\prime} v^{\prime}\rangle_{x,z,t}^+$ become slightly larger than those of the reference LES. For the GQLZ5 case, the inner-scaled mean velocity $U^+$ now provides a much better approximation to that of the reference LES and the GQLZ1 case. As more spanwise Fourier modes are included in the $\mathcal{P}_l$-subspace group, all the turbulence statistics converge to those of the full LES (GQLZ5 and GQLZ25 cases in figure \ref{fig:stat}), similarly to the GQL cases along the streamwise direction investigated in Part 1. Finally, the statistics of the GQLZ25 case give the closest match to those of the reference LES, as expected.

Figure \ref{fig:zspectra} compares the premultiplied spanwise wavenumber spectra of streamwise (left column) and wall-normal (right column) velocities of the reference LES (figures \ref{fig:zspectra}a,b) with those of the GQLZ cases (figures \ref{fig:zspectra}c-h). Typical features of the energy-containing motions in turbulent channel flow can be observed in the spectra of LES \cite[e.g.][]{hwang15}. The spanwise wavenumber spectra of streamwise velocity appear to be aligned along a linear ridge $y \approx 0.1 \lambda_z$ throughout the logarithmic region. At the bottom end of this ridge ($y^+\approx 10$ and $\lambda_z^+\approx 100$), a local maximum with the spanwise spacing of the near-wall streaks (\citealp{kline67}) is found. At the top end ($y/h\approx 0.1$ and $\lambda_z/h\approx 1$), another local maximum can be found, whose spanwise wavelength corresponds to the spanwise length scale of large- and very-large-scale motions (\citealp{kovasznay70,delalamo03}). The spanwise wavenumber spectra of wall-normal velocity are also found to be aligned along a linear ridge $y =0.5 \lambda_z$ (figure \ref{fig:zspectra}a,b). 

\begin{figure}
\begin{minipage}{\textwidth}
\centering
\begin{subfigure}[b]{0.42\textwidth}
  \includegraphics[width=\textwidth]{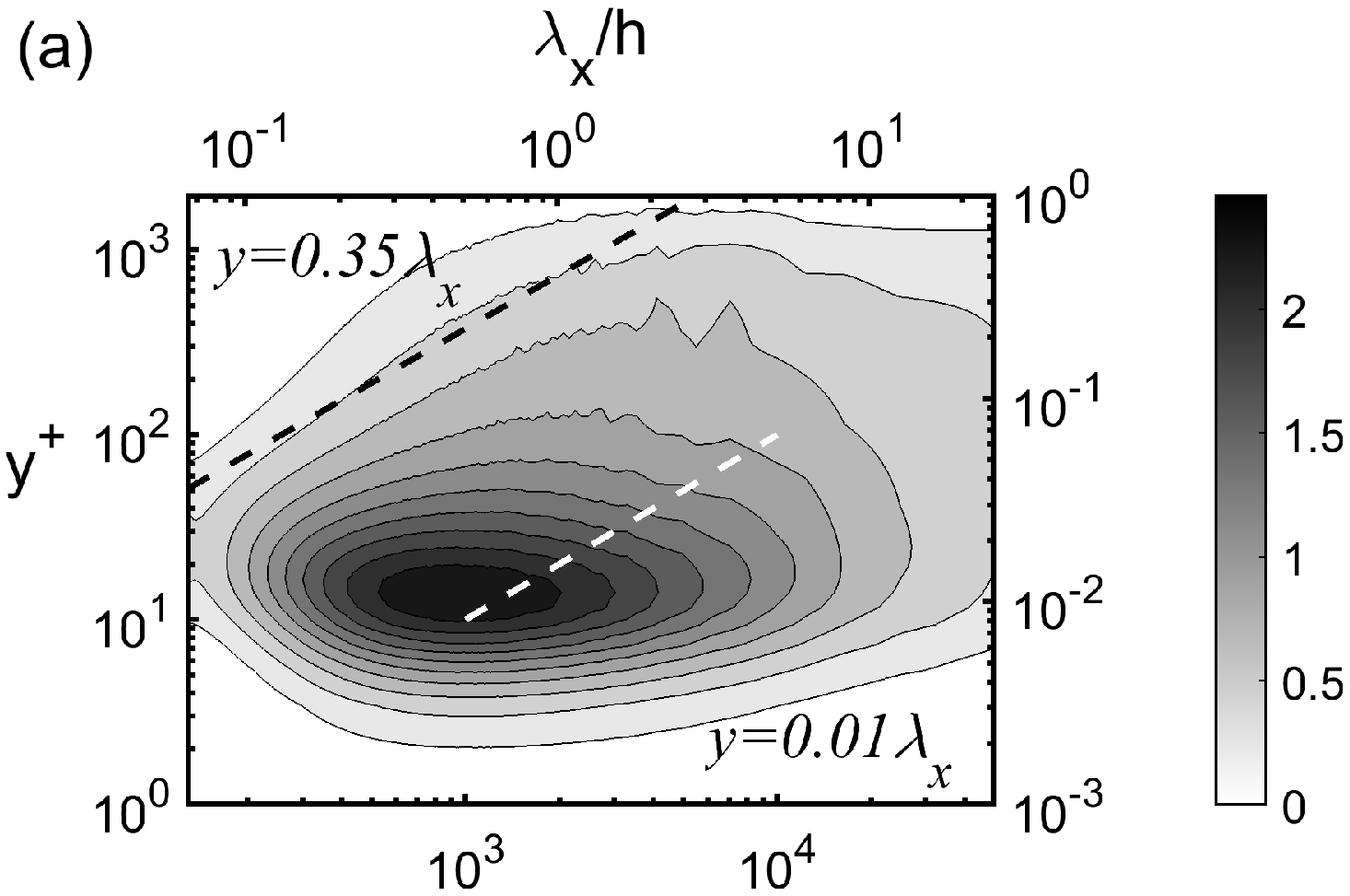}
\label{1}
\vspace{-0.8cm}
\end{subfigure}
\begin{subfigure}[b]{0.42\textwidth}
  \includegraphics[width=\textwidth]{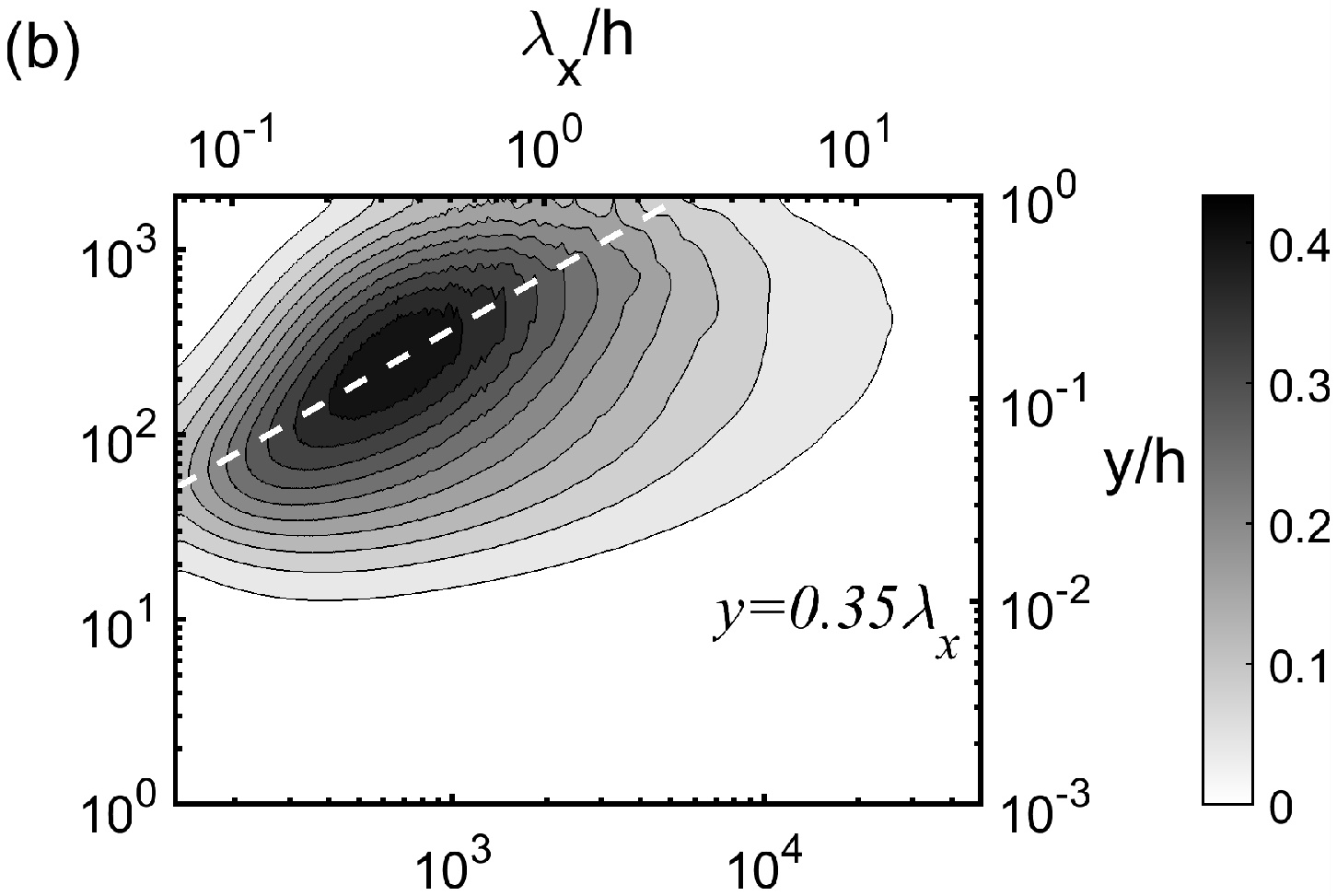}
\label{2}
\vspace{-0.8cm}
\end{subfigure}
\vspace{-0.8cm}
\begin{subfigure}[b]{0.42\textwidth}
  \includegraphics[width=\textwidth]{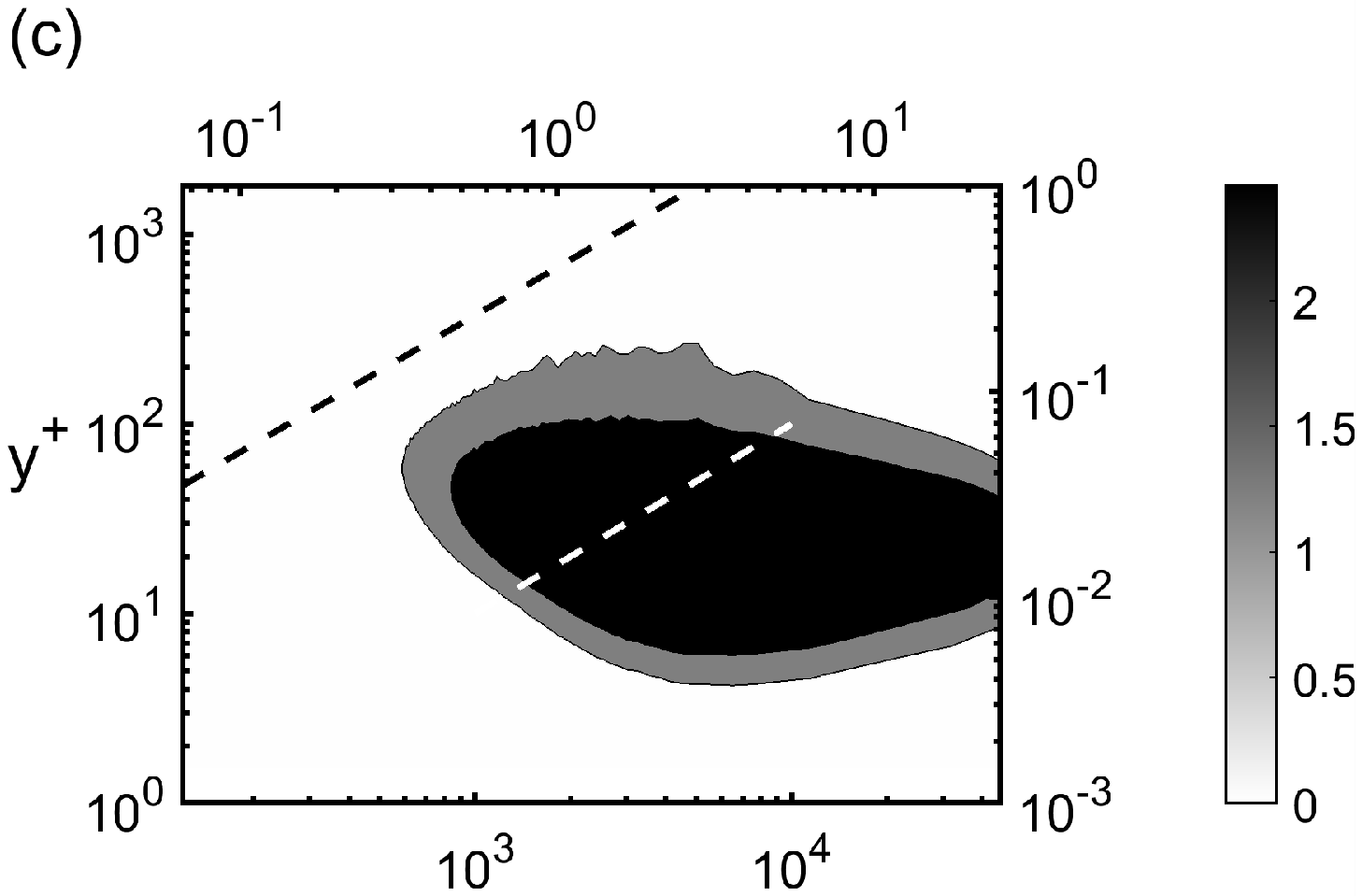}
  \label{3}
\end{subfigure}
\begin{subfigure}[b]{0.42\textwidth}
  \includegraphics[width=\textwidth]{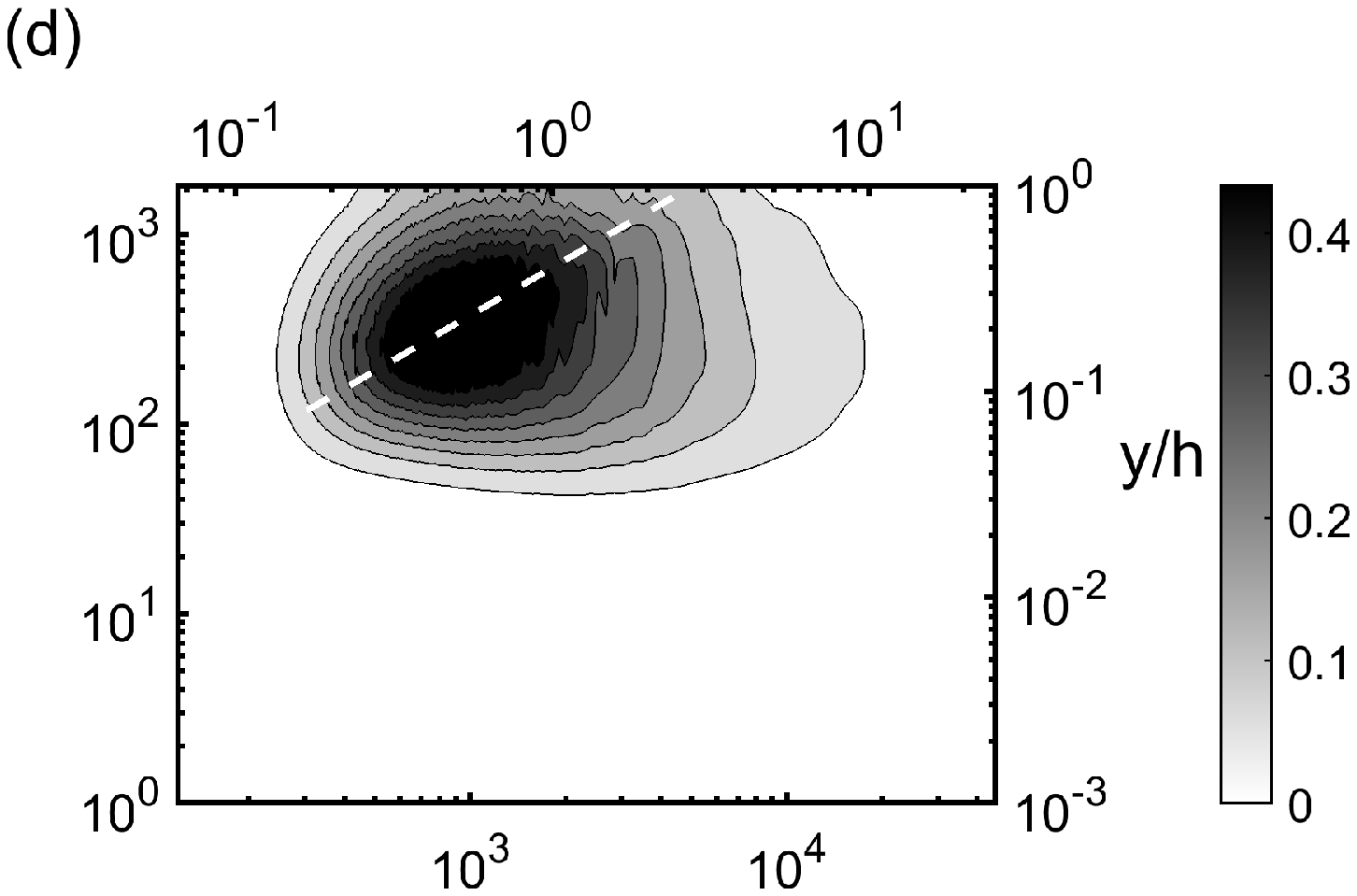}
  \label{4}
\end{subfigure}
\vspace{-0.8cm}
\begin{subfigure}[b]{0.42\textwidth}
  \includegraphics[width=\textwidth]{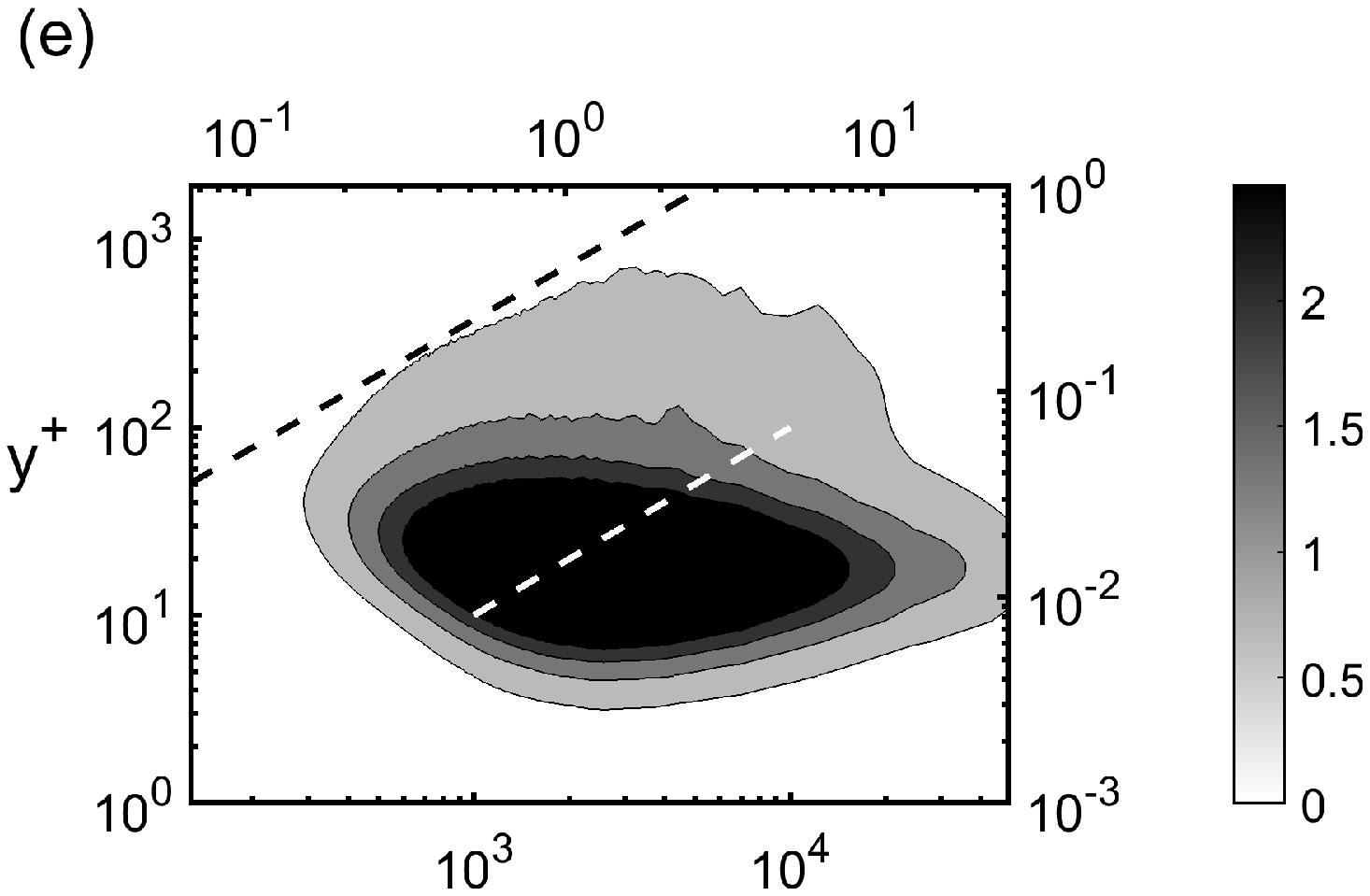}
  \label{5}
\end{subfigure}
\begin{subfigure}[b]{0.42\textwidth}
  \includegraphics[width=\textwidth]{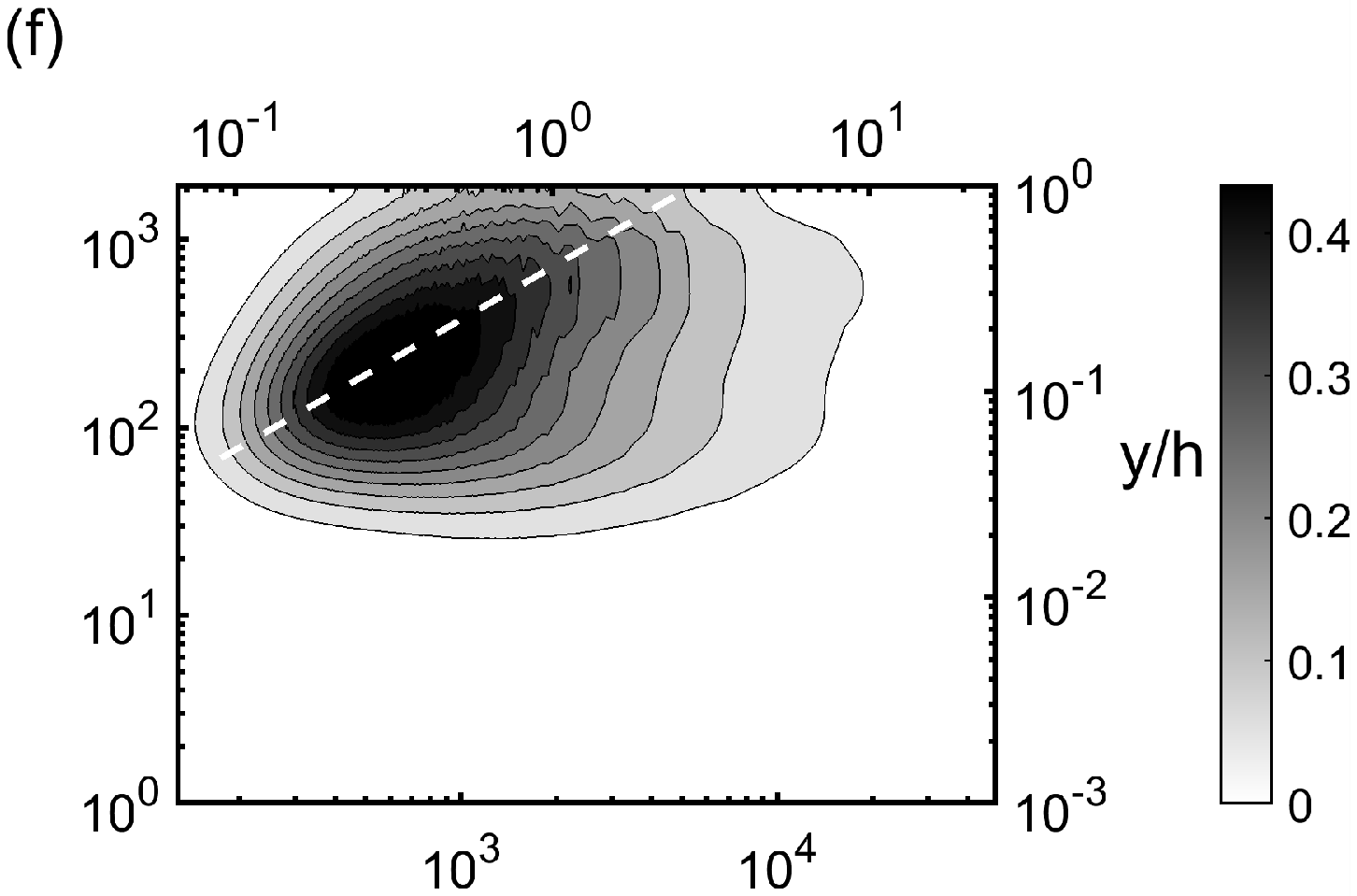}
  \label{6}
\end{subfigure}
\begin{subfigure}[b]{0.42\textwidth}
  \includegraphics[width=\textwidth]{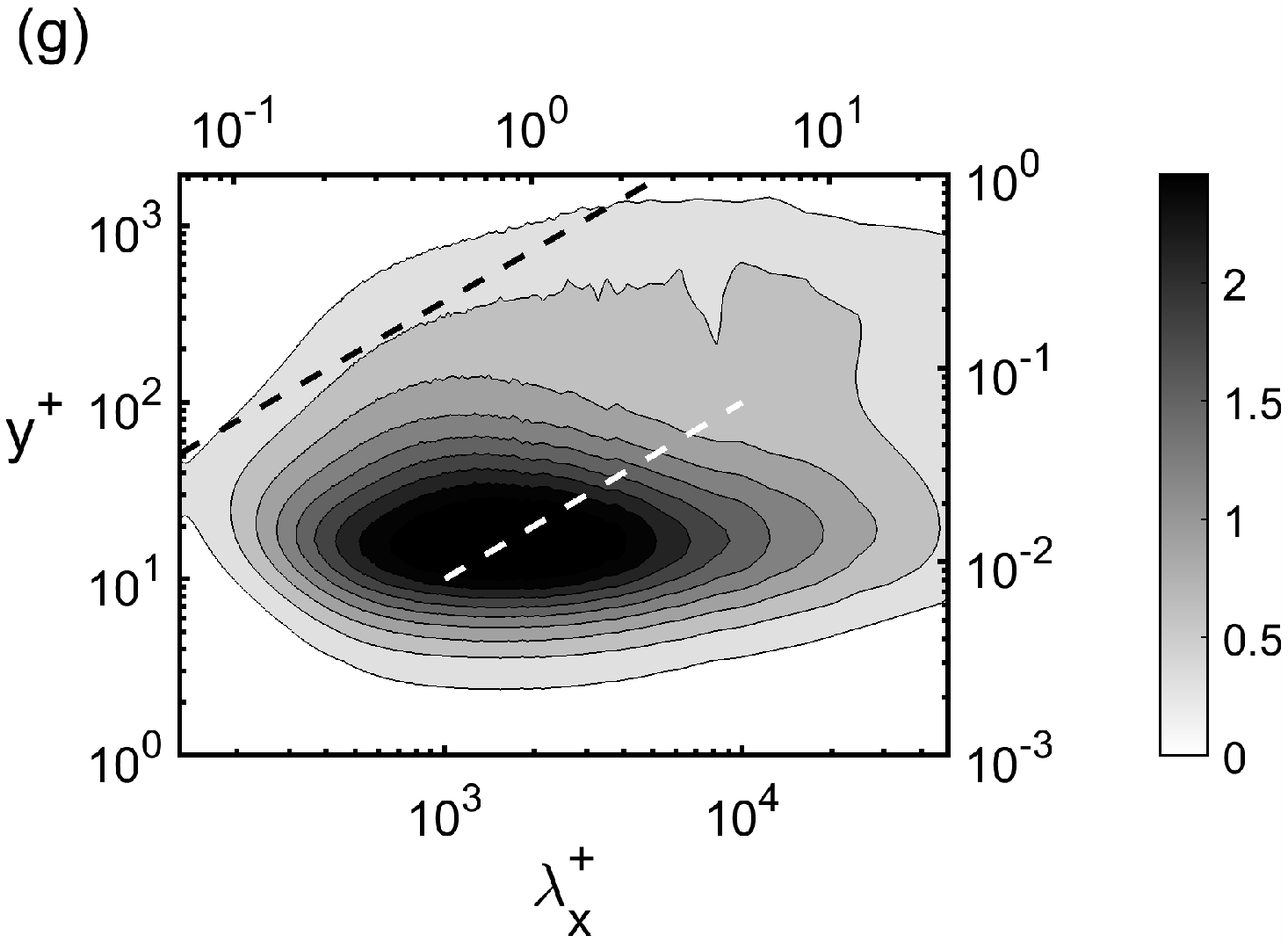}
\end{subfigure}
\begin{subfigure}[b]{0.42\textwidth}
  \includegraphics[width=\textwidth]{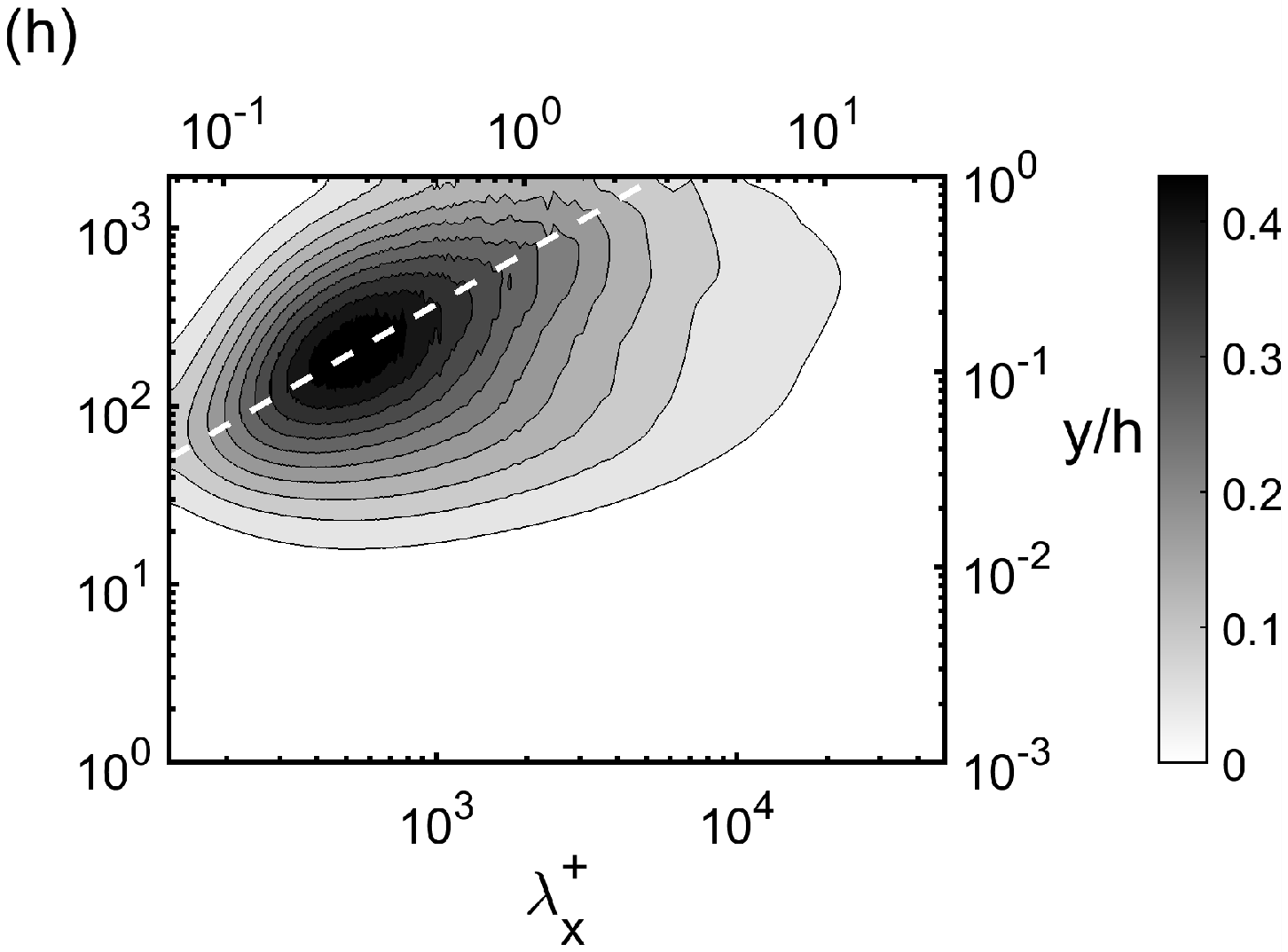}
\end{subfigure}
\end{minipage}
\caption{Premultiplied streamwise wavenumber spectra of streamwise $k_x^+ \Phi_{uu}^+(y^+,\lambda_x^+)$ (left column) and wall-normal $k_x^+ \Phi_{vv}^+(y^+,\lambda_x^+)$ (right column) velocity for (a,b) LES, (c,d) GQLZ1, (e,f) GQLZ5 and (g,h) GQLZ25 cases.}
\label{fig:xspectra}
\end{figure}

The GQLZ1 case shows an increased energy in the region $\lambda_z^+ \approx 500$ $(\lambda_z/h \approx 0.3)$ of the streamwise velocity spectra compared to the LES case, in line with the statistics discussed above. This is in contrast to the reduced energy featured at the other wavelengths (figure \ref{fig:zspectra}c). The spanwise wavenumber spectra of wall-normal velocity show increased energy in the region $\lambda_z^+ \approx 1000$ $(\lambda_z/h \approx 0.5)$, and the spectra span for most of the log region but showing little energy for $\lambda_z^+ \lesssim 200$ (figure \ref{fig:zspectra}d). The reduced energy observed for $\lambda_z^+ \lesssim 200$ explains the lack of Reynolds shear stress of the GQLZ1 case in the near-wall region (figure \ref{fig:stat}d). By incorporating more spanwise Fourier modes in the $\mathcal{P}_l$-subspace group (i.e. GQLZ5 and GQLZ25), the peak location in the streamwise velocity spectra gradually moves towards $(\lambda_z^+,y^+) \approx (100,10)$ along the linear ridge $y=0.1\lambda_z$ and the outer part of the spectra also becomes more energetic. The wall-normal velocity spectra now span the region close to the wall along the linear ridge $y=0.5 \lambda_z$. Moreover, the spectra of the GQLZ25 case are found to be fairly similar to those of the reference LES. An additional case GQLZ64 has been computed (see Appendix \ref{appendix}) with $\lambda_{z,c}^+ \approx 97$, resulting in spanwise wavenumber spectra which do not extend below $\lambda_{z,c}^+$, implying that the velocity field in the $\mathcal{P}_h$-subspace group yields the trivial solution. This issue, investigated in Part 1 for the streamwise GQL models, was explained in terms of the Lyapunov exponent and the instability of the linearised equations in the $\mathcal{P}_h$-subspace group.

The premultiplied streamwise wavenumber spectra of streamwise and wall-normal velocities are shown in figure \ref{fig:xspectra} for the LES and GQLZ cases. The energy-containing motions in turbulent channel flow (LES) at a given spanwise length scale are composed of a bimodal structure with two components: a long streaky structure mainly carrying streamwise turbulent kinetic energy and a short and tall vortex packet carrying turbulent kinetic energy at all the velocity components \citep{hwang15}. The corresponding linear ridges $y \approx 0.01 \lambda_x$ (lower line) and $y \approx 0.35 \lambda_x$ (upper line), along which these structures contribute to the spectra, are plotted in figure \ref{fig:xspectra}. The former corresponds to a long streaky structure and the streamwise velocity spectra show high energy intensity along it (figure \ref{fig:xspectra}a). The latter corresponds to a vortex packet and the streamwise spectra are less energetic along it, whilst the wall-normal velocity spectra are very well aligned with it. The streamwise velocity spectra of the GQLZ1 case are energetic in the region $\lambda_x^+ \gtrsim 600$, whilst the wall-normal velocity spectra do not extend below $\lambda_x^+ \lesssim 200$. The former appear to be around a single point along $y \approx 0.01 \lambda_x$, whilst weak correlation with $y \approx 0.35 \lambda_x$ is observed in the latter. As more spanwise Fourier modes are further incorporated into the $\mathcal{P}_l$-subspace group (figures \ref{fig:xspectra}e-h; i.e. GQLZ5 and GQLZ25), the two spectra begin to show more energy at smaller $\lambda_x$ along the linear ridges. As a consequence, in the GQLZ5 case, the near-wall region is better resolved and the spectra reach out to smaller scales down to $\lambda_x^+ \approx 150$. The spectra of the GQLZ25 case greatly resemble those of the LES case.

It is interesting to observe that the overall behaviours of the GQL models in this study are similar to those of the streamwise QL and GQL models in Part 1 \cite[e.g.][]{thomas14,thomas15,farrell16,tobias17,hernandez}, although the flow decomposition in (\ref{eq:2.0}) implies physically different nature of the GQL models here from those in Part 1. The QL and GQL models in Part 1 are designed to offer a minimal description for the self-sustaining processes at all integral length scales and study the effect of their mutual nonlinear coupling: the full nonlinear equations in the $\mathcal{P}_l$ subspace are primarily to describe the full nonlinear evolution of the long streaks (or streamwise velocity fluctuations), whereas the linearised equations in the $\mathcal{P}_h$ subspace are to resolve the streak instability and the subsequent breakdown process of the streaks. On the contrary, in the GQL models considered here, the full nonlinear equations in the $\mathcal{P}_l$ subspace almost fully resolve every essential elements of the self-sustaining process at least at the integral length scales for $\lambda_z\geq \lambda_{z,c}$ ($\lambda_{z,c}= 2\pi/k_{z,c}$), whereas the linearised equations in the $\mathcal{P}_h$ subspace do not support its presence for $\lambda_z<\lambda_{z,c}$. A further detailed discussion on this issue will be given in \S\ref{sec:sec41} and \S\ref{sec:sec42}. 

\subsection{Spectral energy transfer}\label{sec:sec32}

\begin{figure}
\begin{minipage}{\textwidth}
\centering
\begin{subfigure}[b]{0.42\textwidth}
  \includegraphics[width=\textwidth]{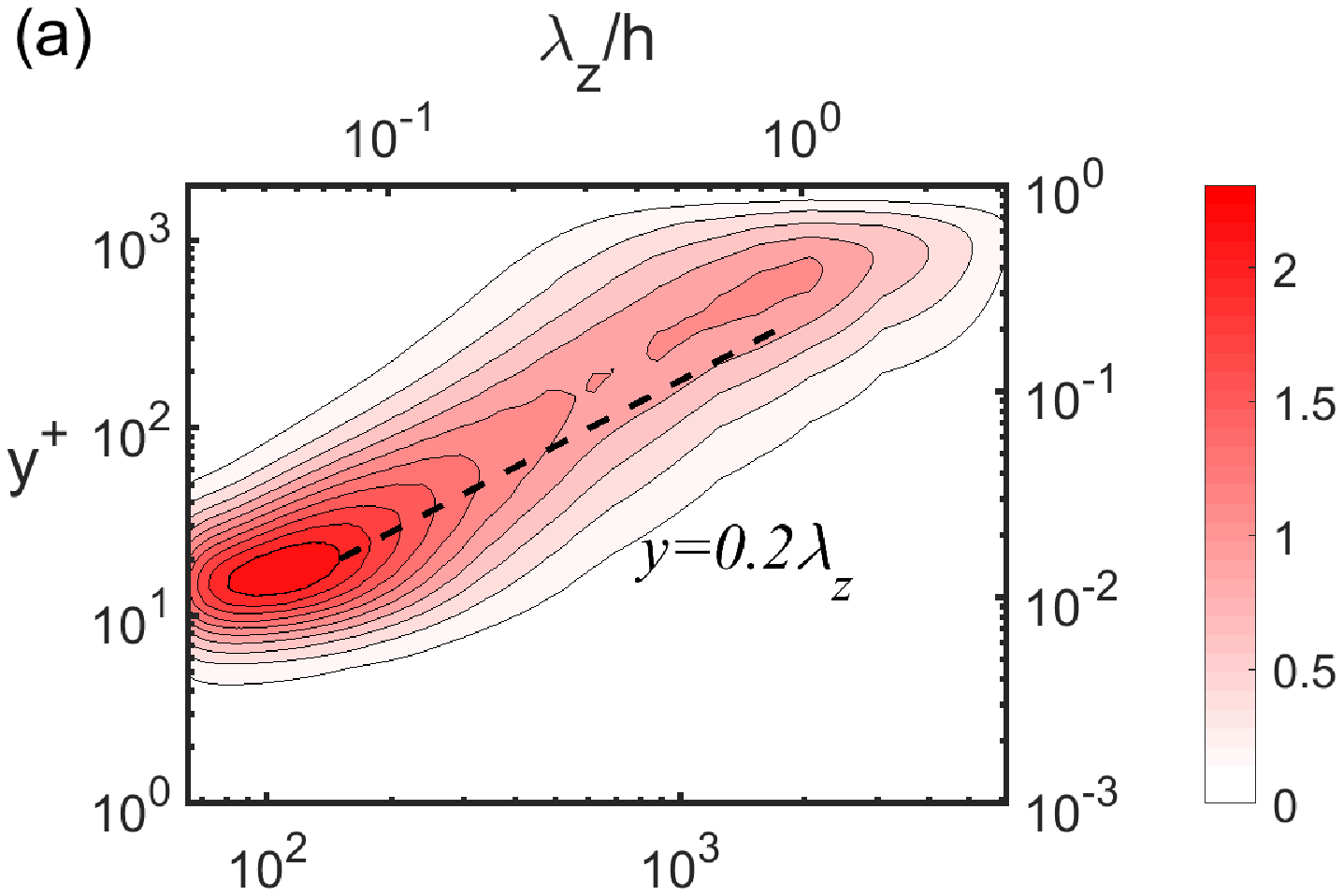}
\label{1}
\end{subfigure}
\vspace{-0.7cm}
\begin{subfigure}[b]{0.42\textwidth}
  \includegraphics[width=\textwidth]{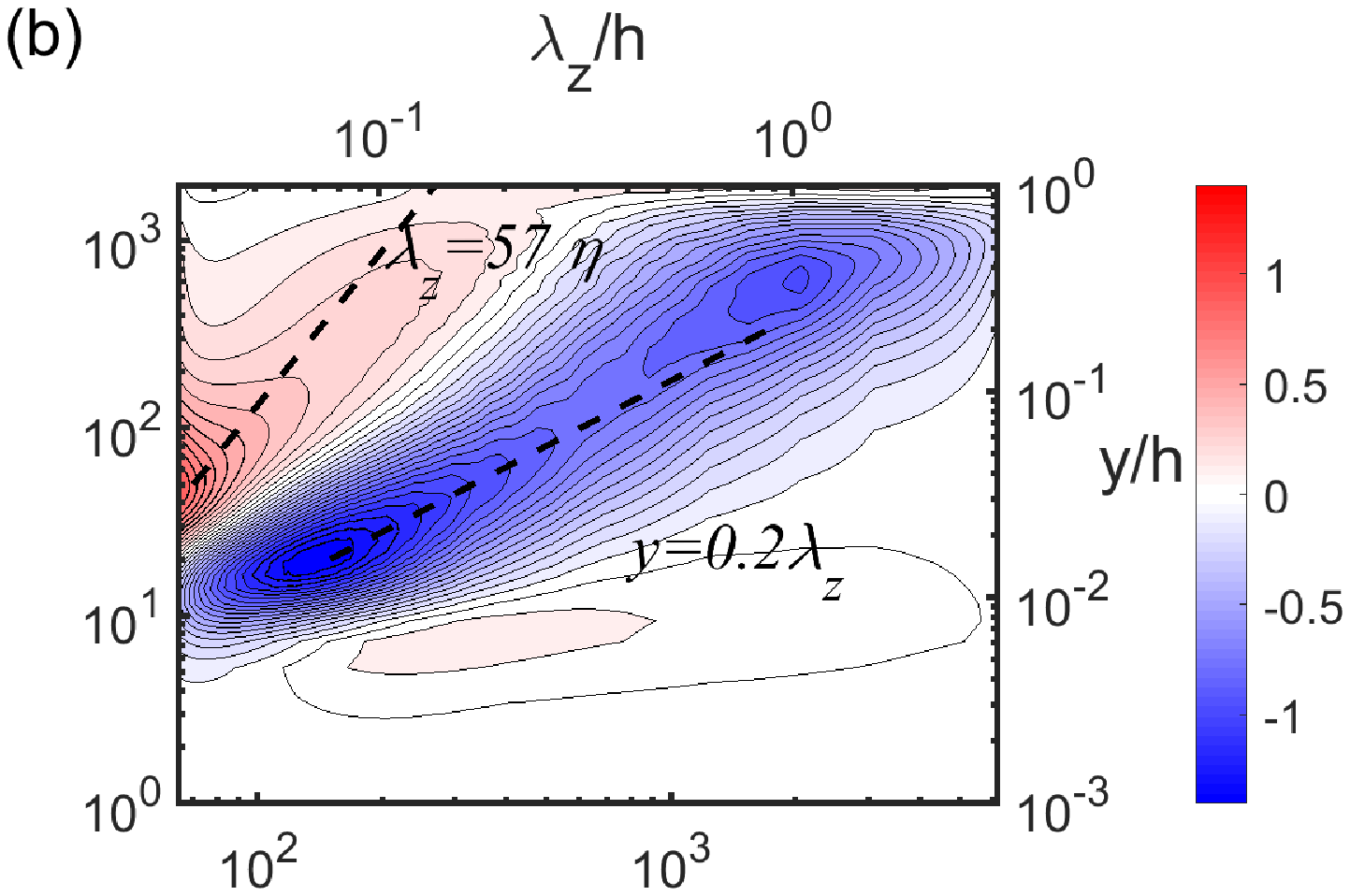}
\label{2}
\end{subfigure}
\begin{subfigure}[b]{0.42\textwidth}
  \includegraphics[width=\textwidth]{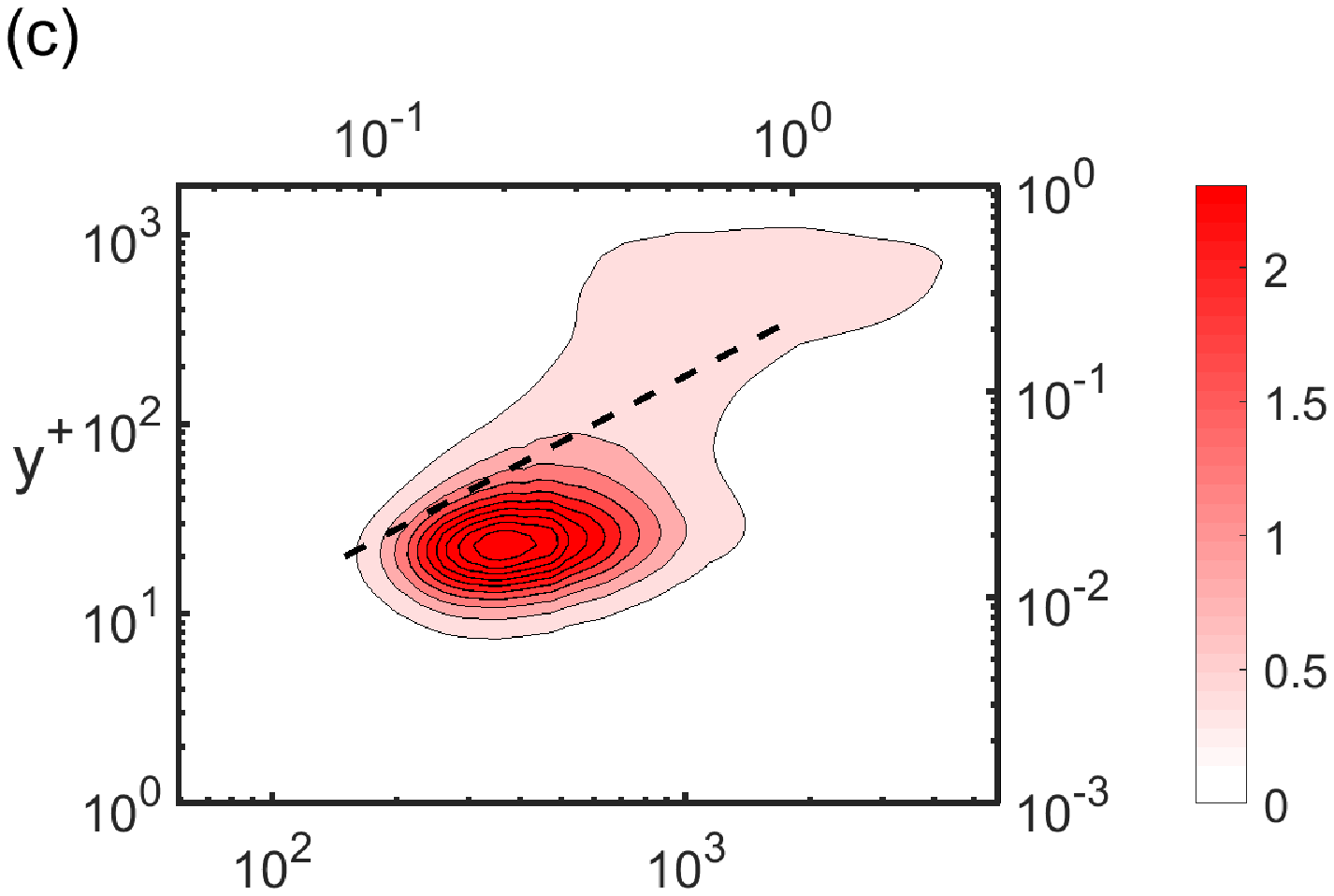}
  \label{3}
\end{subfigure}
\vspace{-0.7cm}
\begin{subfigure}[b]{0.42\textwidth}
  \includegraphics[width=\textwidth]{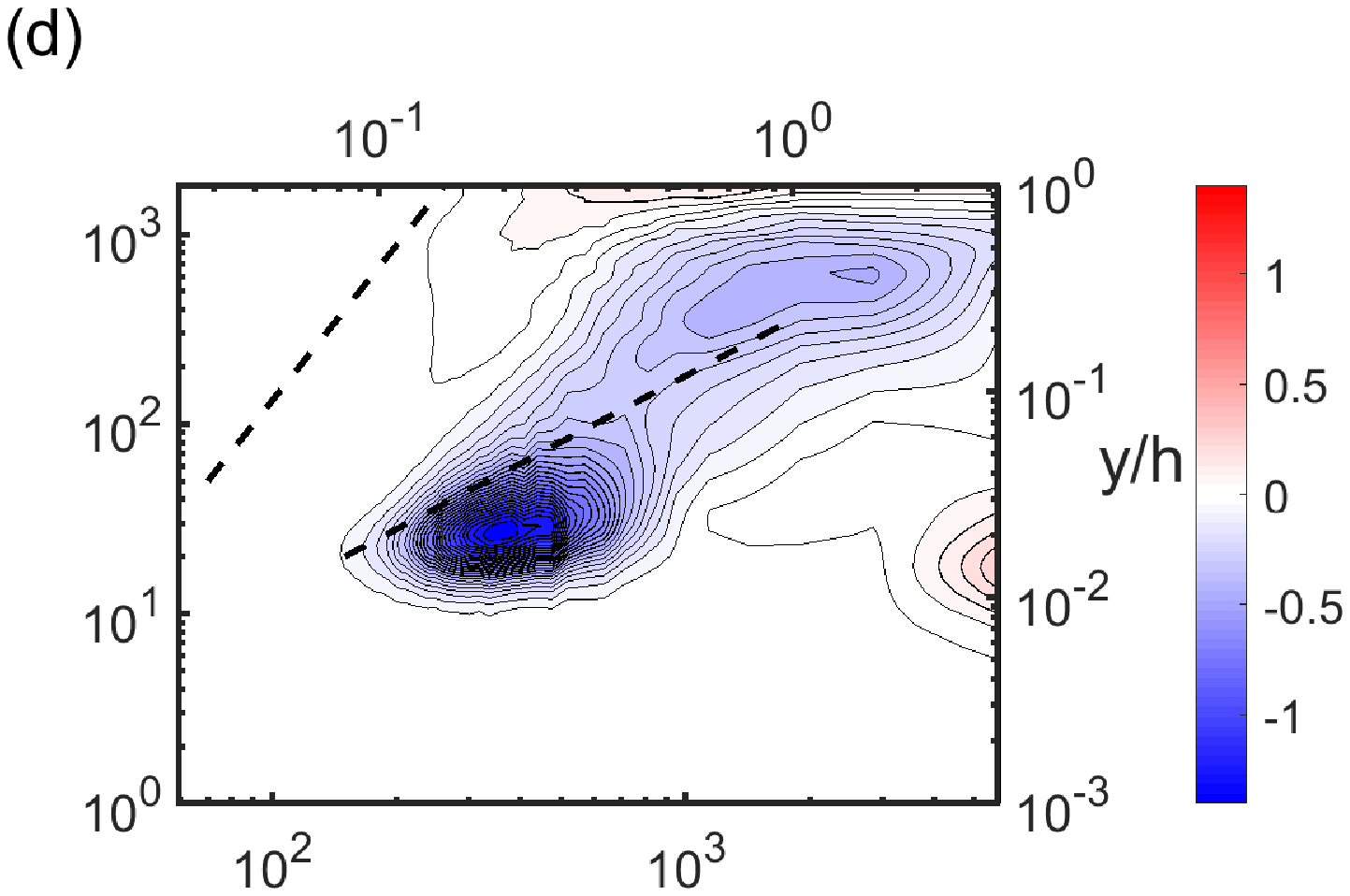}
  \label{4}
\end{subfigure}
\begin{subfigure}[b]{0.42\textwidth}
  \includegraphics[width=\textwidth]{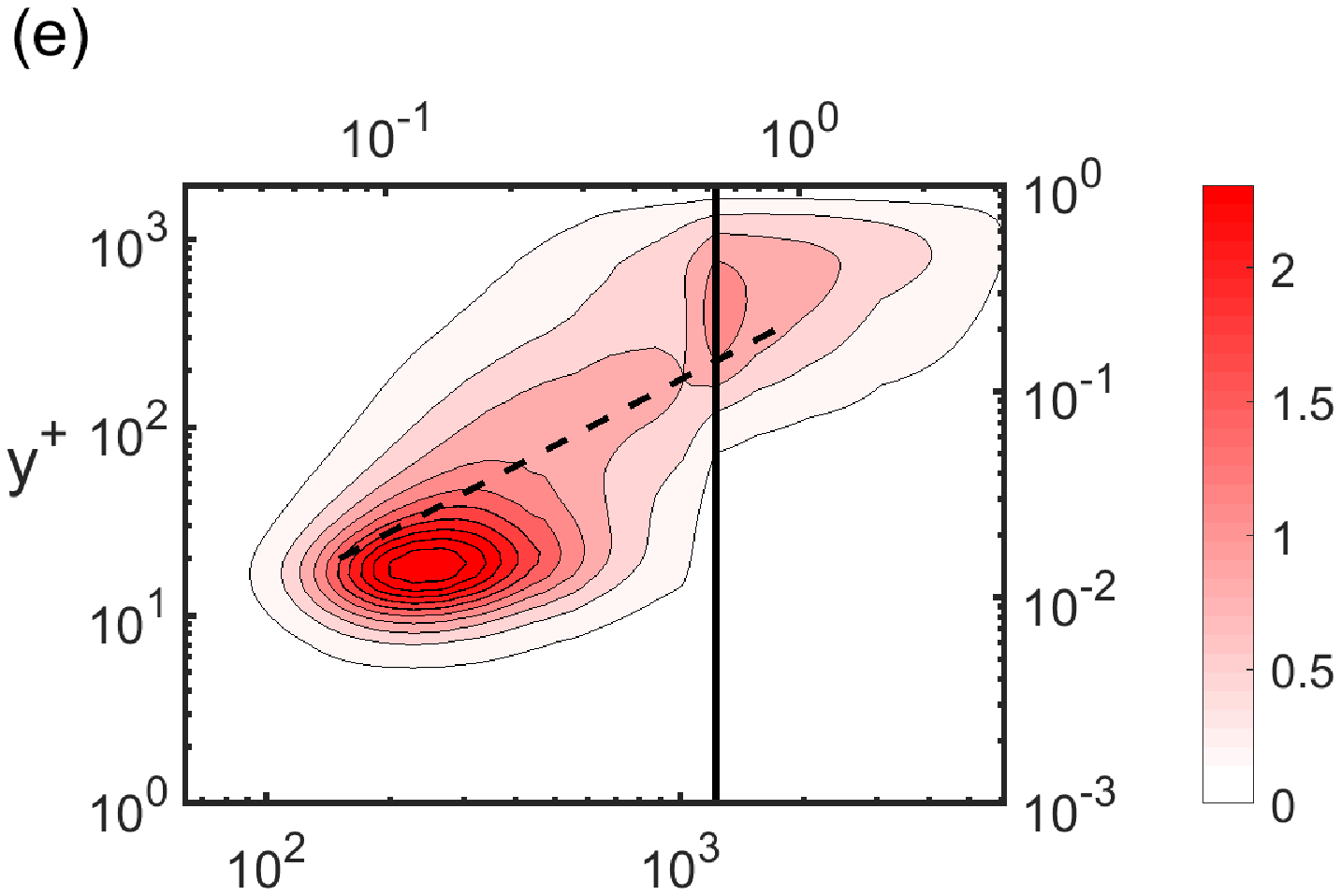}
  \label{5}
\end{subfigure}
\vspace{-0.7cm}
\begin{subfigure}[b]{0.42\textwidth}
  \includegraphics[width=\textwidth]{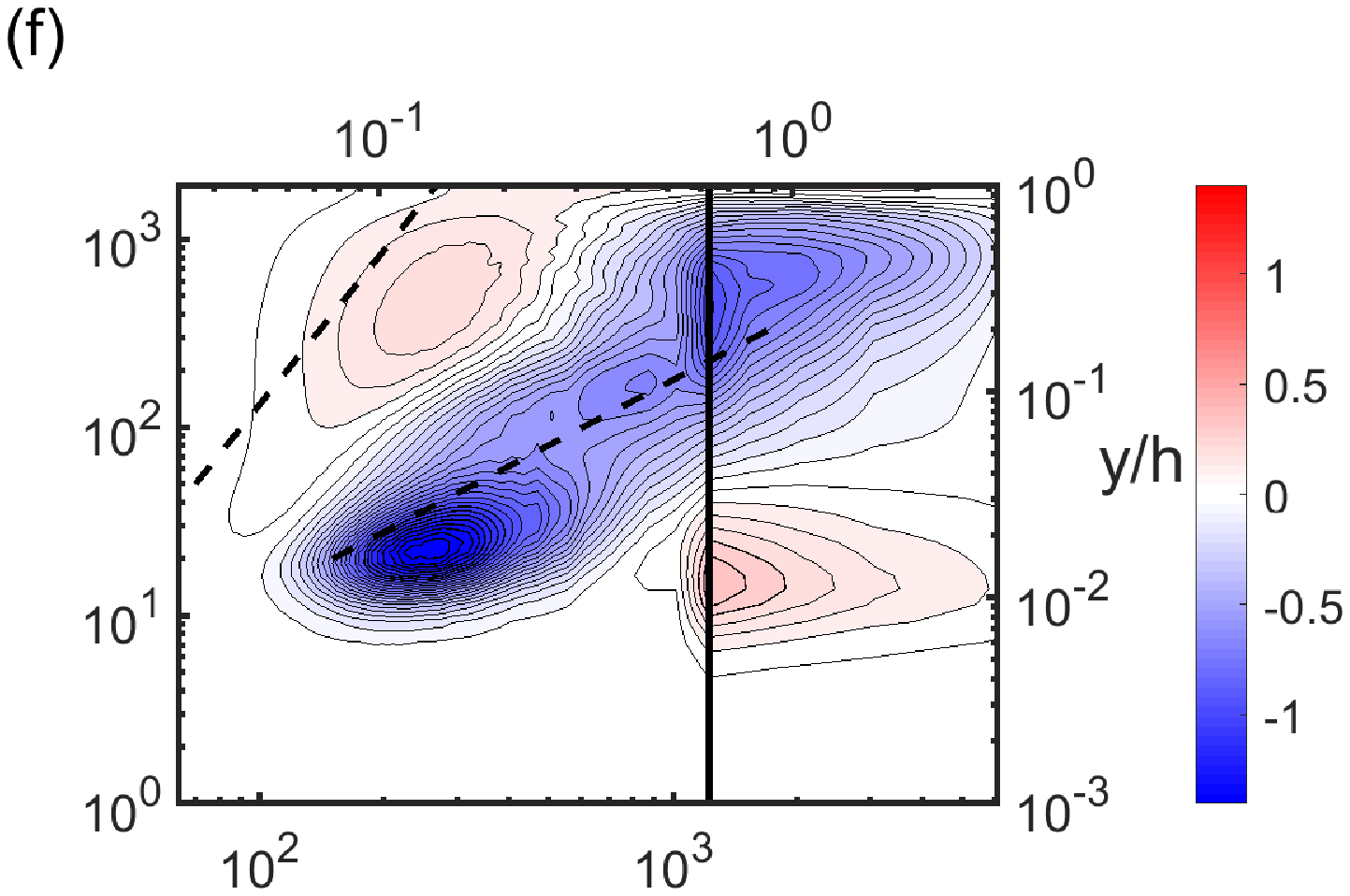}
  \label{6}
\end{subfigure}
\begin{subfigure}[b]{0.42\textwidth}
  \includegraphics[width=\textwidth]{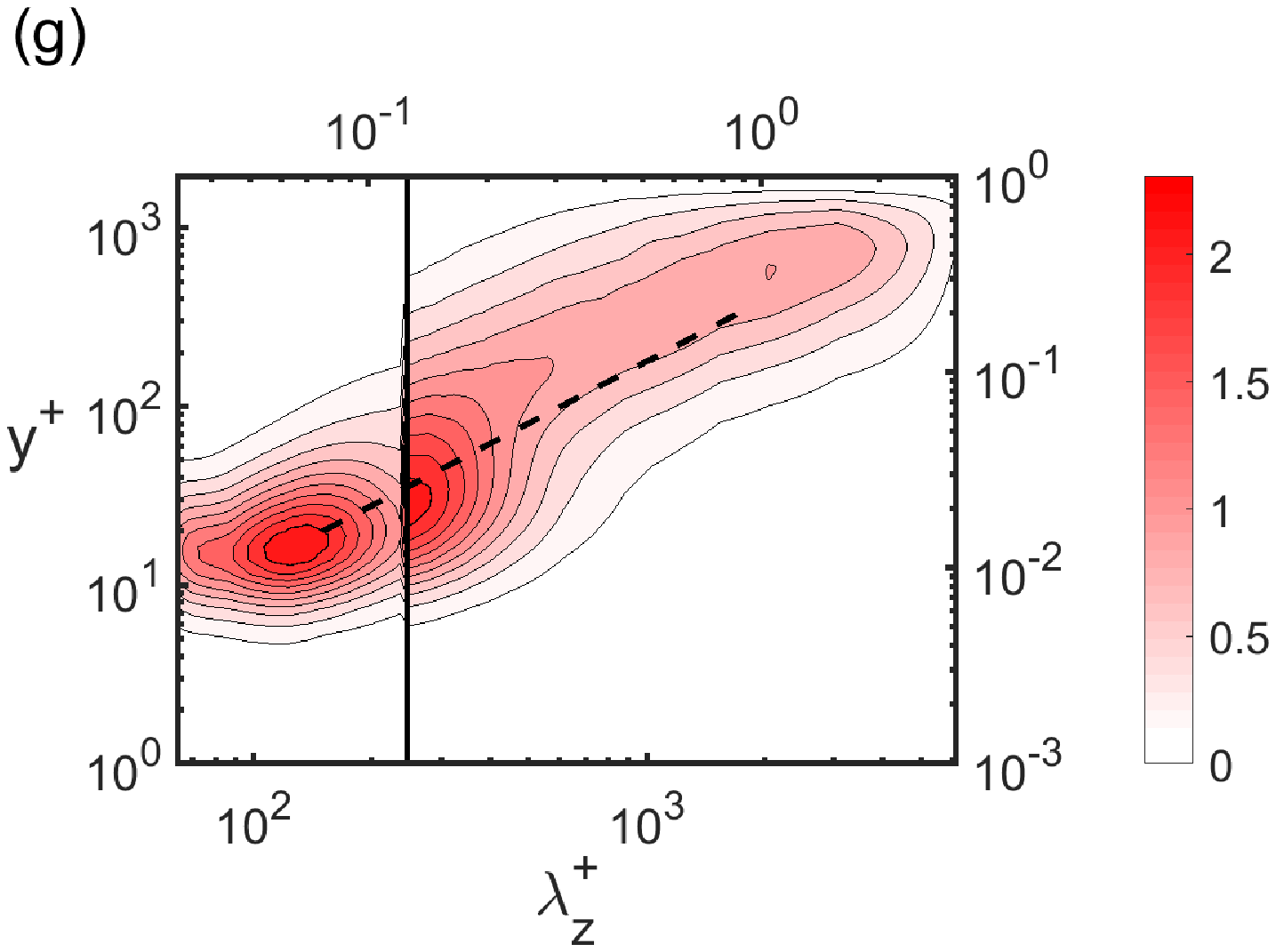}
\end{subfigure}
\begin{subfigure}[b]{0.42\textwidth}
  \includegraphics[width=\textwidth]{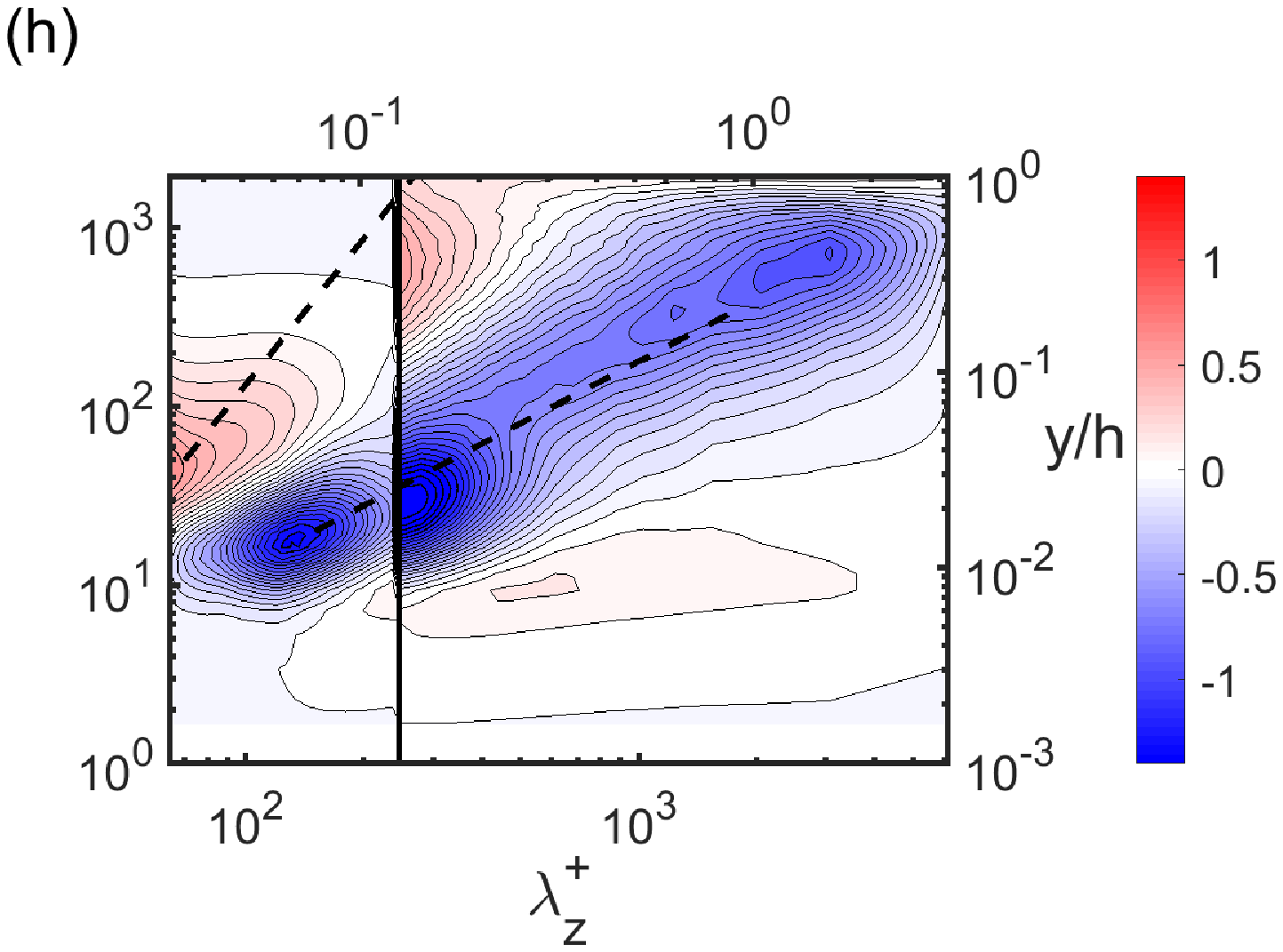}
\end{subfigure}
\end{minipage}
\caption{Premultiplied spanwise wavenumber spectra of production $k_z^+ y^+ \widehat{P}^+(y^+,\lambda_z^+)$ (left column) and turbulent transport $k_z^+ y^+ \widehat{T}_{turb}^+(y^+,\lambda_z^+)$ (right column) for (a,b) LES, (c,d) GQLZ1, (e,f) GQLZ5 and (g,h) GQLZ25 cases. Here, the vertical line represents the spanwise cut-off wavelength ($\lambda_{z,c}$) dividing the $\mathcal{P}_h$- (left) and $\mathcal{P}_l$-subspace (right) regions.}
\label{fig:zenergy}
\end{figure}

\begin{figure}
\begin{minipage}{\textwidth}
\centering
\begin{subfigure}[b]{0.42\textwidth}
  \includegraphics[width=\textwidth]{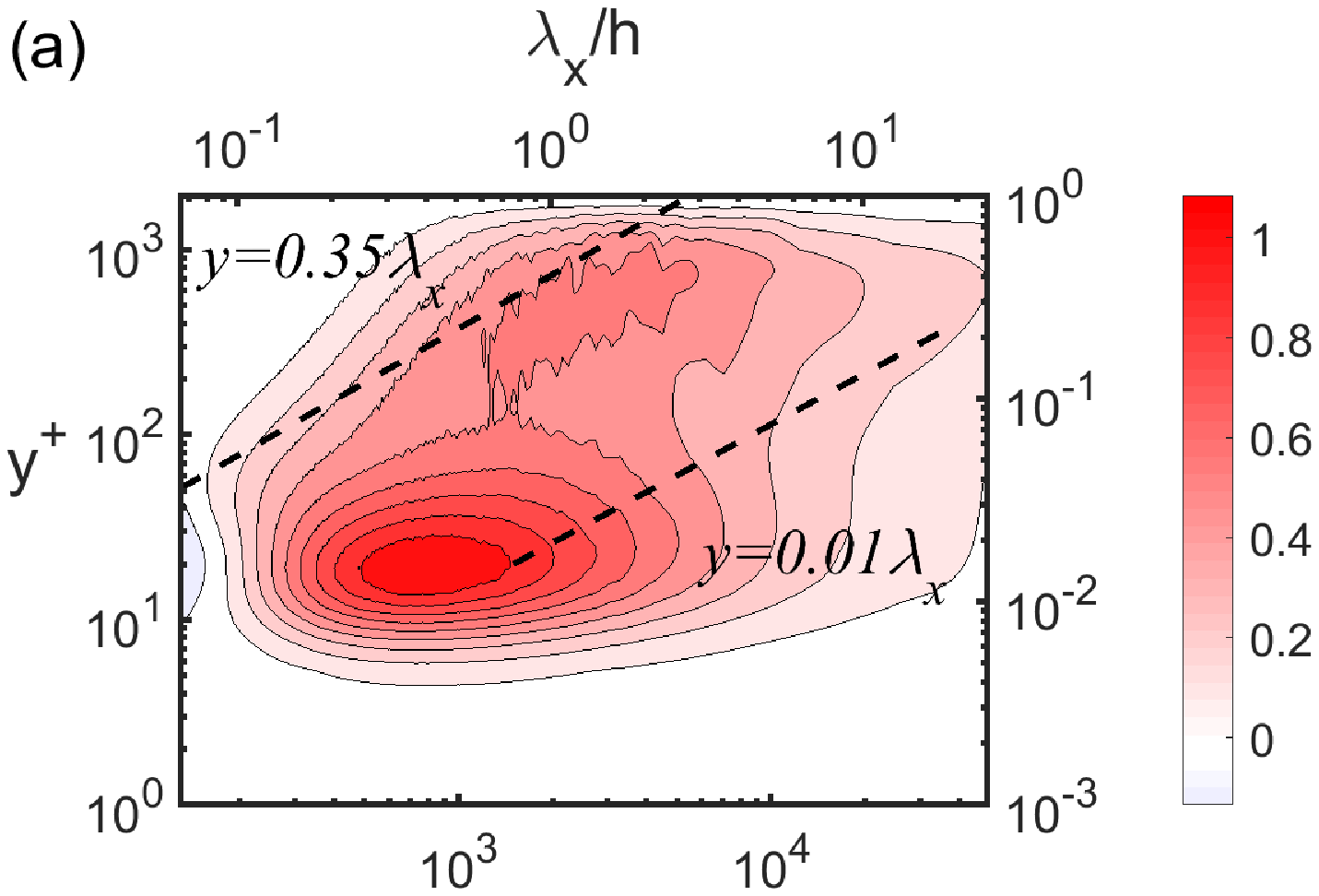}
\label{1}
\end{subfigure}
\vspace{-0.7cm}
\begin{subfigure}[b]{0.42\textwidth}
  \includegraphics[width=\textwidth]{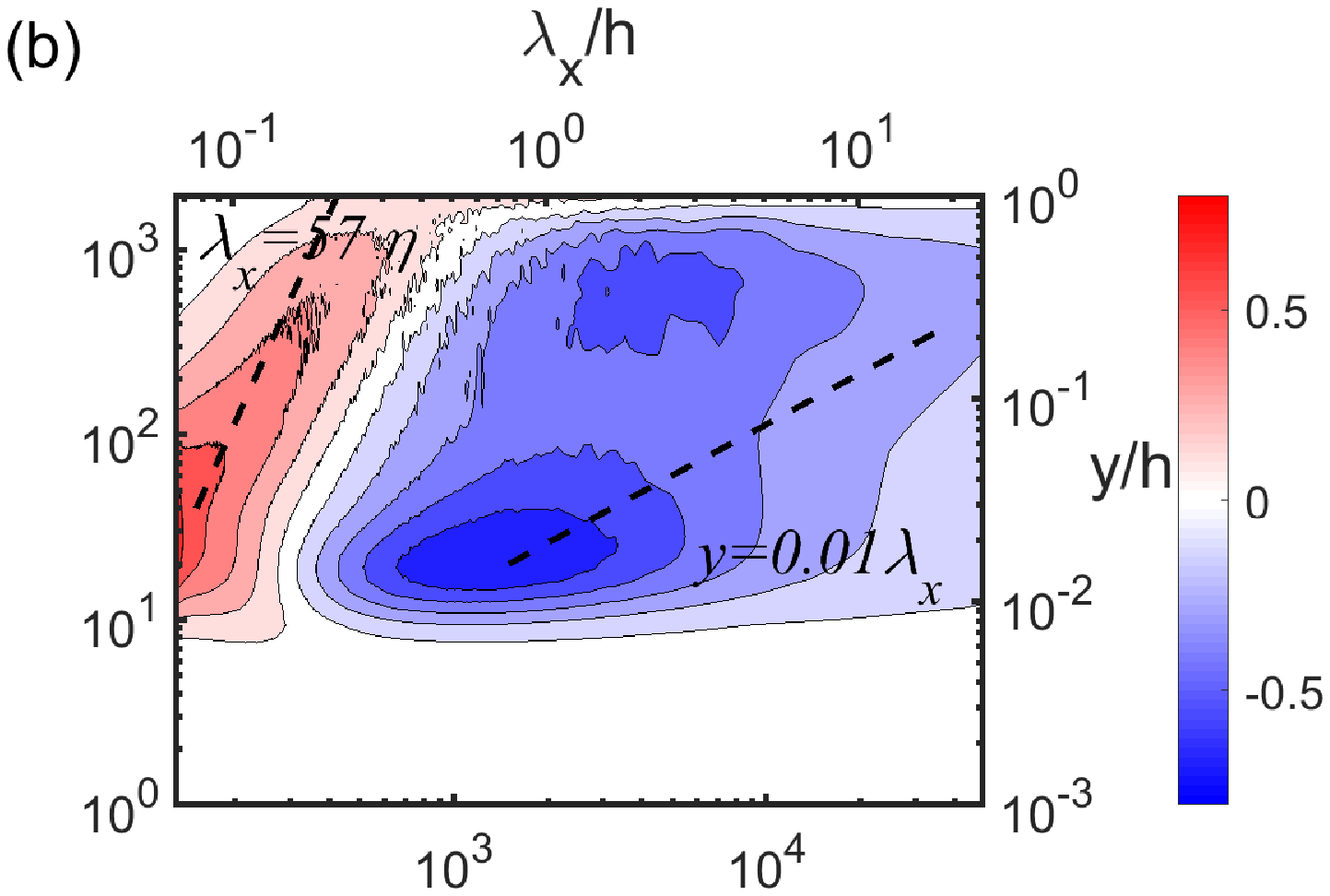}
\label{2}
\end{subfigure}
\begin{subfigure}[b]{0.42\textwidth}
  \includegraphics[width=\textwidth]{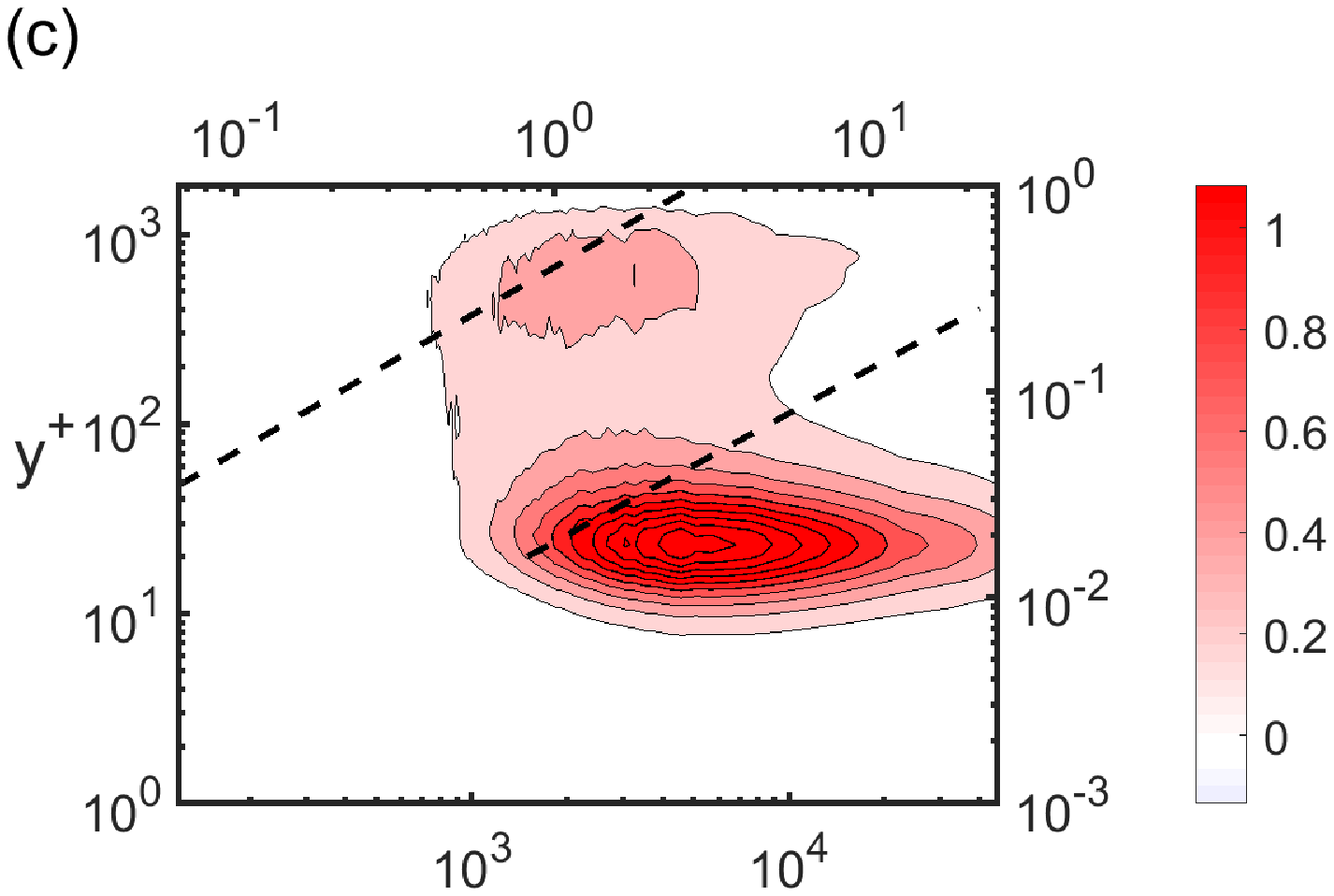}
\label{fig:prodx}
\end{subfigure}
\vspace{-0.7cm}
\begin{subfigure}[b]{0.42\textwidth}
  \includegraphics[width=\textwidth]{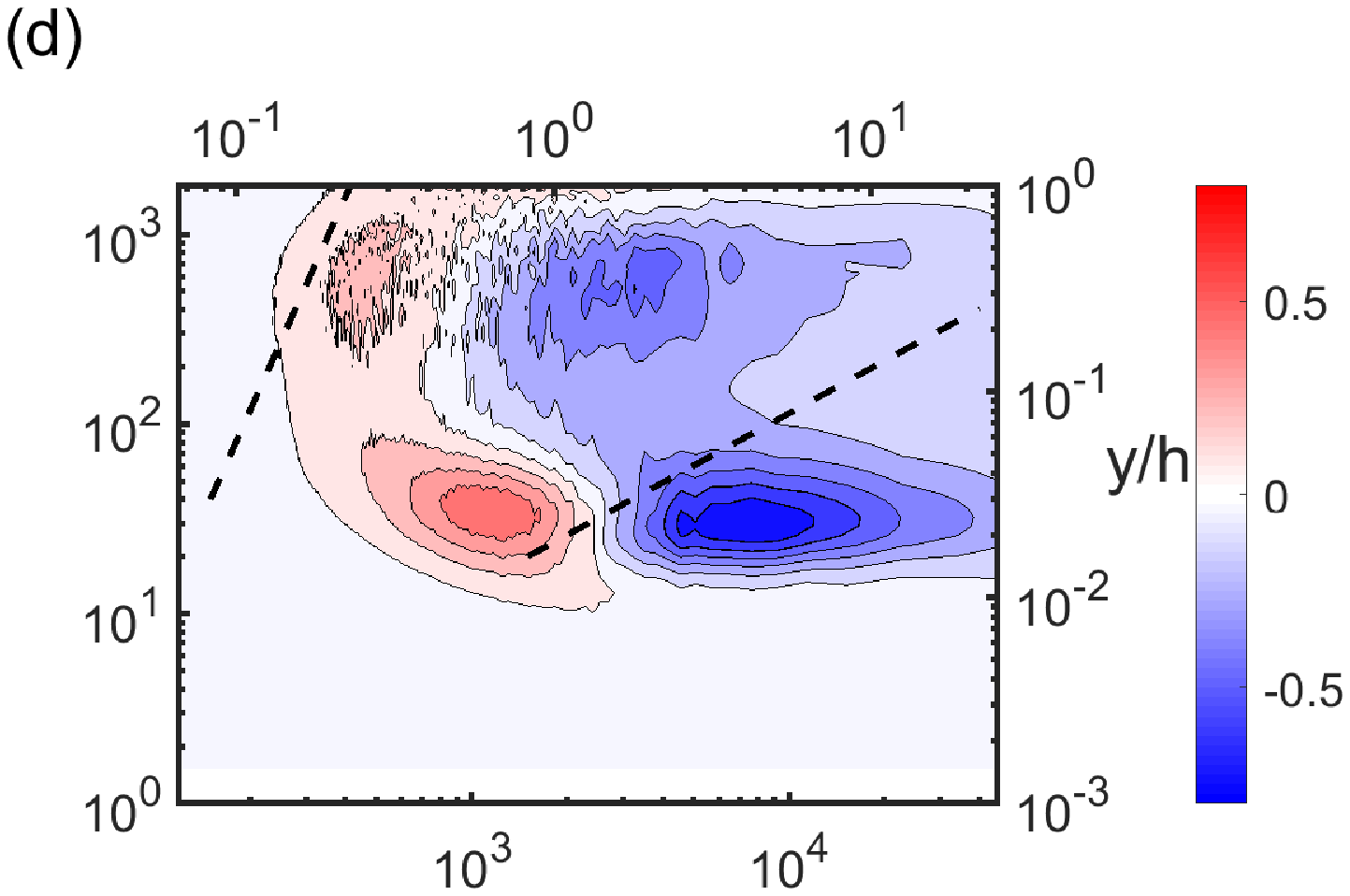}
  \label{4}
\end{subfigure}
\begin{subfigure}[b]{0.42\textwidth}
  \includegraphics[width=\textwidth]{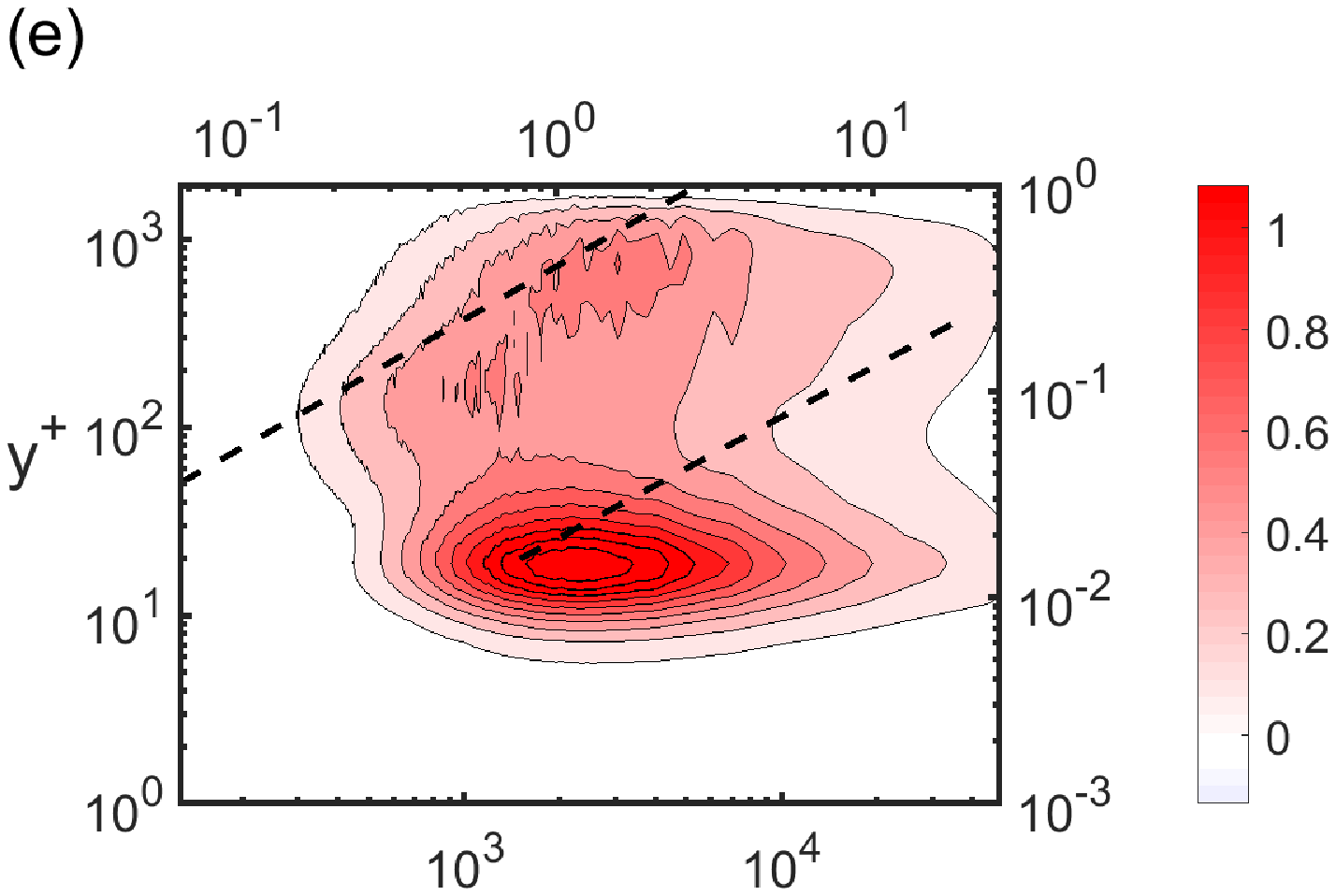}
  \label{5}
\end{subfigure}
\vspace{-0.7cm}
\begin{subfigure}[b]{0.42\textwidth}
  \includegraphics[width=\textwidth]{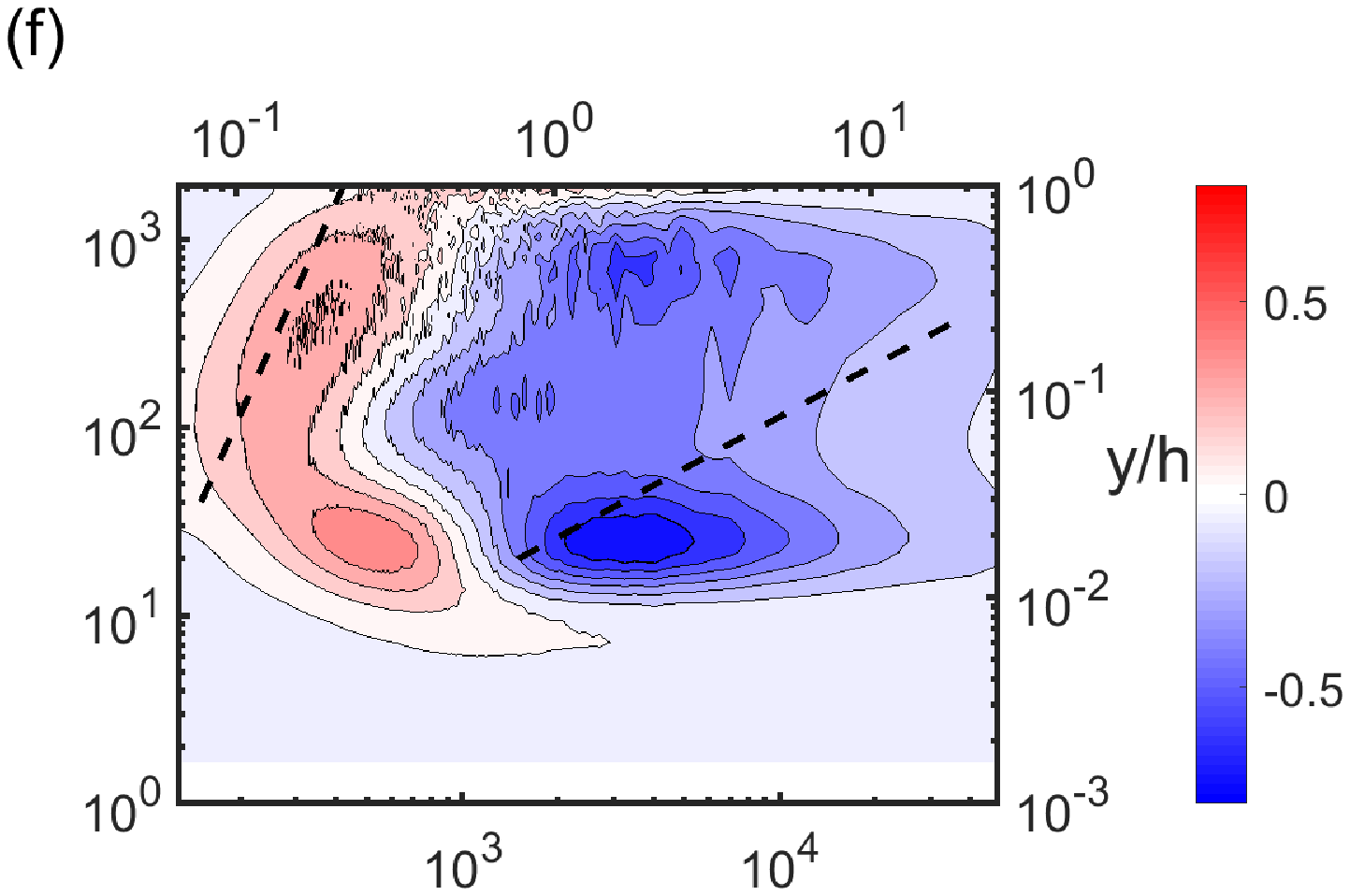}
  \label{6}
\end{subfigure}
\begin{subfigure}[b]{0.42\textwidth}
  \includegraphics[width=\textwidth]{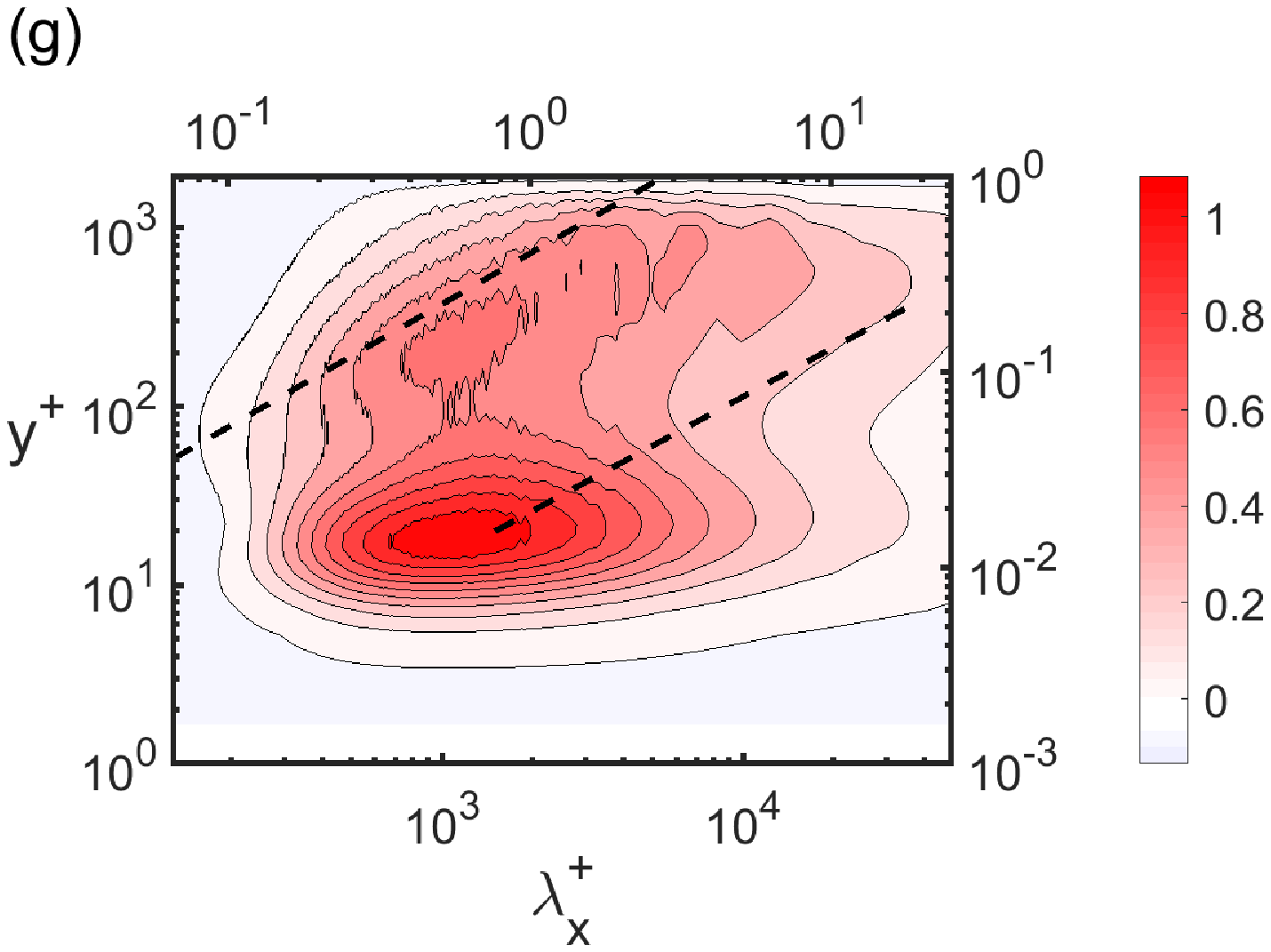}
\end{subfigure}
\begin{subfigure}[b]{0.42\textwidth}
  \includegraphics[width=\textwidth]{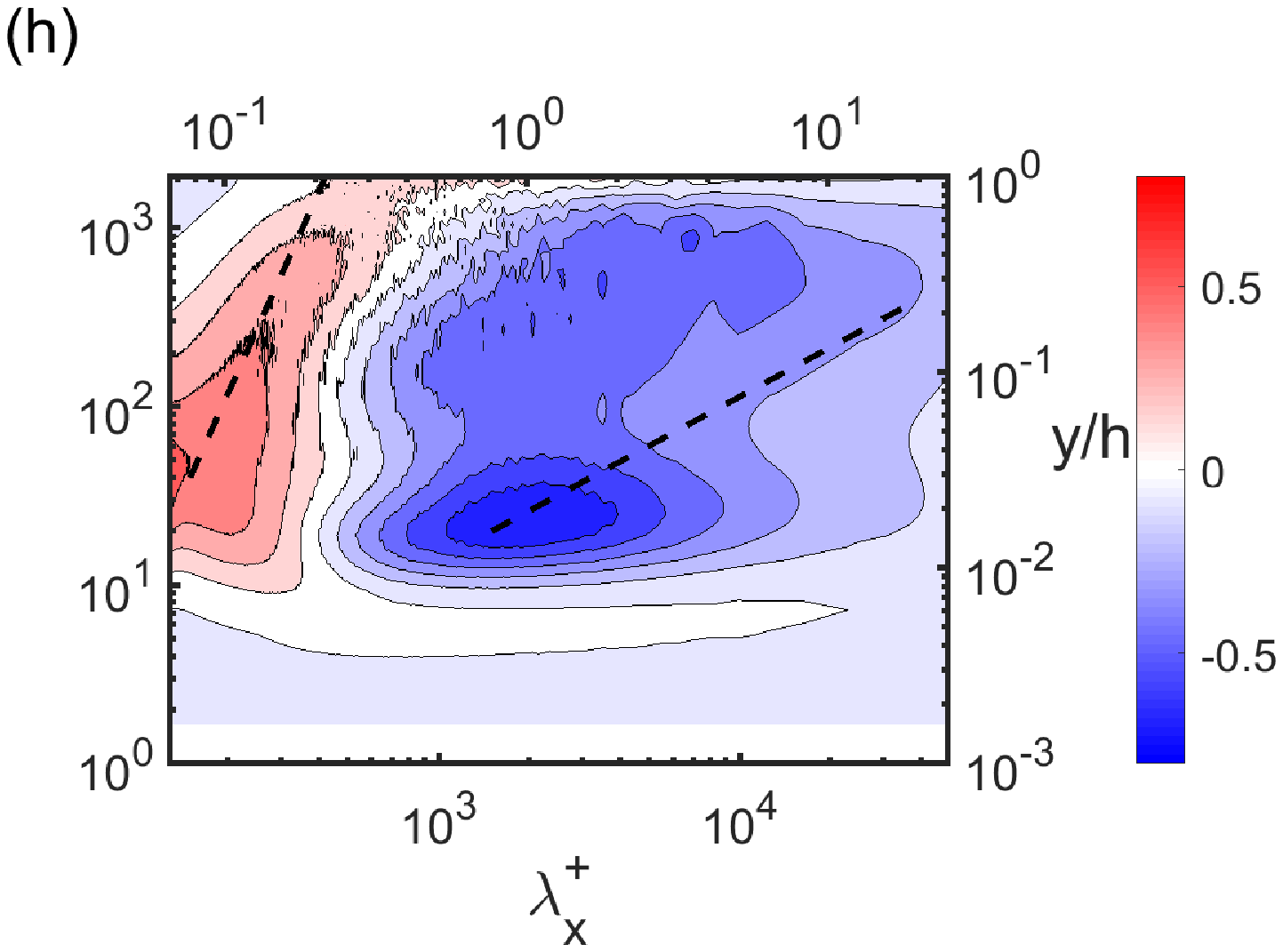}
\end{subfigure}
\end{minipage}
\caption{Premultiplied streamwise wavenumber spectra of production $k_x^+ y^+ \widehat{P}^+(y^+,\lambda_x^+)$ (left column) and turbulent transport $k_x^+ y^+ \widehat{T}_{turb}^+(y^+,\lambda_x^+)$ (right column) for (a,b) LES, (c,d) GQLZ1, (e,f) GQLZ5 and (g,h) GQLZ25 cases.}
\label{fig:xenergy}
\end{figure}

In this section, the spectral energetics of the LES and GQLZ cases are further considered. Given the present study is based an LES, the focus will be given on the production and turbulent spectra only. Note that the pressure and viscous transport spectra are negligibly small in the log and outer regions \citep{cho18,lee19}. The premultiplied one-dimensional spanwise wavenumber spectra of the production and turbulent transport of each case are plotted in figure \ref{fig:zenergy}. The turbulent production spectra in figure \ref{fig:zenergy}(a) appear to be almost uniformly distributed along the ridge $y \approx 0.2 \lambda_z$ especially in the log layer. Owing to the energy conservative nature of the nonlinear terms in the Navier-Stokes equations, the premultiplied turbulent transport spectra in figure \ref{fig:zenergy}(b) show positive and negative regions. The negative turbulent transport (blue contours in figure \ref{fig:zenergy}b) is almost balanced with the (positive) production in the logarithmic and outer layers, and the positive turbulent transport appears along a ridge indicating the Kolmogorov scale (i.e. $\lambda_{z} \approx 57 \eta$, where $\eta$ is the Kolmogorov scale). There can also be observed a region of weakly positive turbulent transport spectra in the region very close to the wall ($y^+<10$) over a wide range of the spanwise wavelength scales ($200 \delta_\nu \lesssim \lambda_z \lesssim 0.8 h$), primarily due to an inverse energy transfer from small to large scales \cite[]{cho18,doohan21}.

The GQLZ1 case shows spectra with increased energy and a production peak centred in $\lambda_z^+ \approx 400$ (figures \ref{fig:zenergy}c,d). The production spectra do not extend for $\lambda_z^+ \lesssim 200$ and they are damaged at large scales ($\lambda_z/h \gtrsim 1$), except in the outer region (figure \ref{fig:zenergy}c). Moreover, similarly to the velocity spectra (figure \ref{fig:zspectra}), the linear scaling of $\lambda_z^+$ with $y^+$ is found to have been weakened. The turbulent transport spectra show positive and negative regions which have been reduced over the entire spanwise wavelengths compared to the reference LES (figure \ref{fig:zenergy}d). Also, the weakly positive spectra in the region close to the wall is confined to the largest spanwise scales $\lambda_z^+ \approx 5000$. This is somehow expected, because the linearised part of the GQLZ1 model would not admit full energy cascade mechanism due to the absence of the self-interacting nonlinear terms in the $\mathcal{P}_h$ subspace -- the spanwise energy cascade should only be admitted through the interaction between $\mathbf{u}_l$ and $\mathbf{u}_h$ \cite[i.e. the `scattering' mechanism proposed by][]{tobias17}. 
By enriching the $\mathcal{P}_l$-subspace group with more spanwise Fourier modes, the resulting spectra start to recover the original range of spanwise wavelengths, the scaling with $y$ and the original position of the peak in the GQLZ5 and GQLZ25 cases. The recovery of the spectra to those of the full LES, however, appears to be in a little delicate manner, and this is seen in the turbulent transport spectra of GQLZ25 (figure \ref{fig:zenergy}h). We note that turbulent transport spectra for $\lambda_{z}<\lambda_{z,c}$ (i.e. the $\mathcal{P}_h$ subspace group) in the outer region (say $y^+\gtrsim 300$) are almost zero, while showing a sharp discontinuity across $\lambda_{z}=\lambda_{z,c}$. This suggests that, in the outer region, the energy cascade from the $\mathcal{P}_l$ to the $\mathcal{P}_h$ subspace does not exist in the GQL25 model. This issue will be discussed further in \S\ref{sec:434}. 

The premultiplied streamwise wavenumber spectra of the energy budget are shown in figure \ref{fig:xenergy}. For the full LES, the production spectra and the negative part of the turbulent transport spectra appear to be in the region bounded by two linear ridges, along which the wall-normal and streamwise velocity spectra are highly energetic (i.e. $y =0.35 \lambda_x$ and $y =0.01 \lambda_x$ in figure \ref{fig:xspectra}). We note that, in the logarithmic region, the premultiplied production spectra, $-k_x^+ y^+ \Phi_{uv}^+(k_x,y)dU^+/dy^+$, are similar to those of Reynolds shear-stress co-spectra, $-k_x^+ \Phi_{uv}^+(k_x,y)$, due to $dU^+/dy^+ \sim 1/y^+$ \cite[]{cho18}. However, the linear scaling of the production spectra with respect to $y$ is not very clear, and this is presumably due to the relatively low Reynolds numbers in the present study (for high Reynolds numbers, e.g. $Re_\tau \approx 5200$, see \citealp{lee19} and \citealp{hwang_lee_2020}, where the linear scaling is clearly visible). 
The turbulent transport spectra in figure \ref{fig:xenergy}(b) show a region of positive values along the ridge corresponding to viscous dissipation ($\lambda_x=57 \eta$) and a region of negative values corresponding to the streamwise production intensity peak in the near-wall region ($\lambda_x^+ \approx 1000$). 

The spectra of GQLZ1 show some resemblance to the LES case but still a very damaged energy cascade compared to that (figure \ref{fig:xenergy}c). In particular, the production spectra exhibit an intense peak highly localised around $\lambda_x^+ \simeq 5000$ and $y^+ \simeq 25$, while a secondary peak is also observed in the outer region. The production spectra show no energy intensity for wavelengths below $\lambda_x^+ = 700$ (figure \ref{fig:xenergy}c). The positive region of the transport spectra appear at approximately the same locations for larger streamwise length scales compared to LES (peaks at $\lambda_x^+ \approx 1000$ and $\lambda_x^+ \approx 500$, respectively), while the negative region is displaced to higher streamwise wavelengths and reduced to the $y^+ \approx 30$ and $y^+ \approx 500$ locations. 
As more spanwise modes are incorporated in the $\mathcal{P}_l$-subspace group, the scalings of both spectra with $y$ and $\eta$ are greatly recovered, i.e. GQLZ5 and GQLZ25 cases (figures \ref{fig:xenergy}e-h). 

\subsection{Componentwise energy transport and pressure strain}

\begin{figure}
\begin{minipage}{\textwidth}
\centering
\begin{subfigure}[b]{0.42\textwidth}
  \includegraphics[width=\textwidth]{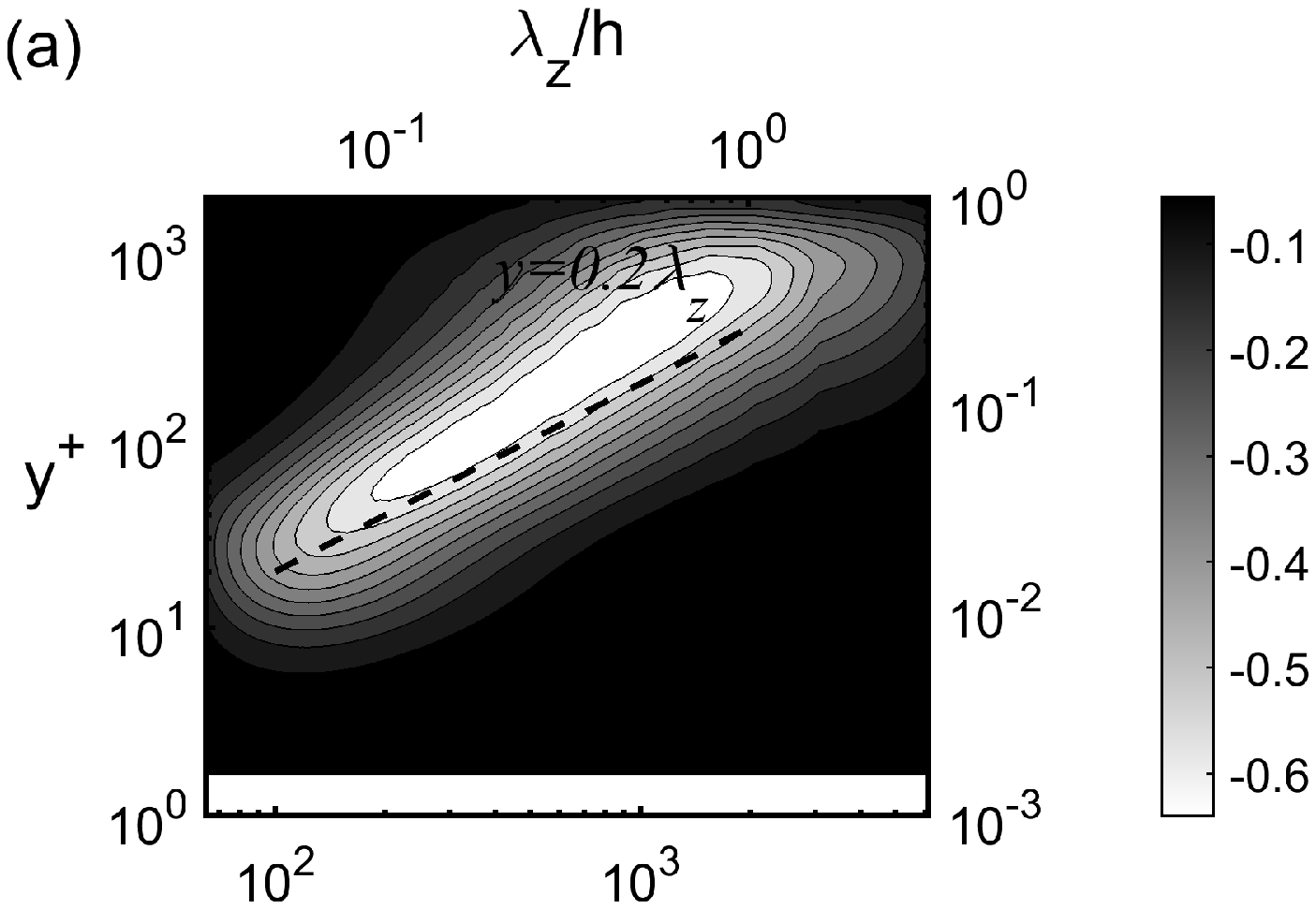}
\label{1}
\end{subfigure}
\vspace{-0.7cm}
\begin{subfigure}[b]{0.42\textwidth}
  \includegraphics[width=\textwidth]{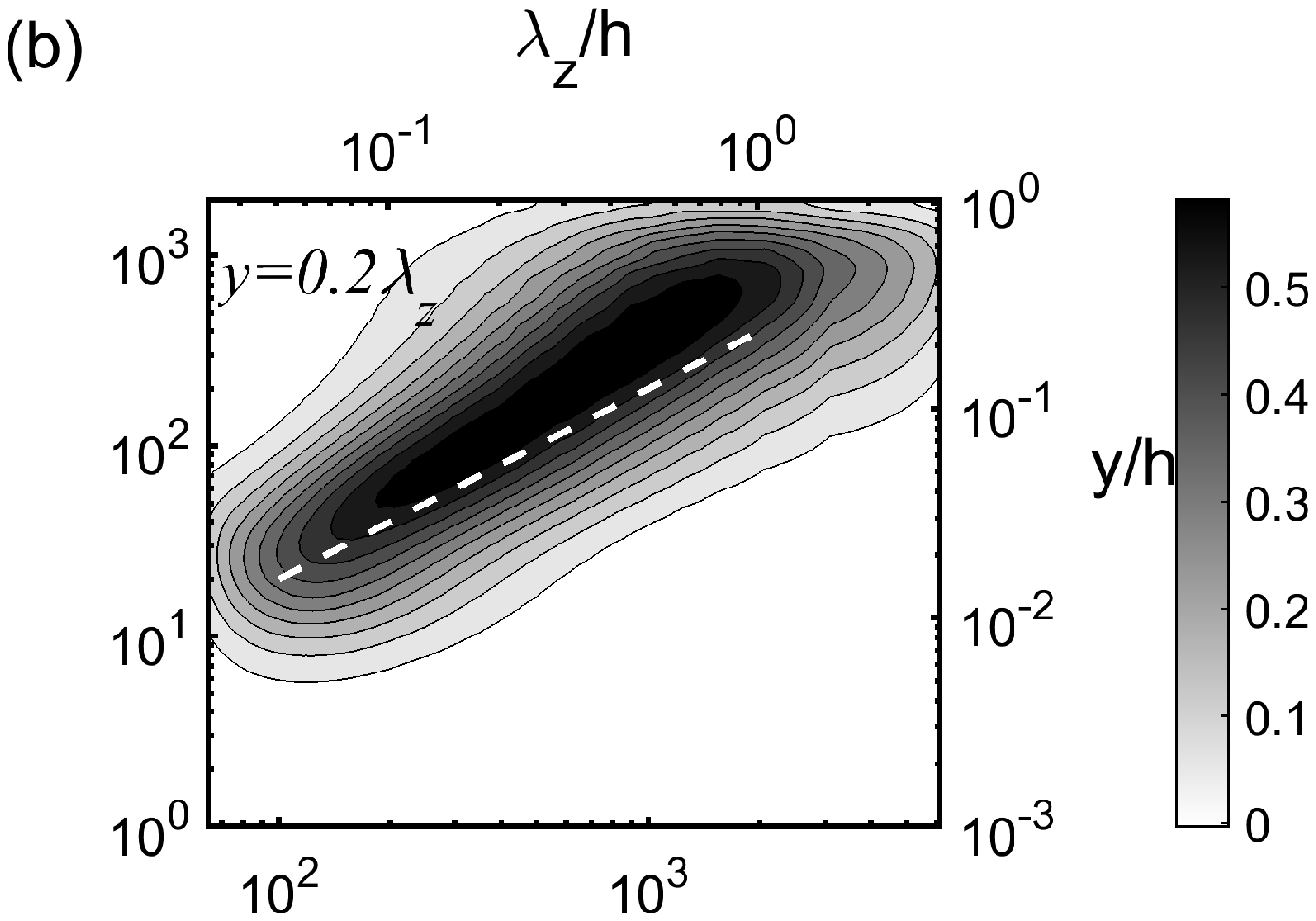}
\label{2}
\end{subfigure}
\begin{subfigure}[b]{0.42\textwidth}
  \includegraphics[width=\textwidth]{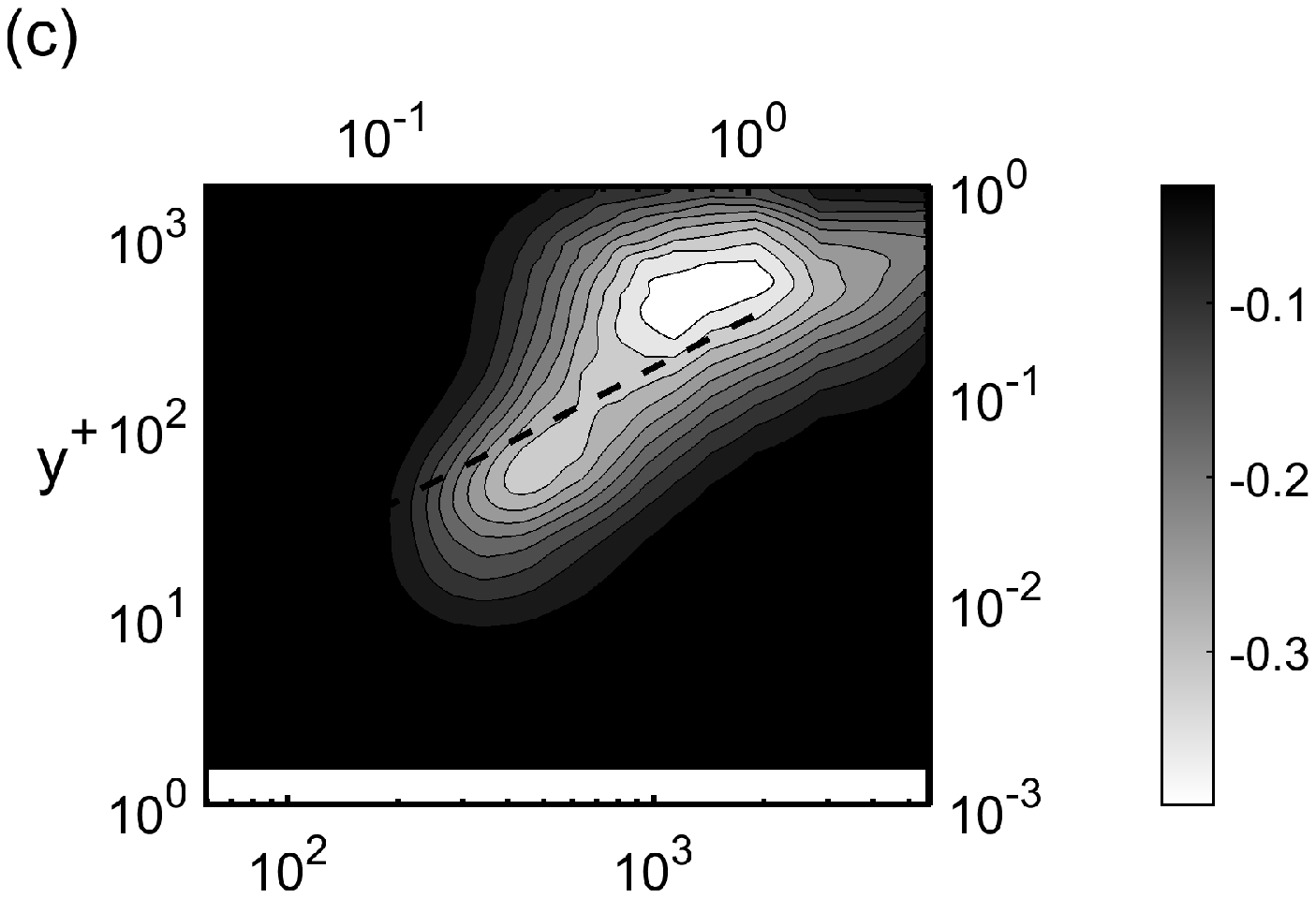}
  \label{3}
\end{subfigure}
\vspace{-0.7cm}
\begin{subfigure}[b]{0.42\textwidth}
  \includegraphics[width=\textwidth]{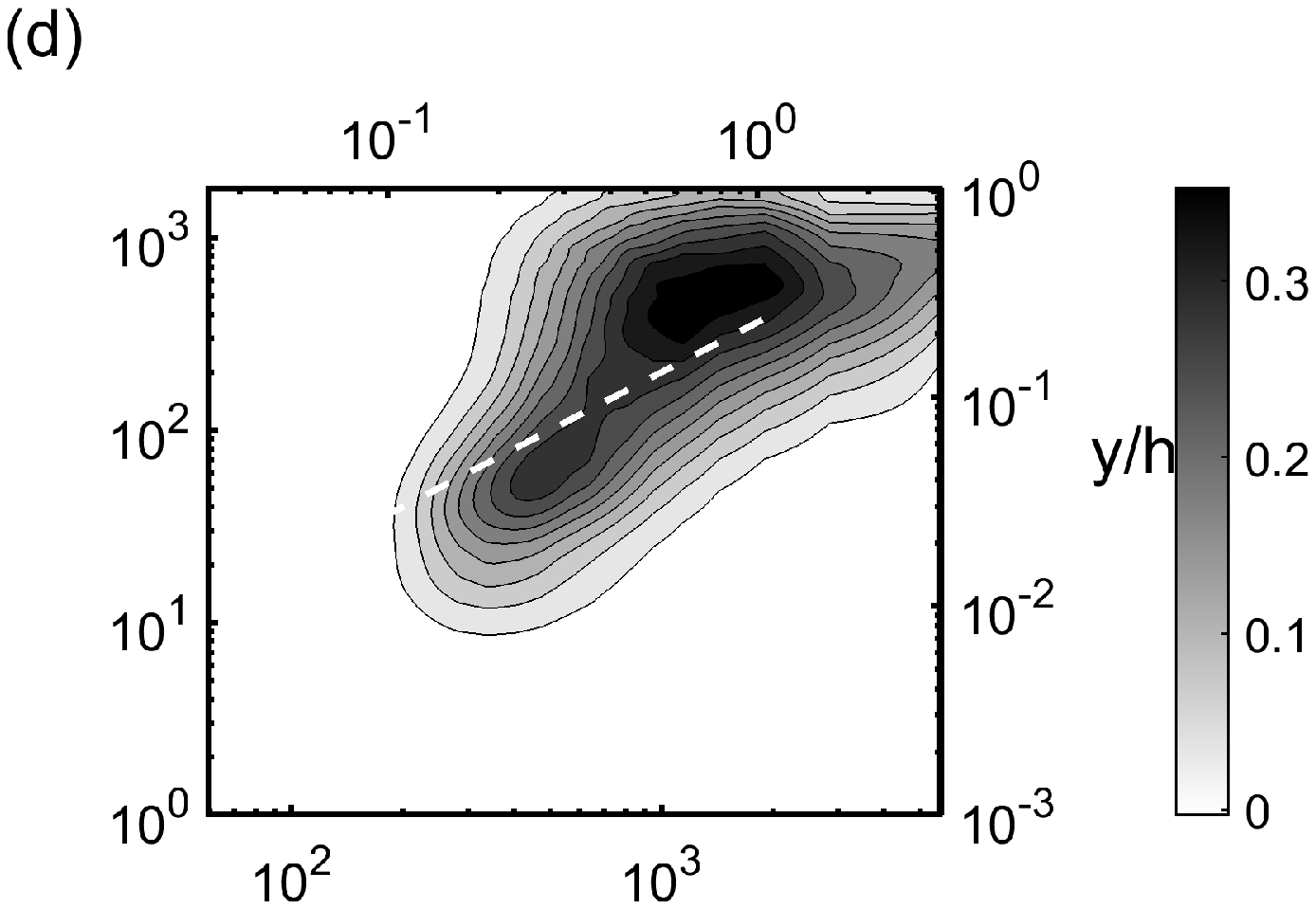}
  \label{4}
\end{subfigure}
\begin{subfigure}[b]{0.42\textwidth}
  \includegraphics[width=\textwidth]{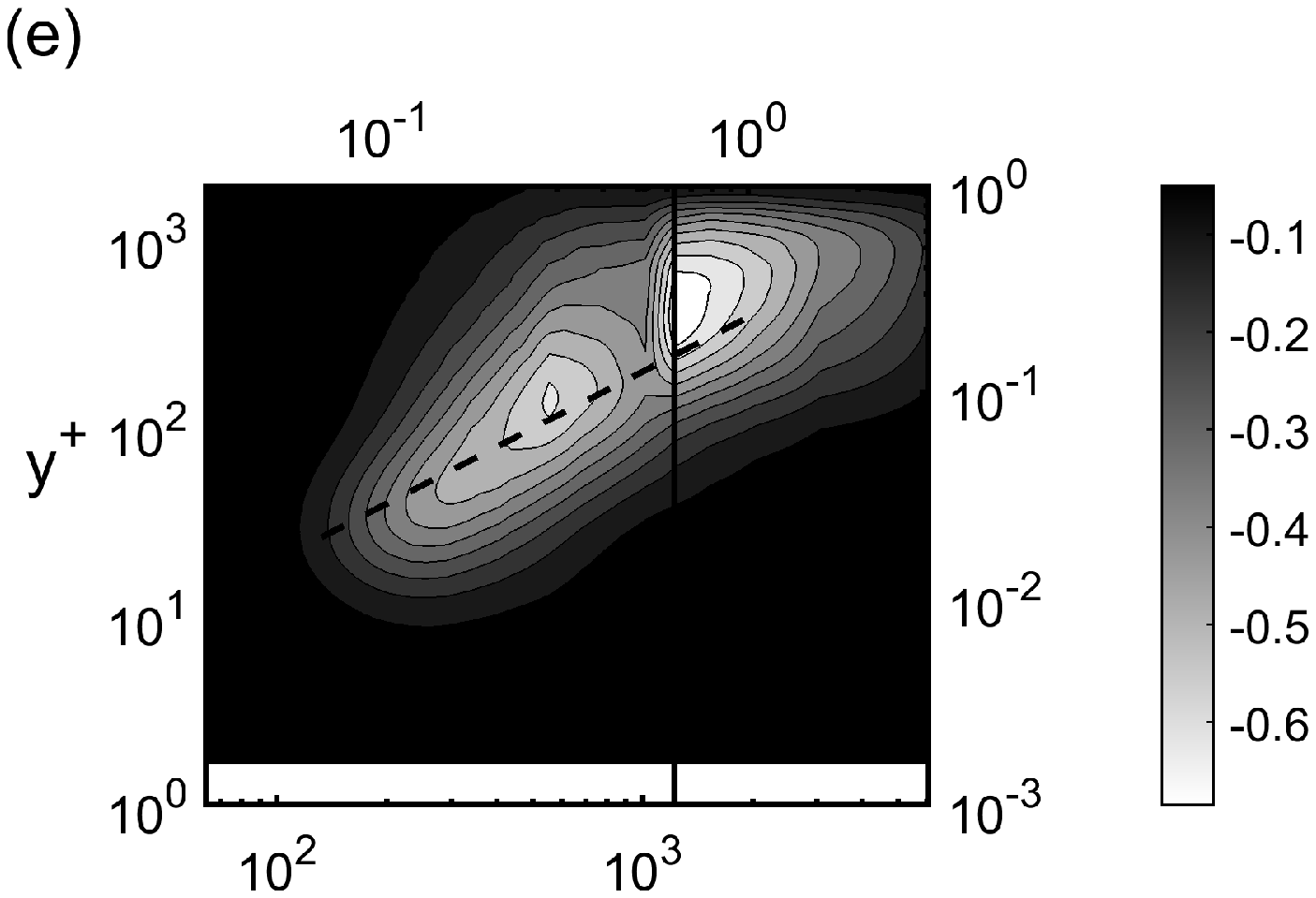}
  \label{5}
\end{subfigure}
\vspace{-0.7cm}
\begin{subfigure}[b]{0.42\textwidth}
  \includegraphics[width=\textwidth]{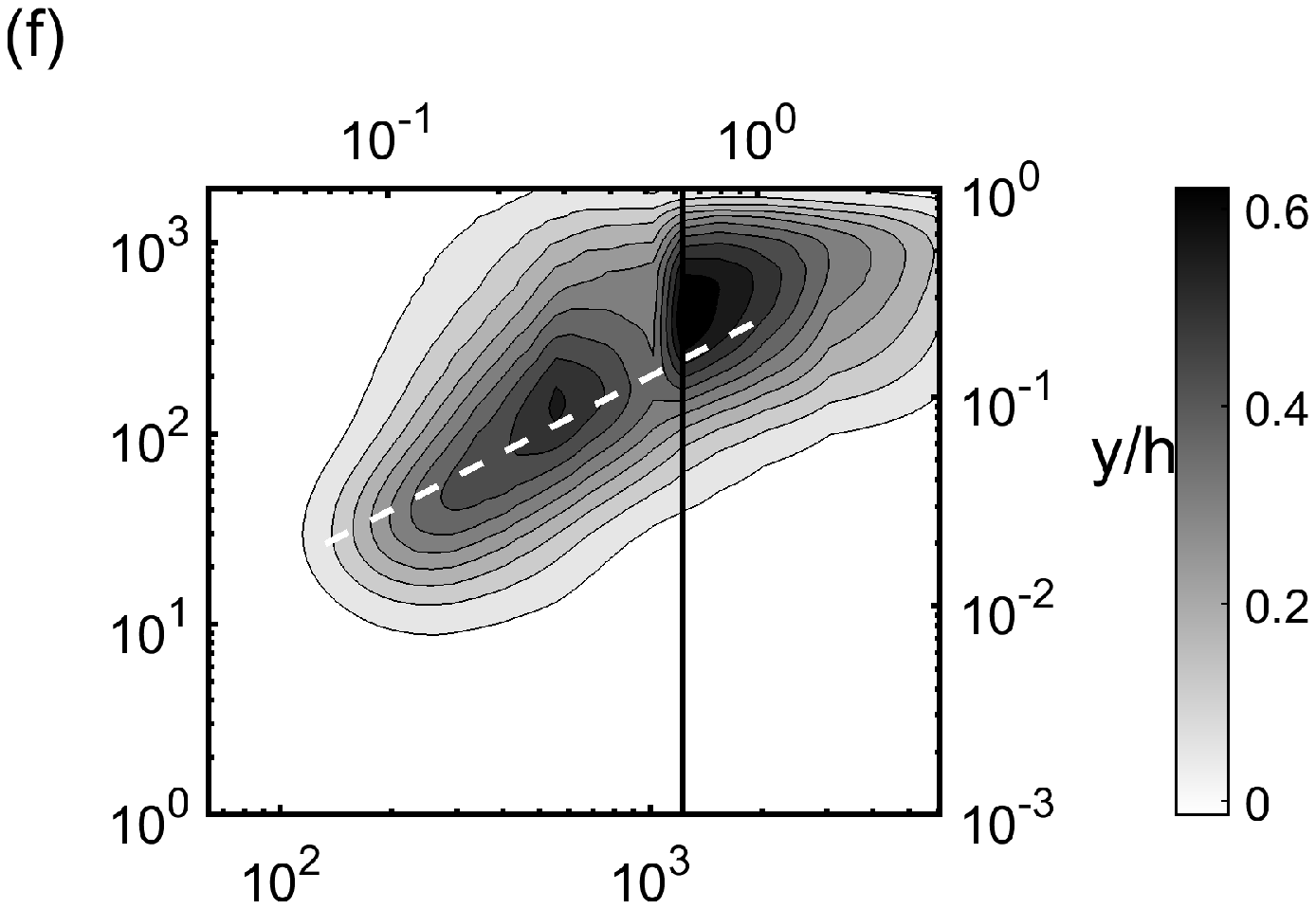}
  \label{6}
\end{subfigure}
\begin{subfigure}[b]{0.42\textwidth}
  \includegraphics[width=\textwidth]{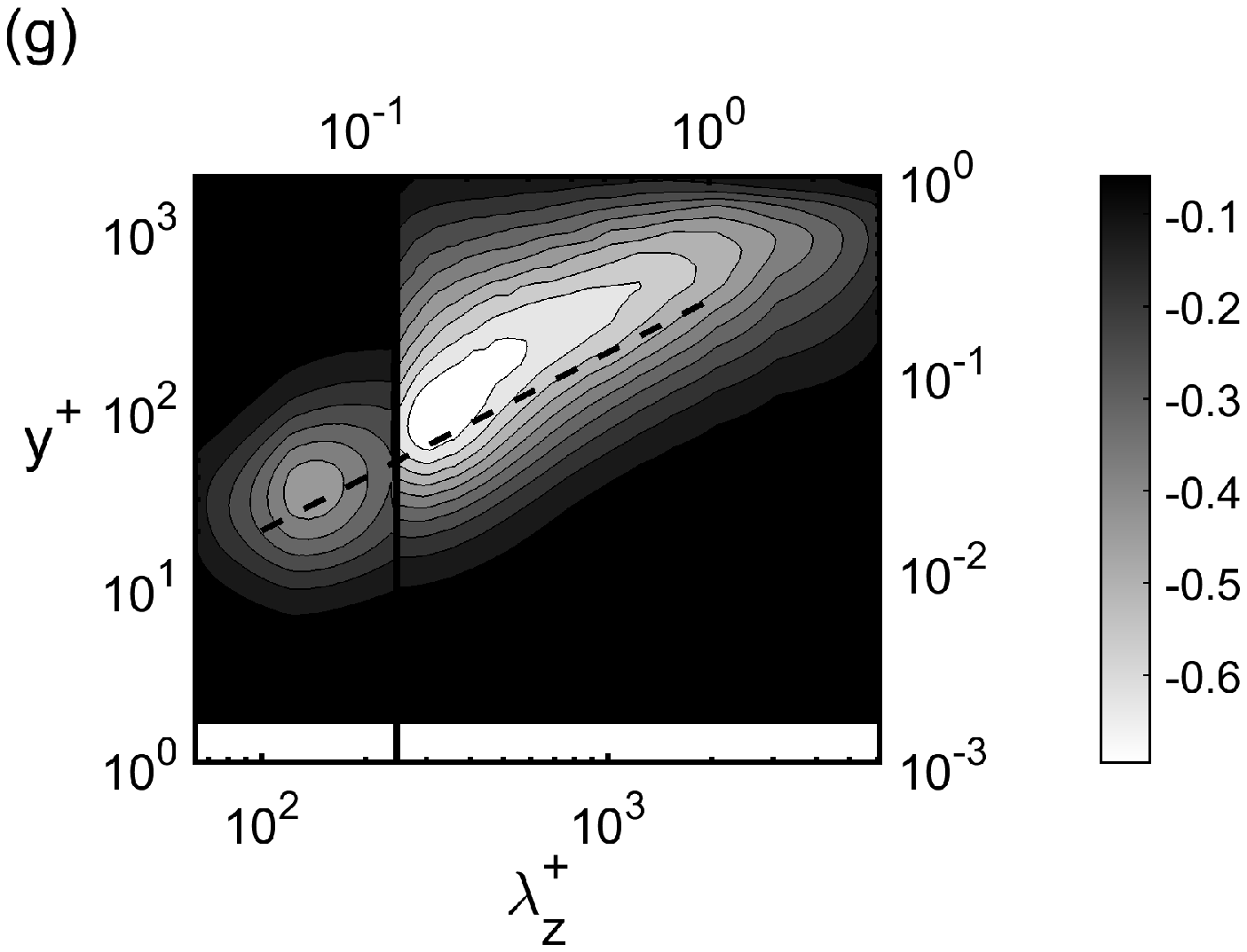}
\end{subfigure}
\begin{subfigure}[b]{0.42\textwidth}
  \includegraphics[width=\textwidth]{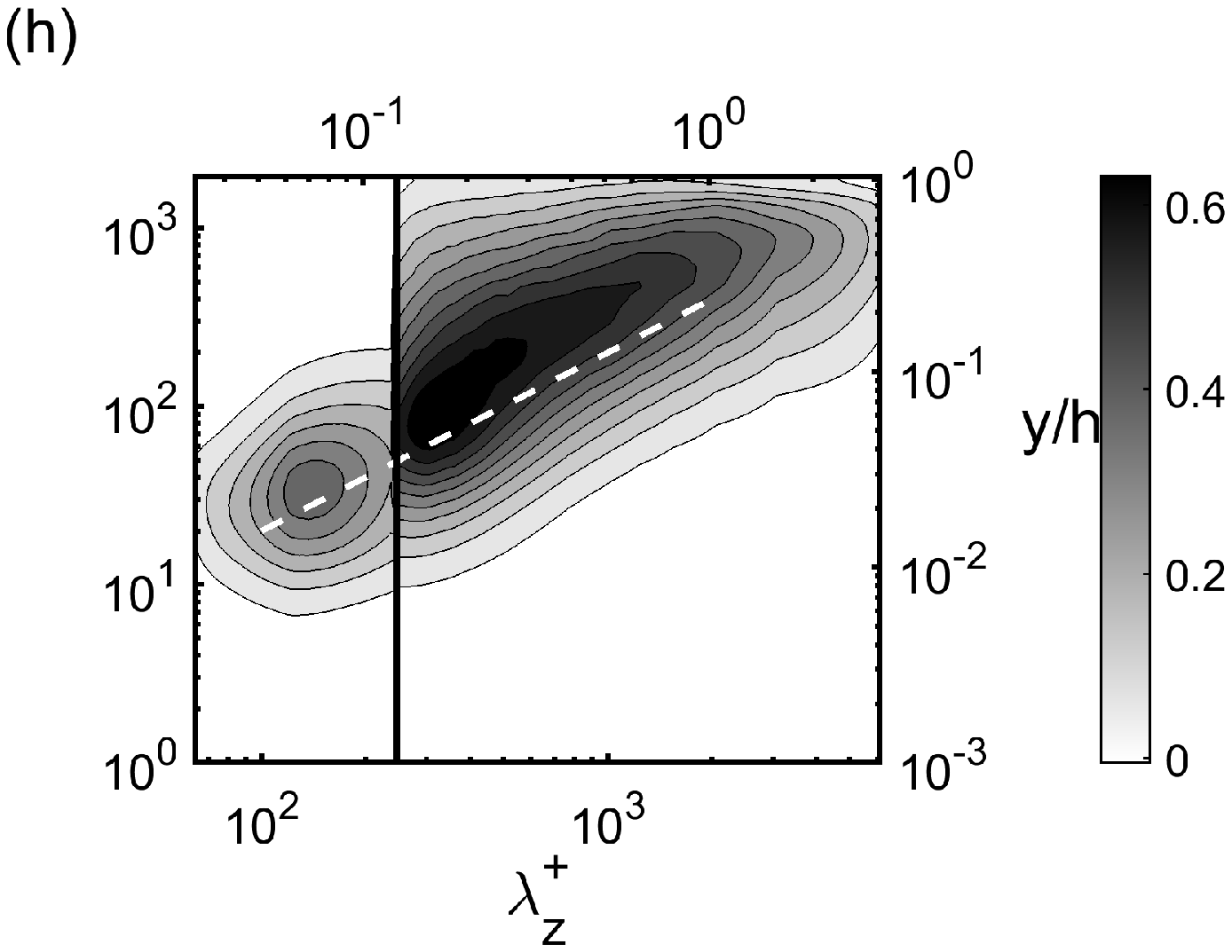}
\end{subfigure}
\end{minipage}
\caption{Premultiplied spanwise wavenumber spectra of $k_z^+ y^+ \widehat{\Pi}_x^+(y^+,\lambda_z^+)$ (left column) and $k_z^+ y^+ \widehat{\Pi}_{yz}^+(y^+,\lambda_z^+)$ (right column) for (a,b) LES, (c,d) GQLZ1, (e,f) GQLZ5 and (g,h) GQLZ25. Here, the vertical line represents the spanwise cut-off wavelength ($\lambda_{z,c}$) dividing the $\mathcal{P}_h$- (left) and $\mathcal{P}_l$-subspace (right) regions. The dashed lines indicate the ridge $y=0.5 \lambda_z$.}
\label{fig:zpi}
\end{figure}

\begin{figure}
\begin{minipage}{\textwidth}
\centering
\begin{subfigure}[b]{0.42\textwidth}
  \includegraphics[width=\textwidth]{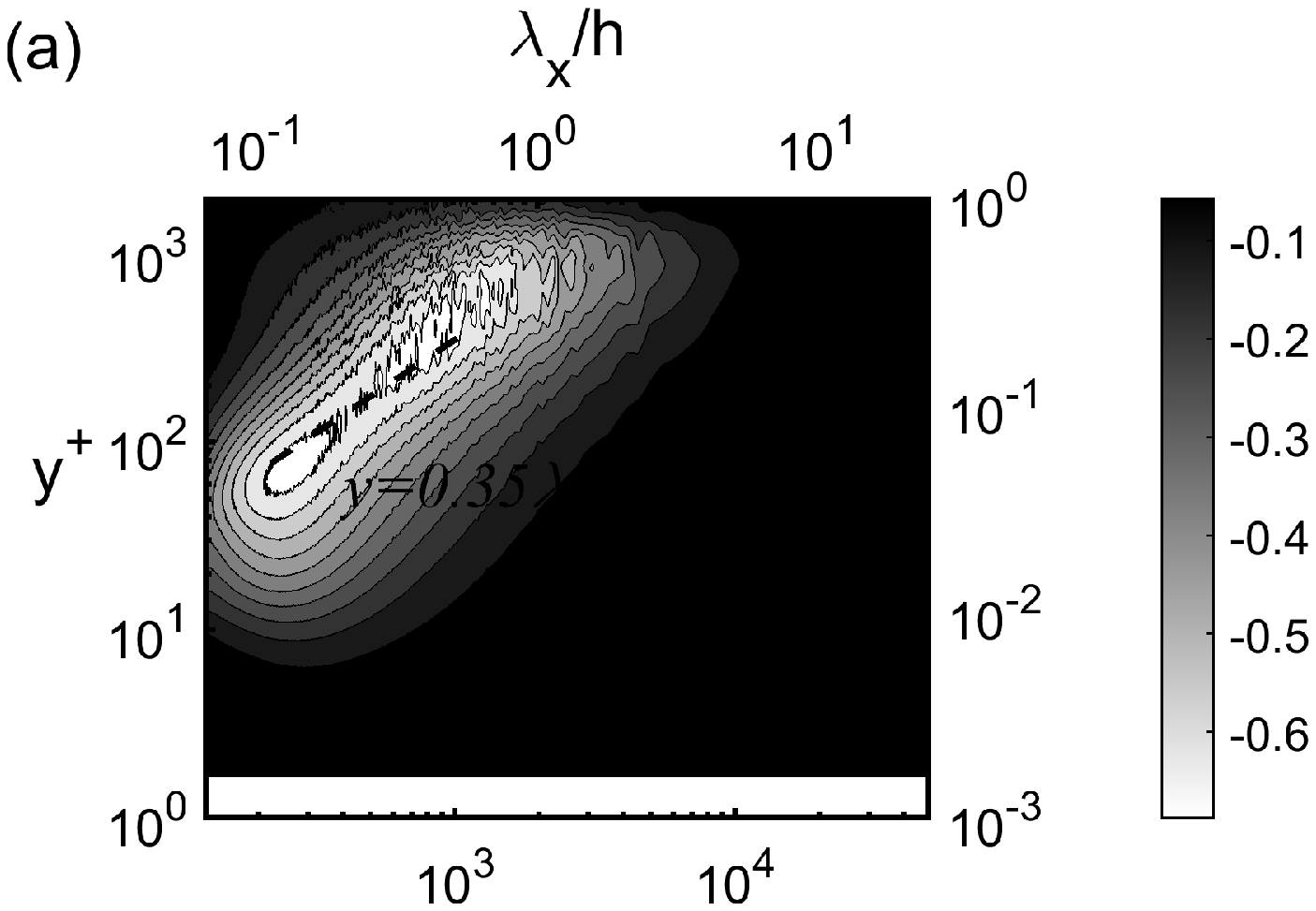}
\label{1}
\end{subfigure}
\vspace{-0.7cm}
\begin{subfigure}[b]{0.42\textwidth}
  \includegraphics[width=\textwidth]{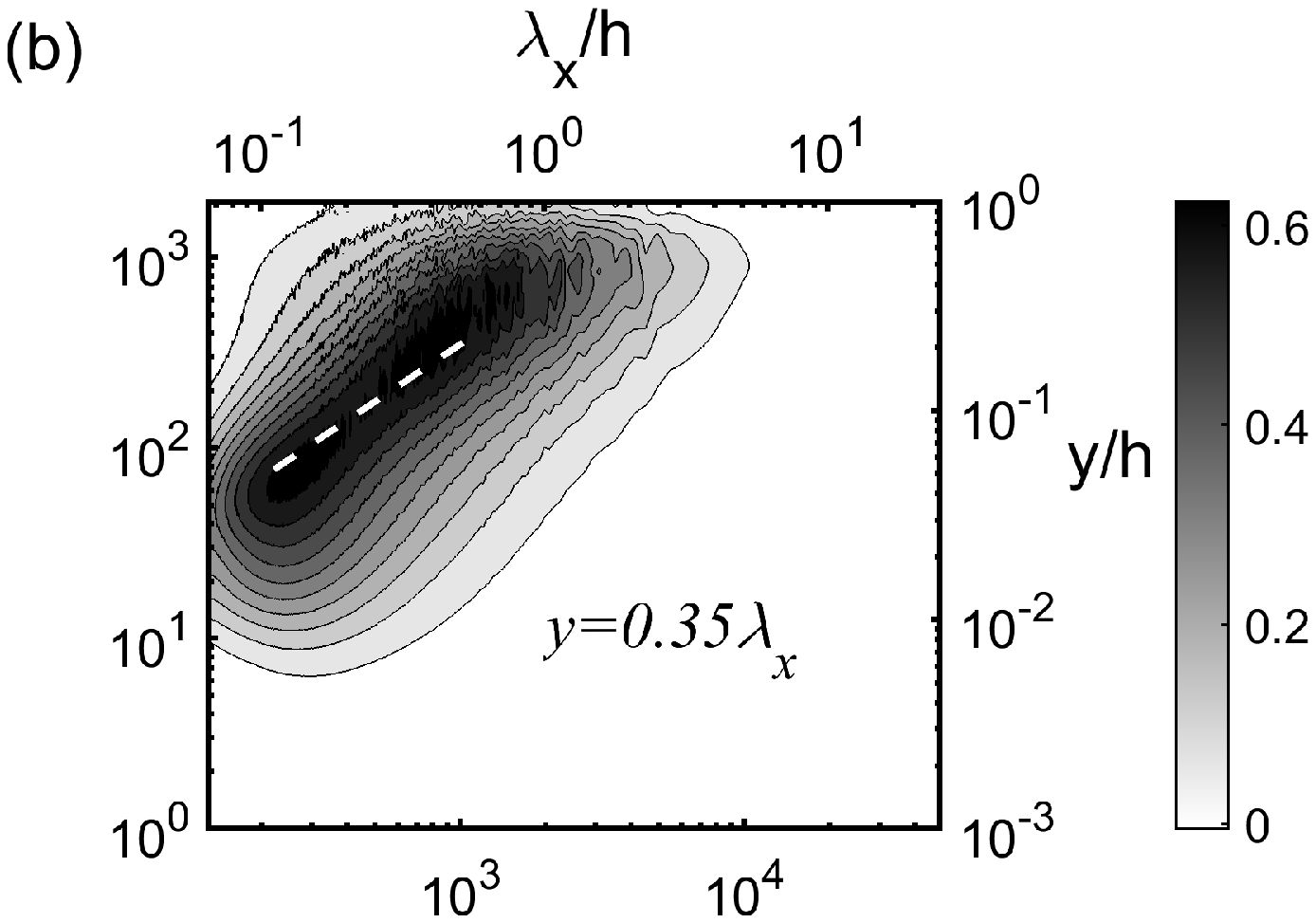}
\label{2}
\end{subfigure}
\begin{subfigure}[b]{0.42\textwidth}
  \includegraphics[width=\textwidth]{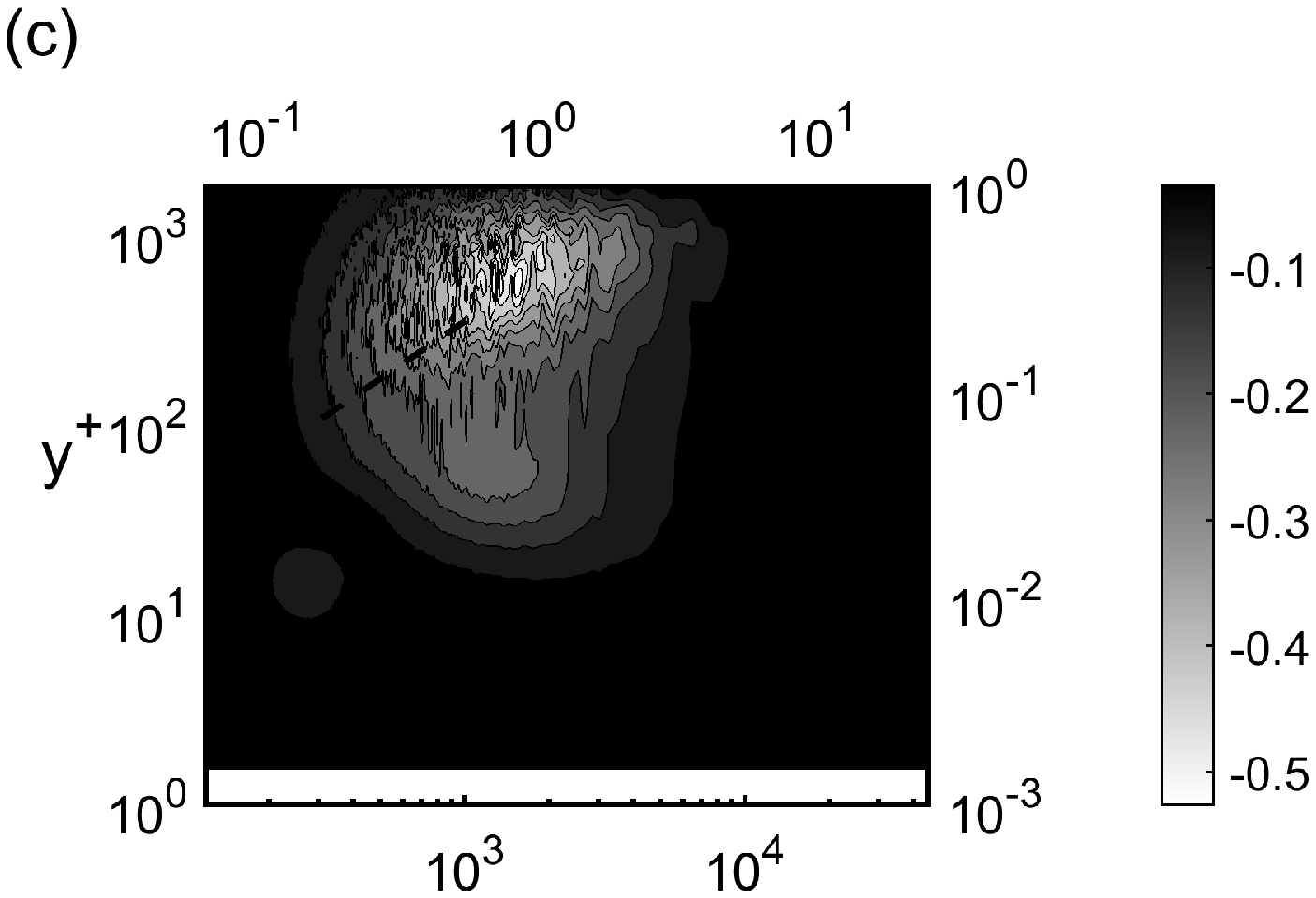}
\label{fig:prodx}
\end{subfigure}
\vspace{-0.7cm}
\begin{subfigure}[b]{0.42\textwidth}
  \includegraphics[width=\textwidth]{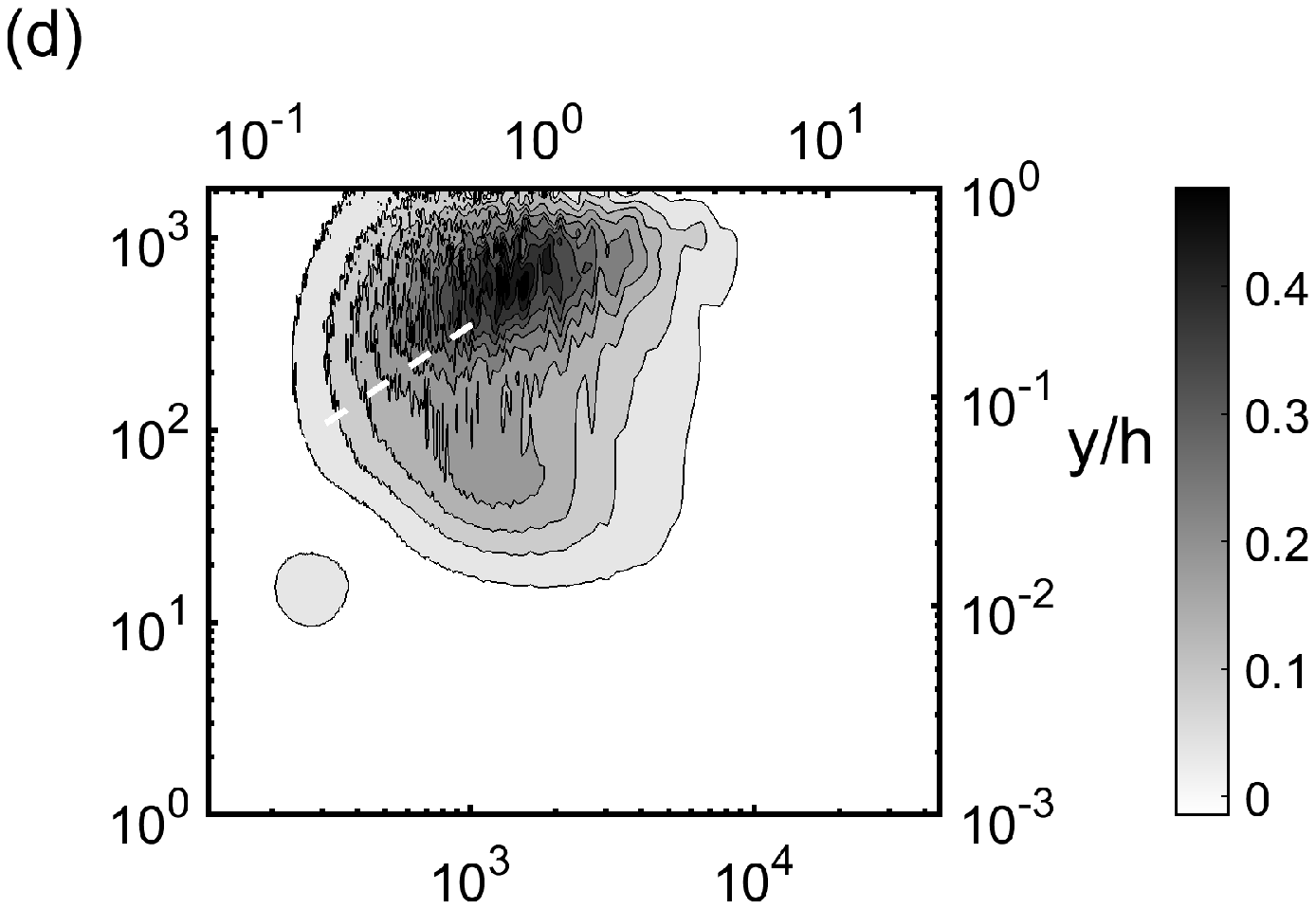}
  \label{4}
\end{subfigure}
\begin{subfigure}[b]{0.42\textwidth}
  \includegraphics[width=\textwidth]{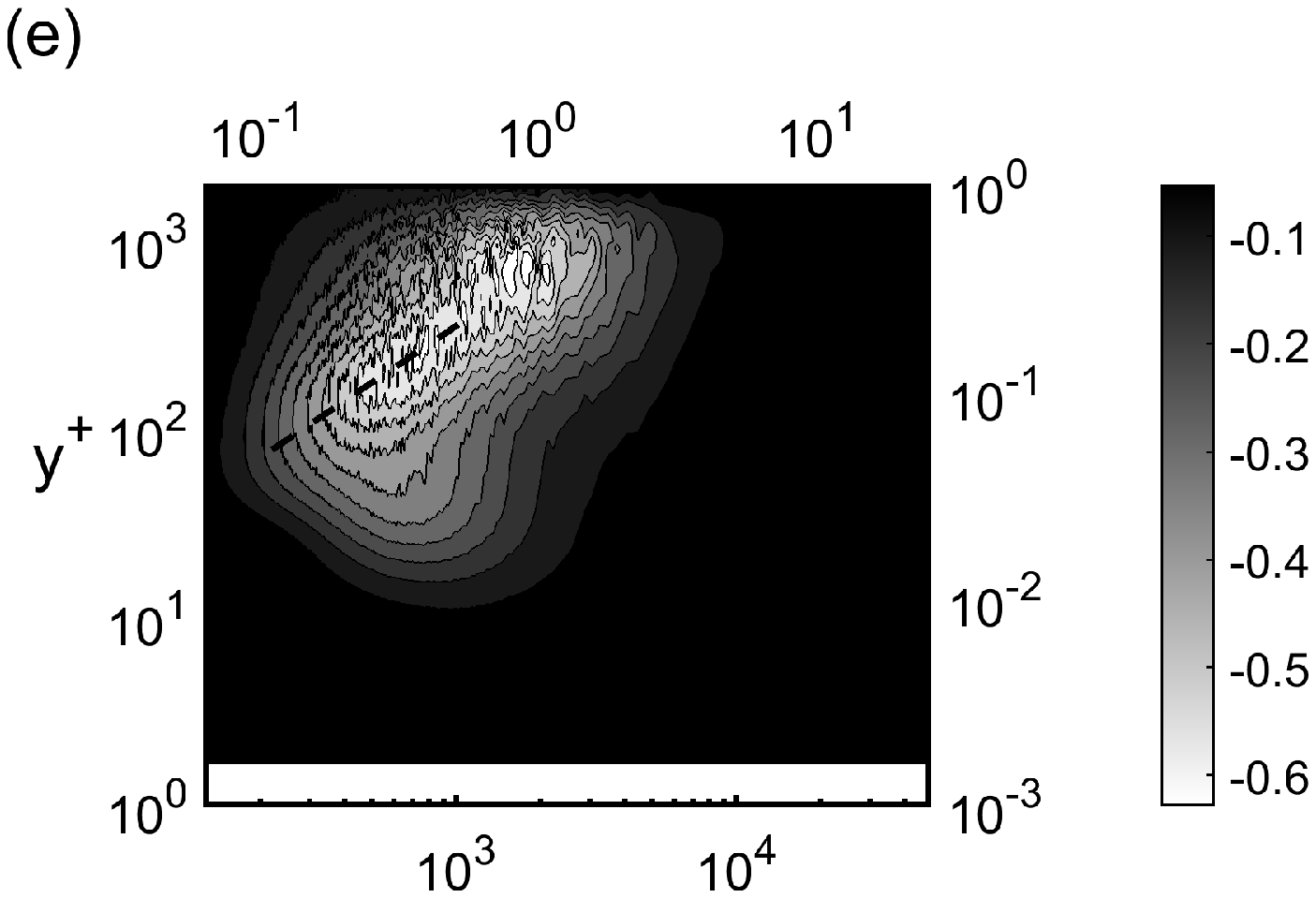}
  \label{5}
\end{subfigure}
\vspace{-0.7cm}
\begin{subfigure}[b]{0.42\textwidth}
  \includegraphics[width=\textwidth]{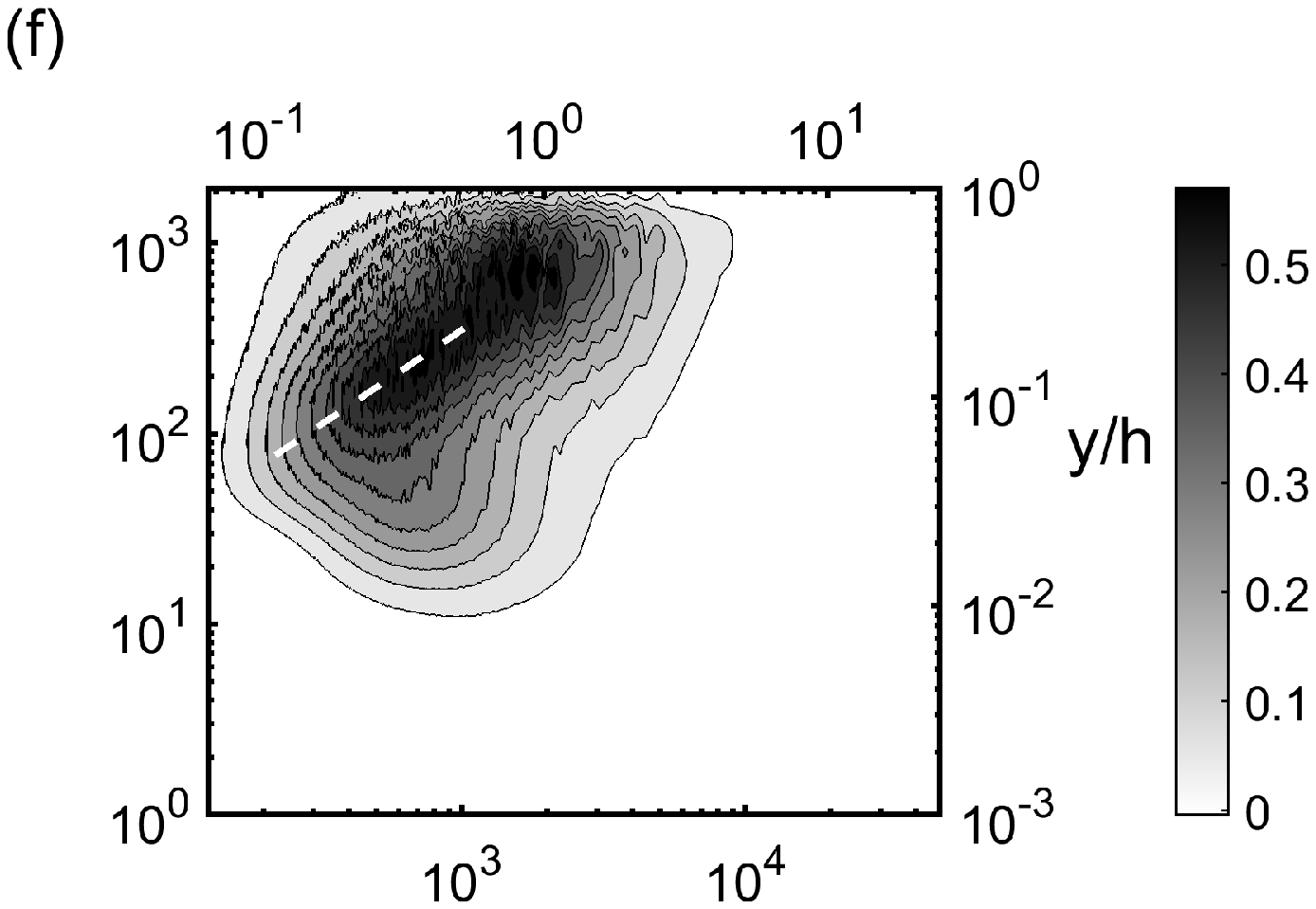}
  \label{6}
\end{subfigure}
\begin{subfigure}[b]{0.42\textwidth}
  \includegraphics[width=\textwidth]{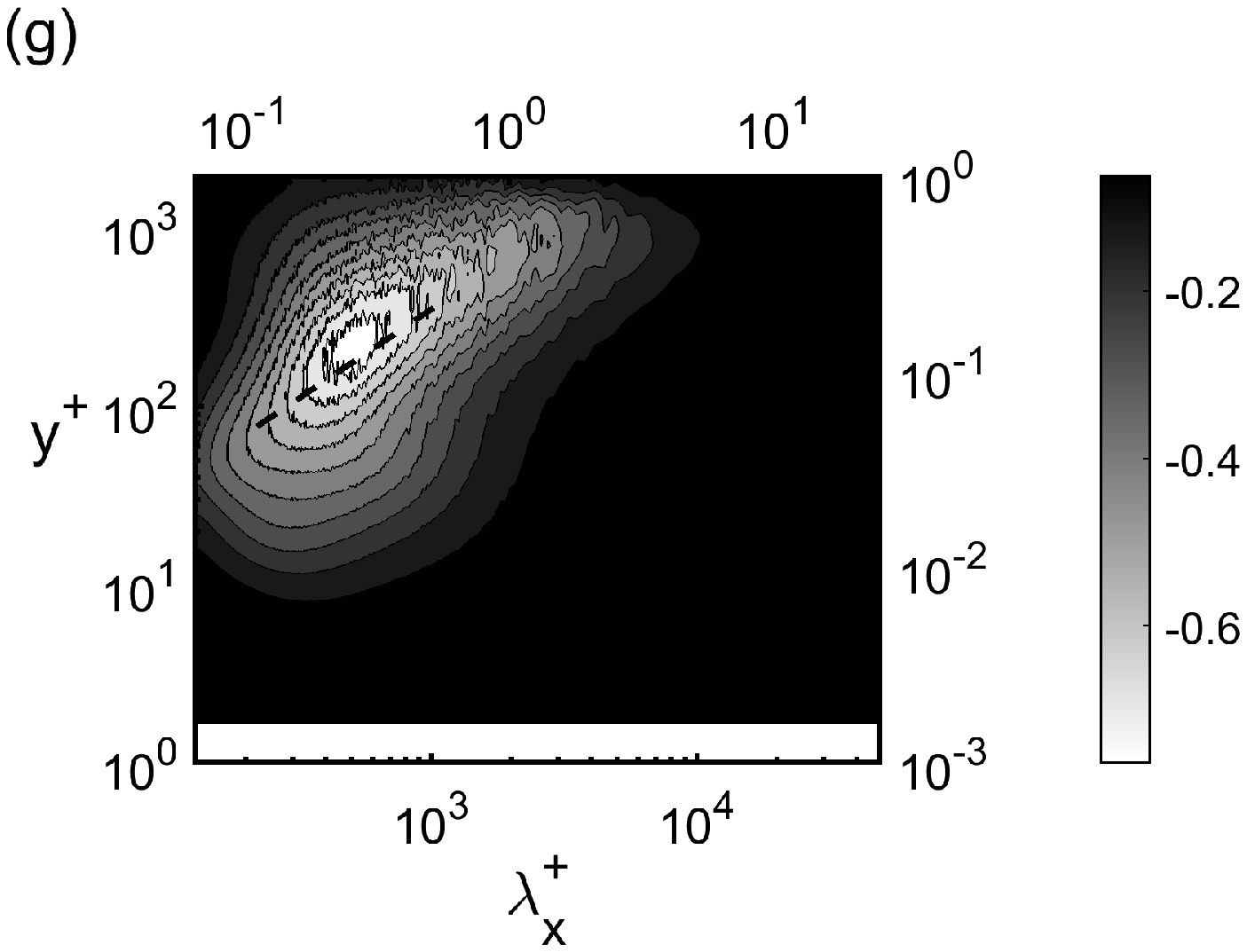}
\end{subfigure}
\begin{subfigure}[b]{0.42\textwidth}
  \includegraphics[width=\textwidth]{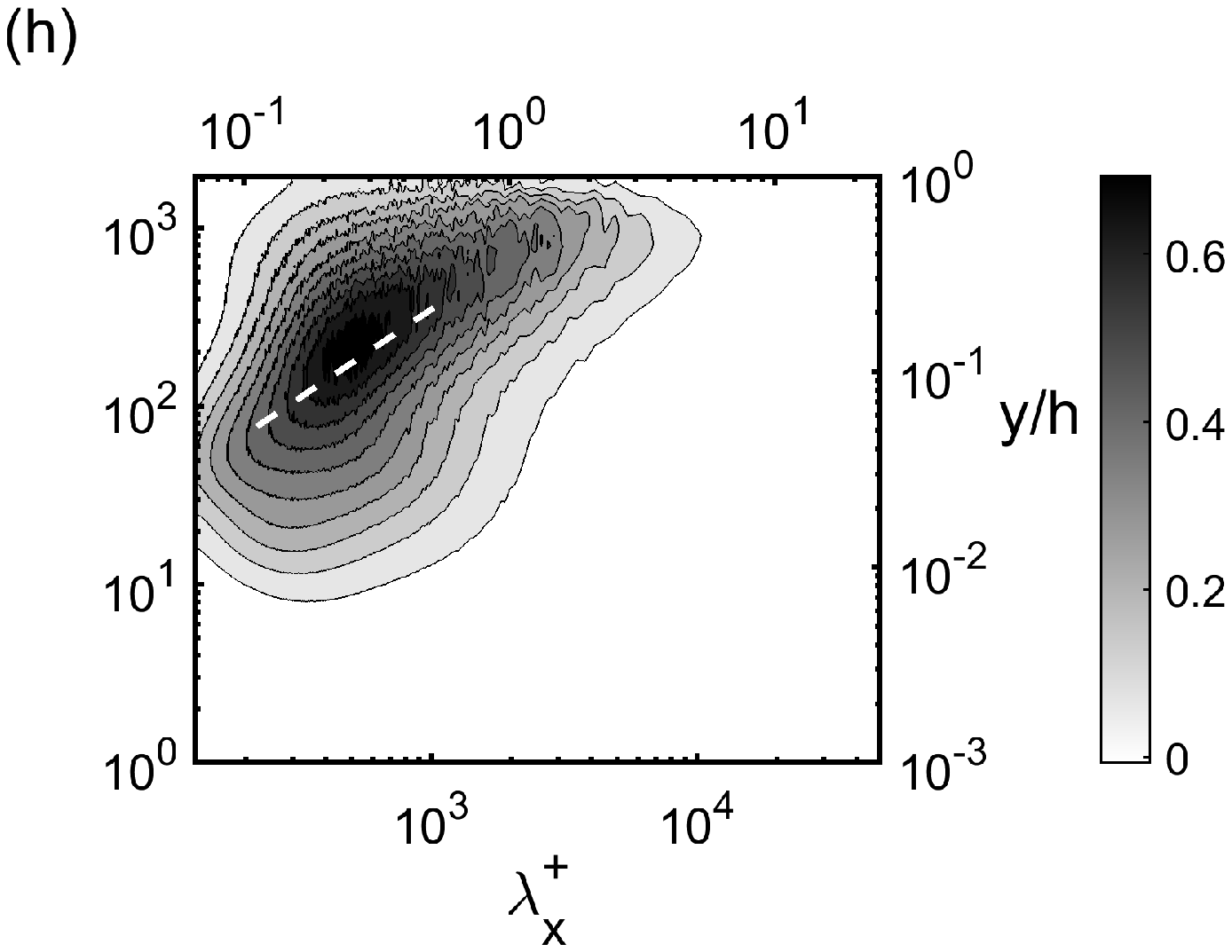}
\end{subfigure}
\end{minipage}
\caption{Premultiplied streamwise wavenumber spectra of $k_x^+ y^+ \widehat{\Pi}_x^+(y^+,\lambda_x^+)$ (left column) and $k_x^+ y^+ \widehat{\Pi}_{yz}^+(y^+,\lambda_x^+)$ (right column) for (a,b) LES, (c,d) GQLZ1, (e,f) GQLZ5 and (g,h) GQLZ25. The dashed lines indicate the ridge $y=0.35 \lambda_x$.}
\label{fig:xpi}
\end{figure}

The pressure-strain spectra are finally explored to understand the mechanism of componentwise TKE distribution in the GQL model. Figure \ref{fig:zpi} shows the spanwise wavenumber spectra of the pressure-strain terms for LES and the GQL models. For all cases, a negative $\widehat{\Pi}_x$ and mainly positive $\widehat{\Pi}_y$ and $\widehat{\Pi}_z$ (combined into $\widehat{\Pi}_{yz}=\widehat{\Pi}_y + \widehat{\Pi}_z$) can be observed in the spanwise wavenumber space. This means that the pressure strain redistributes the energy produced at the streamwise component of the TKE through the production term to the wall-normal and spanwise components. The spectra appear to be distributed approximately along the linear ridge $y=0.2 \lambda_z$. The GQLZ1 case underestimates the spectral energy intensity of all pressure-strain terms, and their spanwise wavenumber spectra do not extend to length scales below $\lambda_z^+ \approx 200$. The two features are improved as $\lambda_{z,c}$ is decreased, the spectra reach smaller spanwise scales in the GQLZ5 case ($\lambda_z^+ \approx 100$). Both spectra of the $\mathcal{P}_h$ subspace group of GQLZ5 and GQLZ25 reproduce with good agreement the spectra of the LES. 
Figure \ref{fig:xpi} shows the streamwise wavenumber spectra of the pressure-strain terms. The pressure-strain intensity appears to be concentrated for wavelengths $\lambda_x^+ \gtrsim 300$ in the GQLZ1 case, but the spectra quickly extend to smaller $\lambda_x$ as the number of spanwise modes in the $\mathcal{P}_l$-subspace group is increased. The streamwise spectra of GQLZ5 and GQLZ25 show excellent agreement with those of the LES.

As was shown in Part 1, the GQL approximations only change the equations for the `slow pressure', $p_l^S$, given by   
\begin{subequations}\label{eq:3.6}
\begin{equation}\label{eq:thel}
\frac{1}{\rho}\nabla^2{p_l^S}=\mathcal{P}_l\Big[-\frac{\partial u_{l,j}}{\partial x_i}\frac{\partial u_{l,i}}{\partial x_j}\Big]+\mathcal{P}_l\Big[-2\frac{\partial u_{l,j}}{\partial x_i}\frac{\partial u_{h,i}}{\partial x_j}\Big]+\mathcal{P}_l\Big[-\frac{\partial u_{h,j}}{\partial x_i}\frac{\partial u_{h,i}}{\partial x_j}\Big],
\end{equation}
in the $\mathcal{P}_l$ subspace and 
\begin{equation}\label{eq:theh}
\frac{1}{\rho}\nabla^2p_h^S=\mathcal{P}_h\Big[-\frac{\partial u_{l,j}}{\partial x_i}\frac{\partial u_{l,i}}{\partial x_j}\Big]+\mathcal{P}_h\Big[-2\frac{\partial u_{l,j}}{\partial x_i}\frac{\partial u_{h,i}}{\partial x_j}\Big]+\mathcal{P}_h\Big[-\frac{\partial u_{h,j}}{\partial x_i}\frac{\partial u_{h,i}}{\partial x_j}\Big],
\end{equation}
\end{subequations}
in the $\mathcal{P}_h$ subspace. In the GQL models, the second term in the right-hand side of (\ref{eq:thel}) and the first and the last terms in the right-hand side of (\ref{eq:theh}) are absent. Therefore, similarly to the GQL approximations considered in the streamwise direction in Part 1, the GQL models here should also experience the lack of pressure-strain transport mediated by such slow pressure, explaining why the GQL models show highly anisotropic turbulent fluctuations biased to the streamwise component especially when $\lambda_{z,c}$ is large (see figure \ref{fig:stat}). 


\section{Discussion} \label{sec:sec4}
\subsection{Comparison with GQL approximations in the streamwise direction}\label{sec:sec41}
Thus far, we have investigated a set of GQL approximations and compared its turbulence statistics, spectra and energy transfer to those of full LES at $Re_\tau \simeq 1700$ in turbulent channel flow. For the GQL approximations in this study, the flow is decomposed into a group of low spanwise wavenumber Fourier modes ($\mathcal{P}_l$ subspace) and the rest ($\mathcal{P}_h$ subspace): see (\ref{eq:2.0}). The former group is solved considering the full nonlinear equations, whereas the latter is obtained by solving the linearised equations about the former. The decomposition employed for the GQL models in the present study evidently has different physical implications to that in Part 1. The QL and GQL models in Part 1 \cite[or RNL from][]{thomas14,thomas15,farrell17} place a minimal model for the self-sustaining process at each of the (spanwise) integral length scales and examine how their nonlinear coupling dynamics and the related streamwise energy cascade play a role in the description of full turbulent dynamics (i.e. the GQL models in Part 1). On the contrary, in the present study, the flow field in the $\mathcal{P}_l$ subspace is supposed to fully incorporate the self-sustaining process at least within the given integral length scales, while that in the $\mathcal{P}_h$ subspace is not expected to support the self-sustaining process (see \S\ref{sec:sec42}). Despite the significant difference in the physical implications between the GQL approximations considered in Part 1 and in this study, it is worth pointing out that the two different types of GQL models exhibit some notable similarities, which would originate from the general nature of the GQL approximation.  

First, the GQL models with large cut-off wavelengths (i.e. $\lambda_{x,c}(\equiv 2\pi/k_{x,c}$) in Part 1 and $\lambda_{z,c}$ in this study) exhibit a considerably reduced multi-scale behaviour. Indeed, the QL model in Part 1 and the GQL1 case in this study show spanwise wavenumber spectra, where the energy density is focused on a particular length scale (figures 3c,d in Part 1 and figures \ref{fig:zspectra}c,d) instead of being spread over a wide range of the spanwise wavenumbers. As discussed in detail in \S4.1 of Part 1, the blockage of nonlinear energy transfer in the GQL approximations with a large cut-off wavelength appears to elevate energy at a particular integral length scale. This subsequently affects the mean-fluctuation dynamics in a way that it does not strongly support turbulent fluctuations in the other wall-normal locations and the corresponding spectral range, resulting in a reduced multi-scale behaviour. It appears that this behaviour is rather common regardless of the nature of the flow decomposition at least in the present channel flow. This implies that a key to the successful prediction of low-order turbulence statistics using the QL/GQL framework would lie in an accurate description of the balanced mean-fluctuation interactions across the wide range of integral length scales. It also explains why much simpler quasilinear models \cite[]{hwang19,skouloudis}, the focus of which was primarily given to modelling of the mean-fluctuation interactions, were reasonably successful in the description of general statistical behaviors of the same flow. 

Second, when $\lambda_{z,c}$ or $\lambda_{x,c}$ is large, the GQL models in Part 1 as well as in the present study exhibit a highly anisotropic turbulent fluctuation which contains significantly elevated energy in the streamwise component. Since the turbulence production in parallel wall-bounded shear flows appears only in the streamwise component, this implies that, in general, GQL approximations prevent the energy distribution mechanism into the other velocity components. In fact, this is exactly the role played by the pressure strain \cite[e.g.][]{cho18,lee19}. Given the nature of any QL/GQL approximations ignoring certain nonlinear fluctuation-fluctuation interactions, it would always damage and/or disrupt the slow pressure originating from the nonlinear fluctuation-fluctuation interactions, explaining the highly anisotropic turbulent fluctuation of the QL/GQL models in Part 1 and in the present study. This observation would be important if one wishes to improve or develop a QL model simpler than the GQL models considered here, as the role of the pressure-strain transport mediated by slow pressure needs to be taken into account for the correct prediction of anisotropy of the given flow.  

\subsection{Statistical description of near-wall turbulence and the self-sustaining process} \label{sec:sec42}
The near-wall turbulence approximately below $y^+ \approx 50-70$ has been understood to be largely independent of the motions in the logarithmic and outer regions. It has consistently been shown that the absence and/or disruption of the log- and outer-region motions in turbulence does not affect the near-wall dynamics which attains the self-sustaining process at the viscous inner length scale \cite[e.g.][]{jimenez91,hamilton95,pinelli99,hwang13,doohan19}. This early observation extends to the logarithmic and outer regions, where similar self-sustaining processes have been found to exist at larger scales in the form of Townsend's attached eddies \cite[]{flores10,hwang10prl,hwang11,hwang15,hwangbengana16}. 

The self-sustaining process is a nonlinear mechanism that drives turbulence utilising the energy from linearly stable mean flow $\mathbf{U}$. For a mathematical description of the self-sustaining process in the $\mathcal{P}_h$ subspace of a full simulation (or the full LES), let us consider the full Navier-Stokes equations with $\mathbf{u}_l=\mathbf{0}$. Employing the velocity decomposition used for the QL model in Part 1 (or the RNL model), $\mathbf{u}_h$ is written as
\begin{equation}\label{eq:4.1}
\mathbf{u}_h=\mathbf{u}_{h,0}  +\mathbf{u}_{h,r},
\end{equation}
where $\mathbf{u}_{h,0}=\langle \mathbf{u}_{h} \rangle_x$. The equations for $\mathbf{u}_{h,0}$ and $\mathbf{u}_{h,r}$ are then given by
\begin{subequations}\label{eq:4.2}
\begin{align}\label{eq:4.2a}
&\frac{\partial \bold{u}_{h,0}}{\partial t}+(\bold{U}\cdot\nabla)\textbf{u}_{h,0}+(\textbf{u}_{h,0}\cdot\nabla)\bold{U}=-\frac{1}{\rho}\nabla p_{h,0}+  \nu \bold{\nabla}^2  \bold{u}_{h,0} \\
& -\mathcal{P}_{h}\left[\left(\textbf{u}_{h,0}\cdot\nabla\right)\textbf{u}_{h,0}\right]-\langle\mathcal{P}_{h}\left[\left(\textbf{u}_{h,r}\cdot\nabla\right)\textbf{u}_{h,r}\right]\rangle_{x}, \nonumber 
\end{align}
and
\begin{align}\label{eq:4.2b}
&\frac{\partial \bold{u}_{h,r}}{\partial t}+\left(\bold{U}\cdot\nabla\right)\textbf{u}_{h,r}+\left(\textbf{u}_{h,r}\cdot\nabla\right)\bold{U}=-\frac{1}{\rho}\nabla p_{h,r}+  \nu \bold{\nabla}^2  \bold{u}_{h,r} \\
&-\mathcal{P}_{h}\left[\left(\textbf{u}_{h,0}\cdot\nabla\right)\bold{u}_{h,r}\right]-\mathcal{P}_{h}\left[\left(\bold{u}_{h,r}\cdot\nabla\right)\textbf{u}_{h,0}\right] \nonumber \\
&-\mathcal{P}_{h}\left[\left(\textbf{u}_{h,r}\cdot\nabla\right)\textbf{u}_{h,r}\right]+\langle\mathcal{P}_{h}\left[\left(\textbf{u}_{h,r}\cdot\nabla\right)\textbf{u}_{h,r}\right]\rangle_x, \nonumber
\end{align}
\end{subequations}
where $p_{h,0}$ and $p_{h,r}$ are defined to enforce $\bold{\nabla} \cdot \bold{u}_{h,0}=0$ and $\bold{\nabla} \cdot \bold{u}_{h,r}=0$, respectively, with $p_h=p_{h,0}+p_{h,r}$, and the eddy viscosity term for LES is ignored for simplicity. Here, $\mathbf{u}_{h,0}$ in (\ref{eq:4.2a}) describes the evolution of the streamwise elongated (or constant) streaks and rolls, whereas $\mathbf{u}_{h,r}$ in (\ref{eq:4.2b}) depicts the streak instability (i.e. instability of a streaky flow, $\mathbf{U}+\mathbf{u}_{h,0}$) -- note that  (\ref{eq:4.2b}) in the absence of its third line containing self-interacting nonlinear terms gives the linearised equations about $\mathbf{U}+\mathbf{u}_{h,0}$ in the $\mathcal{P}_h$ subspace (i.e. the QL model in Part 1). From (\ref{eq:4.2}), it is seen that $\mathbf{u}_{h,0}$ is driven by the presence of $\mathbf{u}_{h,r}$ (i.e. the last term in (\ref{eq:4.2a})), whereas $\mathbf{u}_{h,r}$ physically originates from the inflectional instability, transient growth and/or parametric mechanism primarily depicted by the second term in the second line of (\ref{eq:4.2b}) \cite[]{hamilton95,schoppa02,park11,farrell12,degiovanetti16,degiovanetti17,lozano2021}. This is a two-way interaction between $\mathbf{u}_{h,0}$ (streaks) and $\mathbf{u}_{h,r}$ (streamwise instability wave/quasi-streamwise vortices), which sustains turbulence in the absence of $\mathbf{u}_{l}$. 

Now, we consider the same form of the equations of motions for $\mathbf{u}_h$ from the GQL models in the present study. Further to the velocity decomposition in (\ref{eq:4.1}), $\mathbf{u}_l$ is also decomposed such that
\begin{equation}
\mathbf{u}_l=\mathbf{u}_{l,0}  +\mathbf{u}_{l,r},
\end{equation}
where $\mathbf{u}_{l,0}=\langle \mathbf{u}_{l} \rangle_x$. For the GQL models in this study, the equations for $\mathbf{u}_{h,0}$ and $\mathbf{u}_{h,r}$ are given by
\begin{subequations}\label{eq:4.4}
\begin{align}\label{eq:4.4a}
&\frac{\partial \bold{u}_{h,0}}{\partial t}+(\bold{U}\cdot\nabla)\textbf{u}_{h,0}+(\textbf{u}_{h,0}\cdot\nabla)\bold{U}=-\frac{1}{\rho}\nabla p_{h,0}+  \nu \bold{\nabla}^2  \bold{u}_{h,0} \\
& -\mathcal{P}_{h}[\left(\textbf{u}_{h,0}\cdot\nabla\right)\textbf{u}_{l,0}]-\mathcal{P}_{h}[\left(\textbf{u}_{l,0}\cdot\nabla\right)\textbf{u}_{h,0}] \nonumber \\
&-\langle\mathcal{P}_{h}\left[\left(\textbf{u}_{h,r}\cdot\nabla\right)\textbf{u}_{l}\right]\rangle_{x}-\langle\mathcal{P}_{h}\left[\left(\textbf{u}_{l}\cdot\nabla\right)\textbf{u}_{h,r}\right]\rangle_x, \nonumber 
\end{align}
and
\begin{align}\label{eq:4.4b}
&\frac{\partial \bold{u}_{h,r}}{\partial t}+\left(\bold{U}\cdot\nabla\right)\textbf{u}_{h,r}+\left(\textbf{u}_{h,r}\cdot\nabla\right)\bold{U}=-\frac{1}{\rho}\nabla p_{h,r}+  \nu \bold{\nabla}^2  \bold{u}_{h,r} \\
&-\mathcal{P}_{h}\left[\left(\textbf{u}_{h,r}\cdot\nabla\right)\bold{u}_{l}\right]-\mathcal{P}_{h}\left[\left(\bold{u}_{l}\cdot\nabla\right)\textbf{u}_{h,r}\right] \nonumber \\
& +\langle\mathcal{P}_{h}\left[\left(\textbf{u}_{h,r}\cdot\nabla\right)\textbf{u}_{l}\right]\rangle_{x}+\langle\mathcal{P}_{h}\left[\left(\textbf{u}_{l}\cdot\nabla\right)\textbf{u}_{h,r}\right]\rangle_x. \nonumber 
\end{align}
\end{subequations}
Comparing (\ref{eq:4.2}) with (\ref{eq:4.4}), the differences between the turbulent fluctuations from the self-sustaining process and the GQL models in the present study are now evident. If (\ref{eq:4.4a}) admits a nontrivial $\mathbf{u}_{h,0}$ for $\mathbf{U}$ which does not yield any instability, it should be from the interaction between $\mathbf{u}_l$ and $\mathbf{u}_h$ (the last two lines in (\ref{eq:4.4a})). In such a case, if $\mathbf{u}_l=\mathbf{0}$, there is no way for (\ref{eq:4.4a}) and (\ref{eq:4.4b}) to have a non-trivial statistically stationary solution. Furthermore, the term, which depicts an instability and/or transient growth mechanism from a streaky flow (i.e. the second term in the second line of (\ref{eq:4.2b})), does not exist in (\ref{eq:4.4b}). In this case, an instability may possibly exist from the two second terms respectively in the second and third lines of (\ref{eq:4.4b}), especially when $\mathbf{u}_l$ is non-zero. Importantly, if $\mathbf{u}_l$ is a chaotic velocity field, the non-trivial solution $\mathbf{u}_h$ in (\ref{eq:4.4b}) emerges in the form of the leading neutrally stable Lyapunov vector of (\ref{eq:4.4}), as was discussed in \cite{farrell17} and in Part 1. 

Given the discussion above, it is crucial to have non-zero $\mathbf{u}_l$ for (\ref{eq:4.4}) (i.e. the equations in the $\mathcal{P}_h$ subspace in the present study), whereas (\ref{eq:4.2}) does not need $\mathbf{u}_l$ to create its own dynamics. Despite this important difference between turbulent fluctuations from the self-sustaining process (\ref{eq:4.2}) and from the GQL model (\ref{eq:4.4}), it is remarkable to observe that the GQL model admits velocity spectra fairly similar to those of the full LES (GQLZ25, in particular; compare figures 4a,b with 4g,h). We note that both (\ref{eq:4.2}) and (\ref{eq:4.4}) share the same Navier-Stokes operator linearised about the mean velocity $\mathbf{U}$. In this respect, they can also be viewed as the linearised Navier-Stokes operator driven by chaotic fluctuations, the idea resonating with the previous and on-going linearised Navier-Stokes-based modellling efforts for turbulent flows \cite[e.g.][]{hwang10,mckeon10,moarref13,zare17,mckeon17,hwang19,skouloudis,jovanovic21}. Although such approaches would not be able to fully incorporate some key nonlinear processes described in (\ref{eq:4.2}) and (\ref{eq:4.4}) without additional modelling efforts, the observations here offer an explanation why the linearised Navier-Stokes-based modelling efforts have been so useful to study coherent structures in wall-bounded turbulence.  

\subsection{Scale interactions} \label{sec:sec43}

\begin{table}
  \begin{center}
\def~{\hphantom{0}}
  \begin{tabular}{lccccccccccccc}
\multicolumn{1}{c}{\multirow{2}{*}{~Case~}} &
\multicolumn{1}{c}{\multirow{2}{*}{~(a)~}} & \multicolumn{1}{c}{\multirow{2}{*}{~(b)~}} & 
\multicolumn{1}{c}{\multirow{2}{*}{~(c)~}} & 
\multicolumn{1}{c}{\multirow{2}{*}{~(d)~}}  & \multicolumn{1}{c}{\multirow{2}{*}{~(e)~}}  &
\multicolumn{1}{c}{\multirow{2}{*}{~(f)~}}\\ \\[3pt]
\multicolumn{1}{c}{LES} &  \multicolumn{1}{c}{\cmark} & \multicolumn{1}{c}{\cmark} & \multicolumn{1}{c}{\cmark} &  \multicolumn{1}{c}{\cmark} & \multicolumn{1}{c}{\cmark} & \multicolumn{1}{c}{\cmark} \\ [2pt]
\multicolumn{1}{c}{GQL} &  \multicolumn{1}{c}{\cmark} & \multicolumn{1}{c}{\xmark} & \multicolumn{1}{c}{\cmark} &  \multicolumn{1}{c}{\xmark} & \multicolumn{1}{c}{\cmark} & \multicolumn{1}{c}{\xmark} \\ [2pt]
\multicolumn{1}{c}{TRIAZ25} &  \multicolumn{1}{c}{\cmark} & \multicolumn{1}{c}{\xmark} & \multicolumn{1}{c}{\cmark} &  \multicolumn{1}{c}{\xmark} & \multicolumn{1}{c}{\cmark} & \multicolumn{1}{c}{\cmark} \\ [2pt]
\multicolumn{1}{c}{TRIBZ25} &  \multicolumn{1}{c}{\cmark} & \multicolumn{1}{c}{\xmark} & \multicolumn{1}{c}{\xmark} &  \multicolumn{1}{c}{\cmark} & \multicolumn{1}{c}{\cmark} & \multicolumn{1}{c}{\cmark} \\ [2pt]
\multicolumn{1}{c}{TRICZ25} &  \multicolumn{1}{c}{\cmark} & \multicolumn{1}{c}{\xmark} & \multicolumn{1}{c}{\cmark} &  \multicolumn{1}{c}{\cmark} & \multicolumn{1}{c}{\xmark} & \multicolumn{1}{c}{\cmark} \\ [2pt]
\multicolumn{1}{c}{TRIDZ25} &  \multicolumn{1}{c}{\cmark} & \multicolumn{1}{c}{\xmark} & \multicolumn{1}{c}{\cmark} &  \multicolumn{1}{c}{\cmark} & \multicolumn{1}{c}{\cmark} & \multicolumn{1}{c}{\xmark} \\ [2pt]
  \end{tabular}
\caption{Triadic interactions in the present simulation cases. The tickmark (\cmark) denotes inclusion whereas the crossmark (\xmark) stands for removal of the corresponding triadic interaction. Here, the types of the triadic interactions, indicated by alphabet, are described in figure \ref{fig:triad}.}
  \label{tab:tab2}
  \end{center}
\end{table}

\begin{figure*}
\centering
\begin{subfigure}[b]{0.45\textwidth}
\includegraphics[width=\textwidth]{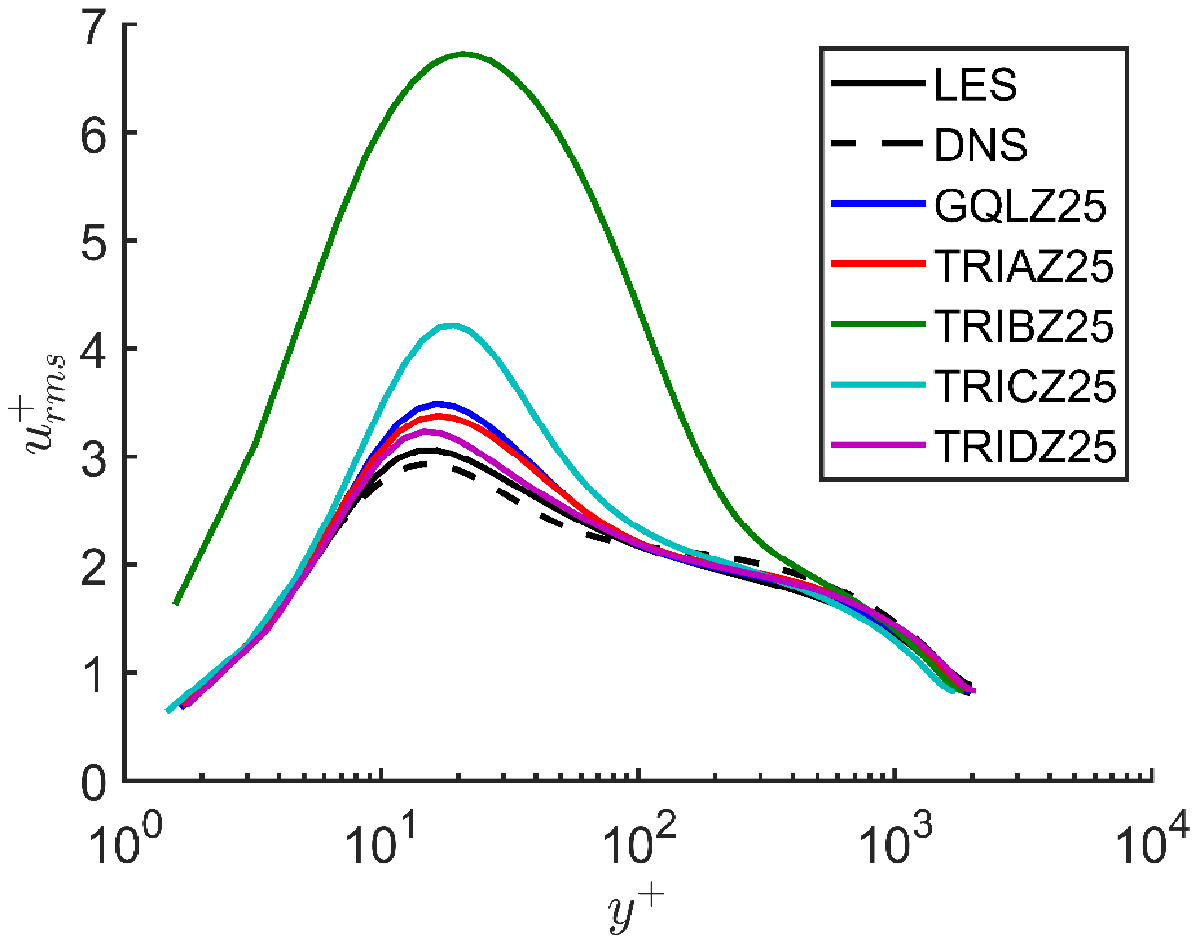}
\caption{$u_{rms}^+(y^+)$}
\label{fig:uu}
\end{subfigure}
\begin{subfigure}[b]{0.45\textwidth}
\includegraphics[width=\textwidth]{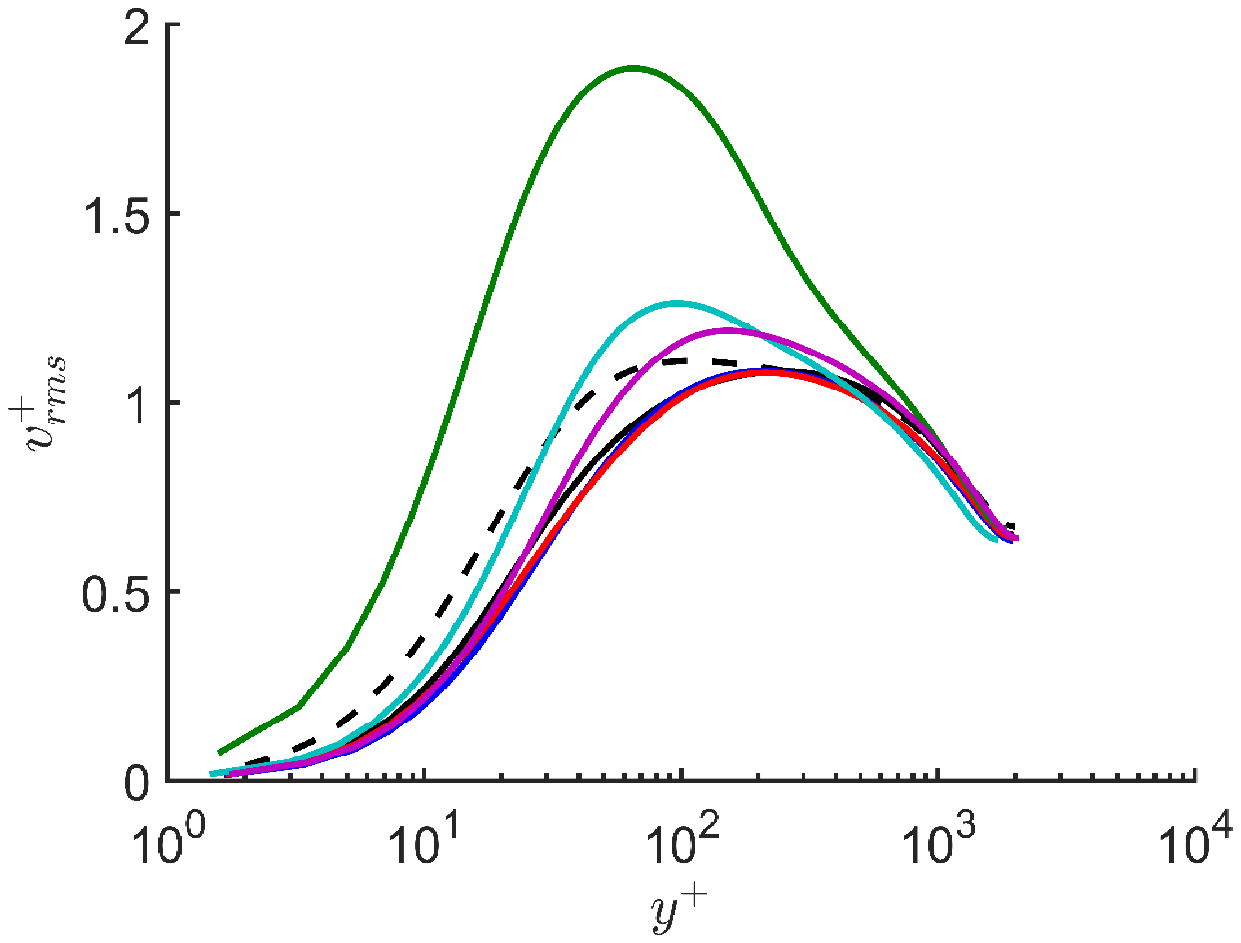}
\caption{$v_{rms}^+(y^+)$}
\label{fig:vv}
\end{subfigure}
\begin{subfigure}[b]{0.45\textwidth}
\includegraphics[width=\textwidth]{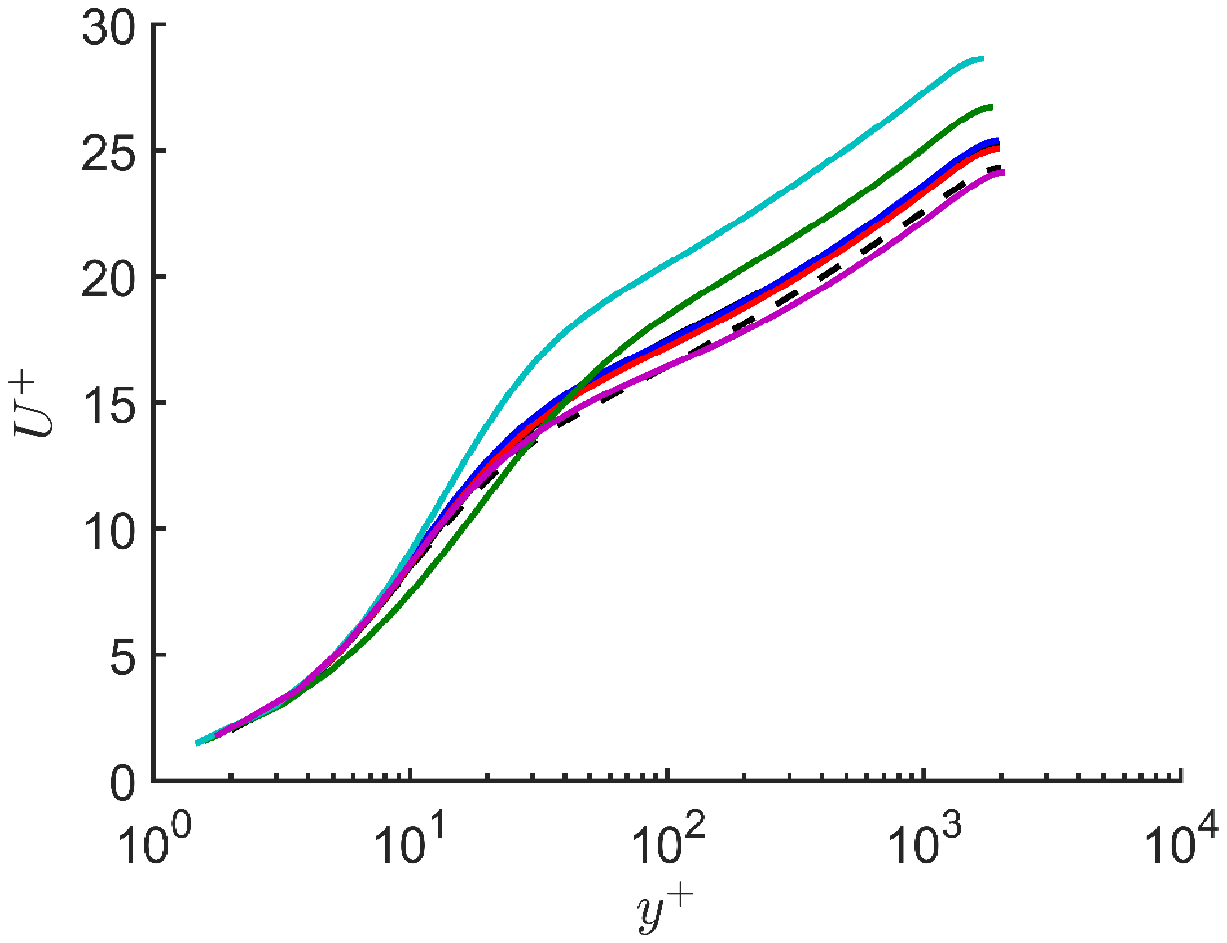}
\caption{$U^+(y^+)$}
\label{fig:ww}
\end{subfigure}
\begin{subfigure}[b]{0.45\textwidth}
\includegraphics[width=\textwidth]{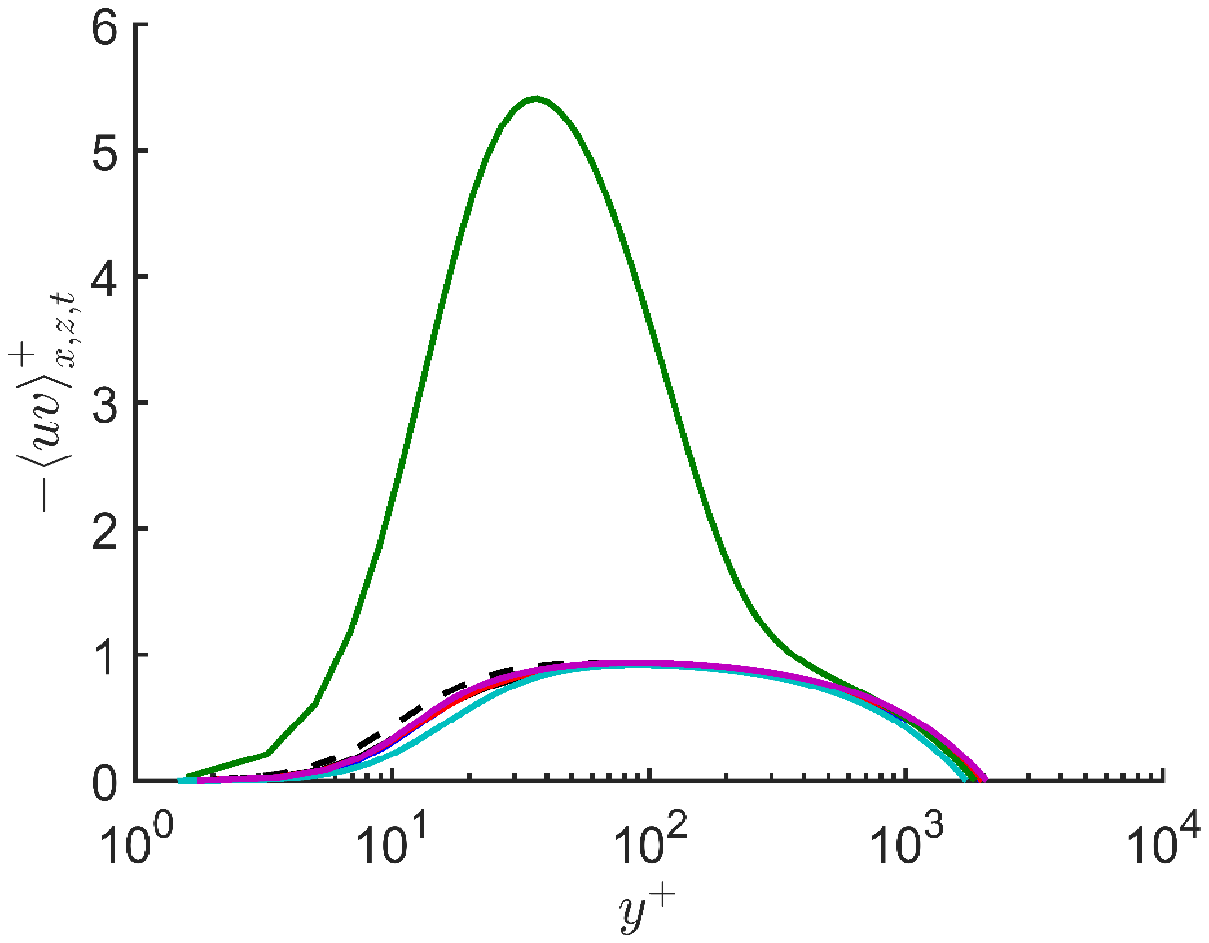}
\caption{$ \langle u^{\prime} v^{\prime}\rangle_{x,z,t} ^+(y^+)$}
\label{fig:uv}
\end{subfigure}
\caption{First- and second-order turbulence statistics for the present LES, DNS at $Re_\tau=2003$ (\citealp{hoyas08}), TRIAZ25, TRIBZ25, TRICZ25 and TRIDZ25 cases: (a) $u_{rms}^+(y^+)$; (b) $v_{rms}^+(y^+)$; (c) $U^+(y^+)$; (d) $ \langle u^{\prime} v^{\prime}\rangle_{x,z,t}^+(y^+)$.}
\label{fig:stat_triad}
\end{figure*}

As discussed in \S\ref{sec:sec1}, the GQL approximations can be utilised as an interventional tool to examine scale interactions, given that it is capable of suppressing particular triadic interactions in a controlled manner (see figure \ref{fig:triad}; see also table \ref{tab:tab2}). Here, we consider the GQLZ25 case as a baseline, where $\lambda_{z,c}^+ \simeq 246$ ($\lambda_{z,c}/h\simeq0.13$) is given for the decomposition of velocity in (\ref{eq:2.0}). We subsequently carry out four additional numerical experiments, in which certain types of triadic interactions between $\mathcal{P}_l$- and $\mathcal{P}_h$-subspace groups are artificially suppressed. In principle, the same numerical experiments can repeatedly be performed by considering different values of $\lambda_{z,c}$. However, in general, this requires a large number of numerical experiments that depends on the choice of $\lambda_{z,c}$, if one wishes to unveil the full triadic interaction dynamics. This is beyond the scope of the present study. Therefore, in the present study, we shall focus on the GQLZ25 case for the purpose of conceptual demonstration. 

The four additional numerical experiments are listed in table \ref{tab:tab2}. Their first- and the second-order turbulence statistics (mean and velocity fluctuations) are reported in figure \ref{fig:stat_triad}. The spanwise and streamwise wavenumber spectra of streamwise and spanwise velocities are shown in figures \ref{fig:zspectra25} and \ref{fig:xspectra25}, and those of production and turbulent transport spectra are given in figures \ref{fig:zenergy25} and \ref{fig:xenergy25}. 

\subsubsection{TRIAZ25: self-interactions within the $\mathcal{P}_h$ subspace}\label{sec:431}
\begin{figure}
\begin{minipage}{\textwidth}
\centering
\begin{subfigure}[b]{0.42\textwidth}
  \includegraphics[width=\textwidth]{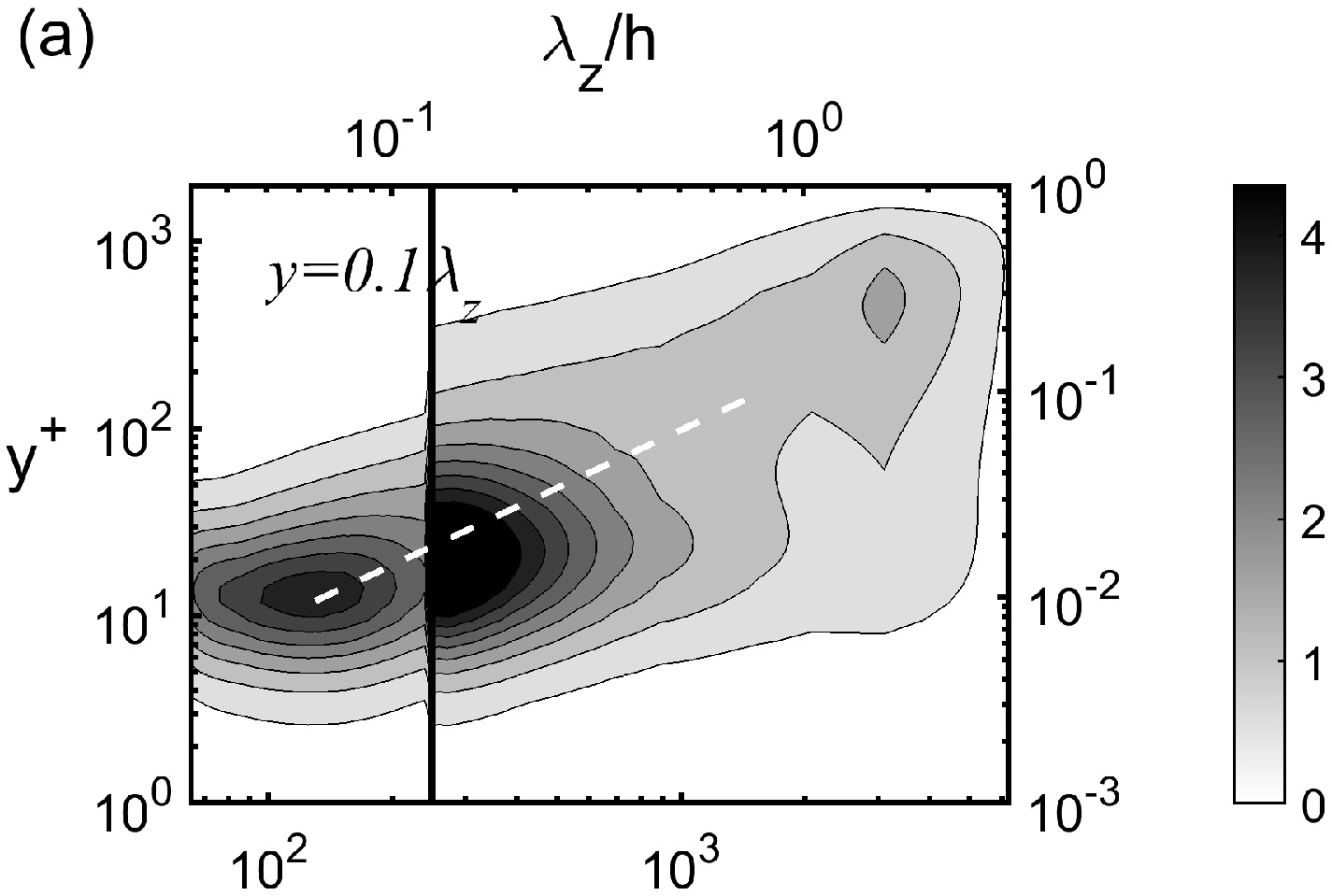}
\label{1}
\vspace{-0.8cm}
\end{subfigure}
\begin{subfigure}[b]{0.42\textwidth}
  \includegraphics[width=\textwidth]{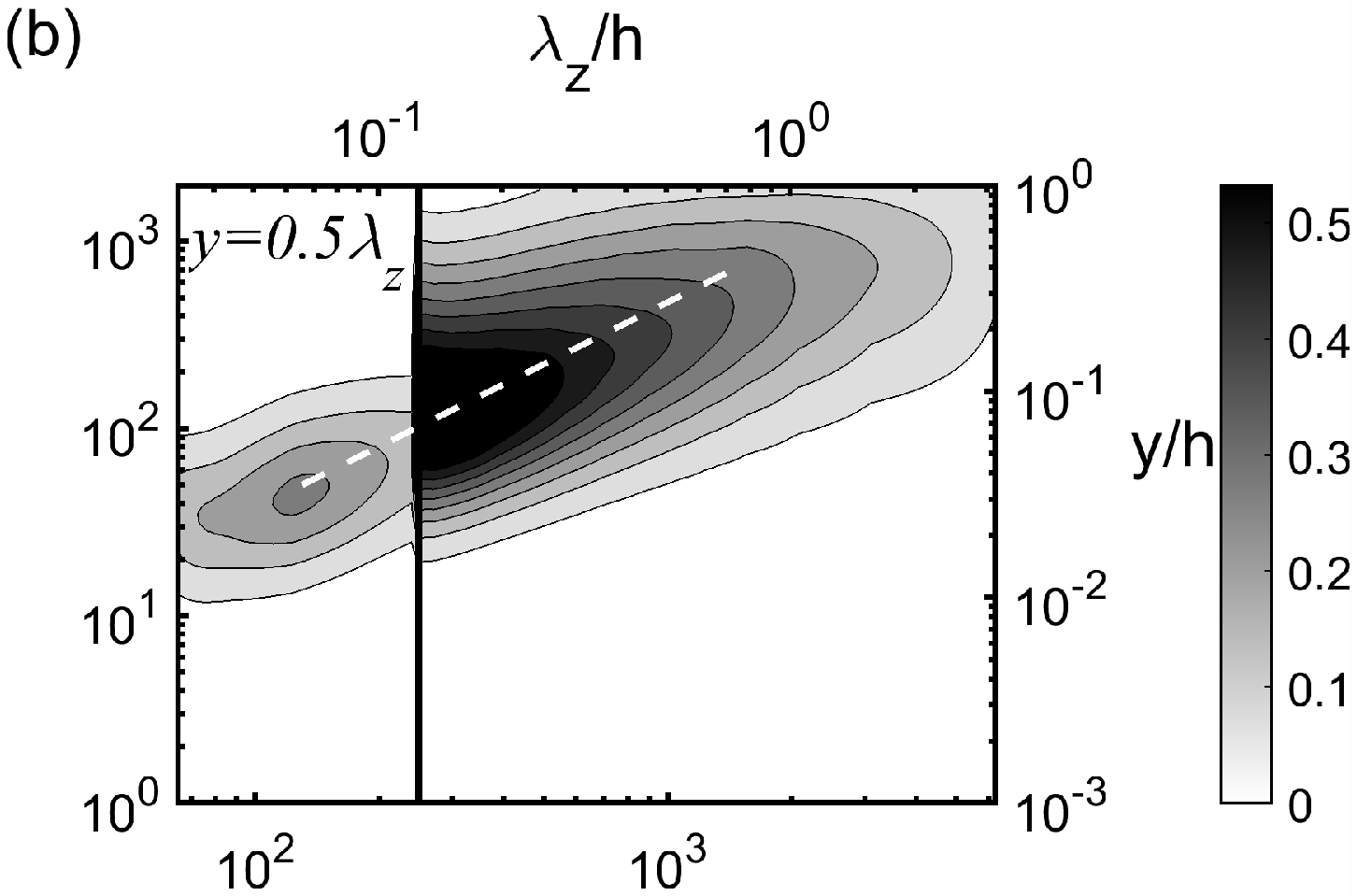}
\label{2}
\vspace{-0.8cm}
\end{subfigure}
\vspace{-0.8cm}
\begin{subfigure}[b]{0.42\textwidth}
  \includegraphics[width=\textwidth]{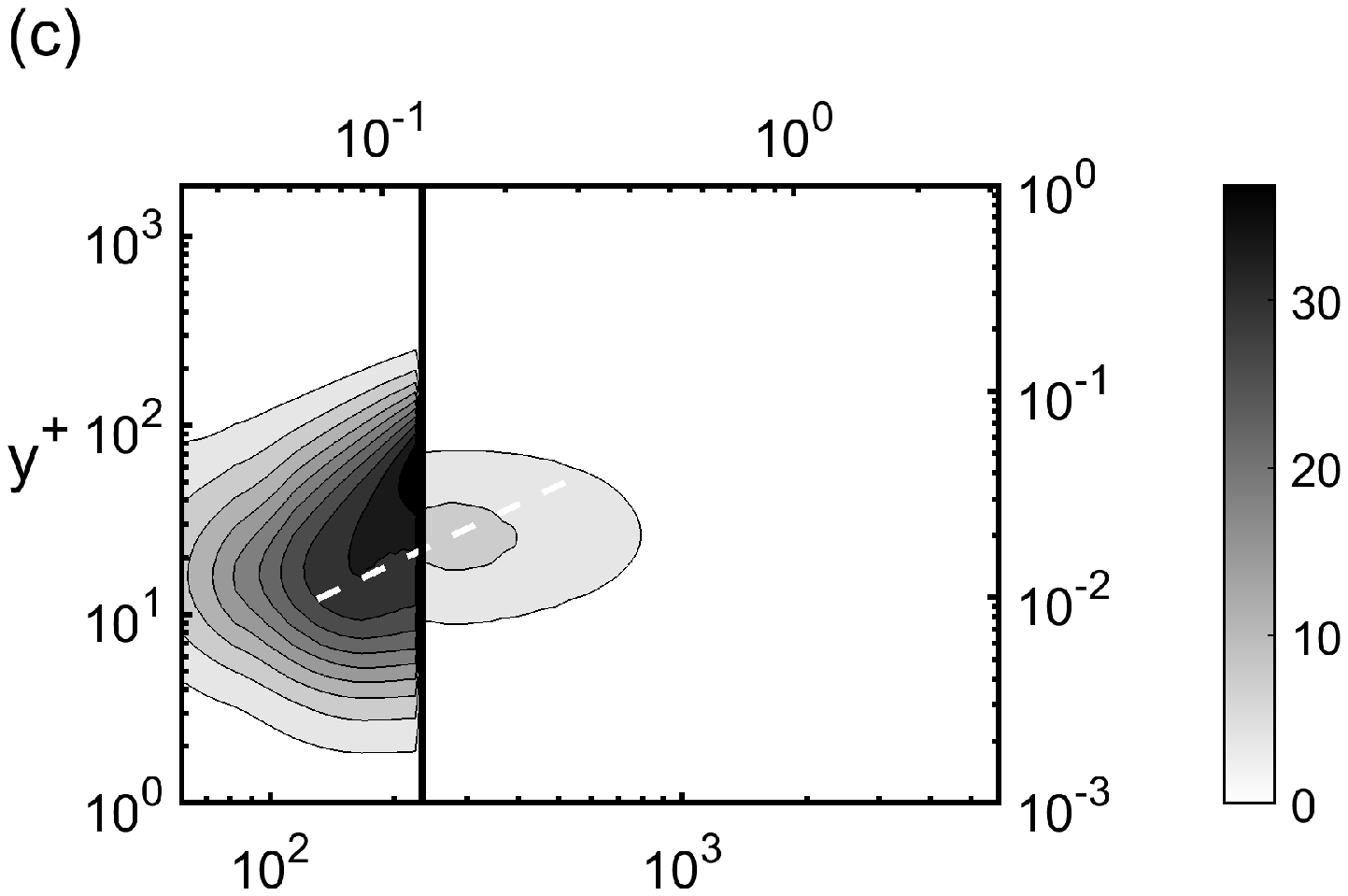}
  \label{3}
\end{subfigure}
\begin{subfigure}[b]{0.42\textwidth}
  \includegraphics[width=\textwidth]{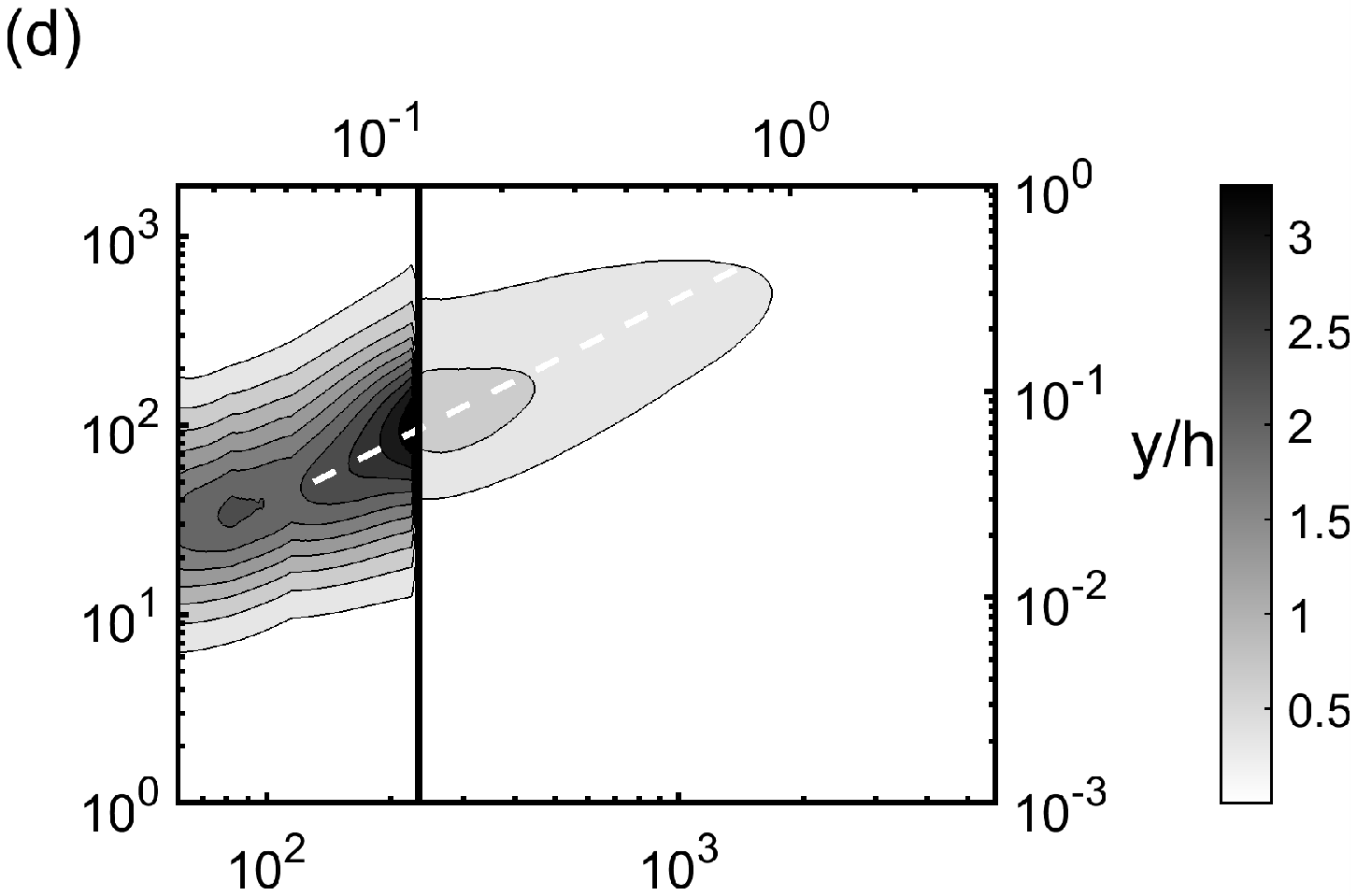}
  \label{4}
\end{subfigure}
\vspace{-0.8cm}
\begin{subfigure}[b]{0.42\textwidth}
  \includegraphics[width=\textwidth]{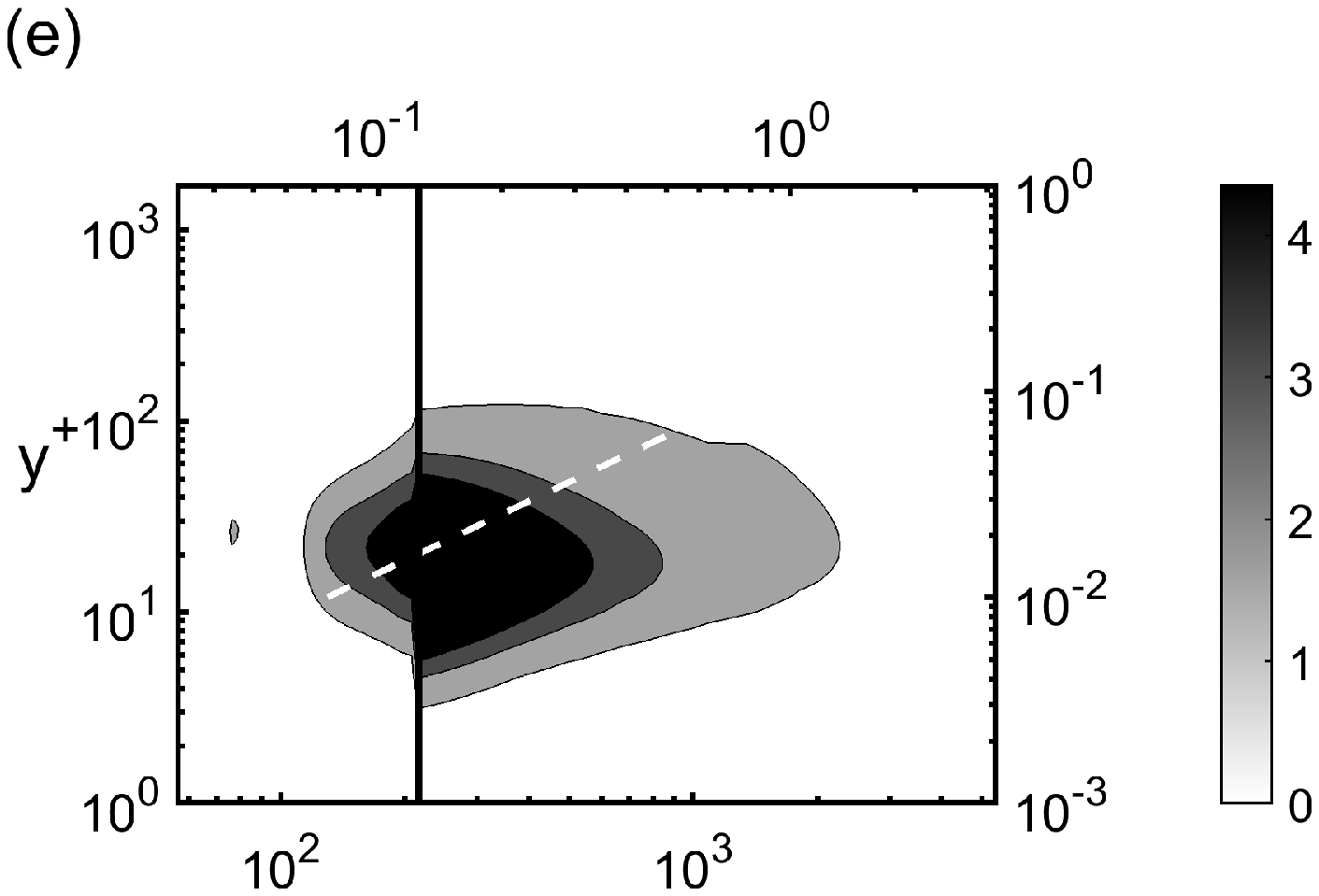}
  \label{5}
\end{subfigure}
\begin{subfigure}[b]{0.42\textwidth}
  \includegraphics[width=\textwidth]{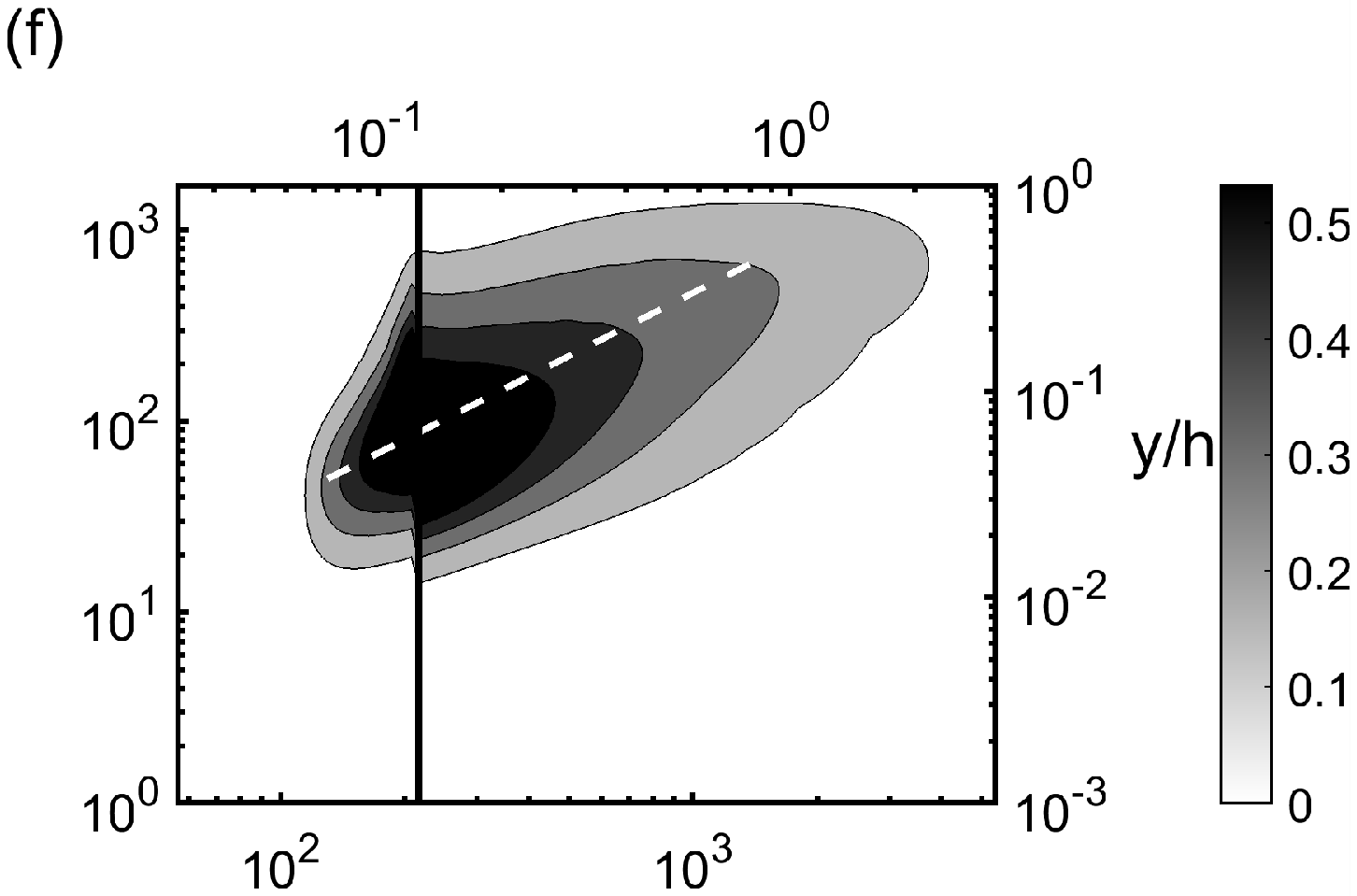}
  \label{6}
\end{subfigure}
\begin{subfigure}[b]{0.42\textwidth}
  \includegraphics[width=\textwidth]{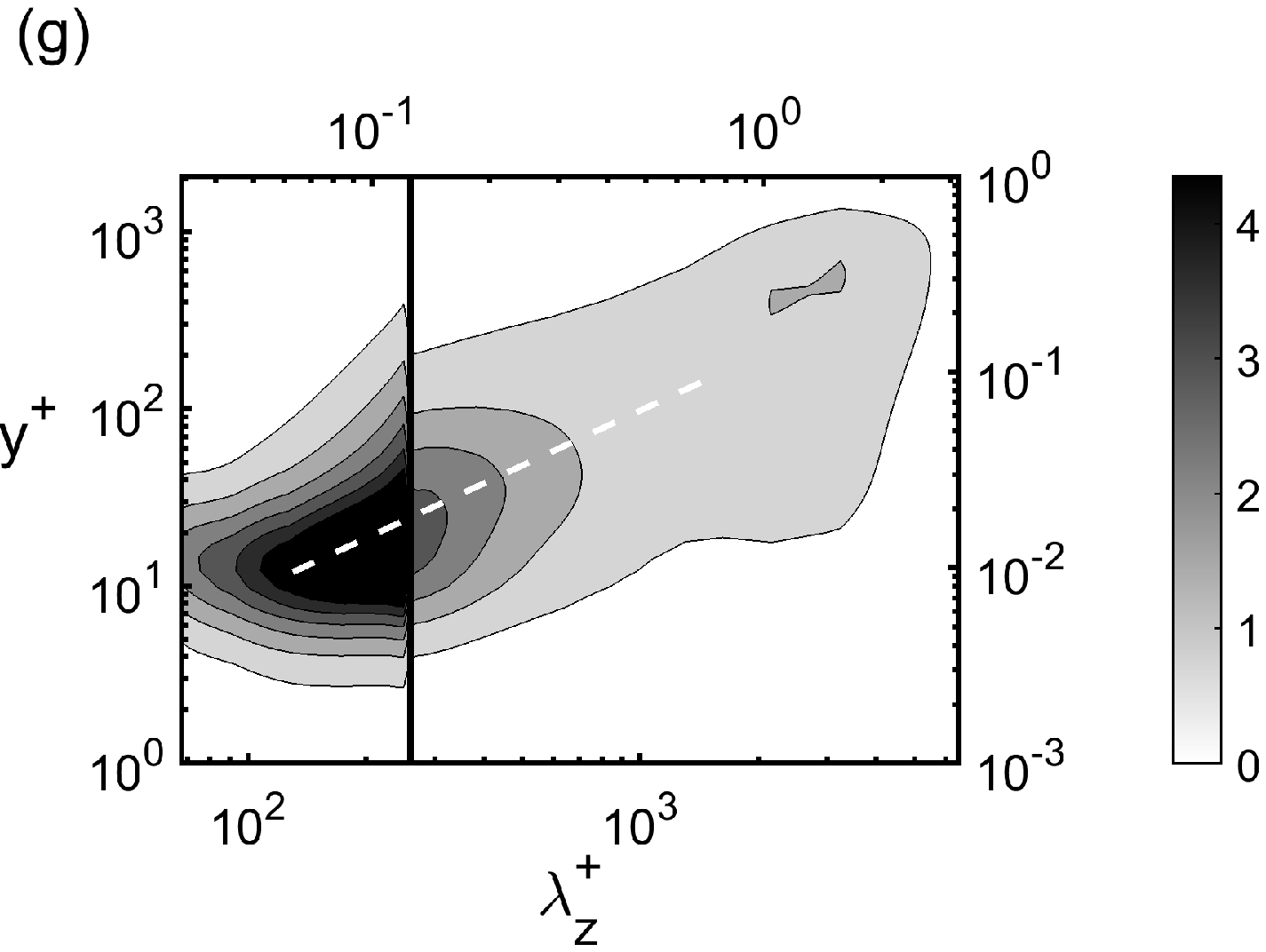}
  \label{5}
\end{subfigure}
\begin{subfigure}[b]{0.42\textwidth}
  \includegraphics[width=\textwidth]{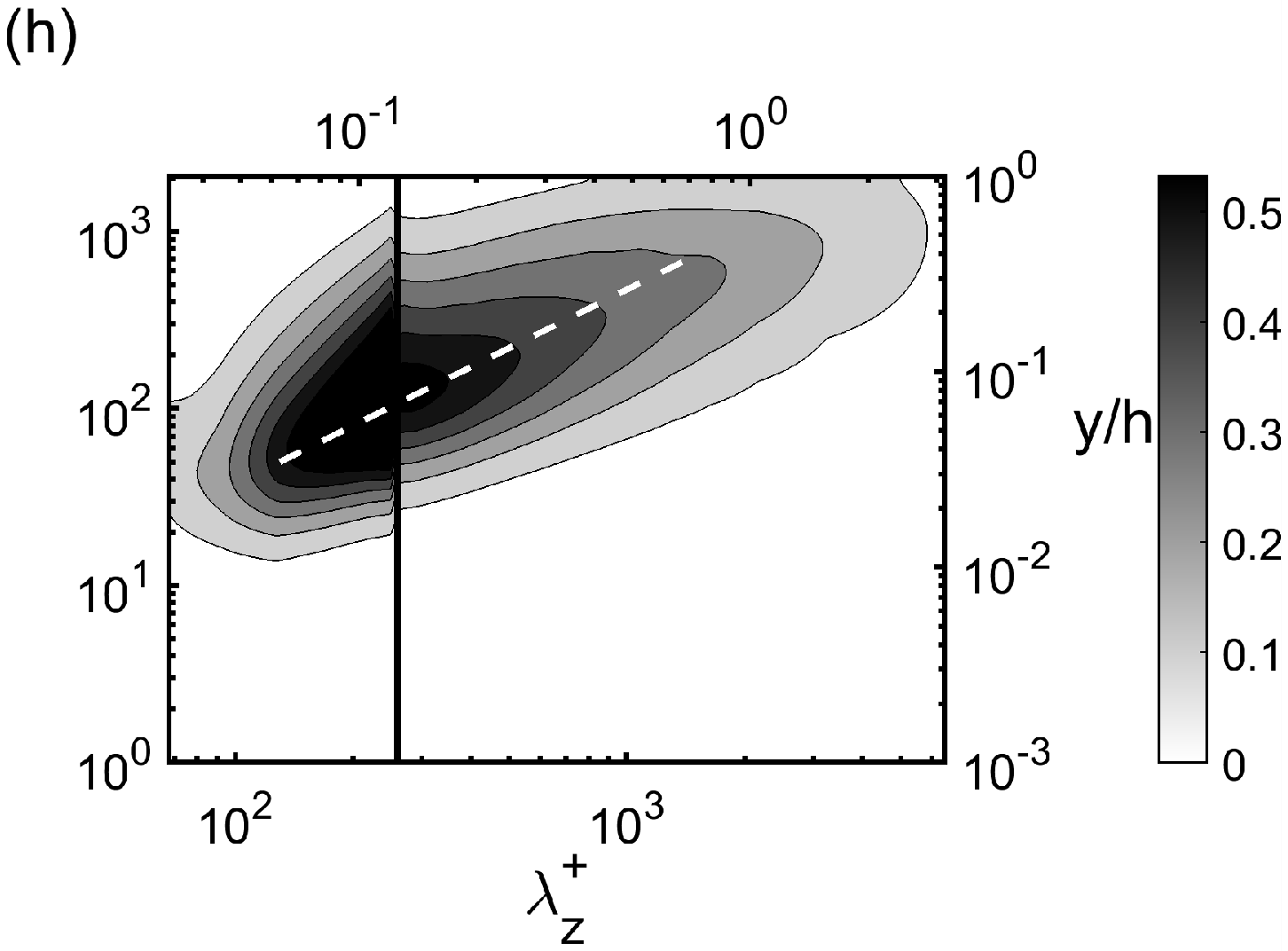}
  \label{6}
\end{subfigure}
\end{minipage}
\caption{Premultiplied spanwise wavenumber spectra of streamwise $k_z^+ \Phi_{uu}^+(y^+,\lambda_z^+)$ (left column) and wall-normal $k_z^+ \Phi_{vv}^+(y^+,\lambda_z^+)$ (right column) velocity for (a,b) TRIAZ25, (c,d) TRIBZ25, (e,f) TRICZ25 and (g,h) TRIDZ25 cases.}
\label{fig:zspectra25}
\end{figure}

\begin{figure}
\begin{minipage}{\textwidth}
\centering
\begin{subfigure}[b]{0.42\textwidth}
  \includegraphics[width=\textwidth]{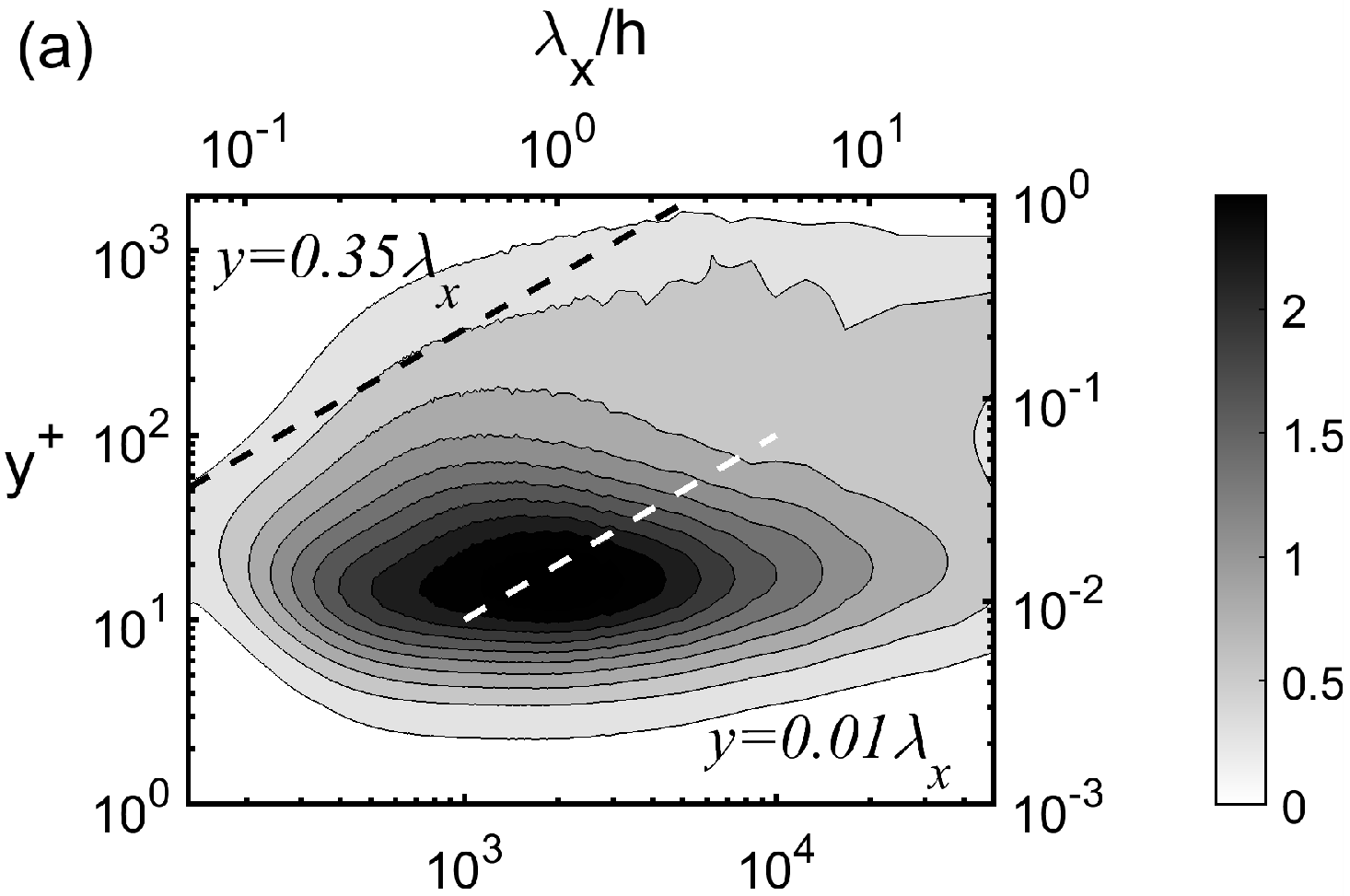}
\label{1}
\vspace{-0.8cm}
\end{subfigure}
\begin{subfigure}[b]{0.42\textwidth}
  \includegraphics[width=\textwidth]{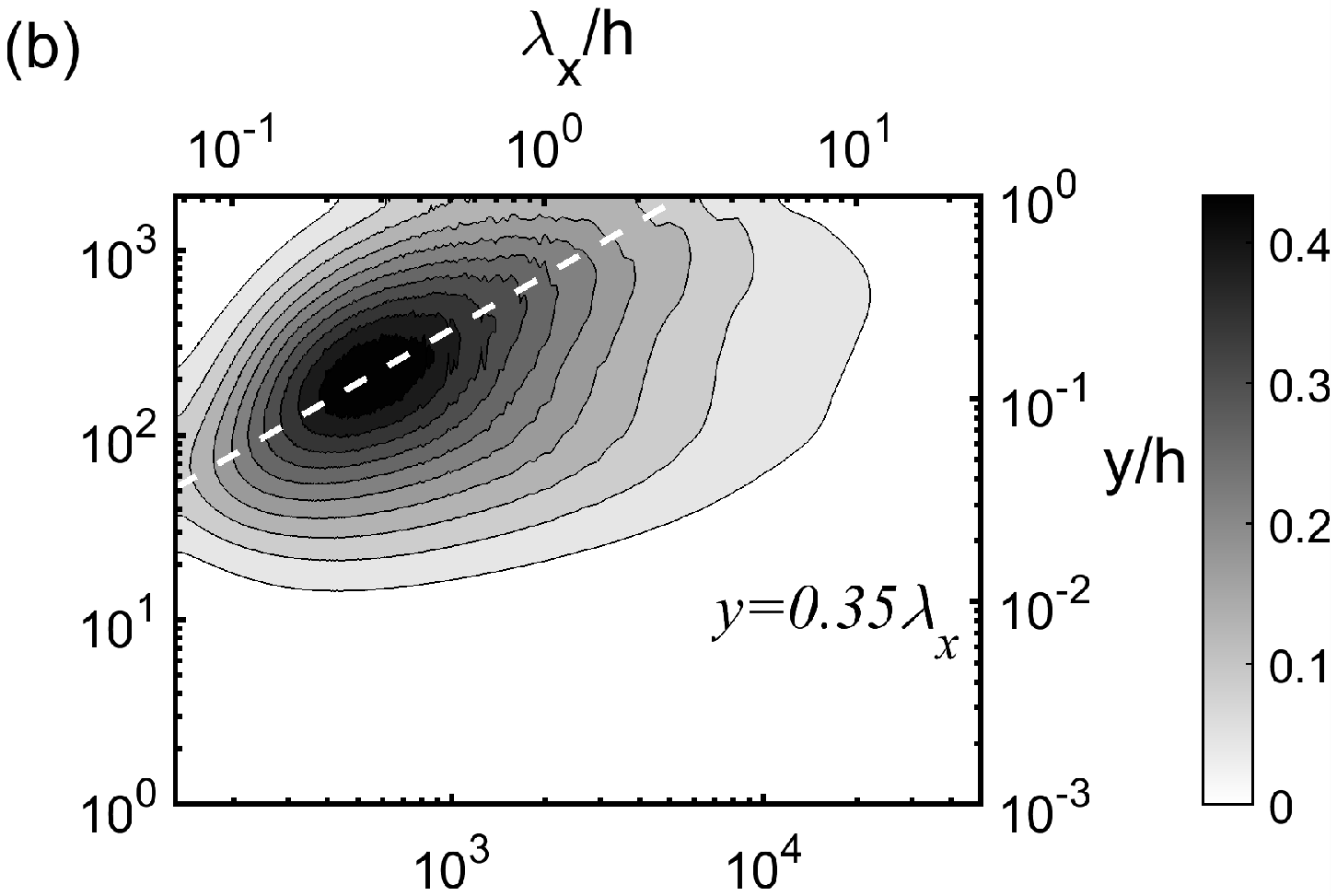}
\label{2}
\vspace{-0.8cm}
\end{subfigure}
\vspace{-0.8cm}
\begin{subfigure}[b]{0.42\textwidth}
  \includegraphics[width=\textwidth]{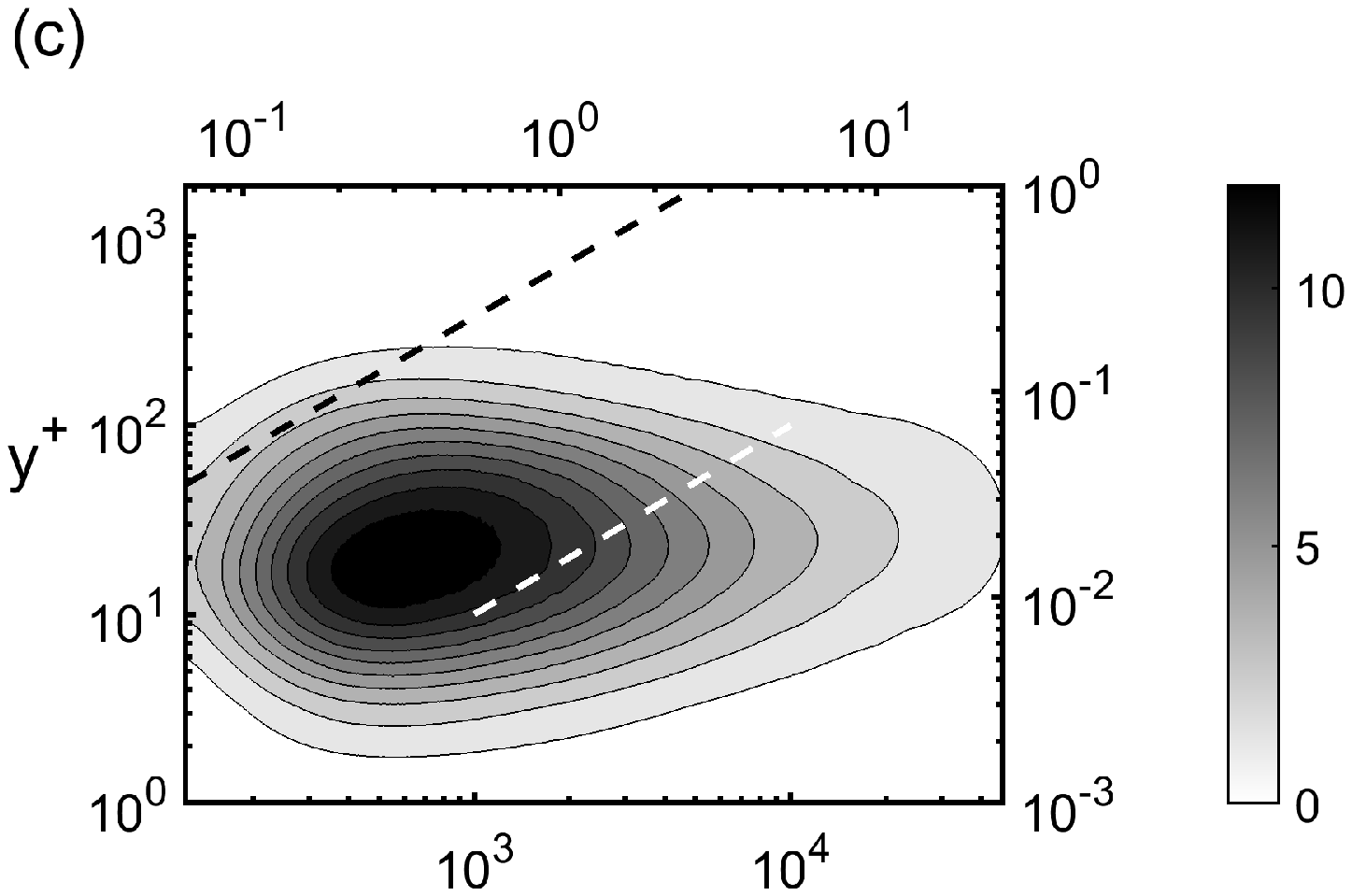}
  \label{3}
\end{subfigure}
\begin{subfigure}[b]{0.42\textwidth}
  \includegraphics[width=\textwidth]{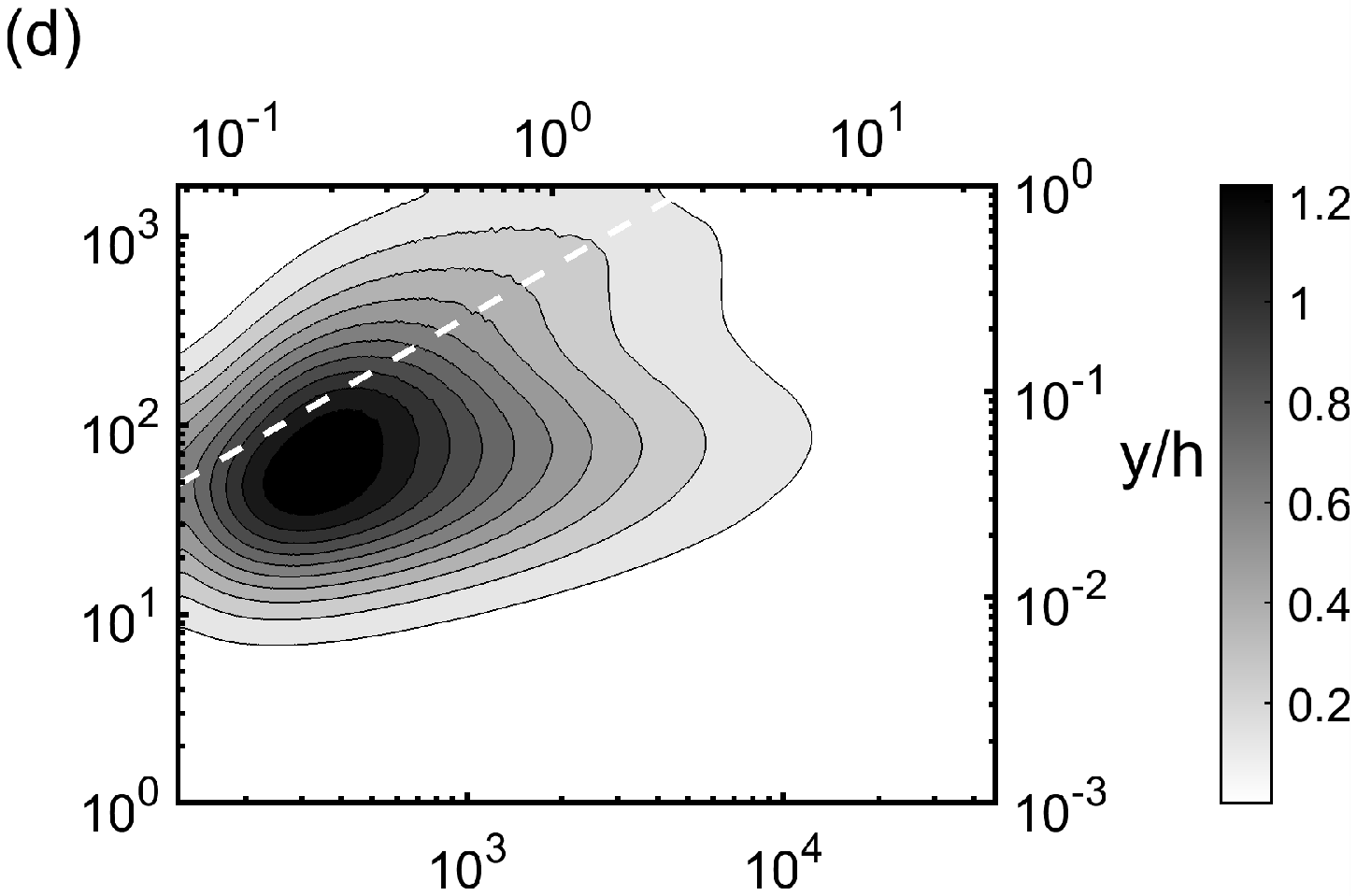}
  \label{4}
\end{subfigure}
\vspace{-0.8cm}
\begin{subfigure}[b]{0.42\textwidth}
  \includegraphics[width=\textwidth]{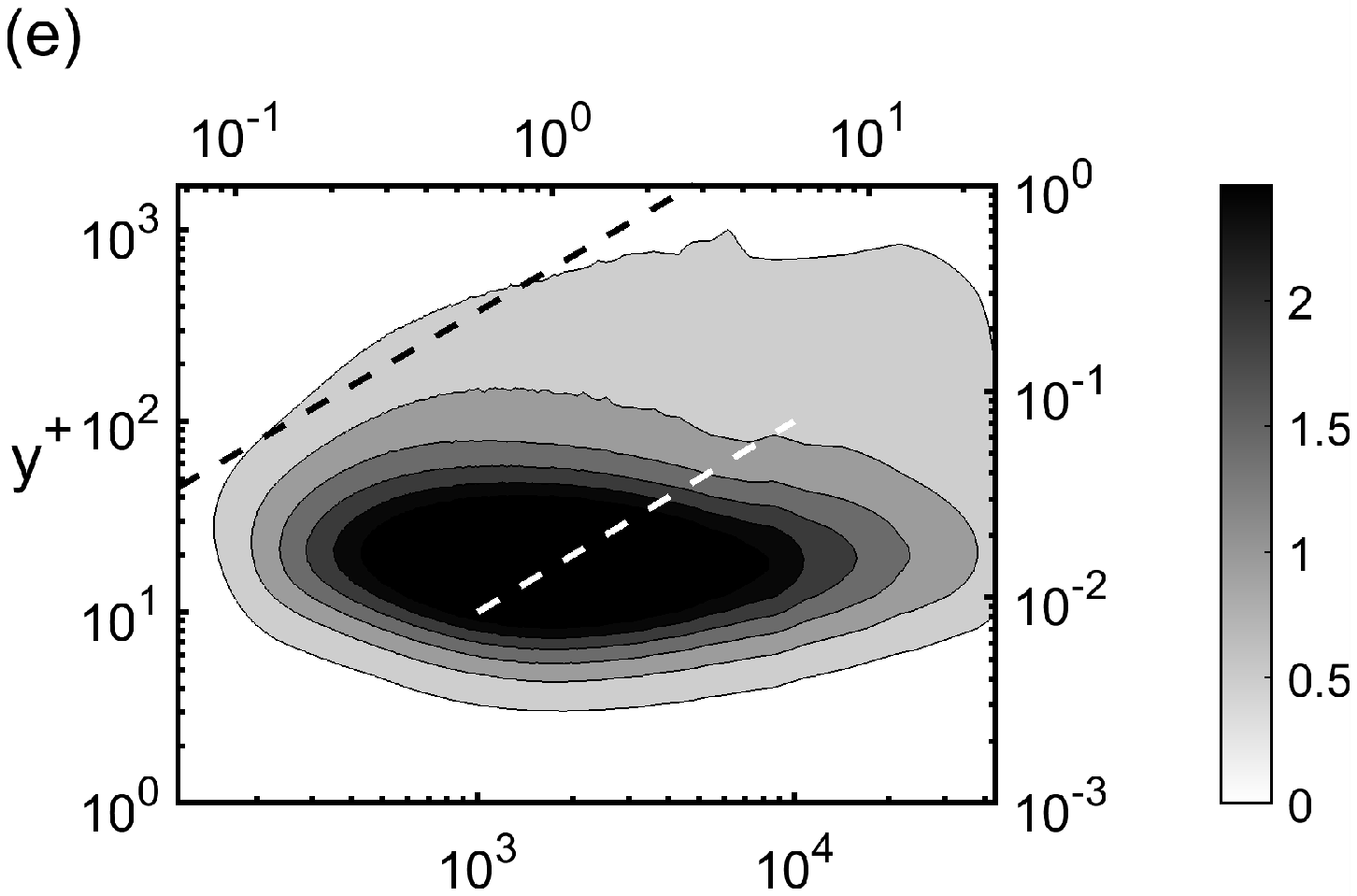}
  \label{5}
\end{subfigure}
\begin{subfigure}[b]{0.42\textwidth}
  \includegraphics[width=\textwidth]{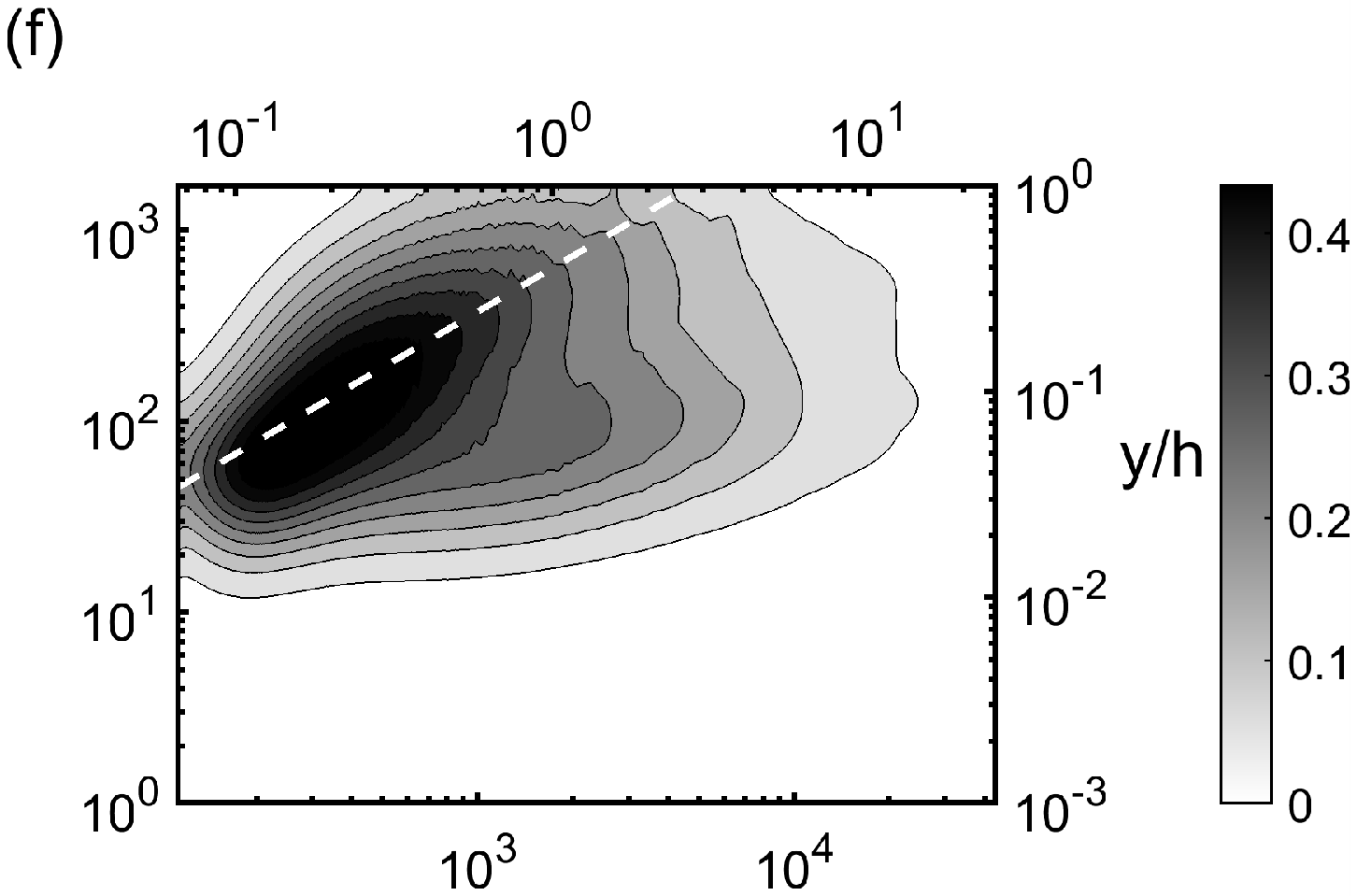}
  \label{6}
\end{subfigure}
\begin{subfigure}[b]{0.42\textwidth}
  \includegraphics[width=\textwidth]{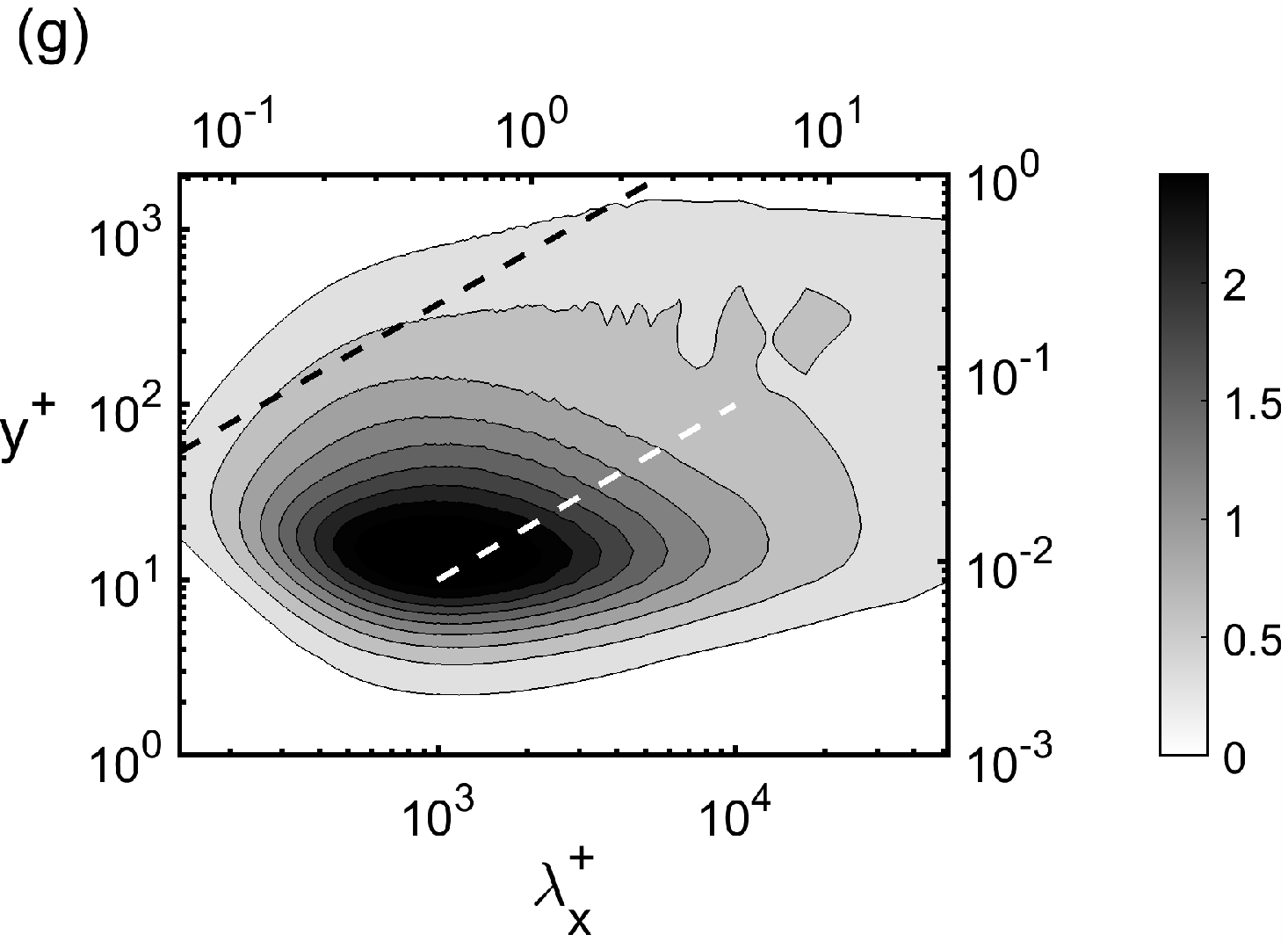}
\end{subfigure}
\begin{subfigure}[b]{0.42\textwidth}
  \includegraphics[width=\textwidth]{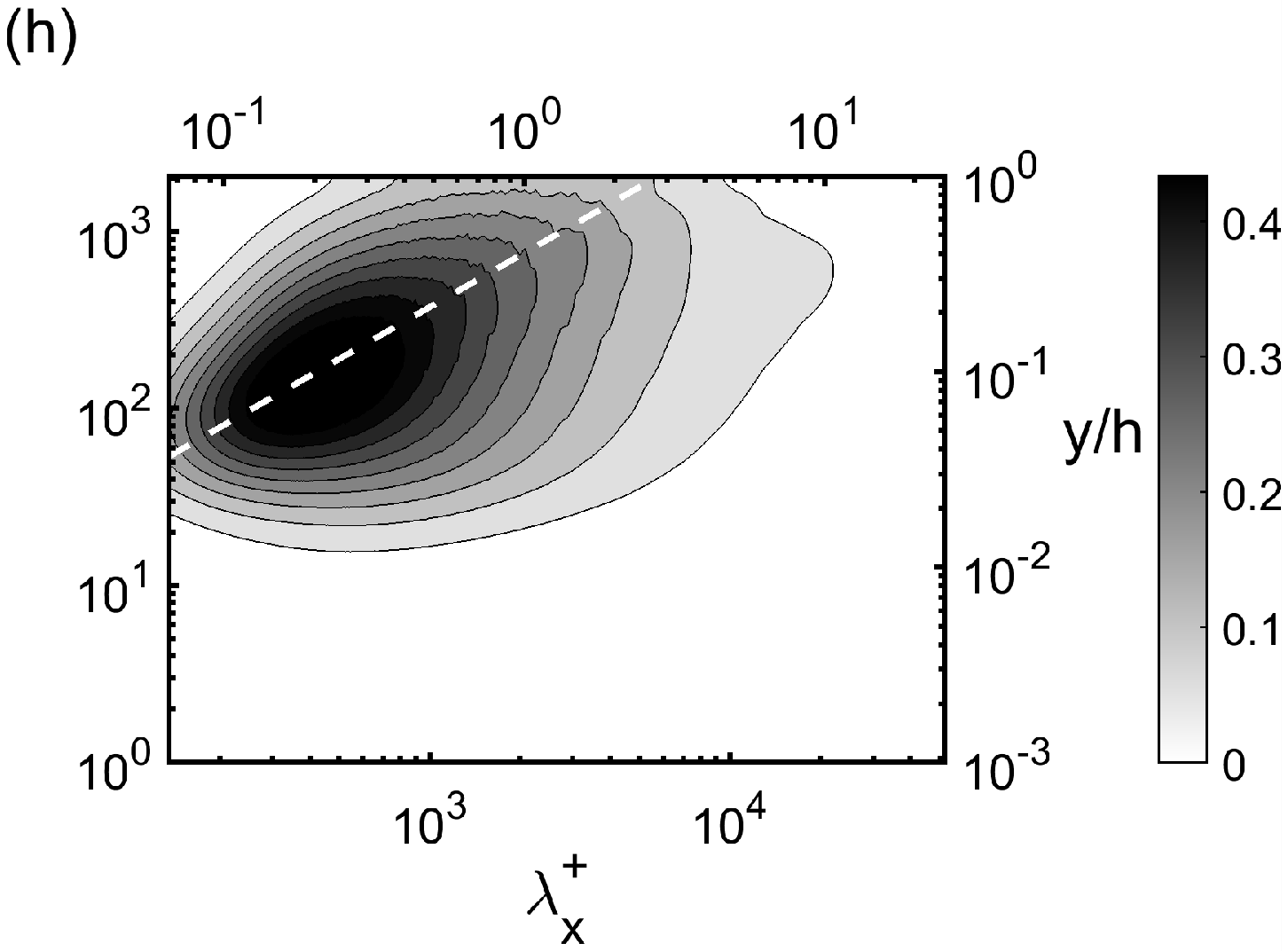}
\end{subfigure}
\end{minipage}
\caption{Premultiplied streamwise wavenumber spectra of streamwise $k_x^+ \Phi_{uu}^+(y^+,\lambda_x^+)$ (left column) and wall-normal $k_x^+ \Phi_{vv}^+(y^+,\lambda_x^+)$ (right column) velocity for (a,b) TRIAZ25, (c,d) TRIBZ25, (e,f) TRICZ25 and (g,h) TRIDZ25 cases. Here, the vertical line represents the spanwise cut-off wavelength ($\lambda_{z,c}$) dividing the $\mathcal{P}_h$- (left) and $\mathcal{P}_l$-subspace  (right) regions.}
\label{fig:xspectra25}
\end{figure}

The TRIAZ25 case explores the contributions of the self-interacting nonlinear terms in the $\mathcal{P}_h$ subspace to its own subspace (i.e. $\mathcal{P}_h$ subspace), and this is represented by the triadic interaction in figure \ref{fig:triad}(f). In the TRIAZ25 case, this interaction term is activated further to the GQL25 model, and it has been designed to understand the importance of such an interaction in the GQL approximation. In fact, figure \ref{fig:stat_triad} shows that the statistics of TRIAZ25 resembles that of GQLZ25 the most, compared to other cases, showing that the effect of this triadic interaction on the low-order statistics of the GQL approximation is moderate. The spanwise wavenumber velocity spectra of the TRIAZ25 case (figures \ref{fig:zspectra25}a,b and \ref{fig:xspectra25}a,b) are also almost identical to those of the GQLZ25 case (figures \ref{fig:zspectra}g,h and \ref{fig:xspectra}g,h). The same holds true for most of the energy spectra (figures \ref{fig:zenergy25}a,b and \ref{fig:xenergy25}a,b), although some minor differences can be observed in the near-wall region. 

It should be mentioned that the addition of the triadic interaction in figure \ref{fig:triad}(f) is directly associated with the activation of the nonlinear terms associated with the self-sustaining process in the $\mathcal{P}_h$ subspace. This becomes evident from (\ref{eq:4.2}), where the contributions of the self-interacting nonlinear terms in the $\mathcal{P}_h$ subspace to its own subspace (i.e. $\mathcal{P}_h$ subspace) were shown to be directly associated with some key sub-processes of the self-sustaining process. A possible explanation to this is that there may exist competition and/or cooperation between the self-sustaining process and the scattering mechanism in driving the near-wall turbulent fluctuations, and in the presence of all scales of turbulent flow, the scattering mechanism may also play an important role. 

\subsubsection{TRIBZ25: inverse energy transfer}
The TRIBZ25 case eliminates the contributions of the self-interacting nonlinear terms in the $\mathcal{P}_h$ subspace to the $\mathcal{P}_l$-subspace group (i.e. the inverse energy transfer from the $\mathcal{P}_h$ to the $\mathcal{P}_l$ subspace), but activates all triadic interactions in the $\mathcal{P}_h$-subspace group. In this case, it is found that its turbulence statistics shows major differences with respect to GQLZ25 and the other cases (figure \ref{fig:stat_triad}). First, $-\langle u^{\prime} v^{\prime}\rangle_{x,z,t}^+$ is greater than one, and $u_{rms}^+$ and $v_{rms}^+$ are also much greater than those of GQLZ25. This strong overestimation of turbulent fluctuations of TRIBZ25 can be explained with the following dimensionless streamwise mean-momentum equation:
\begin{equation}\label{eq:4.5}
    \frac{dU^+}{dy^+}-\frac{\langle u_l v_l \rangle_{x,z,t}}{u_\tau^2}-\frac{\langle u_h v_h \rangle_{x,z,t}}{u_\tau^2}=1-\frac{y}{h}.
\end{equation}
In the TRIBZ25 case, the third term in (\ref{eq:4.5}) is artificially omitted, thus the mean is only balanced with $\langle u_l v_l \rangle_{x,z,t}$. We note that the present simulation imposes a constant mass flow rate and that $\langle u_h v_h \rangle_{x,z,t}$ is not into account in  (\ref{eq:4.5}) in the TRIBZ25 case. Therefore, no matter how strong $-{\langle u_h v_h \rangle_{x,z,t}}$ is, the mean shear has no direct control in reducing its strength, explaining the breakdown of (\ref{eq:4.5}) in the TRIBZ25 case.  

\begin{figure}
\begin{minipage}{\textwidth}
\centering
\begin{subfigure}[b]{0.42\textwidth}
  \includegraphics[width=\textwidth]{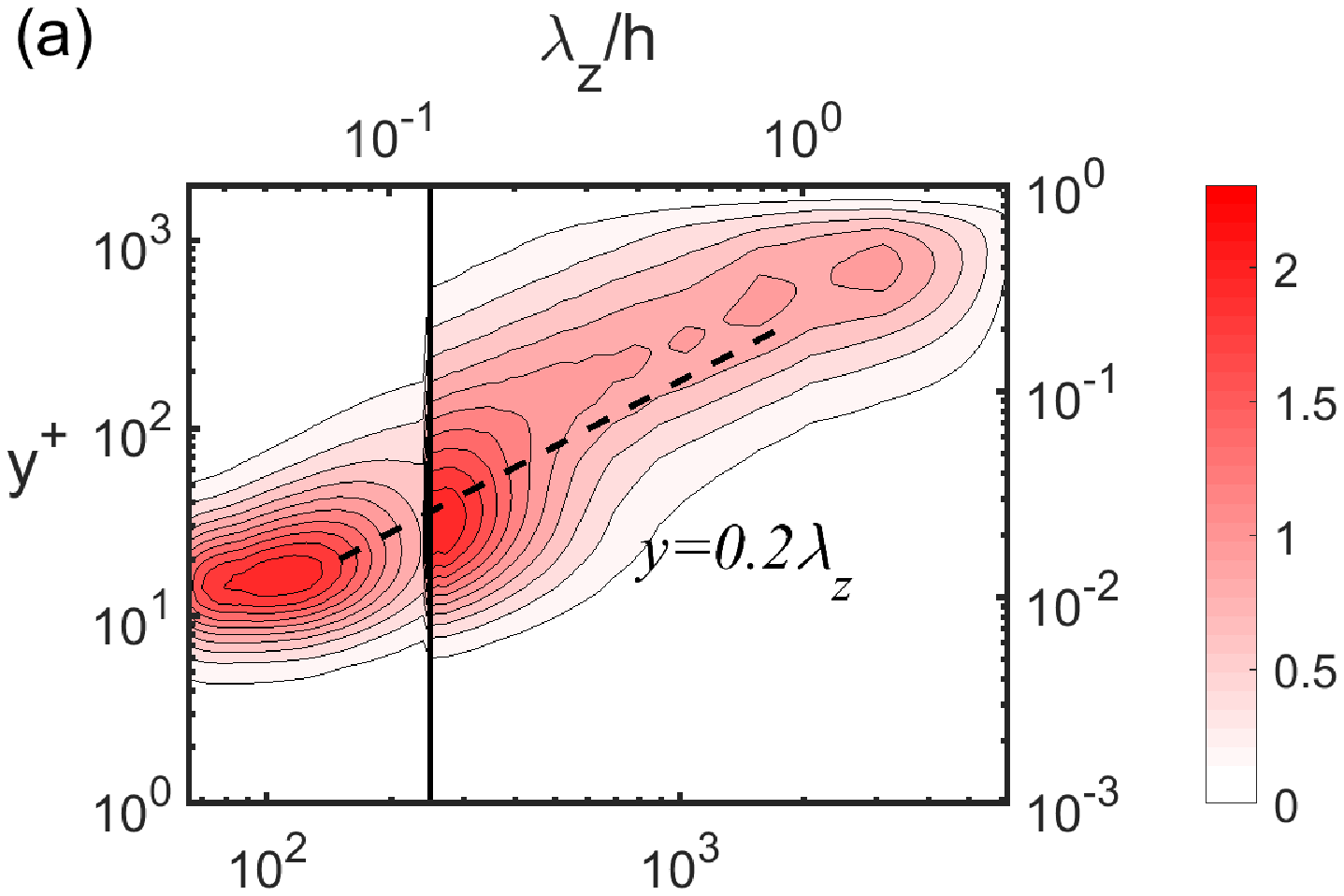}
\label{1}
\end{subfigure}
\vspace{-0.7cm}
\begin{subfigure}[b]{0.42\textwidth}
  \includegraphics[width=\textwidth]{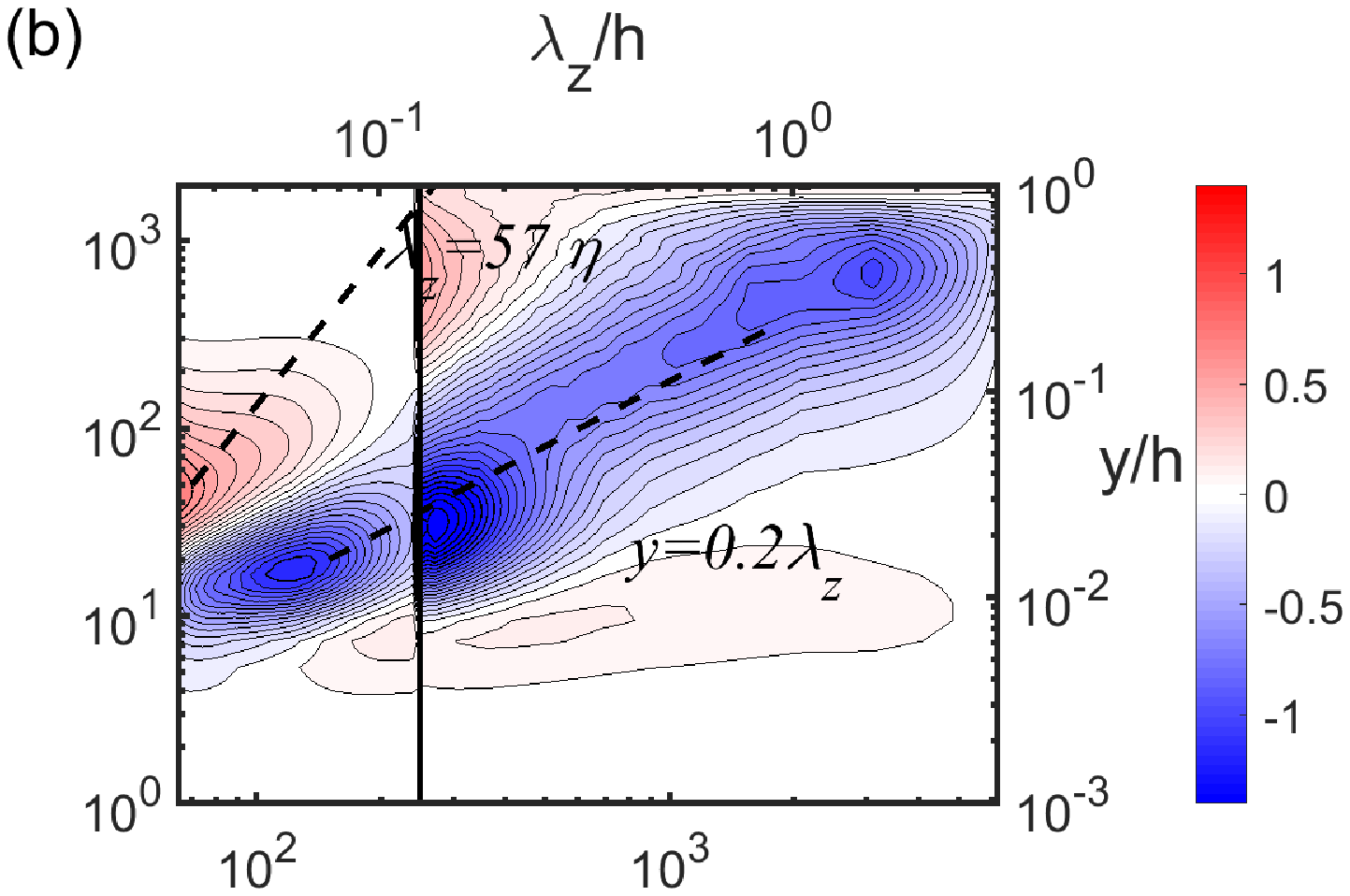}
\label{2}
\end{subfigure}
\begin{subfigure}[b]{0.42\textwidth}
  \includegraphics[width=\textwidth]{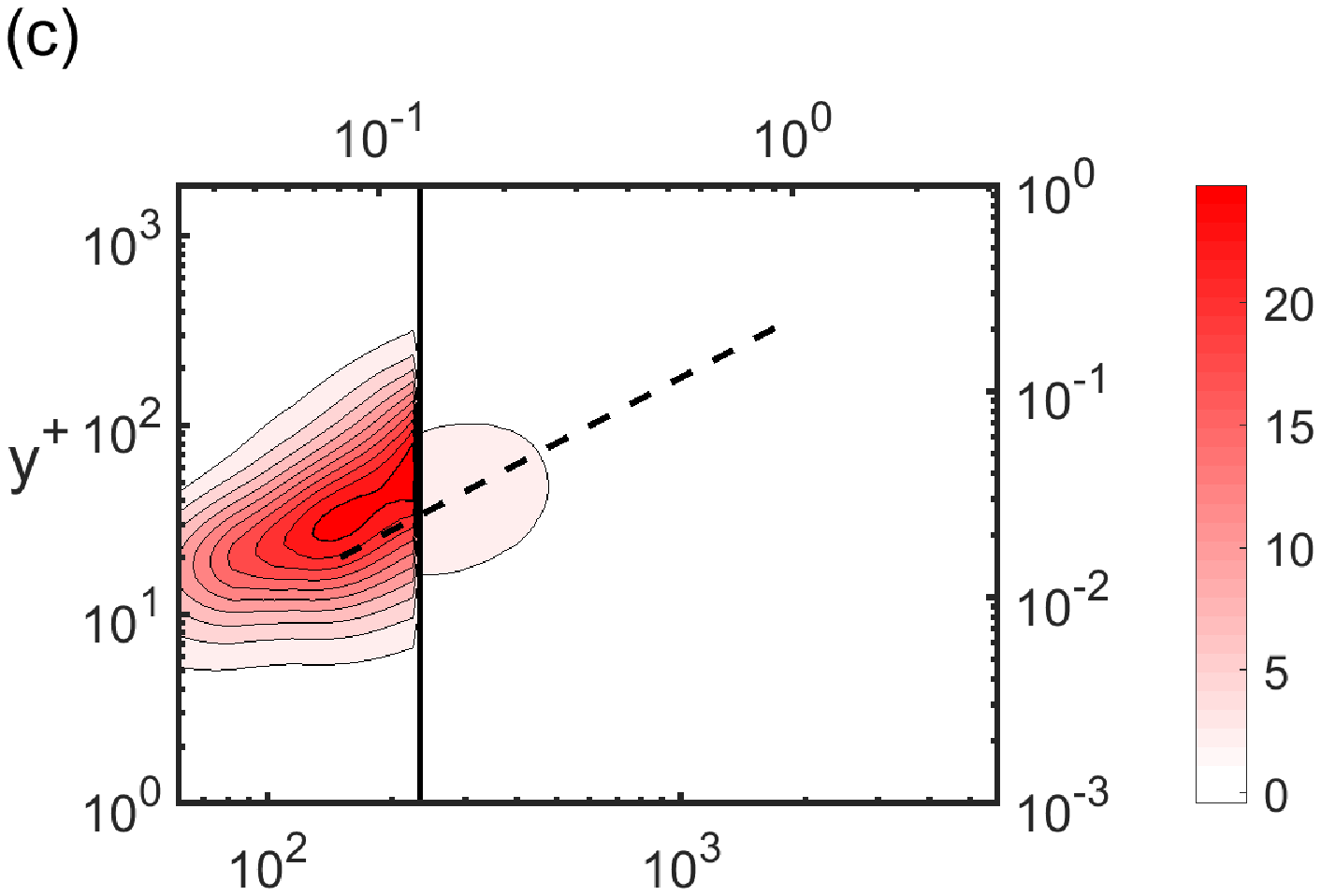}
  \label{3}
\end{subfigure}
\vspace{-0.7cm}
\begin{subfigure}[b]{0.42\textwidth}
  \includegraphics[width=\textwidth]{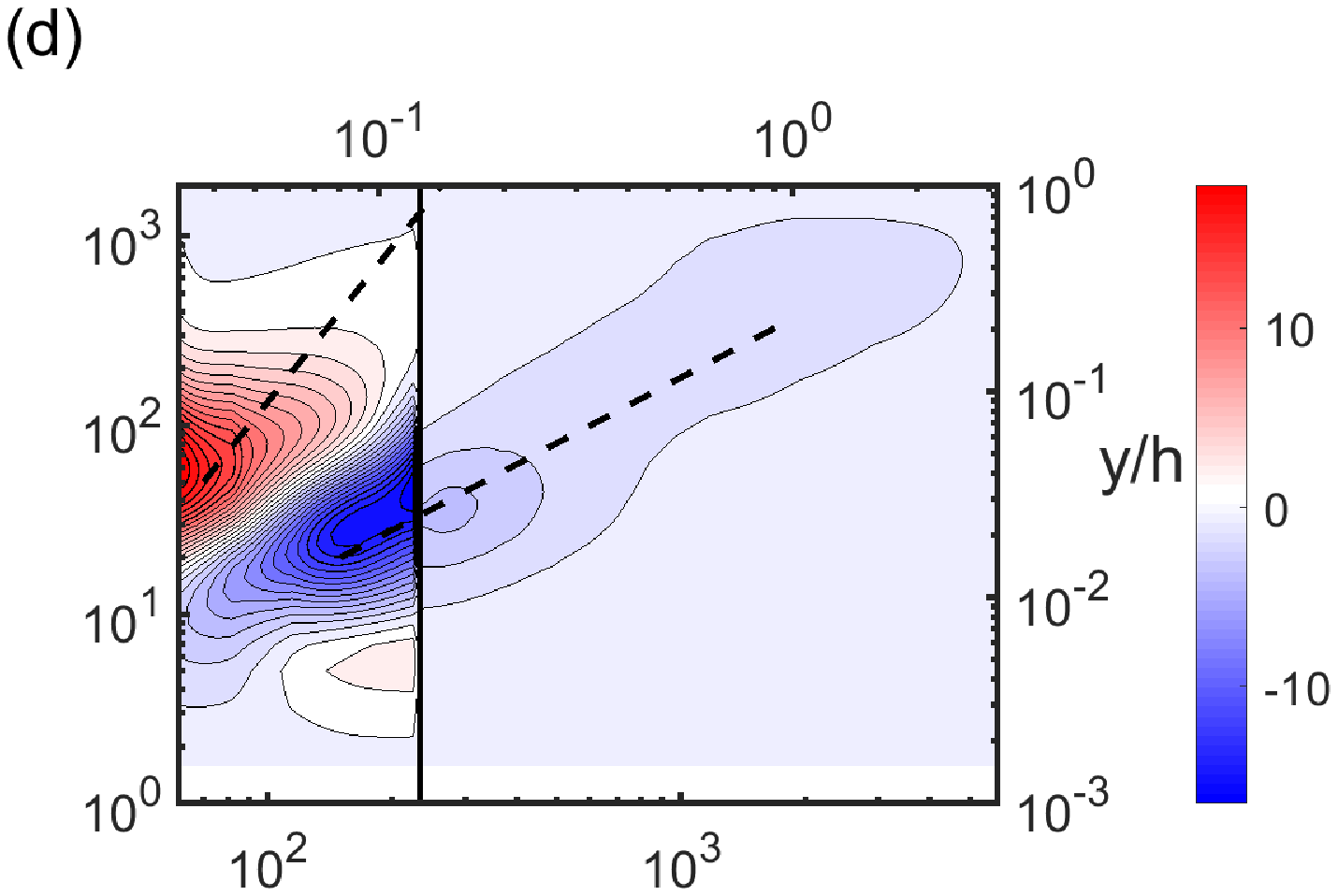}
  \label{4}
\end{subfigure}
\begin{subfigure}[b]{0.42\textwidth}
  \includegraphics[width=\textwidth]{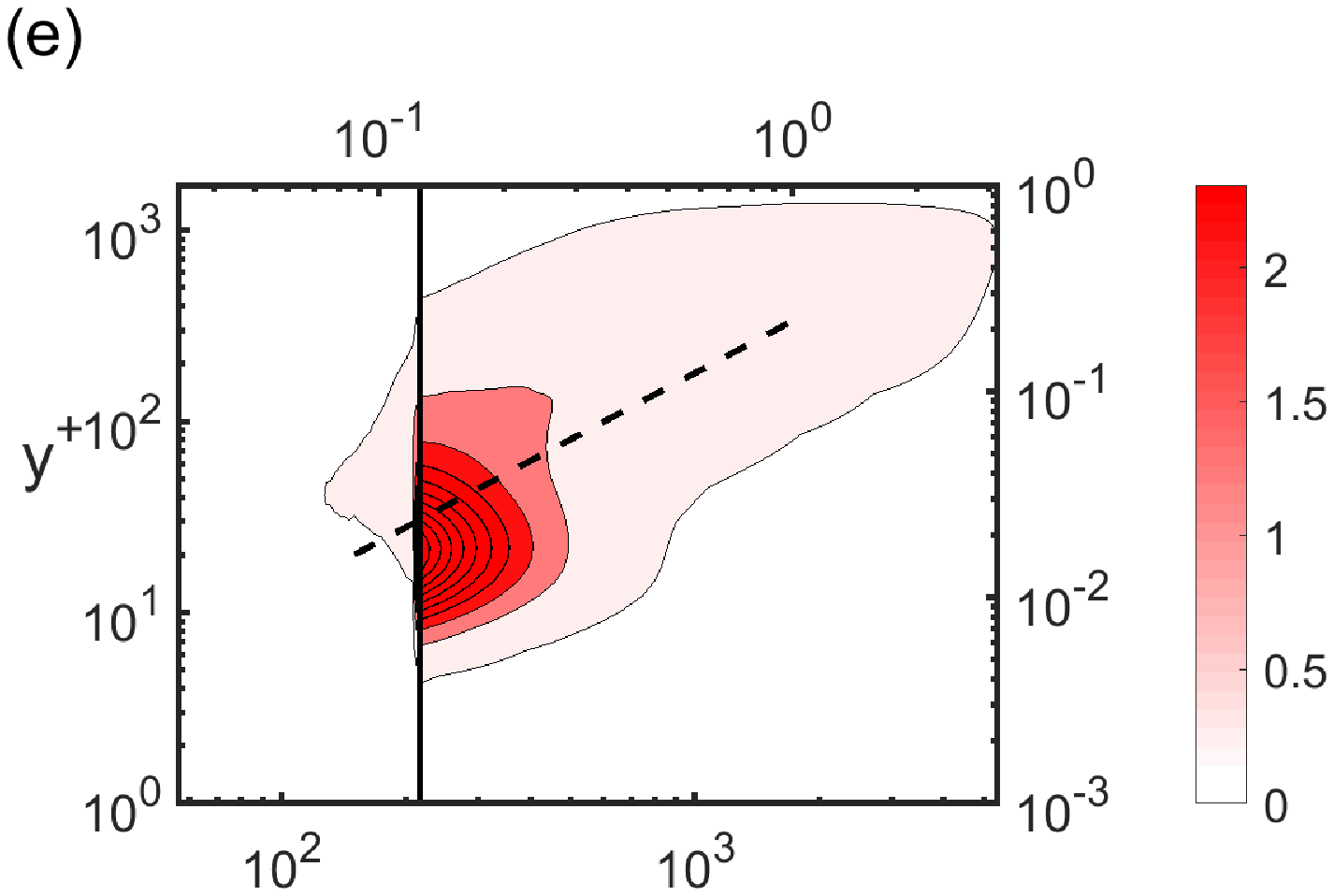}
  \label{5}
\end{subfigure}
\vspace{-0.7cm}
\begin{subfigure}[b]{0.42\textwidth}
  \includegraphics[width=\textwidth]{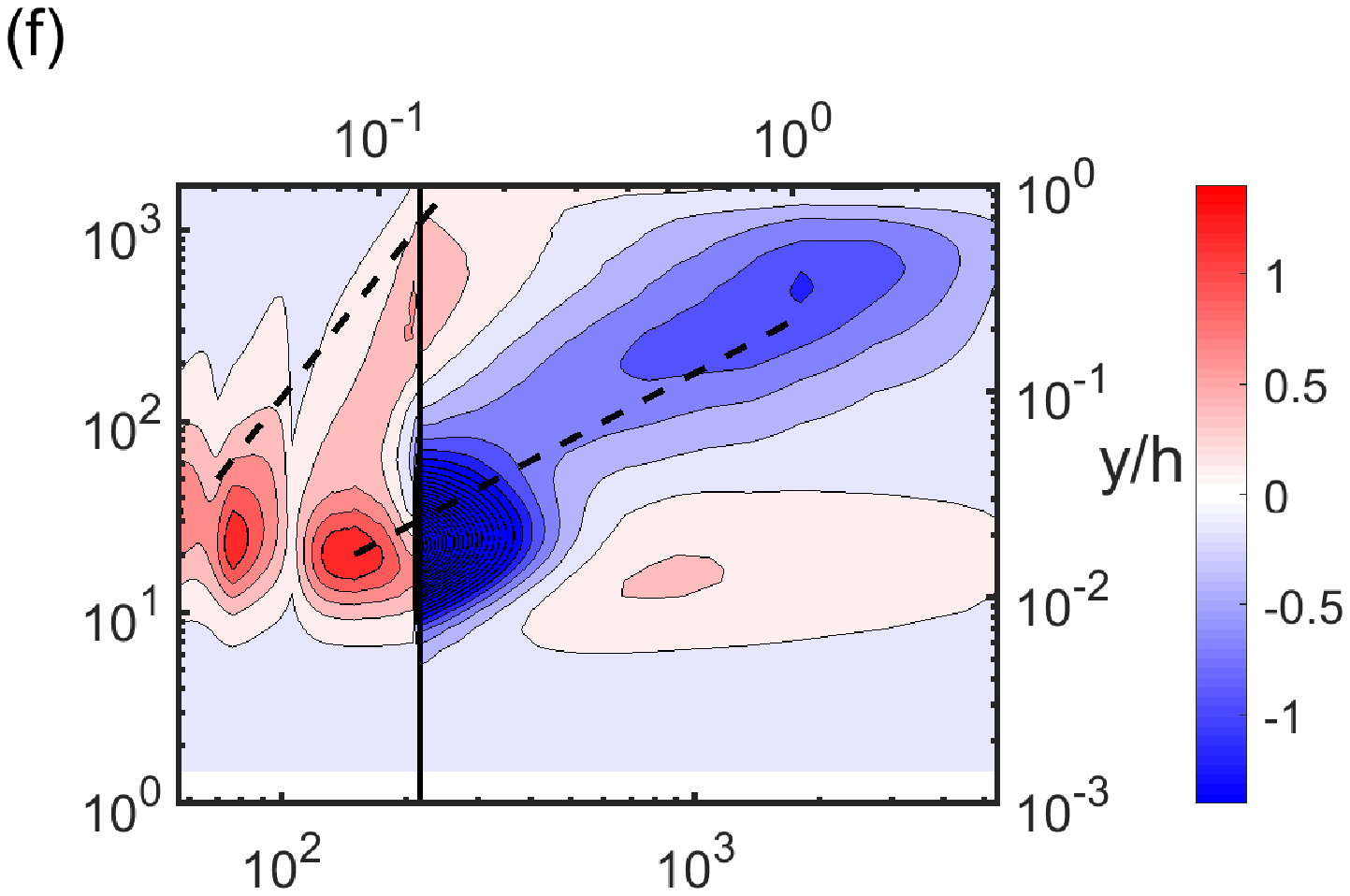}
  \label{6}
\end{subfigure}
\begin{subfigure}[b]{0.42\textwidth}
  \includegraphics[width=\textwidth]{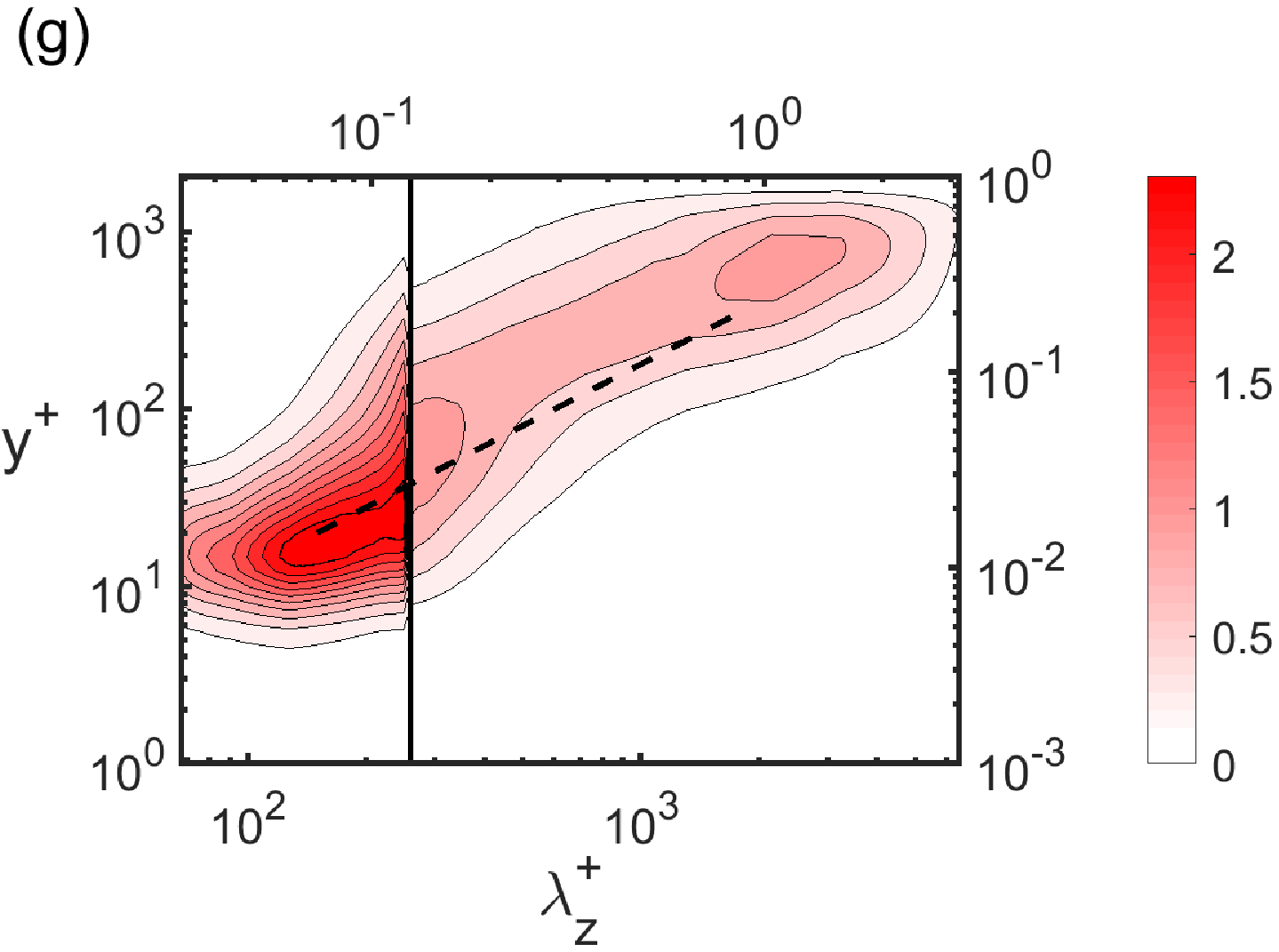}
\end{subfigure}
\begin{subfigure}[b]{0.42\textwidth}
  \includegraphics[width=\textwidth]{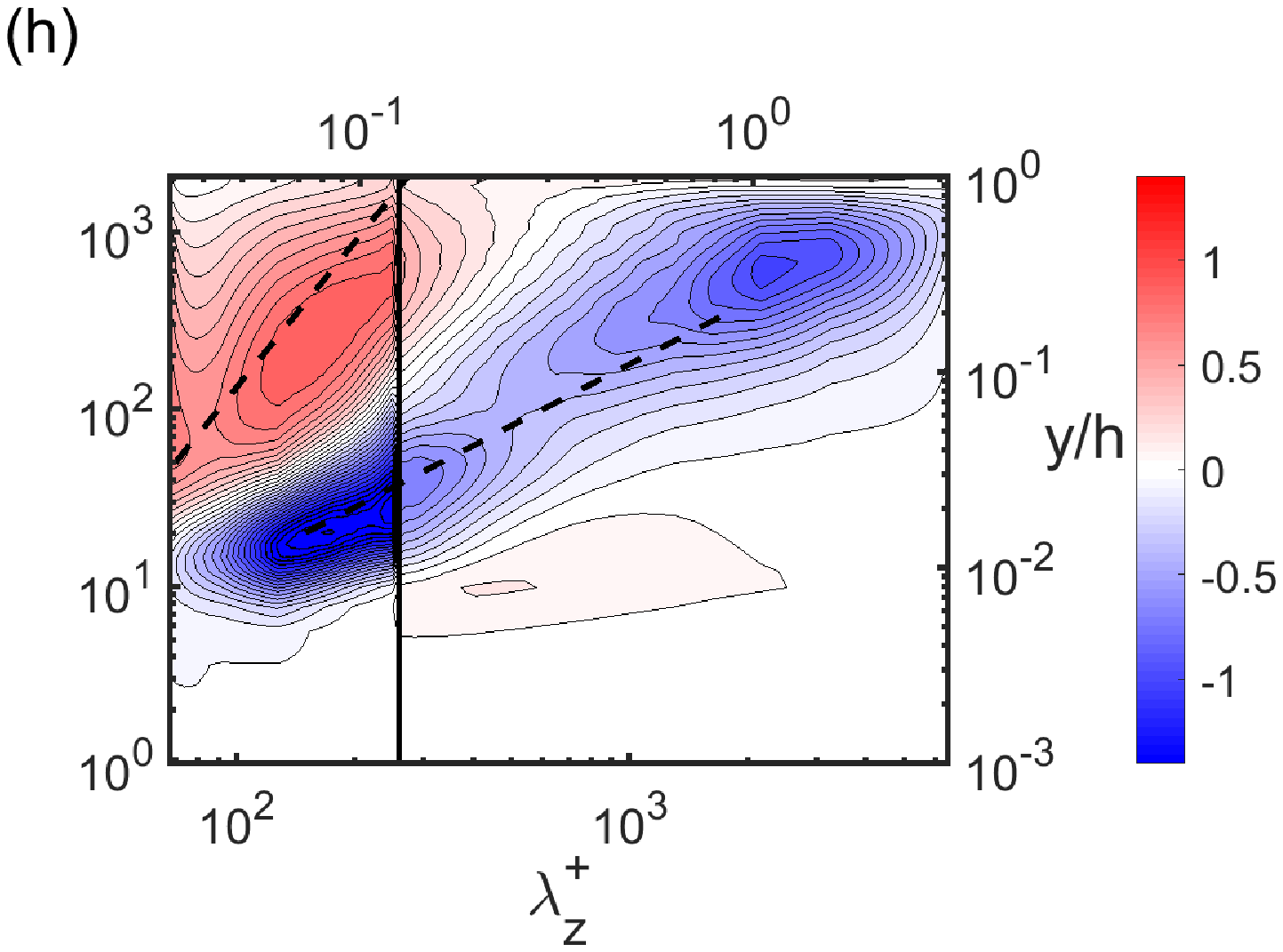}
\end{subfigure}
\end{minipage}
\caption{Premultiplied spanwise wavenumber spectra of production $k_z^+ y^+ \widehat{P}^+(y^+,\lambda_z^+)$ (left column) and turbulent transport $k_z^+ y^+ \widehat{T}_{turb}^+(y^+,\lambda_z^+)$ (right column) for (a,b) TRIAZ25, (c,d) TRIBZ25, (e,f) TRICZ25 and (g,h) TRIDZ25 cases.}
\label{fig:zenergy25}
\end{figure}

\begin{figure}
\begin{minipage}{\textwidth}
\centering
\begin{subfigure}[b]{0.42\textwidth}
  \includegraphics[width=\textwidth]{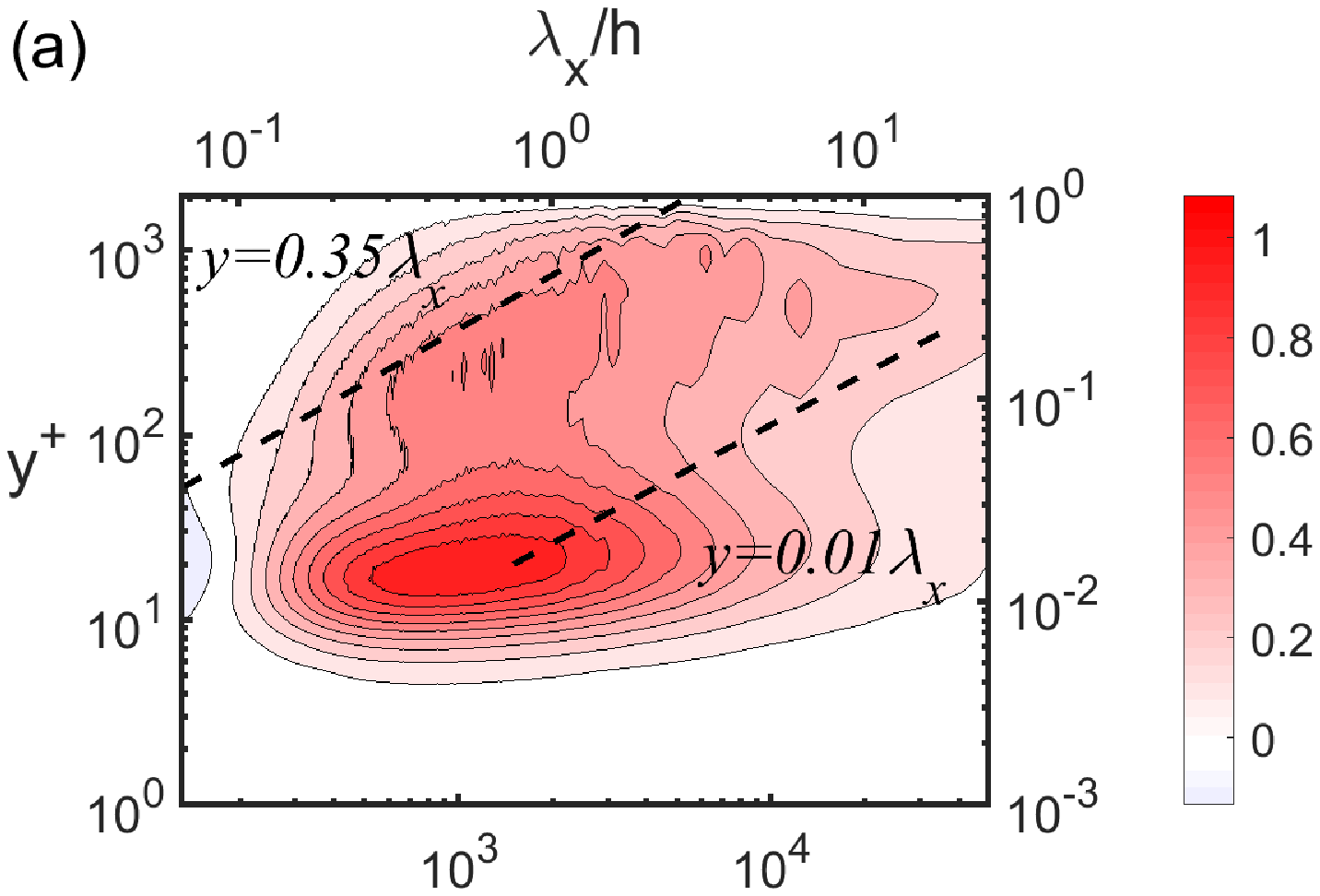}
\label{1}
\end{subfigure}
\vspace{-0.7cm}
\begin{subfigure}[b]{0.42\textwidth}
  \includegraphics[width=\textwidth]{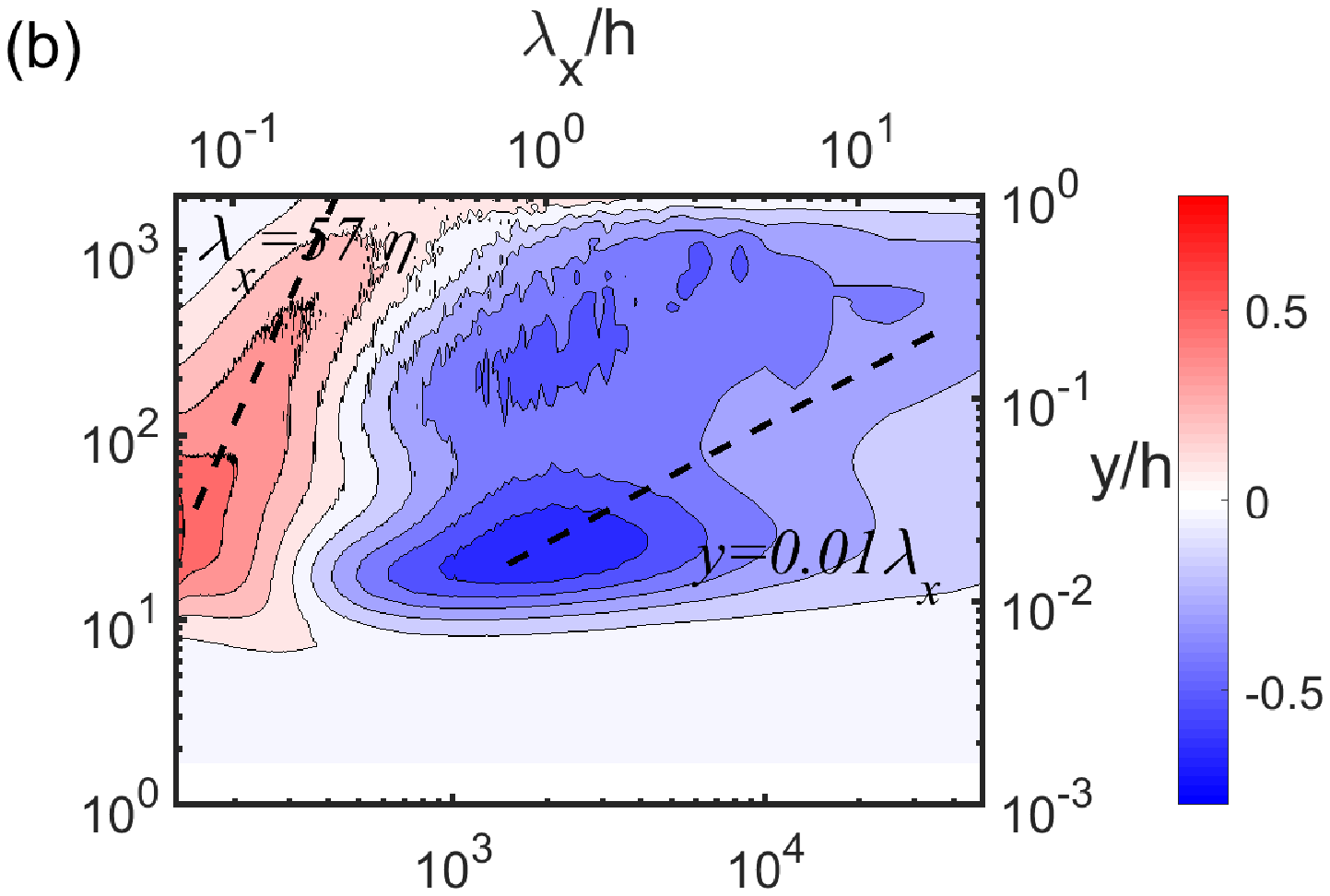}
\label{2}
\end{subfigure}
\begin{subfigure}[b]{0.42\textwidth}
  \includegraphics[width=\textwidth]{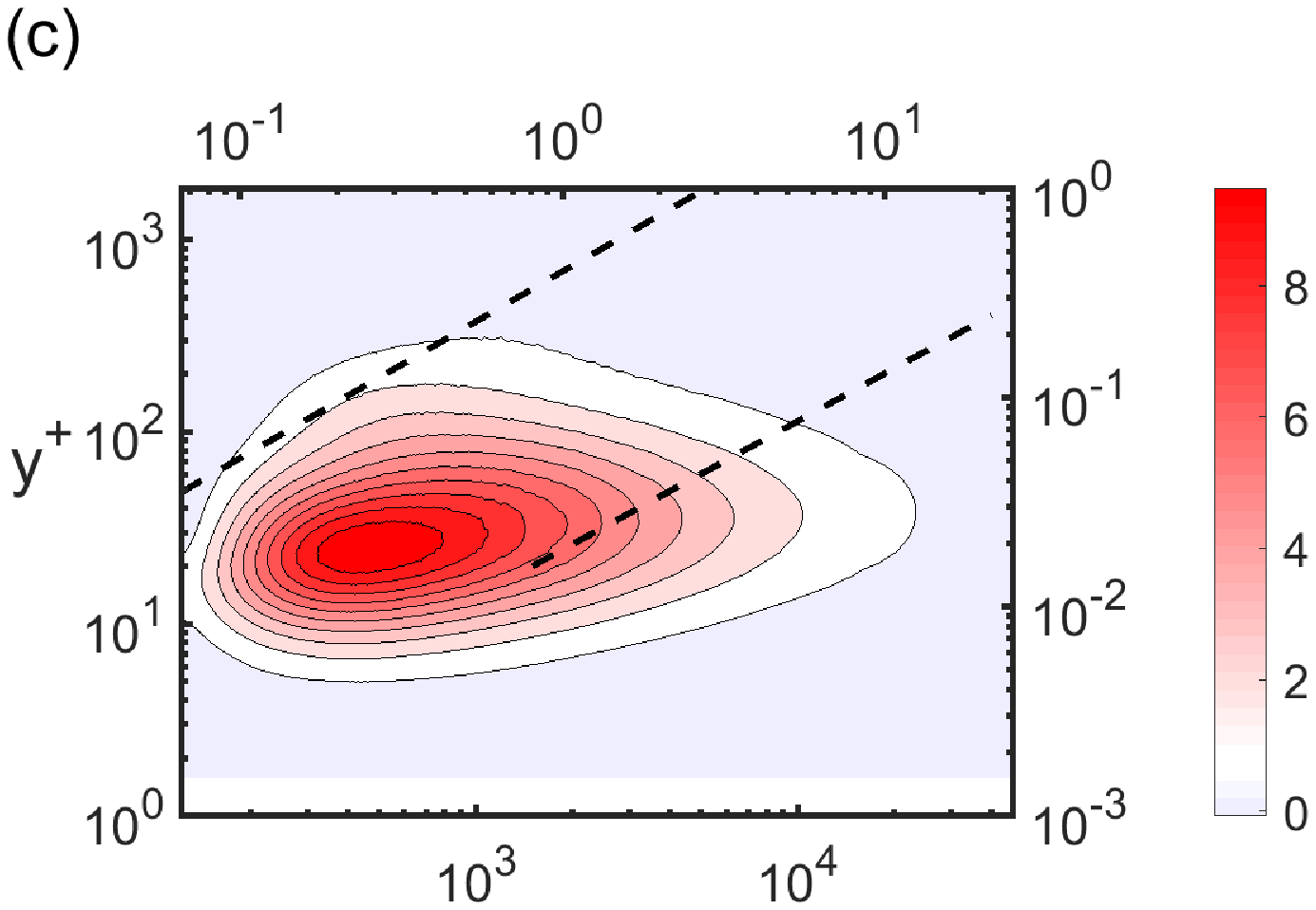}
\label{fig:prodx}
\end{subfigure}
\vspace{-0.7cm}
\begin{subfigure}[b]{0.42\textwidth}
  \includegraphics[width=\textwidth]{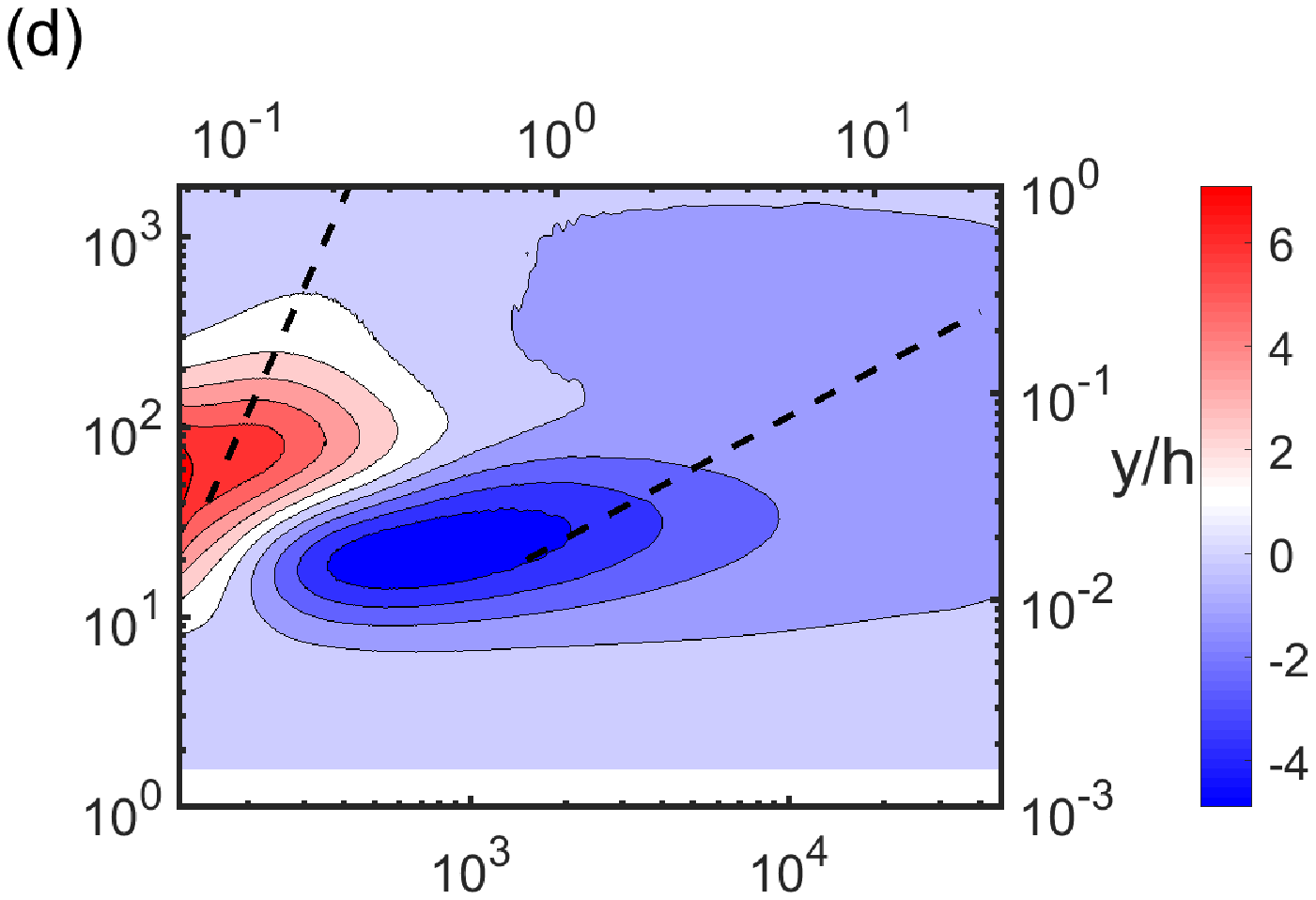}
  \label{4}
\end{subfigure}
\begin{subfigure}[b]{0.42\textwidth}
  \includegraphics[width=\textwidth]{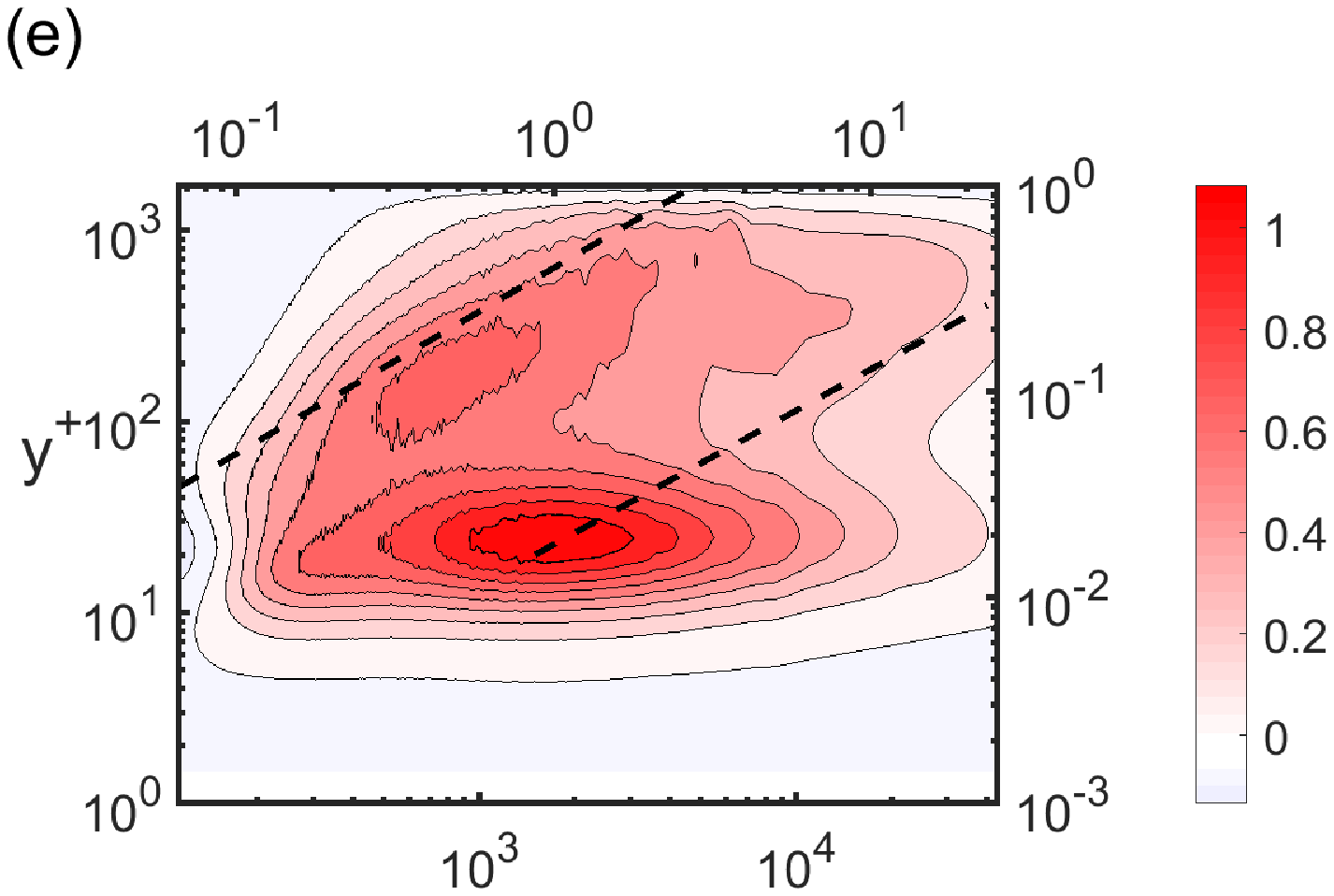}
  \label{5}
\end{subfigure}
\vspace{-0.7cm}
\begin{subfigure}[b]{0.42\textwidth}
  \includegraphics[width=\textwidth]{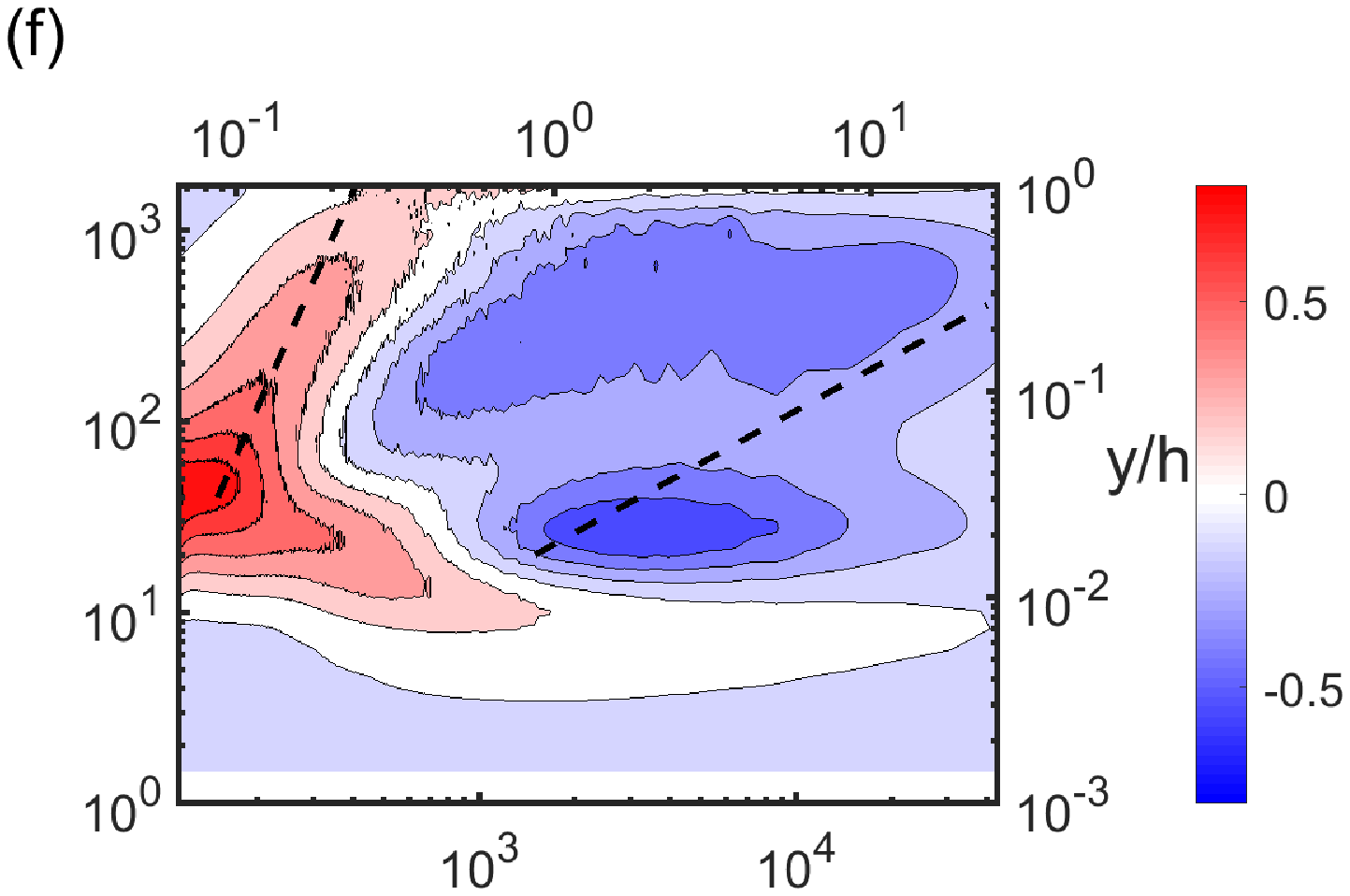}
  \label{6}
\end{subfigure}
\begin{subfigure}[b]{0.42\textwidth}
  \includegraphics[width=\textwidth]{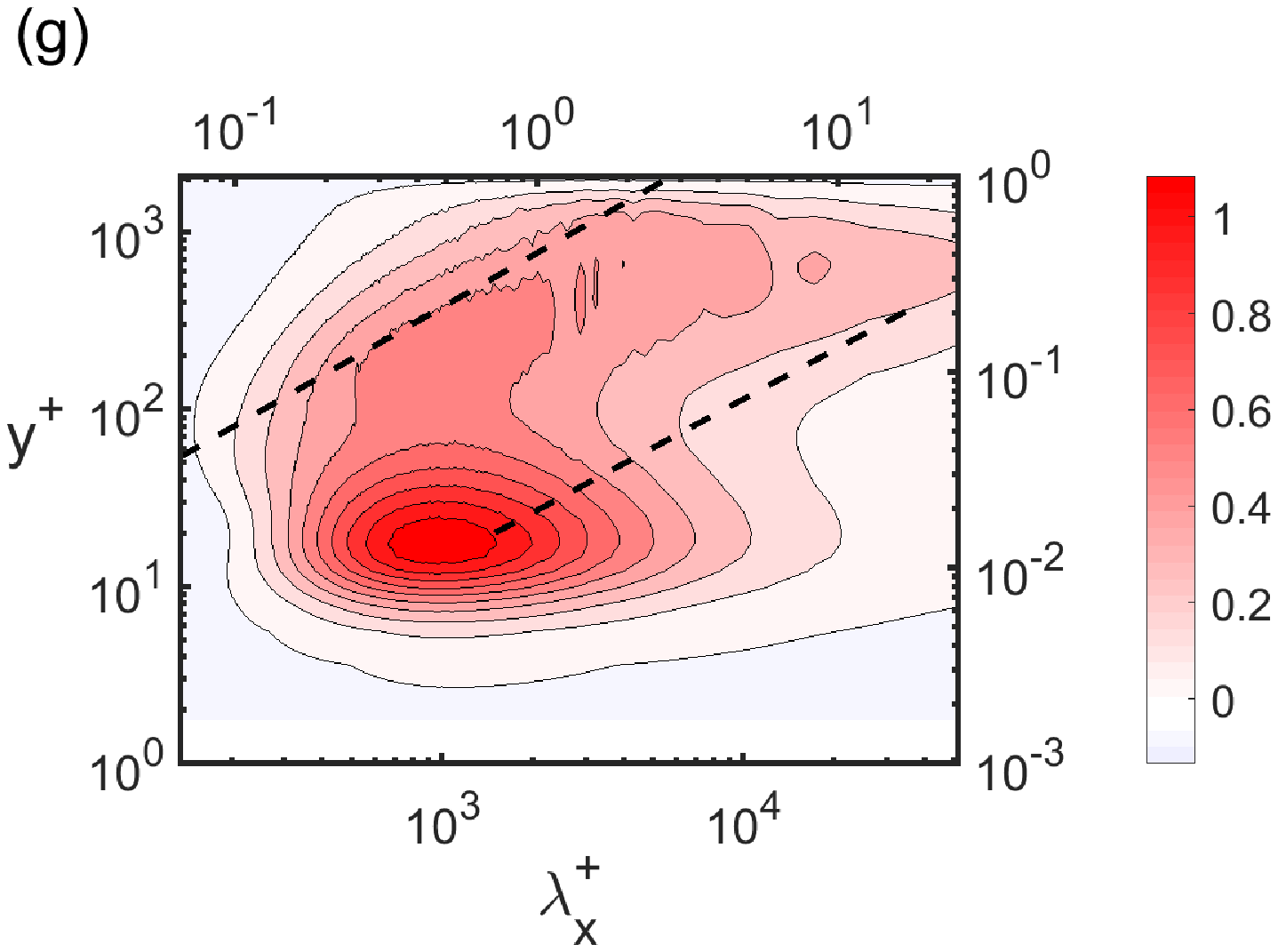}
\end{subfigure}
\begin{subfigure}[b]{0.42\textwidth}
  \includegraphics[width=\textwidth]{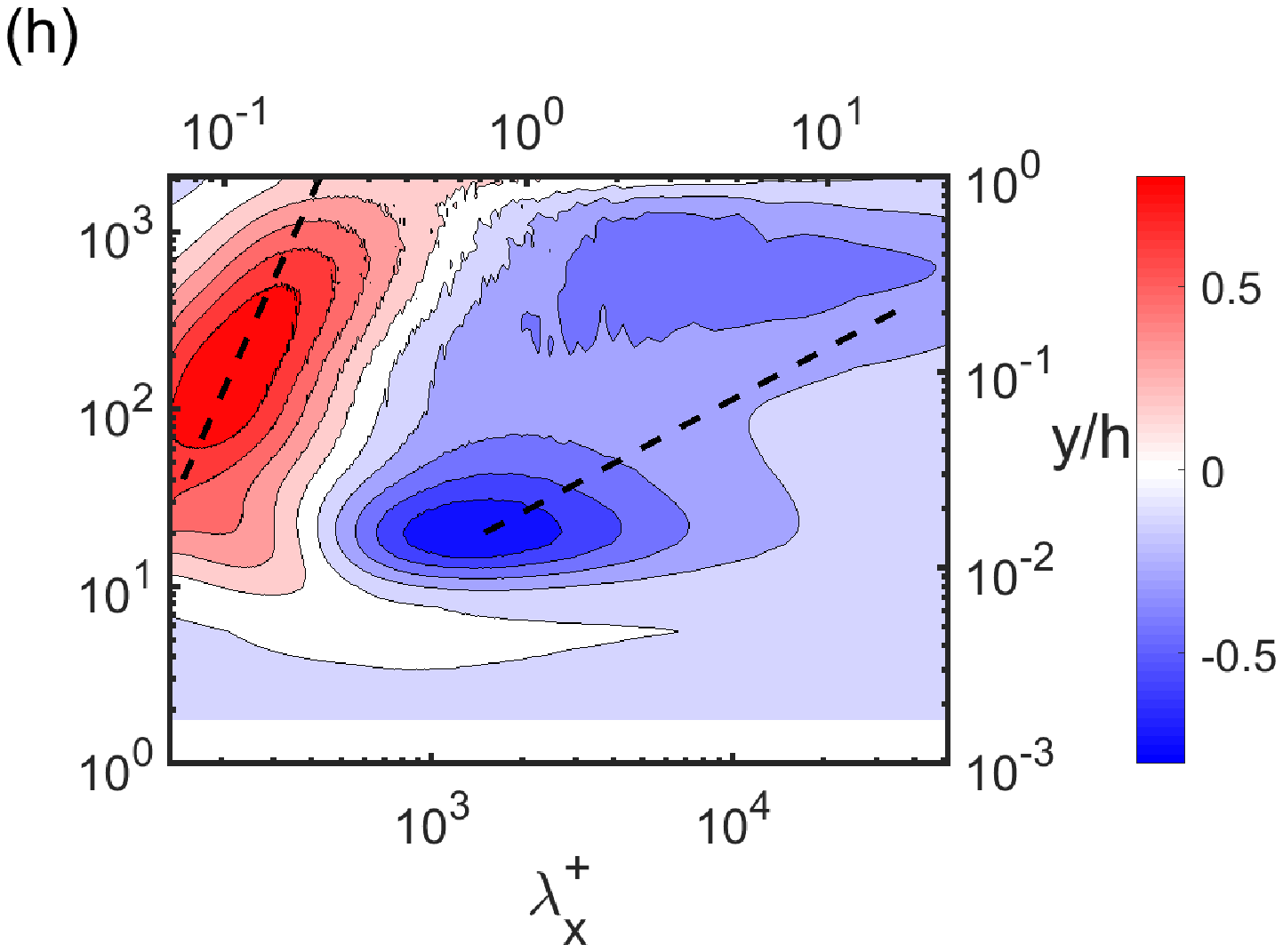}
\end{subfigure}
\end{minipage}
\caption{Premultiplied streamwise wavenumber spectra of production $k_x^+ y^+ \widehat{P}^+(y^+,\lambda_x^+)$ (left column) and turbulent transport $k_x^+ y^+ \widehat{T}_{turb}^+(y^+,\lambda_x^+)$ (right column) for (a,b) TRIAZ25, (c,d) TRIBZ25, (e,f) TRICZ25 and (g,h) TRIDZ25 cases. Here, the vertical line represents the spanwise cut-off wavelength ($\lambda_{z,c}$) dividing the $\mathcal{P}_h$- (left) and $\mathcal{P}_l$-subspace  (right) regions.}
\label{fig:xenergy25}
\end{figure}

In fact, the spanwise wavenumber spectra of velocity and energy budget in figures \ref{fig:zspectra25}(c,d) and \ref{fig:zenergy25}(c,d) show large elevation of the fluctuation energy in the $\mathcal{P}_h$ subspace in the TRIBZ25 case. This effect is also seen in the streamwise wavenumber spectra in figures \ref{fig:xspectra25}(c,d) and \ref{fig:xenergy25}(c,d), where spectral intensity is quite high in the region close to the wall where the spectral intensity of $\mathcal{P}_h$-subspace group would be important. This behaviour might have given an impression that the range of spanwise wavenumbers covered by the spectra is reduced in the TRIBZ25 case. However, it is important to note that turbulent velocity fluctuations in the outer regions remain more or less the same as those of GQLZ25 and LES (see figure \ref{fig:stat_triad}), indicating that the spectral intensity of the outer region in the $\mathcal{P}_l$ subspace is actually not affected significantly. The same behaviour is also observed in all the other spectra in figures \ref{fig:xspectra25}(c,d), \ref{fig:zenergy25}(c,d) and \ref{fig:xenergy25}(c,d) (not shown). This confirms the outer-similarity hypothesis of \cite{townsend76}, who conjectured that turbulence in the outer region remains largely independent of the near-wall dynamics, consistent with many previous studies conducted with surface roughness \cite[e.g.][]{flores07}. 

Finally, it should be mentioned that the spanwise wavenumber spectra of turbulent transport in the TRIBZ25 case exhibit a strong relative suppression of the near-wall positive turbulent transport in the $\mathcal{P}_h$ subspace (figure \ref{fig:zenergy25}d) in comparison to LES (figure \ref{fig:zenergy}b) and GQLZ25 (figure \ref{fig:zenergy}h). The weakly positive turbulent transport observed in GQL25 and L£S  was originally proposed to be the main consequence of inverse energy cascade \citep{cho18}, who visualised the full triadic nonlinear interactions in a turbulent channel flow. However, some recent studies have suggested that it is also generated by the self-sustaining process at each scale \citep{lee19,kawata21}. The TRIBZ25 simulation here shows that the positive turbulent transport in the region close to the wall is significantly weakened by the inhibition of the energy transfer from the $\mathcal{P}_h$ to the $\mathcal{P}_l$ subspace, favouring the proposition by \cite{cho18} on its primary origin. 

\subsubsection{TRICZ25: the scattering mechanism in GQL approximations}\label{sec:433}
The TRICZ25 case enables the contributions of the self-interacting nonlinear terms in both the $\mathcal{P}_l$ and $\mathcal{P}_h$ subspaces to the $\mathcal{P}_h$-subspace group (previously eliminated by the GQL approximation), while it excludes the low/high wavenumber interaction terms in the $\mathcal{P}_h$-subspace group, i.e. the `scattering mechanism' in the GQL models \cite[e.g.][]{tobias17}. The statistics of TRICZ25 shows some subtle differences from that of GQLZ25 (see figure \ref{fig:stat_triad}). In particular, $u_{rms}^+$ (figure \ref{fig:stat_triad}a) appears to be larger, while $\langle u^{\prime} v^{\prime}\rangle_{x,z,t}^+$ depends on the $y$-location: they are greater on the log layer and smaller below and above it. 

The spectra of the TRICZ25 case show more precise picture in the subtle differences (figures \ref{fig:zspectra25}e,f, \ref{fig:xspectra25}e,f, \ref{fig:zenergy25}e,f and \ref{fig:xenergy25}e,f). In particular, the spanwise wavenumber spectra of velocity and energy budget in figures \ref{fig:zspectra25}(e,f) and \ref{fig:zenergy25}(e,f) are found to extend over a narrower range of spanwise wavenumbers. Furthermore, the spectral intensity in the $\mathcal{P}_h$-subspace (for $\lambda_z<\lambda_{z,c}$) is significantly weakened, not covering wavenumbers below $\lambda_z^+ \approx 100$. In particular, the spanwise wavenumber spectra of turbulence production is completely suppressed (figure \ref{fig:zenergy25}e). This suggests that the low/high wavenumber interaction terms in the $\mathcal{P}_h$-subspace group, incorporated in the GQL approximations, are probably the most crucial triadic interactions, because their suppression significantly weakens the energy-containing region of the spanwise wavenumber spectra of velocity. 

It is interesting to note that this observation favours that the scattering mechanism is more important than the self-sustaining process for the generation of near-wall turbulence, when all the other scales are present in the flow -- we note that all the previous numerical experiments demonstrating the self-sustaining process was without the motions larger than the given spanwise length scale \cite[e.g.][]{jimenez91,hamilton95,hwang13}. However, care needs to be taken in understanding this behaviour, because the TRICZ25 case also suppresses all the production terms in (\ref{eq:4.4}) (i.e. the third terms in (\ref{eq:4.4a}) and (\ref{eq:4.4b})), shared by the self-sustaining process (see (\ref{eq:4.2})). In this respect, the issue on the relative importance between the self-sustaining process and the scattering mechanism in the generation of near-wall turbulence still remains an open question, although the numerical experiment carried out with the TRIAZ25 case (e.g. see figure \ref{fig:zspectra25}a,b) appears to support that perhaps the scattering mechanism may be more crucial than the self-sustaining process in the presence of all the scales. Since the existence of the self-sustaining processes and their roles have been well established \cite[]{jimenez91,hamilton95,hwang13}, a more careful examination with additional numerical experiments would certainly be required for the precise understanding on this new observation, and we will leave this for the future investigation.

\subsubsection{TRIDZ25: direct driving from the $\mathcal{P}_l$ to the $\mathcal{P}_h$ subspaces}\label{sec:434}
Finally, the TRIDZ25 case incorporates the contribution of the self-interacting nonlinear term in the $\mathcal{P}_l$ subspace to the $\mathcal{P}_h$-subspace group, and its objective is to analyse the influence of this direct driving mechanism from $\mathcal{P}_l$- to $\mathcal{P}_h$-subspace group. The statistics of TRIDZ25 is found to approximate that of GQLZ25 fairly well (figure \ref{fig:stat_triad}), with a little overestimation of $u_{rms}^+$, $v_{rms}^+$, $\langle u^{\prime} v^{\prime}\rangle_{x,z,t}^+$, showing that the inclusion of the triadic interaction (d) has a moderate effect on the statistics.

The spectra of the TRIDZ25 case offer a more detailed picture on its effect (figures \ref{fig:zspectra25}g,h, \ref{fig:xspectra25}g,h, \ref{fig:zenergy25}g,h and \ref{fig:xenergy25}g,h). The spanwise wavenumber velocity spectra in figures \ref{fig:zspectra25}(g,h) appear to be qualitatively similar to those of the GQLZ25 case. However, the streamwise and wall-normal velocity spectra in the $\mathcal{P}_h$ subspace (for $\lambda_z<\lambda_{z,c}$) are intensified compared to those of GQLZ25 case (figures \ref{fig:zspectra}g,h), especially with a large elevation of the energy in the relatively outer region. This can be explained by the spanwise wavenumber spectra of turbulent transport in figure \ref{fig:zenergy25}(h), where a strong positive energy transport now appears in the outer region of $\mathcal{P}_h$ subspace. This is in contrast to that of the GQLZ25 case (figure \ref{fig:zenergy}h), where such a transport barely exist in the relative outer region of the $\mathcal{P}_h$ subspace (i.e. $y^+\gtrsim 300$).

In Part 1, it was proposed that the $\mathcal{P}_h$ subspace group of a GQL model can admit the trivial solution, if the cut-off wavelength for the decomposition of the flow (i.e. $\lambda_{z,c}$) is sufficiently small \cite[see \S4.2 in][]{paper1}. This was also explained in terms of the Lyapunov exponent of the $\mathcal{P}_h$ subspace group, which can alternatively be interpreted as the activation process of the scattering mechanism proposed by \cite{tobias17}. In the present study, such a case also appears for $\lambda_{z,c}^+ \approx 97$ (GQLZ64 in table \ref{tab:tab} and in Appendix \ref{appendix}), the typical spanwise spacing of the near-wall streaks \cite[]{kline67}. In this case, the simulation result of the TRIDZ25 case suggests that the contribution of the self-interacting nonlinear term in the $\mathcal{P}_l$ subspace to the $\mathcal{P}_h$-subspace group becomes important, when the scattering mechanism discussed in \S\ref{sec:433} plays little role in generating the fluctuation in the $\mathcal{P}_h$ subspace (i.e. when $\lambda_{z,c}$ is sufficiently small). In such a case, the dynamics in the $\mathcal{P}_h$ subspace remains essentially passive to the $\mathcal{P}_l$ subspace in the sense that the $\mathcal{P}_h$ subspace would only admit a non-trivial solution in the presence of the direct driving from the $\mathcal{P}_l$ subspace. In this respect, it is finally worth mentioning that the smallest spanwise computational domain retaining turbulence is $L_z^+ \approx 100$ \cite[][]{jimenez91,hwang13} and the critical spanwise wavelength, below which the GQL model in this study yields the trivial solution in the $\mathcal{P}_h$ subspace, is $\lambda_{z,c}^+ \approx 97$, showing an interesting analogue.

\section{Concluding remarks} \label{sec:sec5}
In the present study, a generalised quasilinear (GQL) approximation has been applied to in turbulent channel flow at $Re_\tau \approx 1700$, continuing from the work of Part 1 \citep{paper1}. The flow is decomposed into two groups, the former of which contains a set of low-wavenumber spanwise Fourier modes and the latter are composed of the rest high-wavenumber modes. The former group is then solved by considering the full nonlinear equations, while the latter group is obtained from the linearised equations about the former. In Part 1, the flow decomposition was based on the streamwise Fourier modes. By doing so, the corresponding GQL models placed a minimal model for self-sustaining process at all integral length scales varying from viscous inner units to outer units, and the nonlinear coupling between the self-sustaining processes from different scales  and the related nonlinear energy transfer were assessed. However, in this study, the flow decomposition gives the low wavenumber group in which the self-sustaining process are fully resolved at least within the given integral length scales, while the linearised high wavenumber group is not able to support the self-sustaining process. Some key conclusions are summarised as follows:
\begin{enumerate}
    \item Despite the important physical difference of the GQL approximation here from that in Part 1, they share some important similarities. First, due to the interference of nonlinear energy transfer given by GQL approximations, they tend to exhibit significantly reduced multi-scale behaviours especially when the cut-off wavelength for the flow decomposition is sufficiently large. Furthermore, turbulent fluctuations have been found to be strongly anisotropic for large cut-off wavelengths, and this is due to damaged slow pressure originating from the fluctuation/fluctuation interactions, some of which are omitted in the GQL approximations. 
    \item While the high wavenumber group of the GQL models in the present study cannot support the self-sustaining process, they can generate some key statistical features at integral scales through the scattering mechanism to certain extent. Although it is not clear how this is exactly possible, this observation could have some potential links to the success of the previous linearised Navier-Stokes operator-based models \cite[e.g.][]{hwang10,mckeon10,moarref13,zare17,mckeon17,hwang19,skouloudis,jovanovic21}.
    
    \item Finally, a set of numerical experiments have been performed to explore the roles of ignored triadic interaction terms in the GQL approximations focusing on scale interactions. These numerical experiments have clarified the role of 1) self-interactions within the high-wavenumber group; 2) inverse energy transfer in the region close to the wall; 3) the scattering mechanism; 4) direct driving mechanism to the the high-wavenumber group. In particular, it has been seen that the scattering mechanism may play a more important role than the self-sustaining process in driving near-wall turbulence in the presence of all scales in the flow. 
\end{enumerate}

The GQL approximations were originally introduced to improve simpler QL models. Here, we have utilised the GQL approximations to explore the coupling dynamics between self-sustaining processes at different length scales, scale interactions and nonlinear energy transfer for wall-bounded turbulence. The knowledge gained here would be useful to develop a simpler and more robust self-consistent description of turbulence using the concepts of the `statistical-state dynamics'. Our on-going efforts have been made along this direction, and some recent works can be found in \cite{hwang19} and \cite{skouloudis}. 

\section*{Acknowledgements}
C. G. H. and Y. H. acknowledge the support of the Leverhulme trust (RPG-123-2019). Q. Y. acknowledges the Natural Science Foundation of China (11802322). Y. H. is also supported by the Engineering and Physical Sciences Research Council (EPSRC; EP/T009365/1) in the UK.

\section*{Declaration of interest}
The authors report no conflict of interest.

\appendix
\section{Velocity and energy-balance spectra of GQLZ64}\label{appendix}

\begin{figure}
\begin{minipage}{\textwidth}
\centering
\begin{subfigure}[b]{0.42\textwidth}
  \includegraphics[width=\textwidth]{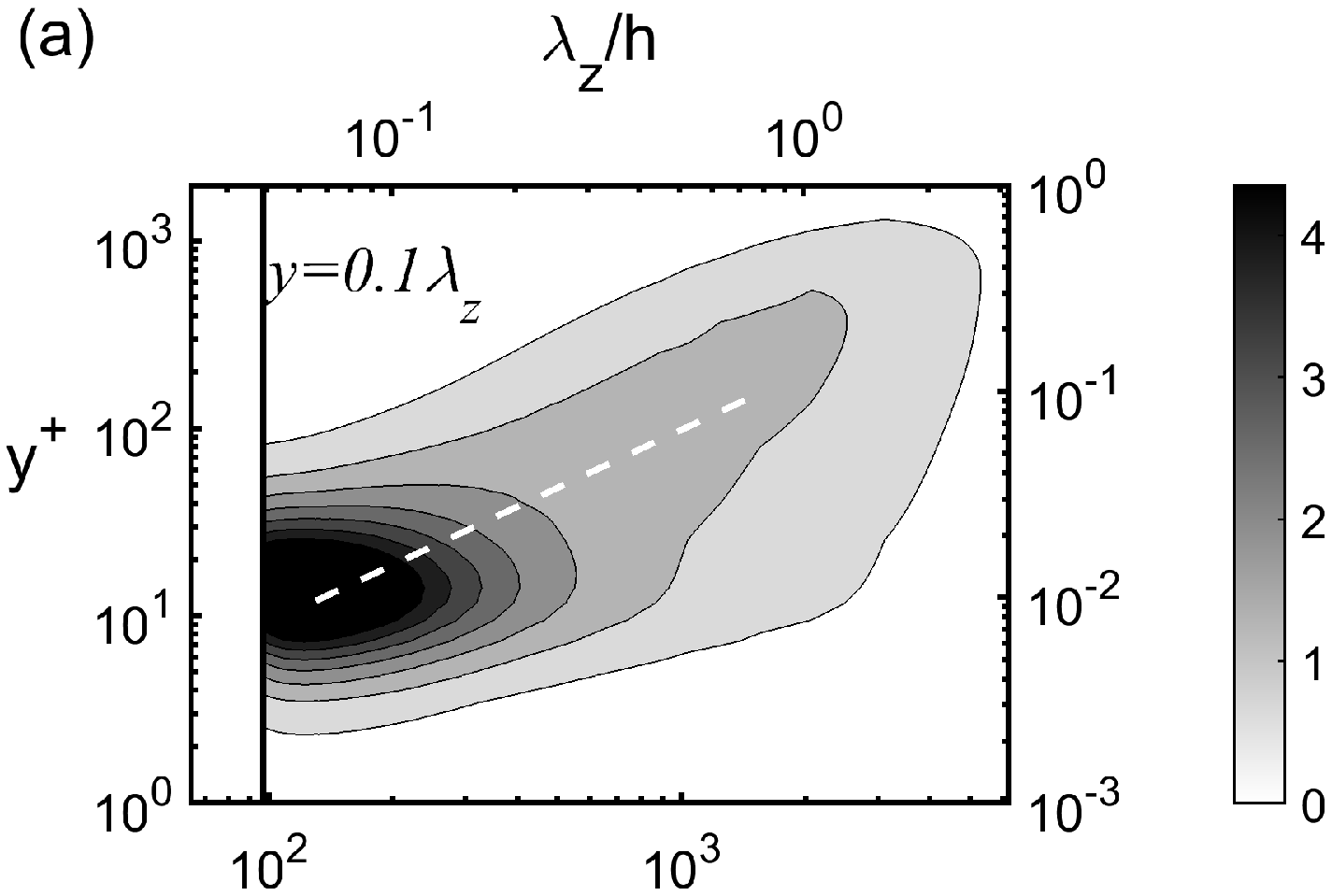}
\label{1}
\end{subfigure}
\vspace{-0.8cm}
\begin{subfigure}[b]{0.42\textwidth}
  \includegraphics[width=\textwidth]{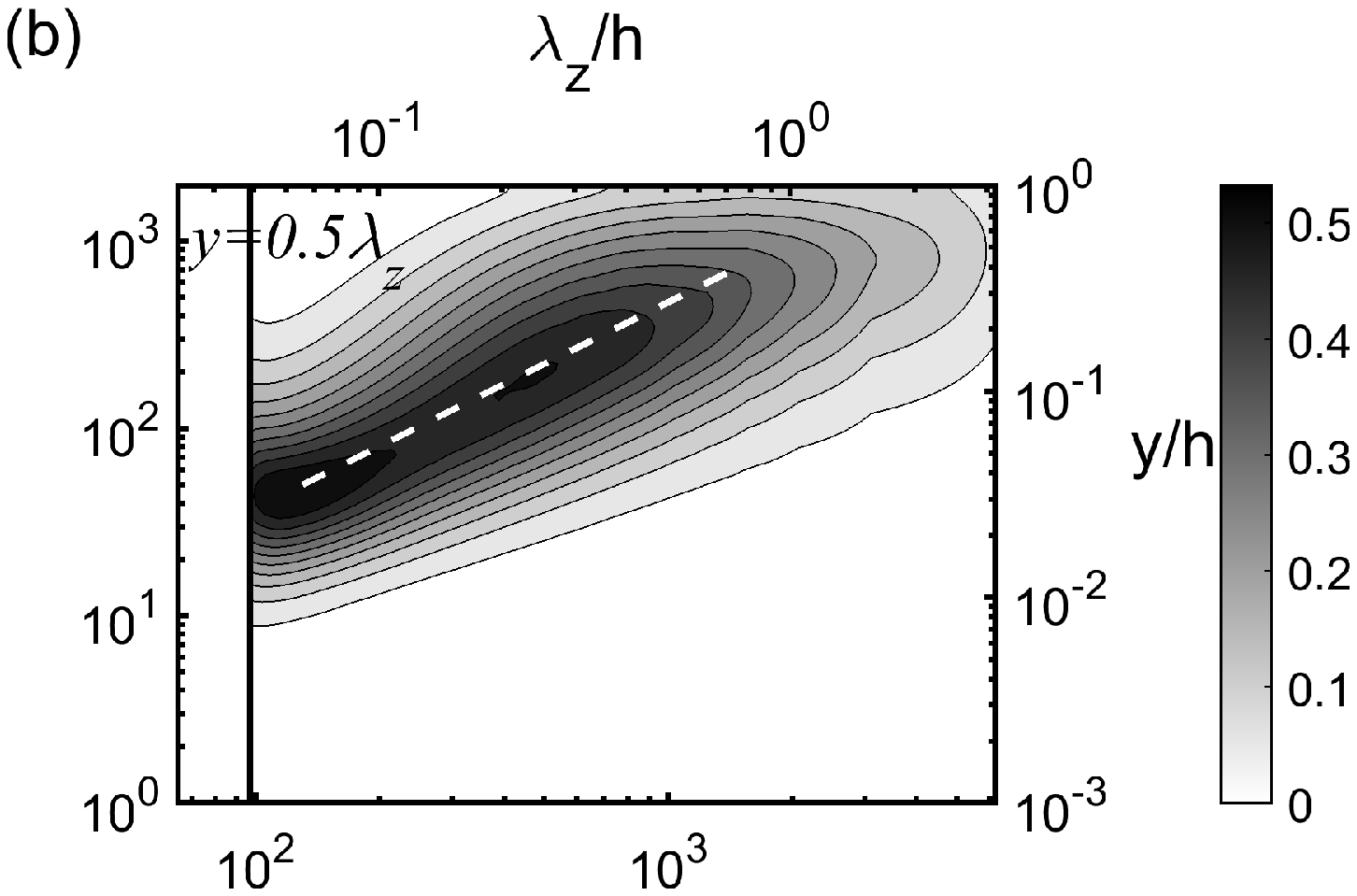}
\label{2}
\end{subfigure}
\begin{subfigure}[b]{0.42\textwidth}
  \includegraphics[width=\textwidth]{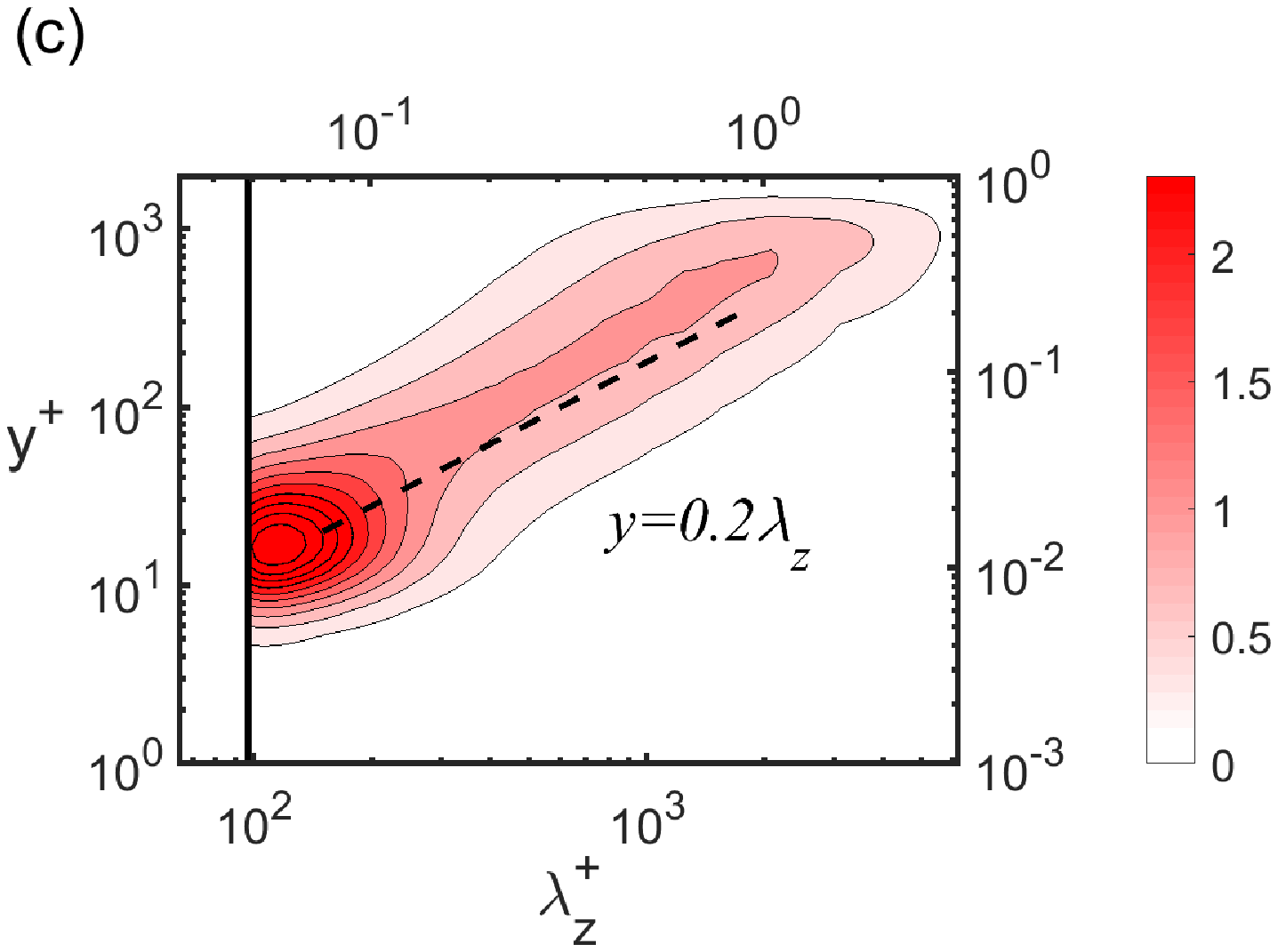}
\label{fig:prodx}
\end{subfigure}
\begin{subfigure}[b]{0.42\textwidth}
  \includegraphics[width=\textwidth]{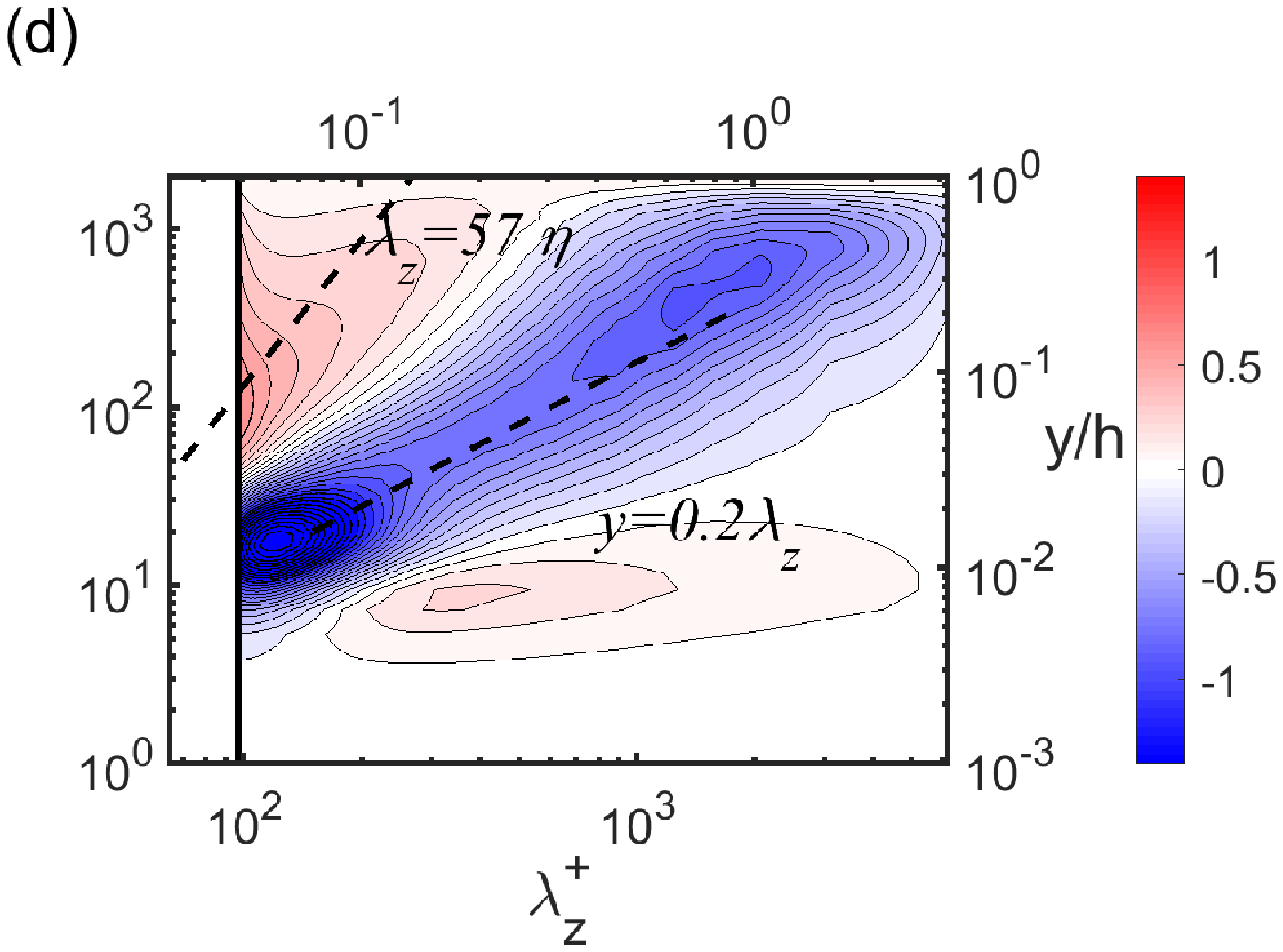}
  \label{4}
\end{subfigure}
\end{minipage}
\caption{Premultiplied spanwise wavenumber spectra of (a) streamwise $k_z^+ \Phi_{uu}^+(y^+,\lambda_z^+)$ and (b) wall-normal $k_z^+ \Phi_{vv}^+(y^+,\lambda_z^+)$ velocity, (c) production $k_z^+ y^+ \widehat{P}^+(y^+,\lambda_z^+)$ and (d) turbulent transport $k_z^+ y^+ \widehat{T}_{turb}^+(y^+,\lambda_z^+)$ for GQLZ64. Here, the vertical line represents the spanwise cut-off wavelength ($\lambda_{z,c}$) dividing the $\mathcal{P}_h$- (left) and $\mathcal{P}_l$-subspace (right) regions.}
\label{fig:z64}
\end{figure}


Figure \ref{fig:z64} shows the premultiplied spanwise wavenumber spectra of the streamwise velocity (figure \ref{fig:z64}a), the wall-normal velocity (figure \ref{fig:z64}b), the production (figure \ref{fig:z64}c) and the turbulent transport (figure \ref{fig:z64}d) of the GQLZ64 case. All spectra greatly resemble those of the reference LES in the $\mathcal{P}_l$ subspace. The trivial solution $\mathbf{u}_h=0$ is however observed in the $\mathcal{P}_h$-subspace group, for a cutoff wavelength $\lambda_{z,c}^+ \approx 97$. This phenomenon, investigated in Part 1 for the streamwise GQL models, has been explained in terms of the Lyapunov exponent and the instability of the linearised equations in the $\mathcal{P}_h$-subspace group \citep{paper1}.

\bibliography{biblio}
\bibliographystyle{jfm}

\end{document}